\documentclass[iop]{emulateapj}

\newcommand{\cha}{\textit{Chandra\/}}
\def\xmm{{XMM-{\it Newton\/}}}

\def\swi{{{\it Swift}-BAT}\/}
\def\xrt{{{\it Swift}-XRT}\/}

\def\flu{{ erg s$^{-1}$} cm$^{-2}$}
\def\leg{{\em COSMOS-Legacy}\/}
\def\nustar{{\it NuSTAR}}

\def\myt{\texttt{MYTorus}}
\def\borus{\texttt{borus02}}

\shorttitle{CT-AGN in the \nustar\ era III: A systematic study of the torus covering factor}
\shortauthors{Marchesi et al.}

\usepackage{natbib}
\usepackage{float}
\usepackage{color}
\usepackage{graphicx}
\usepackage{gensymb}
\usepackage{array}
\usepackage{mathtools}
\usepackage{enumitem}
\usepackage{longtable}
\usepackage{hyperref}

\begin{document}


\title{Compton-thick AGN in the \nustar\ era III: A systematic study of the torus covering factor}

\author{S. Marchesi\altaffilmark{1}, M. Ajello\altaffilmark{1}, X. Zhao\altaffilmark{1}, L. Marcotulli\altaffilmark{1}, M. Balokovi{\'c}\altaffilmark{2}, M. Brightman\altaffilmark{3}, A. Comastri\altaffilmark{4}, G. Cusumano\altaffilmark{5}, G. Lanzuisi\altaffilmark{4,6}, V. La Parola\altaffilmark{5}, A. Segreto\altaffilmark{5}, C. Vignali\altaffilmark{6,4}} 

\altaffiltext{1}{Department of Physics and Astronomy, Clemson University, Clemson, SC 29634, USA}
\altaffiltext{2}{Harvard-Smithsonian Center for Astrophysics, 60 Garden Street, Cambridge, MA 02140, USA}
\altaffiltext{3}{Cahill Center for Astrophysics, California Institute of Technology, 1216 East California Boulevard, Pasadena, CA 91125, USA}
\altaffiltext{4}{Osservatorio di Astrofisica e Scienza dello Spazio di Bologna, Via Piero Gobetti, 93/3, 40129, Bologna, Italy}
\altaffiltext{5}{INAF - Istituto di Astrofisica Spaziale e Fisica Cosmica, Via U. La Malfa 153, I-90146 Palermo, Italy}
\altaffiltext{6}{Dipartimento di Fisica e Astronomia, Alma Mater Studiorum, Universit\`a di Bologna, Via Piero Gobetti, 93/2, 40129, Bologna, Italy}

\begin{abstract}
We present the analysis of a sample of 35 candidate Compton thick (CT-) active galactic nuclei (AGNs) selected in the nearby Universe (average redshift $\langle z\rangle\sim$0.03) with the \swi\ 100-month survey. All sources have available \nustar\ data, thus allowing us to constrain with unprecedented quality important spectral parameters such as the obscuring torus line-of-sight column density ($N_{\rm H, z}$), the average torus column density ($N_{\rm H, tor}$) and the torus covering factor ($f_c$). 
We compare the best-fit results obtained with the widely used \myt\ \citep{murphy09} model with those of the recently published \borus\ model \citep{balokovic18} used in the same geometrical configuration of \myt\ (i.e., with $f_c$=0.5). We find a remarkable agreement between the two, although with increasing dispersion in $N_{\rm H, z}$ moving towards higher column densities.
We then use \borus\ to measure $f_c$. High--$f_c$ sources have, on average, smaller offset between $N_{\rm H, z}$ and $N_{\rm H, tor}$ than low--$f_c$ ones. Therefore, low $f_c$ values can be linked to a ``patchy torus'' scenario, where the AGN is seen through an over-dense region in the torus, while high--$f_c$ objects are more likely to be obscured by a more uniform gas distribution.
Finally, we find potential evidence of an inverse trend between $f_c$ and the AGN 2--10\,keV luminosity, i.e., sources with higher $f_c$ values have on average lower luminosities. 
\end{abstract}


\section{Introduction}
Obscuration in active galactic nuclei (AGNs) has been largely studied over the electromagnetic spectrum, from the optical \citep[e.g.,][]{lawrence91,simpson05}, to the infrared \citep[see, e.g.,][]{jaffe04,nenkova08}, to the X-rays \citep[e.g.,][]{risaliti99,gilli07,marchesi16c}. 
Based on the results of these works, it is commonly accepted that the obscuration is caused by a so-called ``dusty torus'', i.e., a distribution of molecular gas and dust located at $\sim$1--10\,pc from the accreting supermassive black hole (SMBH). While the existence of this obscuring material is universally accepted, its geometrical distribution and chemical composition are still matter of debate, although several works reported that the observational evidence points towards a ``patchy torus'' scenario, where the obscuring material is distributed in clumps formed by optically thick clouds \citep[e.g.,][]{jaffe04,elitzur06,risaliti07,honig07,nenkova08,risaliti11,burtscher13}.

With the launch of the Nuclear Spectroscopic Telescope Array \citep[hereafter \nustar, ][]{harrison13}, the study and characterization of the physics of the obscuring material surrounding accreting SMBHs experienced a significant step forward. \nustar\ is the first telescope with focusing optics at $>$10\,keV, and its sensitivity is almost two orders of magnitude deeper than any of the other previous facilities in the same energy range. Since in the X-rays the observed emission of heavily obscured AGNs peaks at $\sim$30--50\,keV \citep[][]{ajello08a}, where the so-called ``Compton hump'' is observed, while below 5\,keV all the AGN emission is absorbed \citep{koss16}, \nustar, covering the 3--78\,keV energy range, represents the ideal instrument to investigate these otherwise elusive sources. The first years of \nustar\ were dedicated to the analysis of single, well-known CT-AGNs, or to the characterization of small samples of sources \citep[e.g.,][]{balokovic14,puccetti14,annuar15,bauer15,brightman15,koss15,rivers15,masini16,puccetti16}. By the end of 2017, the sample of heavily obscured AGNs observed by \nustar\ was finally large enough to work on a systematic analysis of a statistically significant population of sources.

Consequently, we recently started a project to characterize all the CT-AGNs detected in a volume limited ($d_L$$<$500\,Mpc, Marchesi et al. 2019 in prep.) sample of bright (observed flux $f_{\rm 15-150\,keV}$$\geq$5$\times$10$^{-12}$ erg s$^{-1}$ cm$^{-2}$) AGNs selected in the nearby Universe ($\langle z\rangle$=0.03) using \swi.
As a first step, we analyzed the combined 2--100\,keV spectra of 30 sources having an archival \nustar\ observation and reported to be CT-AGN in previous works, on the basis of their X-ray spectra. Our main goal was to verify how adding the \nustar\ data to the spectral fit improves the general knowledge on the properties of heavily obscured AGNs \citep[][hereafter M18]{marchesi18}. The 2--10\,keV data used in this work were obtained from \xmm, \cha\ and \xrt. The main result of our analysis is the discovery of a systematic offset in the spectral parameters values measured without and with the \nustar\ data. We observed a trend to artificially overestimate the line-of-sight column density ($N_{\rm H, z}$) and the steepness of the spectrum when \nustar\ data are not included in the fit: this effect is only in small part variability--dependent, since only in three out of 30 sources the fit significantly improved allowing for $N_{\rm H, z}$ to vary between the 2--10\,keV and the \nustar\ data. Furthermore, we found that the offset is stronger in sources with low statistics (d.o.f.$<$30) in the 0.3--10\,keV+BAT spectrum, i.e., mostly objects with either a \xrt\ or a short ($<$10\,ks) \cha\ spectrum. In this low-statistics subsample, the intrinsic absorption was overestimated on average by $\sim$40\% in the fits without \nustar\ data. 
As a consequence, less than half (47$^{+10}_{-13}$\%) of the candidate CT-AGNs already reported in the literature are confirmed as bona-fide CT-AGNs in our analysis, and 13 out of 30 sources are found to be Compton thin at the $>$3\,$\sigma$ confidence level. We point out that our analysis was limited to sources previously reported to be CT-AGNs, therefore it is possible that the opposite trend also exists, i.e., there are sources for which the line-of-sight column density is underestimated fitting only the 2--10\,keV+\swi\ data, and that would be found to be CT-AGNs when adding \nustar\ to the fit. 

While the results reported in M18 confirmed the fundamental role of \nustar\ in characterizing heavily obscured AGNs, their effectiveness in constraining the typical geometrical distribution of the obscuring material around the accreting SMBH was limited by the model used in our analysis, \myt\ \citep{murphy09,yaqoob12,yaqoob15}. In fact, while \myt\ has been proven effective in the X-ray spectral fitting of heavily obscured sources, it also assumes a fixed geometry for its torus model, with a torus half-opening angle $\theta_{\rm OA}$=60\degree, i.e., a torus covering factor $f_c$=cos($\theta_{\rm OA}$)=0.5. 
Immediately after the publication of M18, however, \citet{balokovic18} published a new torus model, \borus. This model is an updated version of the extensively used \texttt{BNTorus} model \citep{brightman11}, and has $f_c$ as a free parameter. Notably, $f_c$ is a free parameter also in BNTorus, but \citet{liu15} reported that BNTorus has some issues in the geometry-dependent computation of the torus reprocessed component, and should therefore not be used to derive the torus covering factor. Consequently, we decided to reanalyze our sample, this time using \borus, to measure the torus covering factor and study its trend with different AGN parameters, using a statistically significant sample. As of today, the measurement of the torus covering factor from an X-ray perspective has been limited to single sources, mostly observed with \textit{Suzaku} or \nustar\ \citep[see, e.g.,][]{awaki09,eguchi11,tazaki11,yaqoob12,kawamuro13,farrah16}, or small samples of objects \citep[e.g.,][]{brightman15,masini16}.

This work is organized as follows: in Section \ref{sec:sample} we present the sample used in this work and we describe the data reduction and spectral extraction process for both \nustar\ and the 0.3--10\,keV observations. In Section \ref{sec:model} we describe the models used to perform the spectral fitting. In Section \ref{sec:results} we report the results of the \myt\ spectral fitting of five sources whose data became public recently, which we did not analyze in M18. In Section \ref{sec:myt_vs_borus} we test the recently published \borus\ model, comparing its results with those from the \myt\ one, while in Section \ref{sec:cf} we use \borus\ to measure the torus covering factor ($f_c$) of the sources in our sample, and we study the $f_c$ trend with $N_{\rm H, z}$ and X-ray luminosity. 
Finally, we report our conclusions in Section \ref{sec:concl}. All reported errors are at the 90\% confidence level, if not otherwise stated.

\section{Sample selection and data reduction}\label{sec:sample}
The sources analyzed in this work have been selected from the Palermo BAT 100-month catalog\footnote{\url{http://bat.ifc.inaf.it/100m\_bat\_catalog/100m\_bat\_catalog\_v0.0.htm}}, which reaches a flux limit $f\sim$3.3 $\times$ 10$^{-12}$\,\flu\ in the 15-150 keV band. The public data have been first downloaded from the HEASARC public archive and then processed with the BAT\_IMAGER code \citep{segreto10}. BAT\_IMAGER is used to detect sources in observations made using coded mask instruments. The spectra used in our work are background subtracted and exposure-averaged; the spectral redistribution matrix is the official BAT one\footnote{\url{http://heasarc.gsfc.nasa.gov/docs/heasarc/caldb/data/swift/\\bat/index.html}}. The details of the source counterpart association process are reported in \citet{cusumano10}: we also point out that all the sources in our sample, with the exception of NGC 1358 and ESO 116-G018, have been already reported in previous \swi\ catalogs \citep[see, e.g., ][]{vasudevan13,ricci15,ricci17,oh18}. The 100-month catalog (Marchesi et al. 2019 in prep.) contains 1699 sources, less than 10\% of which (167) are not associated to a counterpart.

30 out of the 35 sources analyzed in this work where first studied in M18. These sources were selected among the 100-month BAT AGNs which were reported to be CT-AGN in previous works and had available \nustar\ archival observation:  a detailed summary of all the papers where these objects were first reported to be CT-AGNs is reported in Table 1 of M18. 14 out of 30 sources had 0.5--10\,keV coverage from \xmm, 2 from \cha, and 14 from \xrt. Notably, for the majority of these sources the line-of-sight column density measurement was originally obtained using only 0.5--10\,keV data, in some cases with the further addition of BAT information in the 15--150\,keV band, but without using \nustar. 

In this paper, we reanalyize all the 30 sources we studied in M18, and we add to the sample five other candidate CT-AGNs from the 100-month BAT catalog with available \nustar\ data that were not studied in the previous work. We report in Table \ref{tab:sample} a summary of these five new sources.

\subsection{Data reduction for the five sources not reported in \citet{marchesi18}}
The five sources not analyzed in M18 are ESO 116-G018 ($z$=0.0185), NGC 1358 ($z$=0.0134), Mrk 3 ($z$=0.0254), MCG--01--30--041 ($z$=0.0188) and NGC 7479 ($z$=0.0079).

NGC 1358 and ESO 116-G018 were first analyzed in \citet{marchesi17a} using \cha\ data. Both sources were found to have $N_{\rm H, z}$$\sim$10$^{24}$ cm$^{-2}$. Therefore, we proposed for a joint \nustar--\xmm\ observation of both these objects, to properly characterize them. Our proposal was accepted (\nustar\ GO Cycle 3, proposal ID: 3258; PI S. Marchesi) and we were granted \nustar\ (50\,ks for both sources) and \xmm\ (50\,ks for ESO 116-G018 and 45\,ks for NGC 1358) time. The results of the spectral analysis have been published in \citet{zhao18a,zhao18b}. 
For these two sources, as well as for Mrk 3 and NGC 7479, we reduced the \xmm\ data using the SAS v16.0.0\footnote{\url{http://xmm.esa.int/sas}} packages and adopting standard procedures. The source spectra were extracted from a 15$^{\prime\prime}$ circular region, corresponding to $\sim$70\% of the encircled energy fraction at 5\,keV for all the three \xmm\ 0.5--10\,keV cameras (MOS1, MOS2 and pn), while the background spectra were obtained from a circle having radius 45$^{\prime\prime}$ located near the source and not contaminated by nearby objects. Each spectrum has been binned with at least 20 counts per bin.

MCG--01--30--041 has only been observed twice with \xrt\ in the 0.5--10\,keV band: since one of these observations was taken simultaneously with the \nustar\ one, we used this in our analysis, to reduce potential variability issues.
We obtained the \xrt\ spectrum using the \textit{Swift} products generator available online \citep[\url{http://www.swift.ac.uk/user\_objects/}; see also][]{evans09}. Due to the spectrum low statistics, we binned it with only 3 counts per bin and we therefore analyzed it using the \texttt{cstat}, rather than the $\chi^2$, statistic (see Section \ref{sec:results}).

Finally, for all five objects the data retrieved for both  \nustar\  Focal Plane Modules \citep[FPMA and FPMB;][]{harrison13} were processed using the  \nustar\  Data Analysis Software (NUSTARDAS) v1.5.1. 
The event data files were calibrated running the {\tt nupipeline} task using the response file from the Calibration Database (CALDB) v. 20180419. With the {\tt nuproducts} script we generated both the source and background spectra, and the ancillary and response matrix files. 
For both focal planes, we used a circular source extraction region with a 30$^{\prime\prime}$ radius, corresponding to $\sim$50\% of the encircled energy fraction over the whole \nustar\ energy range, and centered on the target source; for the background we used the same extraction region positioned far from any source contamination in the same frame. The \nustar\ spectra have then been grouped with at least 20 counts per bin, and cover the energy range from 3 to 50--70\,keV, depending on the quality of the data. 

\begingroup
\renewcommand*{\arraystretch}{1.15}
\begin{table*}
\centering
\scalebox{0.83}{
\begin{tabular}{lccccccccccc}
\hline
\hline
  \multicolumn{1}{c}{4PBC name} & \multicolumn{1}{c}{Source name} &   \multicolumn{1}{c}{R.A.} & \multicolumn{1}{c}{Decl} &  \multicolumn{1}{c}{Type} & \multicolumn{1}{c}{$z$} & \multicolumn{1}{c}{Telescope} & \multicolumn{1}{c}{ObsID} & \multicolumn{1}{c}{Date} &  \multicolumn{1}{c}{Exposure} &  \multicolumn{1}{c}{Rate} &  \multicolumn{1}{c}{Ref.}\\
  & & deg & deg & & & & & & ks & cts s$^{-1}$\\
  \multicolumn{1}{c}{(1)} & (2) & (3) & (4) & (5) & (6) & (7) & (8) & (9) & (10) & (11) & (12)\\
\hline
  J0324.8--6043 & ESO 116-G018 & 51.2210 & --60.7384 & 2 & 0.0185 & \xmm & 0795680201 & 2017-11-01 & 164.8 & 0.015 & (a)\\
      ...                              & ...             &  ...     & ...         &      ...            &      ...       & \nustar & 60301027002 & 2017-11-01 & 90.1 & 0.013 & --\\
  J0333.7--0504 & NGC 1358 & 53.4153 & --5.0894 & 2 & 0.0134 & \xmm  & 0795680101 & 2017-08-01 & 100.6 & 0.014 & (a)\\
    ...                              & ...             &  ...     & ...         &      ...            &      ...       & \nustar & 60301026002 & 2017-08-01& 99.8 & 0.023 & --\\ 
  J0615.5+7102 & Mrk 3 & 93.90129 &  71.037476 & 2 & 0.0254 & \xmm & 0741050501 & 2015-04-06 & 19.9 & 0.089 & (b)\\
  ...                              & ...             &  ...     & ...         &      ...            &      ...       & \nustar & 60002048004 & 2014-09-14 & 66.9  & 0.049 & (c)*\\
  J1152.7-0511 & MCG--01--30--041 & 178.15887 &  --5.206967 & 1.8 & 0.0188 & \xrt & 80062 & 2017-06-14 & 7.1 & 0.065 & (d)\\
    ...                              & ...             &  ...              &      ...   & ...    &      ...       & \nustar & 60061216002 & 2017-06-14 & 53.7 & 0.086 & --\\
  J2304.7+1219 & NGC 7479 & 346.236042  & 12.322889 & 2 & 0.0079 & \xmm & 0025541001 & 2001-06-19  & 29.9 & 0.041 & (e)\\
  ...                              & ...             &  ...     & ...         &      ...            &      ...       & \nustar & 60201037002 & 2016-05-12 & 36.9  & 0.015 & --\\
\hline
\hline
\end{tabular}}\caption{\normalsize \normalsize Candidate CT-AGNs analyzed in this work and not reported in M18. Column (1): ID from the Palermo BAT 100-month catalog (Marchesi et al. 2019 in prep.). (2): source name. (3) and (4): right ascension and declination (J2000 epoch). (5): optical classification (1.8: Seyfert 1.8 galaxy; 2: Seyfert 2 galaxy). (6): redshift. (7): telescope used in the analysis. (8): observation ID. (9): observation date. (10): total exposure, in ks. For \xmm\ and \nustar, this is the sum of the exposures of each camera. (11): average count rate (in cts s$^{-1}$), weighted by the exposure for \xmm\ and \nustar, where observations from multiple instruments are combined. Count rates are computed in the 3--70\,keV band for \nustar\ and in the 2--10\,keV band otherwise. (12): reference for previous assessments of CT nature for the source, as follows. When \nustar\ data were used, the reference is reported on the \nustar\ observation line. Sources previously fitted with a torus model are flagged with a *. a) \citet{marchesi17a}; b) \citet{cappi99}; c) \citet{guainazzi16}; d) \citet{vasudevan13}; e) \citet{severgnini12}.}\label{tab:sample}
\end{table*}
\endgroup

\section{Spectral fitting procedure}\label{sec:model}
The spectral fitting procedure was performed using the XSPEC software \citep{arnaud96}; the Galactic absorption values is the one measured by \citet{kalberla05}. We used \citet{anders89} cosmic abundances, fixed to the solar value, and the \citet{verner96} photoelectric absorption cross-section. Following the same approach described in M18, we fit our data in the 2--150\,keV regime, since in heavily obscured AGNs the 0.5--2\,keV band emission is dominated by non-AGN processes, such as as star-formation and/or diffuse gas emission \citep[see, e.g.,][]{koss15}.

Heavily obscured AGNs have complex spectra, where the contribution of the Compton scattering and of the fluorescent Iron line becomes significant with respect to less obscured AGN spectra. Consequently, these sources should be treated in a  self-consistent way, that allows one to properly measure  $N_{\rm H, z}$, using models developed specifically to this purpose. In M18, we fitted the 30 sources in our sample with a Monte Carlo radiative transfer code: \myt\ \citep{murphy09,yaqoob12,yaqoob15}. In this work we use both \myt\ and \borus\ \citep{balokovic18}, another Monte Carlo radiative transfer code. More in detail, we first fit the spectra of the five sources not reported in M18 using \myt.  We then fit with \borus\ both the 30 sources in M18 and the five new sources, using two distinct model configurations: the first one with \borus\ in the same geometrical configuration than \myt, the second one  allowing the torus covering factor to vary.

\subsection{\myt}\label{sec:myt}
The \myt\ model is divided in three distinct components.

\begin{enumerate}
\item A multiplicative component containing photoelectric absorption and Compton scattering attenuation, with associated equivalent neutral hydrogen column density ($N_{\rm H,z}$). This component is applied to the main power law continuum.
\item A scattered continuum, also known as ``reprocessed component''. This component models those photons that are observed after one or more interactions with the material surrounding the SMBH. The normalization of the reprocessed component with respect to the main continuum hereby denoted as A$_{\rm S}$. 
\item The neutral Fe fluorescent emission lines, more in detail the Fe K$\alpha$ line at 6.4\,keV and the K$\beta$ at 7.06\,keV. We denote the normalization of these lines with respect to the main continuum as A$_{\rm L}$. 
\end{enumerate}

In \myt, the obscuring material surrounding the SMBH is assumed to have a toroidal, azimuthally symmetric shape. The torus covering factor, $f_c$, is not a free parameter and is fixed to  $f_c$=cos($\theta_{\rm OA}$)=0.5, where $\theta_{\rm OA}$=60$\degree$ is the torus half-opening angle. The angle between the torus axis and the observer is free to vary, within the range $\theta_{\rm obs}$=[0--90]$\degree$. In our analysis, both in M18 and here, we use \myt\ in the so-called ``decoupled mode'' \citep{yaqoob15}: for the main continuum, we fix $\theta_{\rm obs}$=90$\degree$, while for the reprocessed component we test two different scenarios, one with $\theta_{\rm obs, AS, AL}$=90$\degree$, the other with $\theta_{\rm obs, AS, AL}$=0$\degree$, checking which one leads to the smaller reduced $\chi^2$, $\chi^2_\nu$=$\chi^2$/(degrees of freedom).
Sources best-fitted with $\theta_{\rm obs, AS, AL}$=90$\degree$  correspond to a scenario where the dense obscuring torus is observed ``edge-on'' and the obscuring material lies between the AGN and the observer. Sources best-fitted with $\theta_{\rm obs, AS, AL}$=0$\degree$, instead, are assumed to describe a patchy torus scenario, in which the reprocessed emission from the inner edge of the torus can reach the observer.

\subsection{\borus}\label{sec:borus}
\borus\ \citep{balokovic18} is an updated and improved version of the widely used \texttt{BNTorus} model \citep{brightman11}. This radiative transfer code models the reprocessed emission component of an AGN X-ray spectrum, i.e., following the \myt\ nomenclature we introduced in the previous section, the ``reprocessed component'' and the neutral Fe emission lines.

In \borus, the obscuring material has a quasi--toroidal geometry, with conical polar cutouts. Both the average torus column density ($N_{\rm H, tor}$) and the torus covering factor are free parameters in the model: the torus covering factor value can vary in the range $f_c$=[0.1--1.0], corresponding to a torus opening angle range $\theta_{\rm OA}$=[0--84]\degree.
In principle, the angle between the torus axis and the observer is also a free parameter of this model, but in our analysis we fix it to $\theta_{\rm obs}$=87\degree, i.e., the upper boundary of the parameter in the model, corresponding to an almost ``edge-on'' configuration. In this work, we decide to fix $\theta_{\rm obs}$ to reduce potential degeneracies between this parameter, $N_{\rm H, tor}$ and $f_c$, particularly in sources with low statistics (i.e., with less than 150--200 degrees of freedom). We are also working on a companion paper (Zhao et al. 2019 in prep.) where we use a sample of nearby AGNs, both obscured and unobscured, to analyze how leaving $\theta_{\rm obs}$ free to vary affects the other spectral parameters.

Finally, since the \borus\ models itself does not take into account line-of-sight absorption, we follow \citet{balokovic18} approach and derive  $N_{\rm H, z}$ in XSPEC using the components \texttt{zphabs $\times$ cabs} to properly model Compton scattering losses out of the line of sight. In the overall fitting model, the $N_{\rm H, z}$ value is a free parameter, independent from $N_{\rm H, tor}$, and assumed to be identical in \texttt{zphabs} and \texttt{cabs}.

\subsection{Additional components to the best-fit model}
Besides using \myt\ and \borus\ in the configurations described in the previous sections, we included the following components to our best-fit model:
\begin{enumerate}
\item A second power law, with photon index $\Gamma_2$=$\Gamma_1$, where $\Gamma_1$ is the photon index of the primary power law. This second power law is introduced to take into account the fraction $f_{\rm scatt}$ of accreting SMBH emission which is scattered, rather than absorbed, by the gas surrounding the SMBH. We assume this component to be unabsorbed. 
\item A constant, $C_{NuS-2-10}$, allowing for a re-normalization of the \nustar\ spectrum with respect to the 2--10\,keV+\swi\ one. Such a component models both cross-calibration offsets between the 2-10\,keV and the \nustar\ data and potential flux variability between the different observations.
\end{enumerate}

\section{\myt\ fitting results for the five sources not reported in Marchesi et al. (2018)}\label{sec:results}
As a first step of our analysis, we fitted with \myt\ the five sources which we did not report in M18: we jointly fitted the 2--10\,keV (from either \xmm, \cha\ or \xrt), \nustar\ and \swi\ data, and the \myt\ configuration is the one described in Section \ref{sec:myt}. 
All sources but NGC 7479 are fitted using the $\chi^2$ statistic: due to its low 2--10\,keV count statistics, we measure the goodness of the fit in NGC 7479 with the W statistic (\texttt{cstat} in XSPEC), which is commonly used when a source does not have enough counts to be fitted with the $\chi^2$ method. However, the \swi\ spectra are already background-subtracted and can therefore not be fitted with \texttt{cstat}: for this reason, we used the multi-statistic approach allowed by XSPEC and fitted the \swi\ data with the $\chi^2$ statistic. The best-fit statistic we report for NGC 7479 in Table \ref{tab:fit_new} is thus the sum of \texttt{cstat} and $\chi^2$.

In Table \ref{tab:fit_new} we also report the best-fit parameters for our new five sources: $N_{\rm H, z}$, $\Gamma$, the 2--10\,keV to \nustar\ cross-normalization constant, $C_{NuS-2-10}$; the main power law component normalization, norm$_{\rm 1}$; the reprocessed and and iron lines relative normalizations, A$_{\rm S}$ and A$_{\rm L}$; the fraction of scattered emission, $f_{\rm scatt}$. The observed flux and the intrinsic luminosity in the 2--10\,keV and in the 15--55\,keV band are also reported. For all five sources, we find that the best-fit model is the one with $\theta_{\rm obs, AS, AL}$=90$\degree$.

Three out of five sources (ESO 116-G018, NGC 1358 and NGC 7479) are confirmed to be CT-AGN at a $>$3\,$\sigma$ level. A fourth source, Mrk 3, is found to have best-fit line-of-sight column density $N_{\rm H, z}$=(7.8$\pm$0.1)$\times$10$^{23}$\,cm$^{-2}$, slightly below the CT threshold. However, this is not an unexpected result, since this source is known to be highly variable, a result confirmed also in our analysis, since we find $C_{NuS-2-10}$=2.07$_{-0.04}^{+0.02}$. A Compton-thin solution for this source was already reported in \citet{yaqoob15}; furthermore, a recent monitoring campaign with \nustar\ performed by \citet{guainazzi16} showed that the line-of-sight column density of Mrk 3 varied in the range $N_{\rm H, z}$=[0.75--0.94]\,$\times$10$^{24}$\,cm$^{-2}$ in a timespan of seven months. Notably, there is an excellent agreement between our $N_{\rm H, z}$ measurement and the one obtained by \citet{guainazzi16} using the same \nustar\ observation, i.e.,  $N_{\rm H, z}$=(7.7$\pm$0.1)$\times$10$^{23}$ cm$^{-2}$. 

Finally, we find that MCG-01-30-041, which was reported to be a CT-AGN ($N_{\rm H, z}$=1.45$_{-0.45}^{+0.74}$\,$\times$10$^{24}$\,cm$^{-2}$) by \citet{vasudevan13}, is in fact an unobscured AGN ($N_{\rm H, z}$$<$10$^{22}$\,cm$^{-2}$) based on our combined \xrt,\swi\ and \nustar\ fit. Notably, we find that even fitting the \xrt\ and \swi\ data only does not produce a CT solution, the best-fit line-of-sight column density being in this case $N_{\rm H, z}$=1.8$_{-1.1}^{+2.1}$\,$\times$\,10$^{23}$\,cm$^{-2}$, with a corresponding power law photon index $\Gamma$=2.13$_{-0.37}^{+0.44}$.
The reason of the discrepancy between our result and the one reported by \citet{vasudevan13} is likely linked to the very low quality of the combined \xrt and \swi\ spectrum used in their work: the \xrt\ spectrum used in their analysis has in fact only $\sim$30 counts in the 2--10\,keV band (our combined \xrt\ and \nustar\ spectrum has $\sim$3850 counts in the same band), and their fit has only 6 degrees of freedom (ours has 297). 

\begin{figure*}
\begin{minipage}[b]{.5\textwidth}
  \centering
  \includegraphics[width=0.75\textwidth,angle=-90]{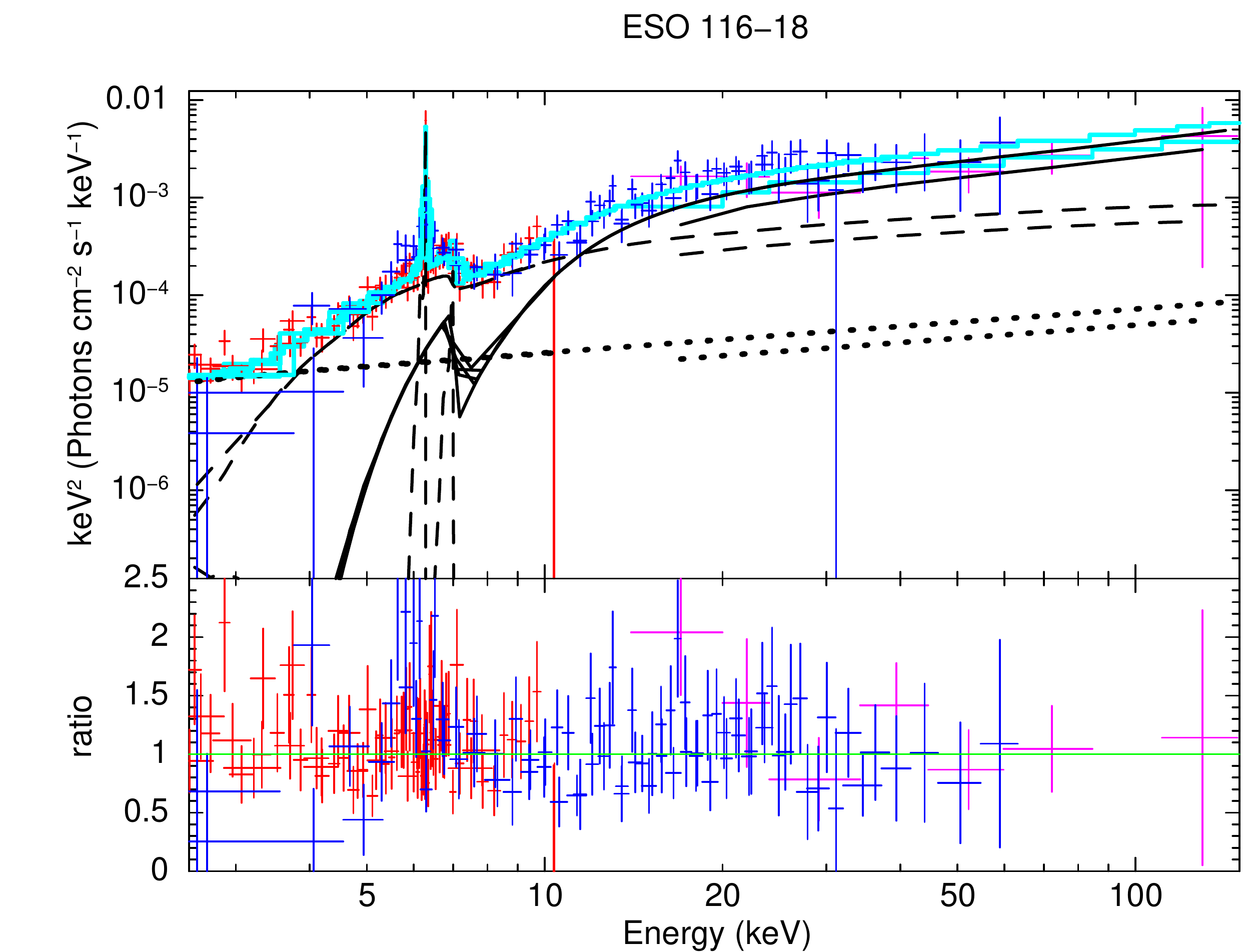}
  \end{minipage}
\begin{minipage}[b]{.5\textwidth}
  \centering
  \includegraphics[width=0.75\textwidth,angle=-90]{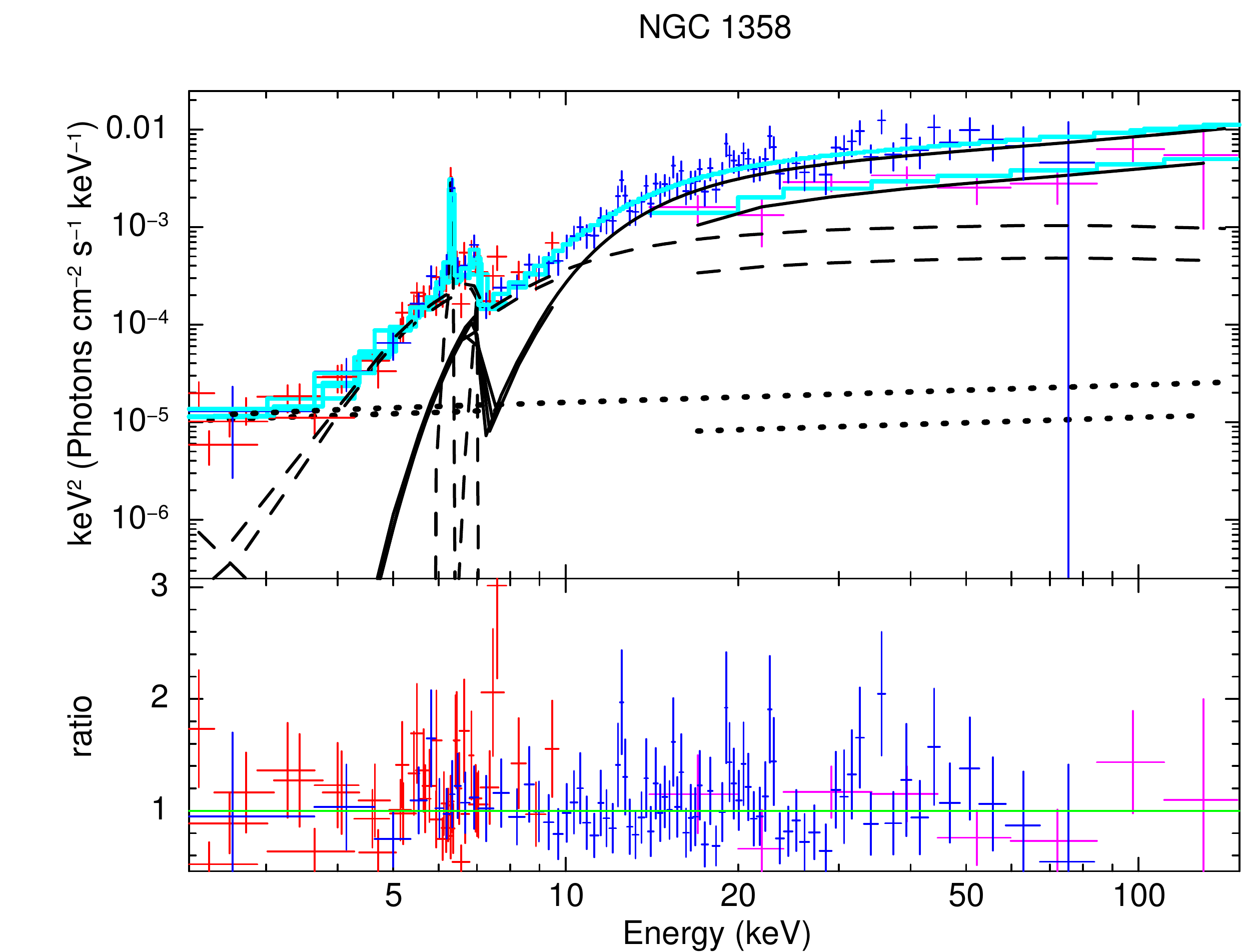}
  \end{minipage}
\begin{minipage}[b]{.5\textwidth}
  \centering
  \includegraphics[width=0.75\textwidth,angle=-90]{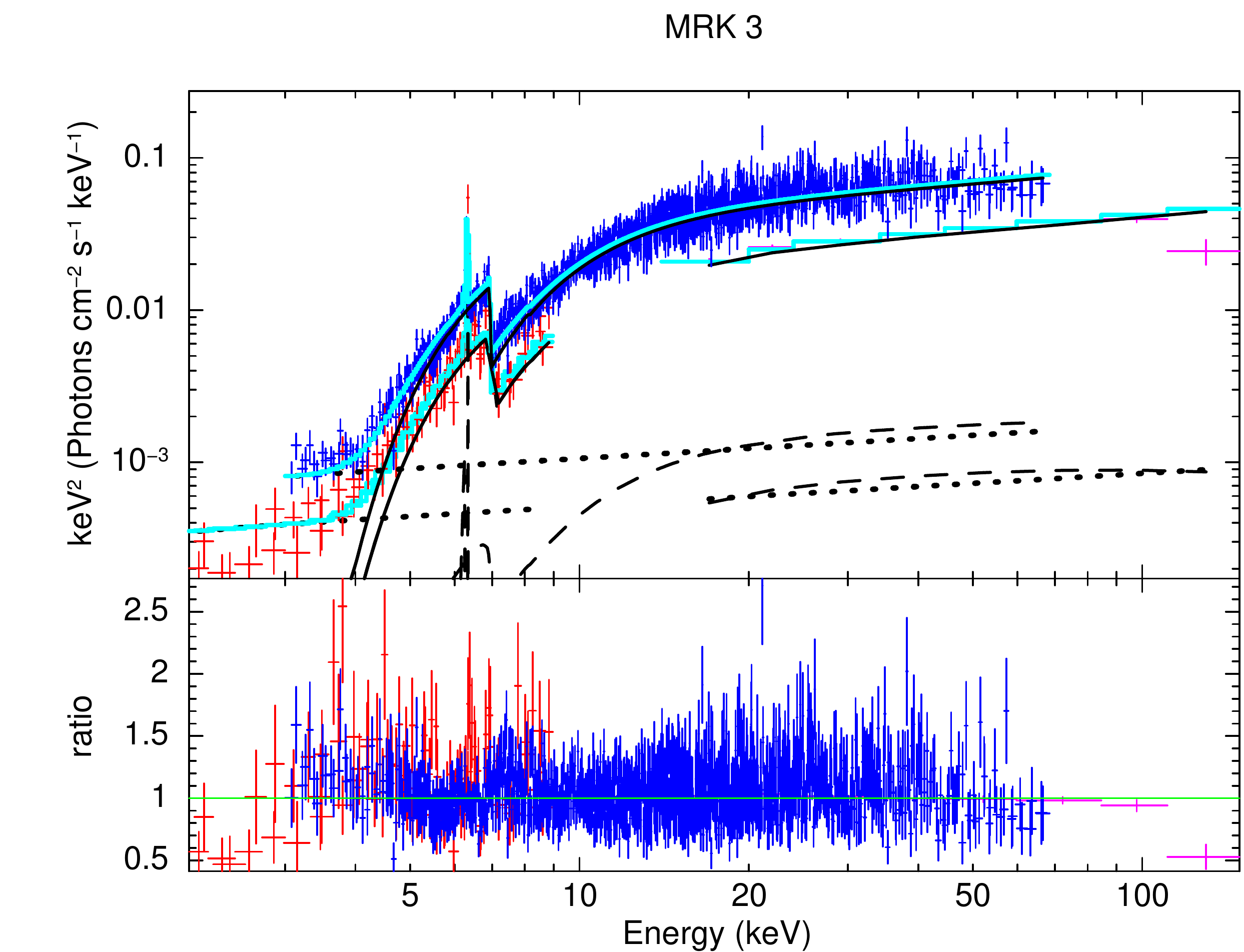}
  \end{minipage}
\begin{minipage}[b]{.5\textwidth}
  \centering
  \includegraphics[width=0.75\textwidth,angle=-90]{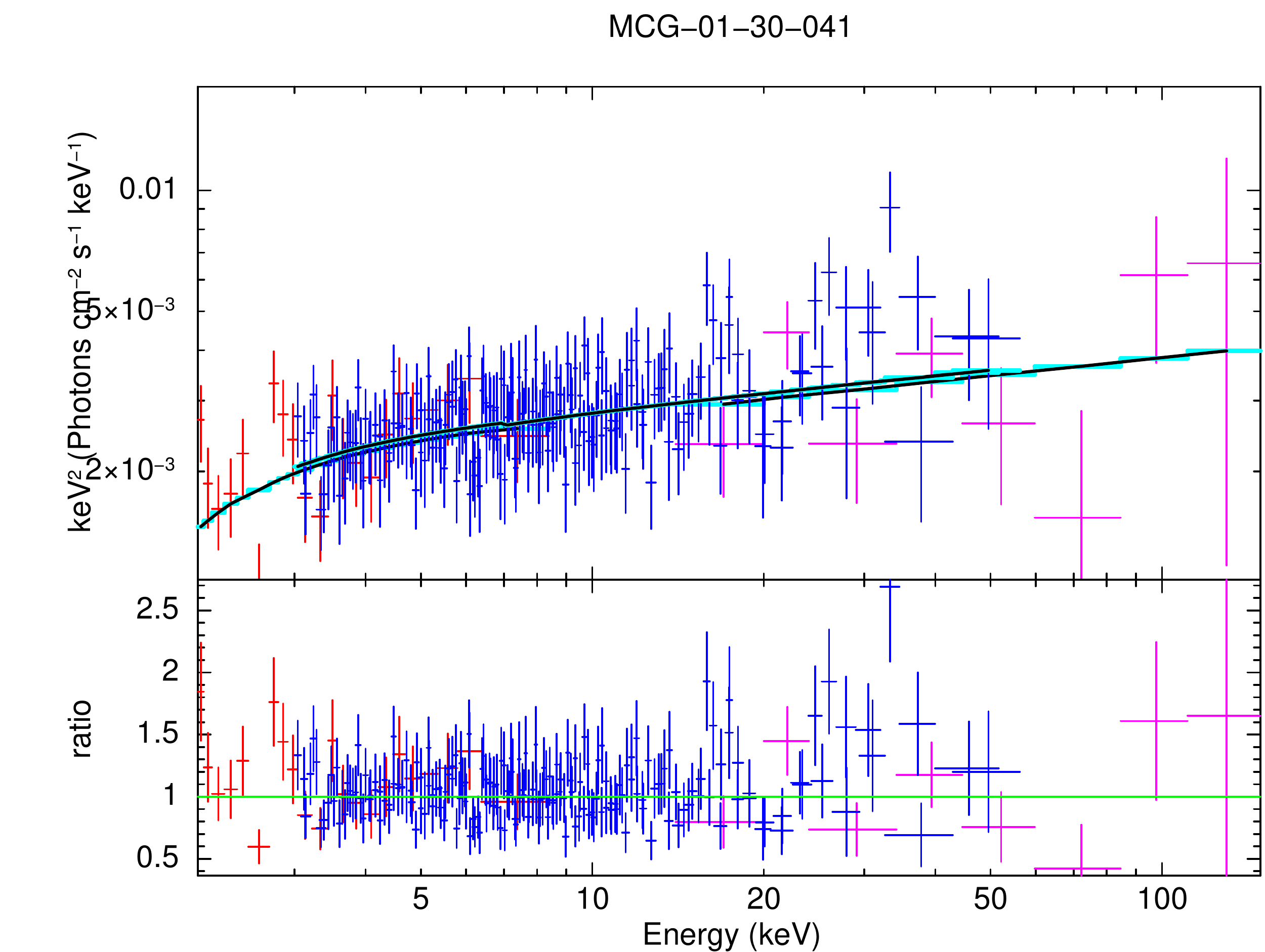}
  \end{minipage}
\begin{minipage}[b]{.5\textwidth}
  \centering
  \includegraphics[width=0.75\textwidth,angle=-90]{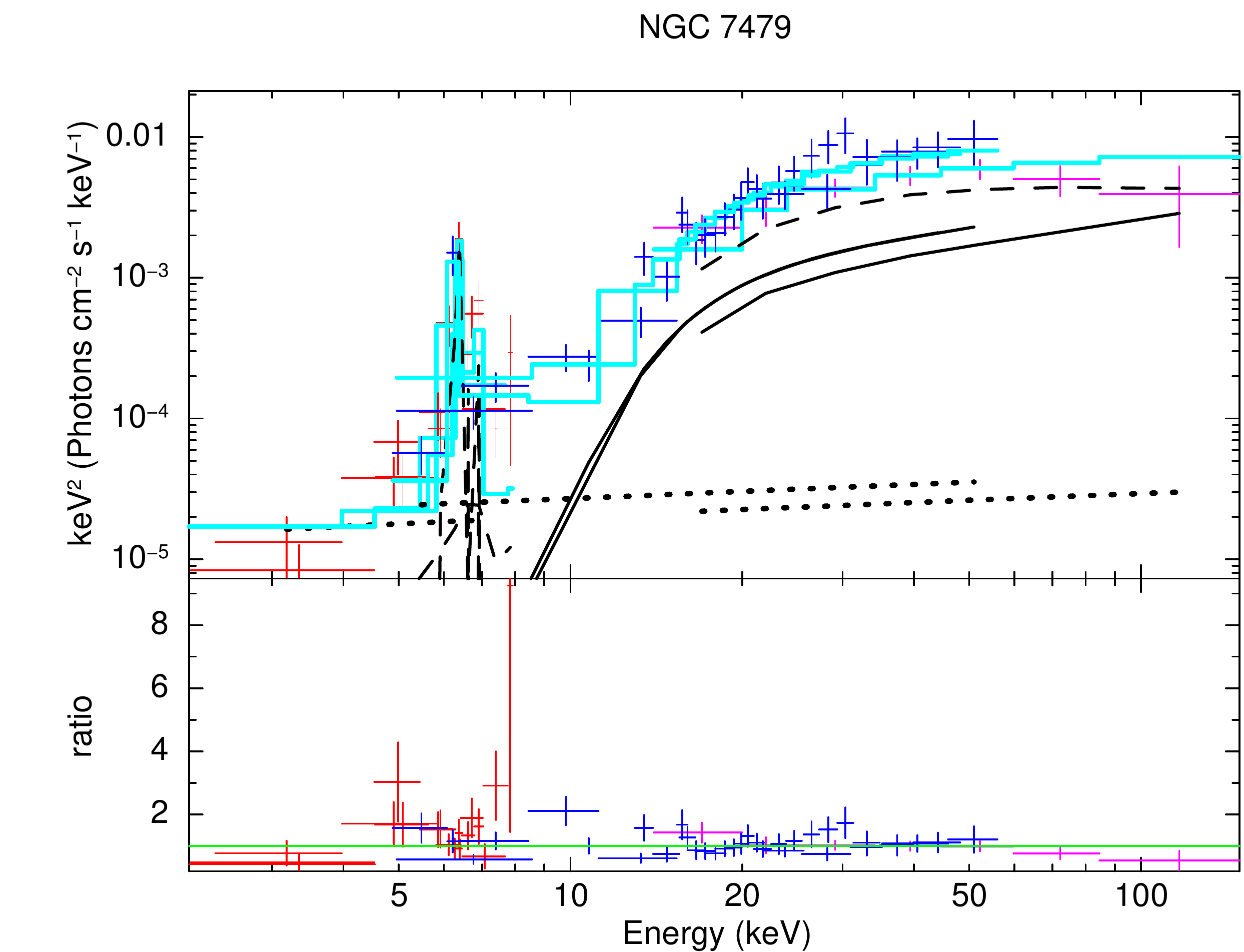}
  \end{minipage}
\caption{\normalsize Background-subtracted spectra (top panel) and data-to-model ratio (bottom) of the five CT-AGNs analyzed in this work and not previously analyzed in M18. 2--10\,keV data are plotted in red, \nustar\ data in blue and \swi\ data in magenta. The best-fitting model is plotted as a cyan solid line, while the single \myt\ components are plotted as black solid (zeroth-order continuum) and dashed (reprocessed component and emission lines) lines. Finally, the main power law component scattered, rather than absorbed, by the torus is plotted as a black dotted line.} \label{fig:spectra_new} 
\end{figure*}

We report in Figure \ref{fig:spectra_new} the 2--100\,keV spectra of the five sources, as well as the corresponding best-fit models.

\begingroup
\renewcommand*{\arraystretch}{1.5}
\begin{table*}
\centering
\scalebox{0.85}{
\begin{tabular}{ccccccccccccc}
\hline
\hline
Source & $N_{\rm H, z}$ & $\Gamma$ & $C_{NuS-2-10}$ & norm$_{\rm 1}$ & A$_{\rm S}$ & $f_{\rm scatt}$ & f$_{\rm 2-10}$ & L$_{\rm 2-10}$ & f$_{\rm 15-55}$ & L$_{\rm 15-55}$ & $\chi^2$/d.o.f.\\
            &                               &                       &                     &   &     \%            &  \\
\hline
ESO 116-G018   & 190.0$_{-33.0}^{+76.0}$ & 1.55$_{-0.15}^{+0.18}$ & 1.05$_{-0.13}^{+0.14}$ & 18.52$_{-6.13}^{+14.89}$.   & 0.84$_{-0.37}^{+0.39}$ & 0.5$_{-0.2}^{+0.3}$ & --12.52$_{-0.09}^{+0.03}$ & 43.30$_{-0.42}^{+0.47}$ & --11.34$_{-0.13}^{+0.02}$ & 43.19$_{-0.19}^{+0.19}$ & 187.0/210\\
NGC 1358          & 236.0$_{-33.0}^{+27.0}$ & 1.82$_{-0.26}^{+0.24}$ & 1.17$_{-0.14}^{+0.15}$ & 141.32$_{-31.33}^{+42.21}$ & 0.25$_{-0.05}^{+0.06}$ & 0.1$_{-0.1}^{+0.1}$ & --12.41$_{-0.05}^{+0.02}$ & 43.39$_{-0.33}^{+0.21}$ & --10.95$_{-0.03}^{+0.02}$ & 43.39$_{-0.10}^{+0.11}$ & 220.0/239\\
Mrk 3                  & 78.9$_{-2.1}^{+2.3}$ & 1.78$_{-0.02}^{+0.02}$ & 2.06$_{-0.05}^{+0.05}$ & 240.64$_{-13.93}^{+25.53}$ & $<$0.30 & 1.3$_{-0.1}^{+0.2}$ & --11.18$_{-0.02}^{+0.03}$ & 44.04$_{-0.04}^{+0.04}$ & --9.91$_{-0.02}^{+0.03}$ & 44.23$_{-0.03}^{+0.03}$ & 1183.4/1098 \\ 
MCG-01-30-041 & $<$1.0 & 1.85$_{-0.04}^{+0.05}$ & 1.03$_{-0.08}^{+0.09}$ & 18.59$_{-1.94}^{+2.87}$       & -- & -- & --11.26$_{-0.07}^{+0.02}$ & 42.64$_{-0.04}^{+0.03}$ & --11.17$_{-0.04}^{+0.03}$ & 42.71$_{-0.01}^{+0.01}$ & 222.2/225\\
NGC 7479          & 363.6$_{-50.3}^{+58.0}$ & 1.83$_{-0.09}^{+0.09}$ & 1.34$_{-0.20}^{+0.22}$ & 110.08$_{-33.92}^{+44.50}$   & 1.00$^f$ & $<$0.5     & {--12.75$_{-0.08}^{+0.04}$} & 42.64$_{-0.18}^{+0.14}$  & --10.96$_{-0.07}^{+0.04}$ & 43.00$_{-0.66}^{+0.26}$ & 200.6/171 \\
\hline
\hline

\end{tabular}}\caption{\normalsize Best fit properties for the five new candidate CT-AGNs analyzed in this work and not reported in M18. The reported parameters have been obtained by fitting all the available data for the given source, including \nustar. $N_{\rm H, z}$ is the line-of-sight column density (in units of 10$^{22}$\,cm$^{-2}$); $\Gamma$ is the power law photon index; $C_{NuS-2-10}$ is the cross-normalization constant between the 2--10\,keV and the \nustar\ data; norm$_{\rm 1}$ is the main power law normalization (in units of ph cm$^2$ s$^{-1}$ keV$^{-1}$$\times$10$^{-4}$), measured at 1\,keV; A$_{\rm S}$ is the intensity of the \myt\ reprocessed component with respect to the main one; $f_{\rm scatt}$ is the percentage of main power law emission scattered, rather than absorbed, by the obscuring material. f$_{\rm 2-10}$, L$_{\rm 2-10}$, f$_{\rm 15-55}$ and L$_{\rm 15-55}$ are the logarithms of the observed flux (in units of erg s$^{-1}$ cm$^{-2}$) and the intrinsic, unabsorbed luminosity (in units of erg s$^{-1}$) measured in the 2--10\,keV and in the 15--55\,keV bands, respectively. Fluxes and luminosities are obtained with XSPEC, using the \texttt{flux} command and the \texttt{clumin} convolution model, respectively. Parameters fixed to a given value are flagged with $^f$. The intensity of the Iron lines, A$_{\rm L}$, is assumed to be $=$A$_{\rm S}$ in all sources but NGC 7479, where is A$_{\rm L}$=4.44$_{-1.02}^{+1.24}$. We find MCG-01-30-041 to be an unobscured AGN, therefore the fit did not required neither the reprocessed component A$_{\rm S}$ nor the scattered component $f_{\rm scatt}$.} 
\label{tab:fit_new}
\end{table*}
\endgroup

\section{Comparison between \texttt{MYTorus} and \texttt{borus02}}\label{sec:myt_vs_borus}
In order to validate the reliability of the \borus\ model and use it to measure the torus covering factor, in this section we compare the results obtained using \myt\ with those obtained using  \borus\ with the same geometrical configuration of \myt, i.e., fixing the torus covering factor to $f_c$=0.5 ($\theta_{\rm OA}$=60$\degree$) and the angle between the observer and the torus axis to $\theta_{\rm obs}$=87$\degree$, i.e., the maximum value allowed by \borus\footnote{While in M18 we used $\theta_{\rm obs}$=90$\degree$, we checked that the difference in the best-fit values of $\Gamma$ and $N_{\rm H, z}$ using $\theta_{\rm obs}$=90$\degree$ and $\theta_{\rm obs}$=87$\degree$ is $<$1\%}. We also fix the strength of the reprocessed component, A$_{\rm S}$, to the best-fit value we obtained using \myt\ (see Table \ref{tab:fit_new} and Table 3 in M18). The overall torus column density, $\log$($N_{\rm H, tor}$), is left free to vary.

\begin{figure*}
\begin{minipage}[b]{.5\textwidth}
  \centering
  \includegraphics[width=1.05\textwidth]{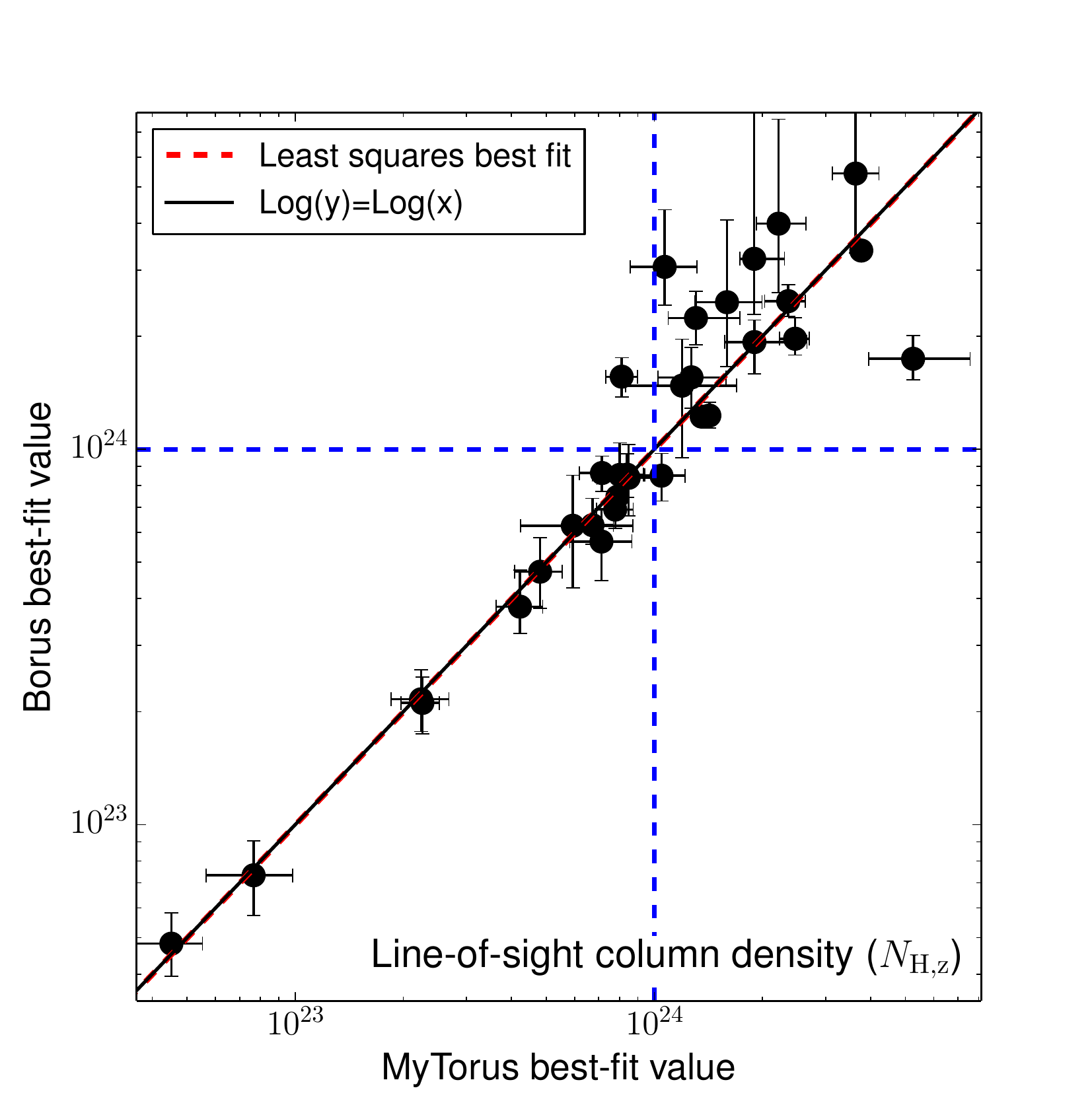}
  \end{minipage}
\begin{minipage}[b]{.5\textwidth}
  \centering
  \includegraphics[width=1.06\textwidth]{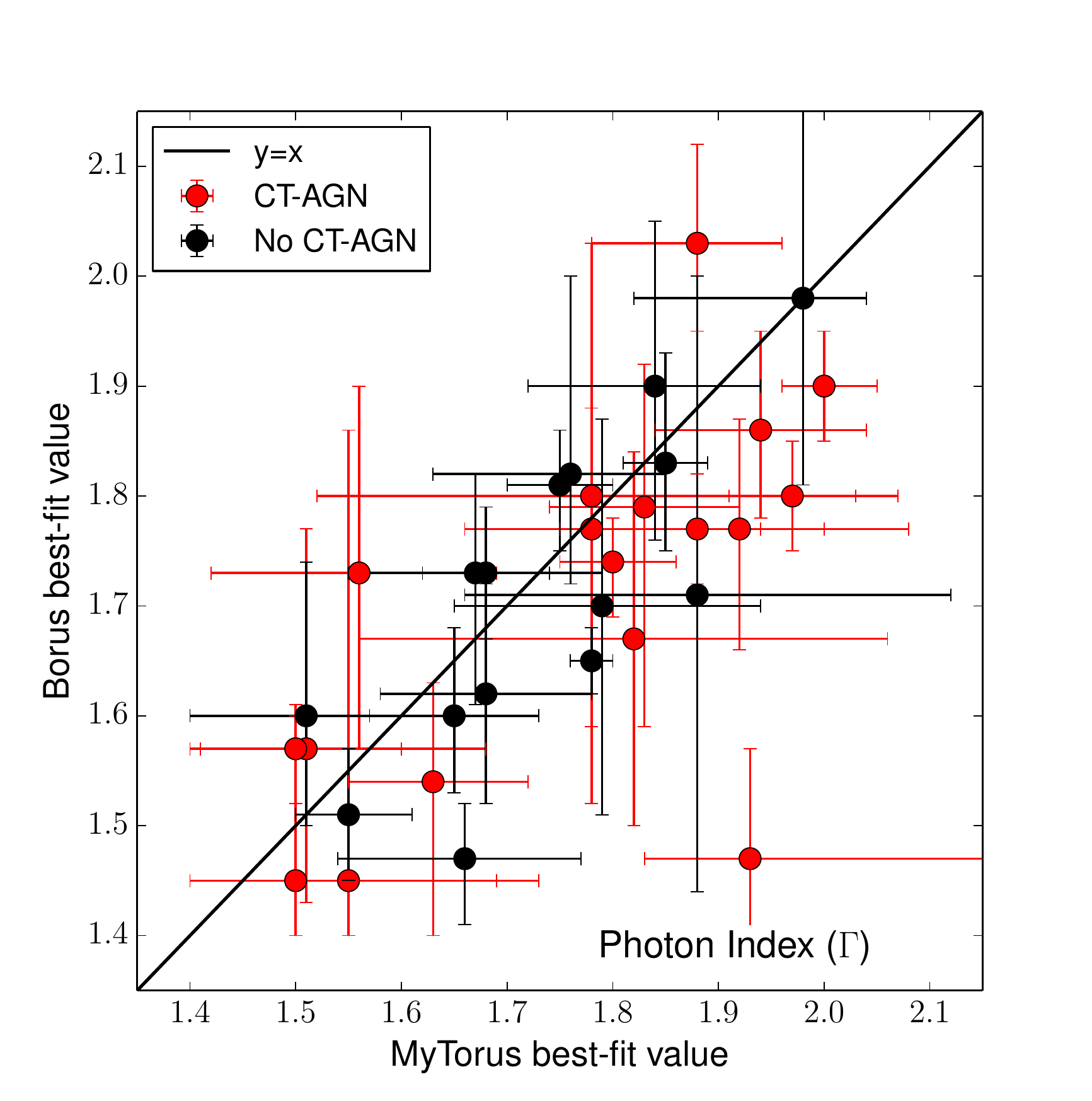}
  \end{minipage}
\caption{\normalsize \textit{Left}: Distribution of the line-of-sight column density $N_{\rm H, z}$ measured using the \borus\ model as a function of the same parameter measured using the \myt\ model. NGC 1068, which has $N_{\rm H, z}>$10$^{25}$ cm$^{-2}$, and RBS 1037 and MCG-01-30-041, which we find to be unobscured AGNs, are not shown in the plot.  The $\log$($y$)=$\log$($x$) relation is plotted as a black solid line, while the best-fit relation $\log$($N_{\rm H, z, Borus}$)=$a$+$b$$\log$($N_{\rm H, z, Borus}$) is plotted as a red dashed line. The blue dashed horizontal and vertical lines mark the CT threshold, $N_{\rm H, z}$=10$^{24}$\,cm$^{-2}$.  \textit{Right}: same as left, but for the photon index $\Gamma$. Source with $N_{\rm H, z}>$10$^{24}$\,cm$^{-2}$ from either the \myt\ or the \borus\ model are plotted in red, while sources with $N_{\rm H, z}<$10$^{24}$\,cm$^{-2}$ in both models are plotted in black.
Here, we do not plot NGC 1229, whose $\Gamma_{\rm NuS}$ value is pegged to $\Gamma$=1.4, the \myt\ model's lower boundary. The $y$=\,$x$ relation is plotted as a black solid line.}\label{fig:trends}
\end{figure*}

In Table \ref{tab:results_myt_vs_borus} we report the best-fit, line-of-sight column density and photon index values for the 35 sources in our sample, using either \myt\ or \borus, while in Figure \ref{fig:trends} we show the trends of these parameters obtained using \borus\ as a function of those obtained using \myt.

As can be seen in Figure \ref{fig:trends}, left panel, there is a general excellent agreement between the $N_{\rm H, z}$ value obtained using \borus\ and the one obtained using \myt: the best-fit slope of the relation $\log$($N_{\rm H, z, Borus}$)=$a$\,$\log$($N_{\rm H, z, MyT}$)+b is $a$=0.97$\pm$0.05, and the Spearman rank order correlation coefficient of the distribution is $\rho$=0.90, and the p-value for such a $\rho$ value to be derived by an uncorrelated population is $p$=1.2$\times$10$^{-12}$. However, the high overall correlation observed in the whole sample hides the existence of two clear separate trends: sources having $N_{\rm H, z}$$\leq$10$^{24}$ cm$^{-2}$ based on both the \myt\ and the \borus\ best-fit have excellent correlation, the Spearman rank order correlation coefficient being $\rho$=0.92 (p-value $p$=1.1$\times$10$^{-6}$). On the opposite, we find a weaker correlation for sources being CT-- according to at least one of the two models: in this second scenario, the sample has $\rho$=0.49 and p-value $p$=0.04. We also find that one source (NGC 1194), is found to be CT using \borus\ and has instead $N_{\rm H, z, MyT}$$<$10$^{24}$ cm$^{-2}$, while another object (MCG+06-16-028) is a CT-AGN with \myt\ but has $N_{\rm H, z, Borus}$$<$10$^{24}$ cm$^{-2}$, although for this source the two  $N_{\rm H, z, MyT}$ measurements are consistent at the 90\% confidence level.

The larger dispersion observed in the Compton thick regime is not unexpected, since above the CT threshold the $N_{\rm H, z}$ measurement becomes more complex, because of the slightly different geometries of the models. Nonetheless, it is worth noticing that for 11 out of the 16 sources being confirmed CT-AGNs based on both \myt\ and \borus\ have $N_{\rm H, z}$ values in agreement at the 90\% confidence level.

In Figure \ref{fig:trends}, right panel, we plot the best-fit photon index obtained using \borus, as a function of the same parameter obtained using \myt. Sources classified as CT-AGNs by either \myt\ or \borus\ are plotted in red, while sources with $N_{\rm H, z}<$10$^{24}$\,cm$^{-2}$ in both models are plotted in black. As can be seen, the dispersion of the distribution is quite large, the Spearman rank order correlation coefficient being $\rho$=0.64 and the p-value for such a $\rho$ value to be derived by an uncorrelated population is $p$=6.8$\times$10$^{-5}$. However, the overall agreement between the two models is remarkable, the average photon index measured with \myt\ being  $\langle \Gamma_{\rm MyT} \rangle$=1.74, with associated standard deviation $\sigma_{\rm \Gamma Myt}$=0.16, while the average photon index measured using \borus\ is $\langle \Gamma_{\rm Borus} \rangle$=1.69, with associated standard deviation $\sigma_{\rm \Gamma Borus}$=0.15. Furthermore, 29 out of 35 sources (i.e., 83\% of the objects in our sample) have $\Gamma_{\rm Borus}$ consistent with $\Gamma_{\rm MyT}$ within the 90\% confidence uncertainty. We point out that measuring $\Gamma$ with such an high accuracy in CT-AGNs is a remarkable result, since in such obscured objects the intrinsic power-law continuum is not directly observable in any part of the spectrum below 5--10\,keV. 

In conclusion, our analysis shows that, when using \borus\ in a geometrical configuration consistent with the \myt\ one, the best-fit results from the two models are in good agreement. Such an evidence supports the main goal of this work, i.e., using \borus\ to measure the obscuring material covering factor for the 35 objects in our sample.

\begingroup
\renewcommand*{\arraystretch}{1.5}
\begin{table*}
\centering
\scalebox{1}{
\begin{tabular}{c|ccc|ccc}
\hline
\hline
&  \multicolumn{3}{c|}{\texttt{MYTorus}} & \multicolumn{3}{c}{\texttt{borus02}}\\ 
Source & $N_{\rm H, z}$ & $\Gamma$ & $\chi^2$/DOF & $N_{\rm H, z}$ & $\Gamma$ & $\chi^2$/DOF\\
&10$^{22}$ cm$^{-2}$ & & & 10$^{22}$ cm$^{-2}$ &\\
\hline
  NGC 424                & 130.6$_{-21.3}^{+42.5}$ & 1.51$_{-0.11l}^{+0.17}$ & 429.8/335 & 223.7$_{-33.9}^{+39.5}$ & 1.57$_{-0.14}^{+0.20}$ & 398.6/334  \\  
  MCG+08-03-018          & 47.9$_{-7.2}^{+7.4}$ & 1.84$_{-0.12}^{+0.10}$ & 176.9/151 & 46.9$_{-9.5}^{+10.9}$ & 1.90$_{-0.14}^{+0.15}$ & 169.5/150  \\  
  NGC 1068$^N$               & 1000.0$_{-142.8}^{+0.0u}$ & 1.88$_{-0.10}^{+0.08}$ & 602.5/552 & 1000.0$^f$ & 2.03$_{-0.08}^{+0.09}$ & 664.7/551  \\  
  NGC 1194               & 81.1$_{-7.9}^{+8.6}$ & 1.50$_{-0.09}^{+0.10}$ & 307.5/244 & 156.1$_{-18.1}^{+19.4}$ & 1.57$_{-0.05}^{+0.04}$ & 272.9/243  \\  
  NGC 1229               & 42.3$_{-6.0}^{+6.6}$ & 1.40$^f$ & 98.5/97  & 38.1$_{-5.8}^{+9.7}$  & 1.47$_{-0.07l}^{+0.14}$ & 95.4/96   \\  
  ESO 116-G018             & 190.0$_{-33.0}^{+76.0}$ & 1.55$_{-0.15l}^{+0.18}$ & 187.0/210 & 193.0$_{-28.0}^{+34.0}$ & 1.45$_{-0.05l}^{+0.41}$ & 190.0/209  \\  
  NGC 1358               & 236.0$_{-33.0}^{+27.0}$ & 1.82$_{-0.26}^{+0.24}$ & 220.0/239 & 248.0$_{-46.0}^{+40.0}$ & 1.67$_{-0.17}^{+0.17}$ & 230.0/238 \\  
  ESO 201-IG004          & 71.3$_{-13.1}^{+15.4}$ & 1.51$_{-0.11l}^{+0.14}$ & 108.7/84  & 56.8$_{-12.1}^{+12.7}$ & 1.60$_{-0.10}^{+0.14}$ & 104.7/83   \\  
  2MASXJ03561995-6251391 & 83.9$_{-10.5}^{+9.4}$ & 1.98$_{-0.16}^{+0.06}$ & 134.5/138 & 85.1$_{-11.2}^{+11.8}$ & 1.98$_{-0.17}^{+0.18}$ & 132.3/137  \\  
  CGCG 420-15            & 71.5$_{-9.7}^{+8.5}$ & 1.66$_{-0.12}^{+0.11}$ & 268.6/219 & 86.4$_{-9.3}^{+9.5}$  & 1.47$_{-0.06}^{+0.05}$ & 260.0/218  \\  
  MRK 3                  & 78.9$_{-2.1}^{+2.3}$ & 1.78$_{-0.02}^{+0.02}$ & 1183.4/1098 & 74.7$_{-1.7}^{+2.2}$  & 1.65$_{-0.03}^{+0.03}$ & 1142.4/1097 \\  
  ESO 005-G004           & 106.9$_{-21.2}^{+24.7}$ & 1.63$_{-0.08}^{+0.09}$ & 86.0/72  & 306.1$_{-64.2}^{+128.7}$& 1.54$_{-0.14l}^{+0.09}$ & 87.2/71   \\  
  MCG+06-16-028          & 104.7$_{-17.3}^{+17.0}$ & 1.56$_{-0.14}^{+0.13}$ & 82.3/87  & 85.0 $_{-12.3}^{+12.5}$ & 1.73$_{-0.16}^{+0.17}$ & 76.3/86   \\  
  2MASXJ09235371-3141305 & 67.3$_{-9.6}^{+9.6}$ & 1.76$_{-0.13}^{+0.09}$ & 193.7/145 & 62.8 $_{-7.0}^{+11.2}$ & 1.82$_{-0.10}^{+0.18}$ & 182.2/144  \\  
  NGC 3079               & 246.7$_{-23.5}^{+23.5}$ & 1.94$_{-0.10}^{+0.10}$ & 206.9/182 & 197.0$_{-18.7}^{+27.2}$ & 1.86$_{-0.08}^{+0.09}$ & 204.1/181  \\  
  NGC 3393               & 189.7$_{-16.5}^{+40.5}$ & 1.78$_{-0.12}^{+0.22}$ & 66.7/92  & 321.6$_{-93.1}^{+678.4}$& 1.77$_{-0.18}^{+0.11}$ & 54.8/91   \\  
  2MASXJ10523297+1036205 & 7.7$_{-2.0}^{+2.2}$ & 1.55$_{-0.05}^{+0.06}$ & 545.8/515 & 7.3$_{-1.6}^{+1.7}$  & 1.51$_{-0.06}^{+0.06}$ & 534.9/514  \\  
  RBS 1037               & $<$1.0 & 1.75$_{-0.05}^{+0.05}$ & 315.6/313 & $<$0.2  & 1.81$_{-0.06}^{+0.05}$ & 319.7/312  \\  
  MCG-01-30-041          & $<$1.0 &  1.85$_{-0.04}^{+0.05}$ & 222.2/225 & $<$1.5  & 1.83$_{-0.09}^{+0.10}$  & 220.1/224  \\  
  NGC 4102               & 77.8$_{-8.8}^{+9.5}$ & 1.67$_{-0.12}^{+0.12}$ & 217.5/191 & 69.1$_{-7.5}^{+7.3}$  & 1.73$_{-0.12}^{+0.09}$ & 204.1/190  \\  
  B2 1204+34             & 4.5$_{-0.9}^{+1.0}$ & 1.68$_{-0.06}^{+0.06}$ & 222.5/248 & 4.8$_{-0.9}^{+1.0}$  & 1.73$_{-0.06}^{+0.06}$ & 217.3/247  \\  
  NGC 4945               & 377.0$_{-15.7}^{+16.6}$ & 1.97$_{-0.06}^{+0.06}$ & 1508.8/1486& 338.3$_{-9.2}^{+11.1}$ & 1.80$_{-0.05}^{+0.05}$ & 1543.5/1485 \\  
  NGC 5100               & 22.6$_{-2.9}^{+2.6}$ & 1.68$_{-0.10}^{+0.10}$ & 191.4/197 & 21.1$_{-3.7}^{+3.7 }$ & 1.62$_{-0.10}^{+0.10}$  & 191.8 /196  \\  
  IGR J14175-4641        & 80.1$_{-12.9}^{+14.0}$ & 1.79$_{-0.14}^{+0.15}$ & 84.3/80  & 85.4$_{-17.4}^{+18.6}$ & 1.70$_{-0.19}^{+0.17}$ & 83.6/79   \\  
  NGC 5643               & 159.4$_{-29.8}^{+40.2}$ & 1.93$_{-0.10}^{+0.22}$ & 154.0/137 & 246.4$_{-80.2}^{+161.2}$& 1.47$_{-0.07l}^{+0.10}$  & 153.9/136  \\  
  MRK 477                & 22.4$_{-3.9}^{+4.4}$ & 1.65$_{-0.08}^{+0.08}$ & 245.9/227 & 21.6$_{-3.9}^{+4.3}$  & 1.60$_{-0.07}^{+0.08}$ & 260.2/226  \\  
  NGC 5728               & 142.3$_{-8.7}^{+8.6}$ & 1.88$_{-0.06}^{+0.06}$ & 362.4/329 & 123.0$_{-8.9}^{+10.3}$ & 1.77$_{-0.05}^{+0.05}$ & 363.6/328  \\  
  CGCG 164-019           & 119.5$_{-36.2}^{+50.0}$ & 1.78$_{-0.26}^{+0.29}$ & 59.9/58  & 147.7$_{-52.7}^{+48.6}$ & 1.80$_{-0.28}^{+0.23}$ & 59.6/57   \\  
  NGC 6232               & 59.3$_{-16.9}^{+27.9}$ & 1.44$_{-0.04l}^{+0.34}$ & 35.9/34  & 62.6$_{-19.8}^{+22.7}$ & 1.40$^f$  & 36.1/33   \\  
  NGC 6240               & 135.5$_{-6.4}^{+6.5}$ & 1.80$_{-0.05}^{+0.06}$ & 533.8/496 & 122.2$_{-6.6 }^{+7.2}$  & 1.74$_{-0.05}^{+0.04}$ & 520.1/495  \\  
  ESO 464-G016           & 84.8$_{-15.6}^{+17.3}$ & 1.88$_{-0.22}^{+0.24}$ & 67.9/76  & 83.9$_{-17.5}^{+18.9}$ & 1.71$_{-0.27}^{+0.29}$ & 67.0/75   \\  
  NGC 7130               & 221.8$_{-29.4}^{+42.4}$ & 1.50$_{-0.10l}^{+0.19}$ & 61.3/83  & 399.0$_{-137.2}^{+358.0}$& 1.45$_{-0.05l}^{+0.15}$ & 67.9/82   \\  
  NGC 7212               & 126.9$_{-24.5}^{+31.4}$ & 1.92$_{-0.17}^{+0.16}$ & 129.0/121 & 155.5$_{-27.0}^{+31.4}$ & 1.77$_{-0.11}^{+0.10}$  & 123.4/120  \\  
  NGC 7479               & 363.6$_{-50.3}^{+58.0}$ & 1.83$_{-0.09}^{+0.09}$ & 200.6/171* & 542.5$_{-209.2}^{+411.3}$ & 1.79$_{-0.20}^{+0.13}$ & 206.8/170* \\  
  NGC 7582               & 525.6$_{-130.7}^{+231.8}$ & 2.00$_{-0.04}^{+0.05}$ & 340.7/320 & 174.2$_{-21.0}^{+26.8}$ & 1.90$_{-0.05}^{+0.05}$ & 353.6 /319  \\   
\hline
\hline
\end{tabular}}\caption{\normalsize Best fit properties for the 35 candidate CT-AGN analyzed in this work, using either \myt\ or \borus\ in the \myt\ configuration (covering factor $f_c$=0.5, observing angle $\theta_{\rm Obs}$=87\degree). $N_{\rm H, z}$ is the AGN line-of-sight column density, in units of 10$^{22}$\,cm$^{-2}$; $\Gamma$ is the power law photon index. In NGC 4102, leaving $N_{\rm H, z, 2-10}$ free to vary with respect to $N_{\rm H, z, NuS}$ leads to a significant improvement of the fit: the reported value is the \nustar\ one. textbf{Sources flagged with $^N$  are objects where we fitted the \nustar\ data alone.} Parameters fixed to a given value are flagged with $^f$. 90\% confidence errors flagged with $l$ and $u$ indicate that the value is pegged at either the lower ($\Gamma$=1.4 in both \myt\ and \borus) or the upper ($N_{\rm H, z}$=10$^{25}$\,cm$^{-2}$ in \myt) boundary  of the parameter in the \myt\ model. For these sources, the reported values should therefore be treated as lower limits on the actual 90\% confidence uncertainties. Given their low quality, in NGC 7479 the \xmm\ and \nustar\ spectra are fitted using Cstat, rather than than $\chi^2$. The \swi\ data are instead already background subtracted and are therefore fitted using the $\chi^2$ statistic. The best-fit statistic values we report for NGC 7479, flagged with $^*$, are therefore the sum of the two.}\label{tab:results_myt_vs_borus}
\end{table*}
\endgroup

\section{Constraining the torus covering factor}\label{sec:cf}
As a preliminary caveat, we point out that the whole analysis reported in this Section is based on single-epoch \nustar\ observation: this represents a partial limitation to our approach, since it has been shown \citep[see, e.g.,][]{balokovic18}, that the analysis of multi-epoch \nustar\ observations allows one to place stronger constraints on the covering factor, simultaneously reducing potential degeneracies between parameters. Nonetheless, this work represents, as of today, the most complete analysis of single-epoch \nustar\ spectra of nearby, heavily obscured AGNs.

In using \borus\ with $f_c$ as a free parameter, we followed the approach adopted by \citet{balokovic18} when fitting single-epoch \nustar\ observations. Therefore, for each object we performed a set of 36 fits, in each of which we kept $N_{\rm H, tor}$ fixed: in each iteration, we increased  the log($N_{\rm H, tor}$) value by 0.1, starting from log($N_{\rm H, tor}$)=22 and stopping at log($N_{\rm H, tor}$)=25.5, i.e., the lower and upper boundaries of the parameter in \borus. We then assumed as best-fit $N_{\rm H, tor}$ the one corresponding to the minimum $\chi^2$ fit. Furthermore, we also report in Appendices \ref{app:nhz_vs_fc} and \ref{app:nhtor_vs_fc} the confidence contours of $f_c$ against the line-of-sight column density and log($N_{\rm H, tor}$), respectively. As can be seen, we find no evidence of degeneracy between $f_c$ and either of the two column densities, and in the vast majority of the cases we are able to reliably constrain all three parameters.

To avoid complications in the spectral fitting caused by potential degeneracies between the torus covering factor, $f_c$, the torus average column density, $N_{\rm H, tor}$, and the angle between the observer and the torus axis, $\theta_{\rm Obs}$, we also decided to fix this last parameter to its upper boundary, $\theta_{\rm Obs}$=87\degree, i.e.,  we assume that the torus is observed almost edge-on. Notably, this same approach was adopted in \citet{brightman15}, and allows one to explore all the possible covering factor solutions.

We report the best-fit spectra obtained using \borus\ having $f_c$ as a free parameter in Appendix \ref{app:borus_spectra}, and the corresponding best-fit parameters in Table \ref{tab:cf}: as can be seen, RBS 1037 is the only source in our sample for which $f_c$ is completely unconstrained. This is not an unexpected result, since RBS 1037 is one of the two unobscured AGNs in our sample. Notably, we find a similar result for the other unobscured AGN in our sample, MCG-01-30-041, for which we obtain a $f_c$ best-fit value, but with large uncertainties ($f_c$=0.62$_{-0.52l}^{+0.28}$). We also find that NGC 424 and NGC 1068 are best-fitted with two reprocessed components, having same covering factor but different $N_{\rm H, tor}$: notably, for both sources it has already been proposed in the literature either a reprocession-dominated scenario \citep[for NGC 424, see][]{balokovic14} or a multi-reprocessed component scenario \citep[for NGC 1068, see][]{bauer15}.

We also point out that NGC 6240 is a well known dual AGN \citep[see, e.g.][]{puccetti16}: in our analysis we use \xmm\ and \nustar\ data for this source, and we therefore do not resolve the two nuclei, thus implying that our measurement of $f_c$ for NGC 6240 should be treated as an average of $f_c$ for the two nuclei. Notably, our best-fit parameters are in excellent agreement with those obtained by \citet{puccetti16} fitting the combined \cha, \nustar\ and \xmm\ spectrum. Furthermore, to test the reliability of our assumption, we simulate two different \xmm+\nustar\ spectra of NGC 6240, using the best-fit parameters of the two cores, as observed with \cha\ and reported in \citet{puccetti16}. In one simulation, we fix the covering factor of the Southern core (three times more luminous than the the Northern one) to $f_{c,S}$=1, and that of the Northern core to $f_{c,N}$=0.11; in the second, we do the opposite (i.e., $f_{c,S}$=0.11 and $f_{c,N}$=1). In both simulations, we fix the average torus column density to log($N_{\rm H, tor}$)=24.2, i.e., the best-fit value we obtain in our analysis, We then fit the simulated spectra with a single-$f_c$ model: we find that in both cases the spectra are best-fitted by a model having $f_c$$\sim$0.5--0.6$\pm$0.1. This test suggests that the $f_c$ measurement we obtain for NGC 6240 can be treated as the average of the covering factors of the two nuclei: since for NGC 6240 we measure a covering factor  $f_c$=0.75$_{-0.24}^{+0.25}$, it is therefore likely that both AGNs have large covering factors.

\begin{figure*}
\begin{minipage}[b]{.33\textwidth}
  \centering
  \includegraphics[width=1.06\textwidth]{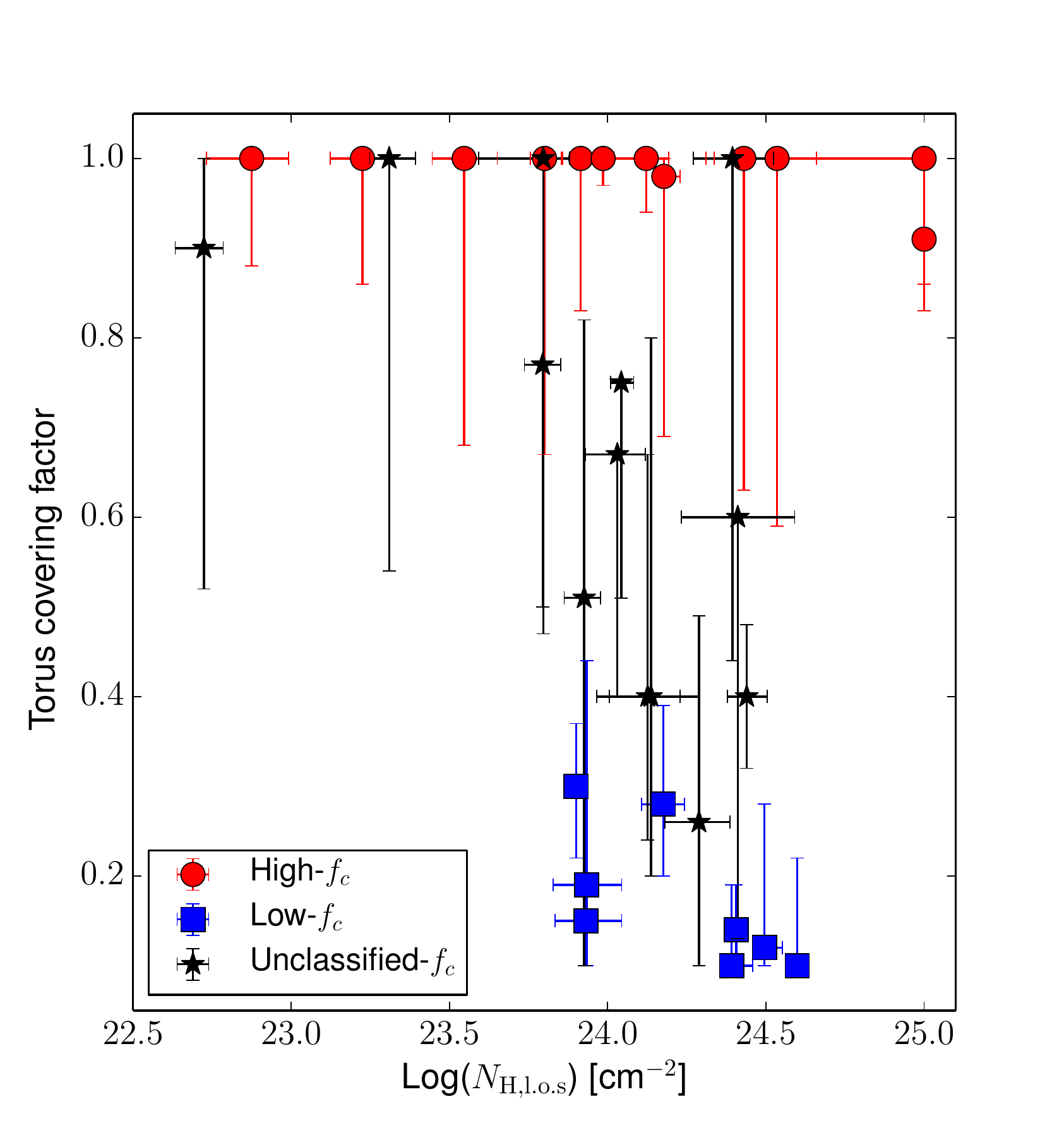}
  \end{minipage}
\begin{minipage}[b]{.33\textwidth}
  \centering
  \includegraphics[width=1.09\textwidth]{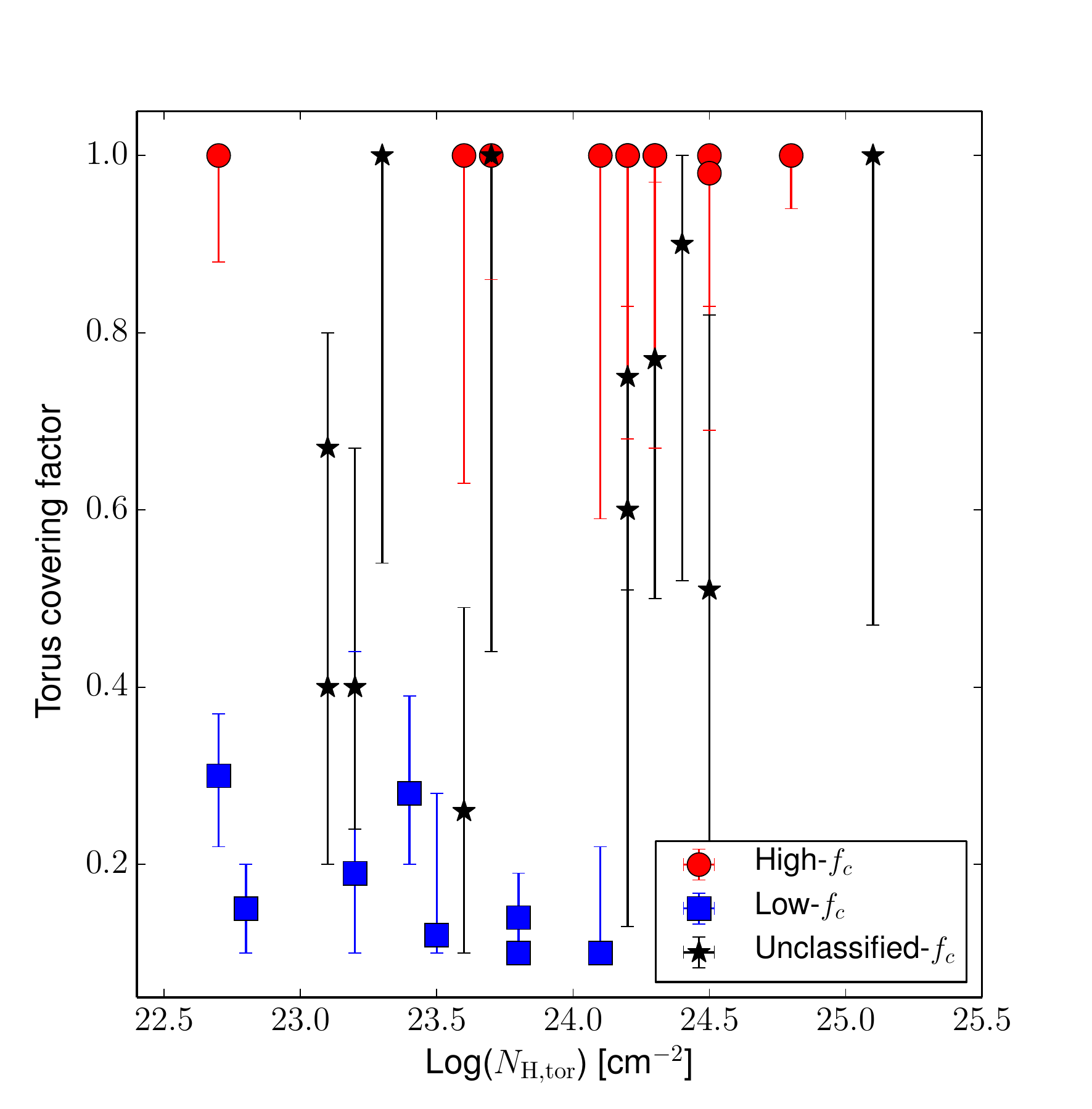}
  \end{minipage}
\begin{minipage}[b]{.33\textwidth}
  \centering
  \includegraphics[width=1.05\textwidth]{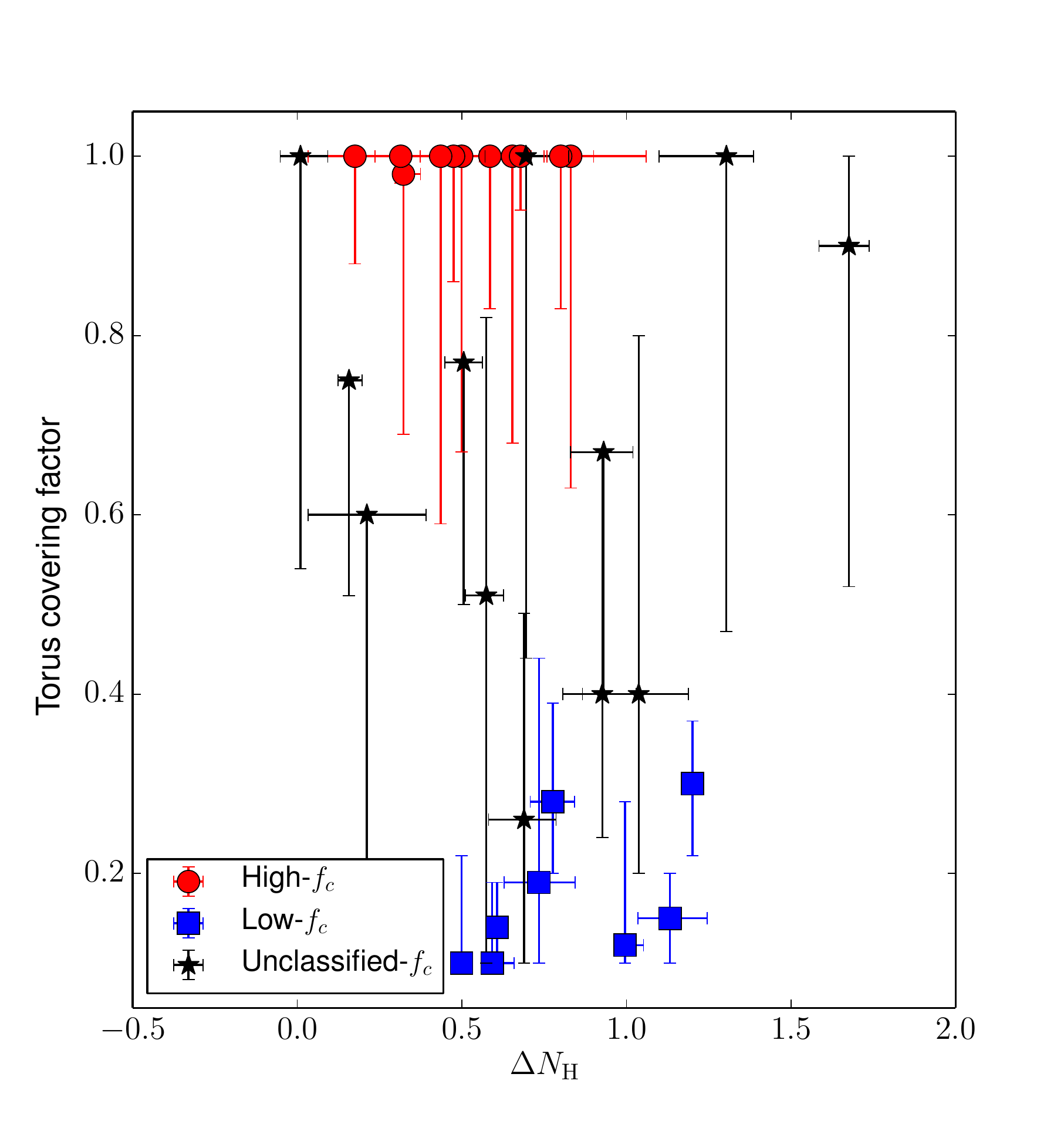}
  \end{minipage}
\caption{\normalsize Torus covering factor as a function of the line-of-sight column density ($N_{\rm H, z, l.o.s.}$, left), the torus average column density ($N_{\rm H, tor}$, center) and the difference between the logarithms of the two quantities ($\Delta$$N_{\rm H}$) for the sources in our sample having $N_{\rm H, z, l.o.s.}$$>$10$^{22}$\,cm$^{-2}$. Since NGC 424 and NGC 1068 are best fitted with two reprocessed components, having different $N_{\rm H, tor}$, we do not report them in the central and right panel. Sources with covering factor 90\% confidence lower boundary $>$0.55 are plotted as red circles, sources with with covering factor 90\% confidence upper boundary $<$0.45 are plotted as blue squares, and sources not belonging to either of the other two classes are plotted as black stars.}\label{fig:nh_vs_cf}
\end{figure*}

To further investigate the relation between $f_c$ and both $N_{\rm H, z, l.o.s.}$ and $N_{\rm H, tor}$, we divide the 33 obscured AGNs in our sample in three different classes, based on their covering factor best-fit value and 90\% confidence uncertainties:
\begin{enumerate}
\item High--$f_c$ sources (red circles in Figure \ref{fig:nh_vs_cf}): objects having 90\% confidence lower boundary $>$0.55. 12 sources belong to this group.
\item Low--$f_c$ sources (blue squares): objects having 90\% confidence upper boundary $<$0.45. 8 sources belong to this group.
\item Undefined--$f_c$ sources (black stars): objects which do not belong to any of the two previous classes, mostly because of their large uncertainties on $f_c$. 13 sources belong to this group.
\end{enumerate}

In Figure \ref{fig:nh_vs_cf}, left panel, we show the covering factor as a function of the line-of-sight column density: no clear trend can be immediately identified, especially around and above the Compton thick regime, where we observe both low-- and high-$f_c$ sources. Particularly, it is worth noting that all low-$f_c$ objects have $\log$($N_{\rm H, z, l.o.s.}$)$\geq$23.9.  However, as shown in Figure \ref{fig:nh_vs_cf}, central panel, all eight low--$f_c$ sources have best-fit $\log$($N_{\rm H, tor}$)$<$24.1, i.e., the average column density of the torus is not Compton thick in all sources but NGC 4945, where $N_{\rm H, tor}$ is just above the CT-threshold. 8 out of 11 high--$f_c$ sources (NGC 1068  is not included in this computation, since it has two different best-fit $N_{\rm H, tor}$) have instead  $\log$($N_{\rm H, tor}$)$>$24, i.e., their obscuring torus is on average Compton thick.

We parameterize the difference between the torus average column density and the line-of sight column density with 
\begin{equation}
\Delta N_{\rm H}=\lvert \log(N_{\rm H, tor})-\log(N_{\rm H, z, l.o.s.})\lvert.
\end{equation}
We find that in the low--$f_c$ sample the offset between the two column densities is large, being almost one order of magnitude ($\langle$$\Delta N_{\rm H,low-f_c}$$\rangle$=0.82, with standard deviation $\sigma$=0.25). In all eight sources, the average column density is smaller than the line-of-sight column density. Consequently, the low covering factor values measured in these objects can be linked to a ``patchy torus'' scenario, where the accreting SMBH is observed through an over-dense (with respect to the overall gas distribution) obscuring region.

On the basis of this result, the eight low--$f_c$ objects can be promising candidates for long-term monitoring campaigns, with the aim of detecting significant flux and $N_{\rm H, z, l.o.s.}$ variability. In fact, in a patchy torus the line-of-sight obscuration is caused by an over-dense, Compton-thick cloud located in a less-dense, Compton-thin environment. In such a scenario, a monitoring campaign can allow one to observe a significant flux and/or line-of-sight column density variation \citep[see, e.g., the case of NGC 1365 in][]{risaliti05}.
Notably, one of these eight sources, NGC 4945, is already known to show strong $>$10\,keV variability (see Section \ref{sec:cf_compare} for a more detailed discussion).

While all the low--$f_c$ sources have best-fit $\log$($N_{\rm H, tor}$)$<$24.1, eight out of 11 high--$f_c$ sources (i.e., 73\% of the high--$f_c$ subsample)  have best-fit $\log$($N_{\rm H, tor}$)$>$24. Interestingly, two out of the three objects with high $f_c$ and $\log$($N_{\rm H, tor}$)$<$24 are among the least obscured sources in our sample, namely 2MASXJ10523297+1036205 and MRK 477, thus suggesting a tighter correlation between $N_{\rm H, tor}$ and $N_{\rm H, z, l.o.s.}$ than the one observed in the low--$f_c$ subsample. In fact, the average offset between the two column densities is $\langle$$\Delta N_{\rm H,high-f_c}$$\rangle$=0.53, with standard deviation $\sigma$=0.20.

In conclusion, we find potential evidence of correlation between the torus covering factor and the difference between the average torus column density and the line-of-sight column density, supporting a scenario where sources with low $f_c$ are more likely to have a patchy torus, while sources with high $f_c$ are more likely to be obscured by a more uniform distribution of gas.

\subsection{Effects on $N_{\rm H, z, l.o.s.}$ and $\Gamma$ with respect to the \myt\ configuration}\label{sec:nh_change}
In 27 out of 35 sources (77.1\% of the whole sample) we find that leaving $f_c$ free to vary does not affect significantly the measurement of the line-of-sight column density, i.e., the measurement of $N_{\rm H, z, l.o.s.}$ with \borus\ in the free--$f_c$ configuration is consistent, within the 90\% confidence uncertainties, with the one obtained using \myt.

Six out of the eight sources for which instead we find a significant disagreement between the two $N_{\rm H, z, l.o.s.}$ measurements show a significant improvement in the fit statistics ($\Delta$$\chi^2$=[8.5--35.3]): we flag these sources with $^B$ in Table \ref{tab:cf}, and we assume that for these objects the measurements obtained with \borus\ should be treated as the best-fit solutions. This is particularly interesting for three of these six objects, namely MCG+08-03-018, ESO 201-IG004 and CGCG 420--15, which were found to have $\log$($N_{\rm H, z, l.o.s.}$)$<$24 using \myt\ and are instead re-classified as CT-AGNs using \borus, although only CGCG 420-15 has $\log$($N_{\rm H, z, l.o.s.}$)$>$24 at a $>$3\,$\sigma$ confidence level. Based on these new measurements, we find that 19 out of 35 candidate CT-AGNs (i.e., 54\% of the sources in our sample) are confirmed CT-AGNs, while other three sources (IGR J14175-4641, NGC 5728 and ESO 464-G016) have $\log$($N_{\rm H, z, l.o.s.}$)$<$24 but their 90\% confidence upper uncertainty lie above the CT threshold. In none of the eight sources with a significant discrepancy in the $N_{\rm H, z, l.o.s.}$ measurement the difference in line-of-sight column density can be explained exclusively by a discrepancy in the $\Gamma$ measurement.

Finally, there is a general excellent consistency in the photon indices measurements, independently from the geometrical configuration assumed in \borus:  in fact, 32 out of 35 sources (91.4\% of the whole sample) have $\Gamma$ in agreement within the 90\% confidence uncertainty. In one of the remaining three sources, NGC 4945, the disagreement is actually fairly small, being $\Delta$$\Gamma$/$\Gamma_{\rm min}$=0.06, where $\Gamma_{\rm min}$ is the smallest of the two photon index measurements. The remaining two objects (NGC 1194 and CGCG 420--15) also have different $N_{\rm H, z, l.o.s.}$, and in both cases the fit is significantly improved ($\Delta$$\chi^2$=10.3 and $\Delta$$\chi^2$=34.1, respectively) in the fit where $f_c$ is left free to vary.

\begingroup
\renewcommand*{\arraystretch}{1.5}
\begin{table*}
\centering
\scalebox{1}{
\begin{tabular}{ccccccc}
\hline
\hline
Source & $N_{\rm H, z, l.o.s.}$ & $\Gamma$ & $\log$($N_{\rm H, tor}$) & $f_c$ & $\log$(L$_{\rm 2-10keV}$) & $\chi^2$/DOF\\
&10$^{22}$ cm$^{-2}$ \\
\hline
  NGC 424                & 265.1$_{-35.8}^{+44.4}$ & 1.81$_{-0.05}^{+0.07}$ & 23.3, 24.3   & 0.40$_{-0.08}^{+0.08}$ & 42.74$_{-0.02}^{+0.02}$ & 398.6/335 \\
  MCG+08-03-018$^B$          & 107.3$_{-22.2}^{+24.4}$ & 2.17$_{-0.19}^{+0.26}$ & 23.1 & 0.67$_{-0.27}^{+0.33u}$  & 43.27$_{-0.26}^{+0.16}$ & 150.9/148 \\
  NGC 1068$^N$                & 1000$^f$ & 2.15$_{-0.05}^{+0.04}$ & 23.1, 24.7   & 0.91$_{-0.05}^{+0.08}$ & 42.38$_{-0.02}^{+0.02}$ &  574.0/551 \\
  NGC 1194$^B$               & 246.6$_{-8.3 }^{+40.6}$ & 1.86$_{-0.05}^{+0.09}$ & 23.8 & 0.10$_{-0.00l}^{+0.09}$ & 43.61$_{-0.09}^{+0.07}$ & 262.6/243 \\
  NGC 1229               & 35.2$_{-7.3 }^{+9.6 }$ & 1.57$_{-0.16}^{+0.14}$ & 24.2   & 1.00$_{-0.32}^{+0.00u}$  & 42.83$_{-0.47}^{+0.22}$ &  91.8/95  \\
  ESO 116-G018             & 313.0$_{-14.0}^{+43.0}$ & 1.59$_{-0.19l}^{+0.22}$ & 23.5   & 0.12$_{-0.02l}^{+0.16}$ & 43.30$_{-0.42}^{+0.47}$ & 189.0/209 \\
  NGC 1358                 & 255.0$_{-10.0}^{+14.0}$ & 1.60 $_{-0.07}^{+0.10}$ & 23.8 & 0.14$_{-0.04l}^{+0.05}$ & 43.40$_{-0.33}^{+0.21}$ & 231.0/238\\
  ESO 201-IG004$^B$          & 133.8$_{-32.4}^{+35.7}$ & 1.69$_{-0.21}^{+0.21}$ & 23.2   & 0.40$_{-0.16}^{+0.27}$ &  43.53$_{-0.42}^{+0.21}$ & 96.2/83  \\
  2MASXJ03561995-6251391 & 84.3$_{-11.4}^{+10.7}$ & 1.99$_{-0.26}^{+0.20}$ & 24.5   & 0.51$_{-0.41l}^{+0.31}$ & 44.55$_{-0.26}^{+0.16}$ & 132.2/137 \\
  CGCG 420-15$^B$            & 150.0$_{-21.9}^{+24.8}$ & 1.92$_{-0.13}^{+0.14}$ & 23.4   & 0.28$_{-0.08}^{+0.11}$ & 41.83$_{-0.12}^{+0.09}$ & 225.9/218 \\
  MRK 3                  & 79.5$_{-1.9}^{+2.9}$  & 1.68$_{-0.02}^{+0.03}$ & 22.7   & 0.30$_{-0.08}^{+0.07}$ & 44.04$_{-0.07}^{+0.06}$ & 1120.1/1097\\
  ESO 005-G004           & 248.1$_{-61.7}^{+85.9}$ & 1.47$_{-0.07l}^{+0.12}$ & 23.7   & 1.00$_{-0.56}^{+0.00u}$  & 41.87$_{-0.89}^{+0.27}$ & 86.1/71  \\
  MCG+06-16-028          & 82.2$_{-10.9}^{+10.9}$ & 1.83$_{-0.15}^{+0.15}$ & 24.5   & 1.00$_{-0.17}^{+0.00u}$  & 42.84$_{-0.31}^{+0.18}$ & 69.4/86  \\
  2MASXJ09235371-3141305 & 63.2$_{-6.1}^{+8.9}$  & 2.01$_{-0.13}^{+0.10}$ & 24.3   & 1.00$_{-0.33}^{+0.00u}$  & 43.57$_{-0.62}^{+0.41}$ & 178.5/144 \\
  NGC 3079$^B$               & 150.5$_{-9.0}^{+19.1}$ & 1.93$_{-0.10}^{+0.06}$ & 24.5   & 0.98$_{-0.29}^{+0.02u}$  & 41.83$_{-0.03}^{+0.03}$ & 194.6/181 \\
  NGC 3393               & 257.8$_{-86.7}^{+132.3}$ & 1.72$_{-0.13}^{+0.22}$ & 24.2   & 0.60$_{-0.47}^{+0.40u}$  & 42.75$_{-0.09}^{+0.07}$ & 55.3/91  \\
  2MASXJ10523297+1036205 & 7.5$_{-2.1}^{+2.3}$  & 1.51$_{-0.04}^{+0.04}$ & 22.7   & 1.00$_{-0.12}^{+0.00u}$  & 43.90$_{-0.11}^{+0.09}$ & 521.1/513 \\
  RBS 1037               & $<$0.1  & 1.85$_{-0.07}^{+0.04}$ & 23.6   & --  & 42.71$_{-0.02}^{+0.02}$ & 317.2/312 \\
  MCG-01-30-041          & $<$1.7  & 1.85$_{-0.09}^{+0.09}$ & 24.8   & 0.62$_{-0.52l}^{+0.30}$ & 43.90$_{-0.04}^{+0.03}$ & 219.9/224 \\
  NGC 4102               & 62.3$_{-7.7}^{+8.7}$  & 1.75$_{-0.11}^{+0.09}$ & 24.3   & 0.77$_{-0.27}^{+0.23u}$  & 41.29$_{-0.06}^{+0.06}$ & 201.3/190 \\
  B2 1204+34             & 5.3$_{-1.0}^{+0.8}$  & 1.86$_{-0.13}^{+0.05}$ & 24.4   & 0.90$_{-0.38}^{+0.10u}$  & 43.65$_{-0.19}^{+0.13}$ & 214.5/247 \\
  NGC 4945$^N$               & 397.2$_{-15.2}^{+15.2}$  & 1.95$_{-0.06}^{+0.06}$ & 24.1   & 0.10$_{-0.00l}^{+0.12}$  & 42.33$_{-0.04}^{+0.04}$ &  1404.1/1398\\
  NGC 5100               & 20.4$_{-2.7}^{+4.3}$  & 1.57$_{-0.07}^{+0.11}$ & 23.3   & 1.00$_{-0.46}^{+0.00u}$  & 42.99$_{-0.54}^{+0.23}$ & 189.7/196 \\
  IGR J14175-4641        & 85.9$_{-18.6}^{+24.9}$ & 1.70$_{-0.19}^{+0.21}$ & 23.2   & 0.19$_{-0.09l}^{+0.25}$ & 43.96$_{-0.73}^{+0.34}$ & 83.6/79  \\
  NGC 5643               & 269.4$_{-65.0}^{+187.5}$ & 1.55$_{-0.15}^{+0.13}$ & 23.6   & 1.00$_{-0.37}^{+0.00u}$  & 41.42$_{-0.06}^{+0.06}$ & 150.0/137 \\
  MRK 477                & 16.8$_{-3.5}^{+3.7}$  & 1.60$_{-0.06}^{+0.06}$ & 23.7   & 1.00$_{-0.14}^{+0.00u}$  & 43.09$_{-0.05}^{+0.04}$ & 223.0/223 \\
  NGC 5728$^B$               & 96.8$_{-3.3}^{+5.2}$  & 1.81$_{-0.04}^{+0.07}$ & 24.3   & 1.00$_{-0.03}^{+0.00u}$  & 42.74$_{-0.12}^{+0.09}$ & 327.1/328 \\
  CGCG 164-019           & 137.1$_{-44.6}^{+57.1}$ & 1.79$_{-0.32}^{+0.36}$ & 23.1   & 0.40$_{-0.20}^{+0.40}$  & 42.43$_{-0.28}^{+0.17}$ & 59.3/57  \\
  NGC 6232               & 62.6$_{-23.5}^{+13.2}$ & 1.40$^f$ & 25.1   & 1.00$_{-0.53}^{+0.00u}$  & 41.89$_{-0.30}^{+0.18}$ & 32.8/33  \\
  NGC 6240               & 110.4$_{-8.3}^{+10.4}$ & 1.75$_{-0.04}^{+0.03}$ & 24.2   & 0.75$_{-0.24}^{+0.25u}$  & 43.58$_{-0.02}^{+0.02}$ & 518.7/495 \\
  ESO 464-G016           & 85.5$_{-17.2}^{+25.3}$ & 1.67$_{-0.27}^{+0.29}$ & 22.8   & 0.15$_{-0.05l}^{+0.05}$  & 43.17$_{-0.98}^{+0.70}$ & 67.2/74  \\
  NGC 7130               & 343.1$_{-126.3}^{+656.9}$ & 1.40$^f$ & 24.1   & 1.00$_{-0.41}^{+0.00u}$  & 42.30$_{-0.06}^{+0.06}$ & 66.0/82  \\
  NGC 7212               & 194.4$_{-42.7}^{+48.8}$ & 1.99$_{-0.22}^{+0.30}$ & 23.6   & 0.26$_{-0.16l}^{+0.23}$ & 43.63$_{-0.09}^{+0.08}$ & 120.6/120 \\
  NGC 7479               & 132.4$_{-27.7}^{+23.8}$ & 1.64$_{-0.19}^{+0.14}$ & 24.8   & 1.00$_{-0.06}^{+0.00u}$  & 42.02$_{-0.19}^{+0.15}$ & 167.2/170  \\
  NGC 7582               & 1000.0$^f$ & 1.96$_{-0.05}^{+0.03}$ & 24.2   & 1.00$_{-0.17}^{+0.00u}$  & 42.53$_{-0.08}^{+0.07}$ & 316.6/317 \\
\hline
\hline
\end{tabular}}\caption{\normalsize Best fit properties for the 35 candidate CT-AGNs analyzed in this work, using \borus\ allowing the covering factor, $f_c$, to vary. $N_{\rm H, z}$ is the torus line-of-sight column density, in units of 10$^{22}$\,cm$^{-2}$; $\Gamma$ is the power law photon index. $\log$($N_{\rm H, tor}$) is the logarithm of the torus overall column density (in units of cm$^{-2}$), i.e., the $N_{\rm H, tor}$ value for which we obtain the smallest $\chi_\nu$ value (see the text for more details), $f_c$ is the torus covering factor  and $\log$(L$_{\rm 2-10keV}$ is) the logarithm of the intrinsic, absorption corrected 2--10\,keV luminosity (in units of erg s$^{-1}$. Sources flagged with $^B$ are objects for which the $N_{\rm H, z}$ measurement differs significantly from the one obtained using \borus\ in the \myt\ geometrical configuration, and for which the $\chi^2$ of the fit is significantly improved (see Section \ref{sec:nh_change}), while sources flagged with $^N$  are objects where we fitted the \nustar\ data alone.
Parameters fixed to a given value are flagged with $^f$. 90\% confidence errors flagged with $l$ and $u$ indicate that the value is pegged at either the lower ($\Gamma$=1.4; $f_c$=0.1) or the upper ($f_c$=0.91) boundary of a given parameter. For these sources, the reported values should therefore be treated as lower limits on the actual 90\% confidence uncertainties. Finally, $f_c$ is unconstrained in RBS 1037, one of the two unobscured AGN in our sample.}\label{tab:cf}
\end{table*}
\endgroup

\subsection{Comparison with previous results}\label{sec:cf_compare}
NGC 4945 is one of the sources in our sample for which several measurements of the covering factor are reported in literature. Particularly, several works \citep{madejski00,done03,yaqoob12,puccetti14} explained the significant variability observed in NGC 4945 above $>$10\,keV, as well as the source weak reprocessed component, as two indicators of a low--$f_c$ scenario for this object. However, \citet{brightman15}, using the \texttt{BNTorus} model \citep{brightman11} and its modified version with $f_c$=1, \texttt{sphere}, found a high covering factor for NGC 4945, even consistent with $f_c$=1. However, as pointed out by \citet{liu15}, the \texttt{BNTorus} presents some issues in properly treating the AGN reprocessed component, therefore potentially leading to unreliable $f_c$ measurements. The existence of this issue was then confirmed in \citet{balokovic18}, and properly taken into account when developing \borus.
 
\begin{figure*}
\begin{minipage}[b]{.5\textwidth}
  \centering
  \includegraphics[width=1.05\textwidth]{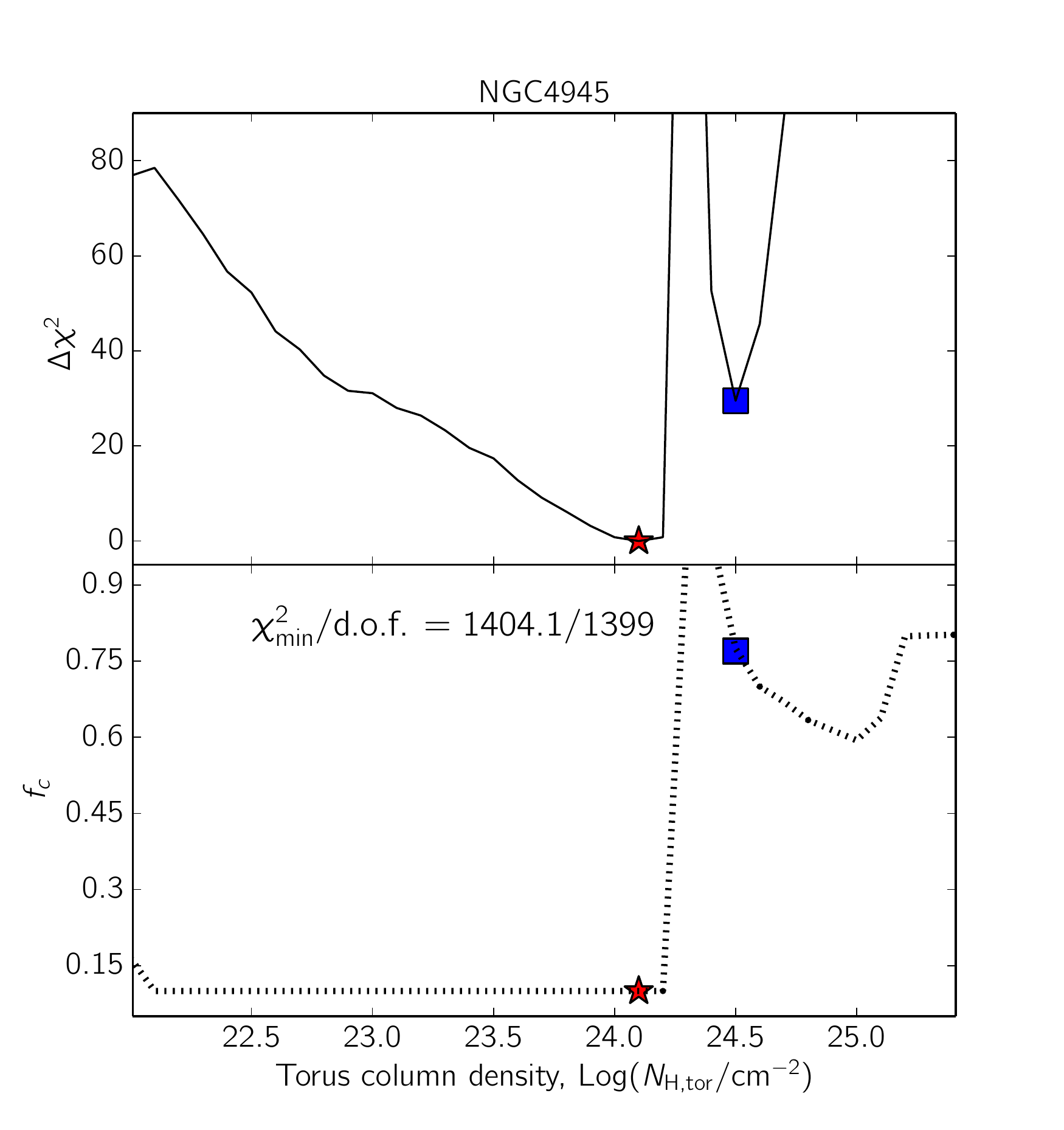}
  \end{minipage}
\begin{minipage}[b]{.5\textwidth}
  \centering
  \includegraphics[width=1.06\textwidth]{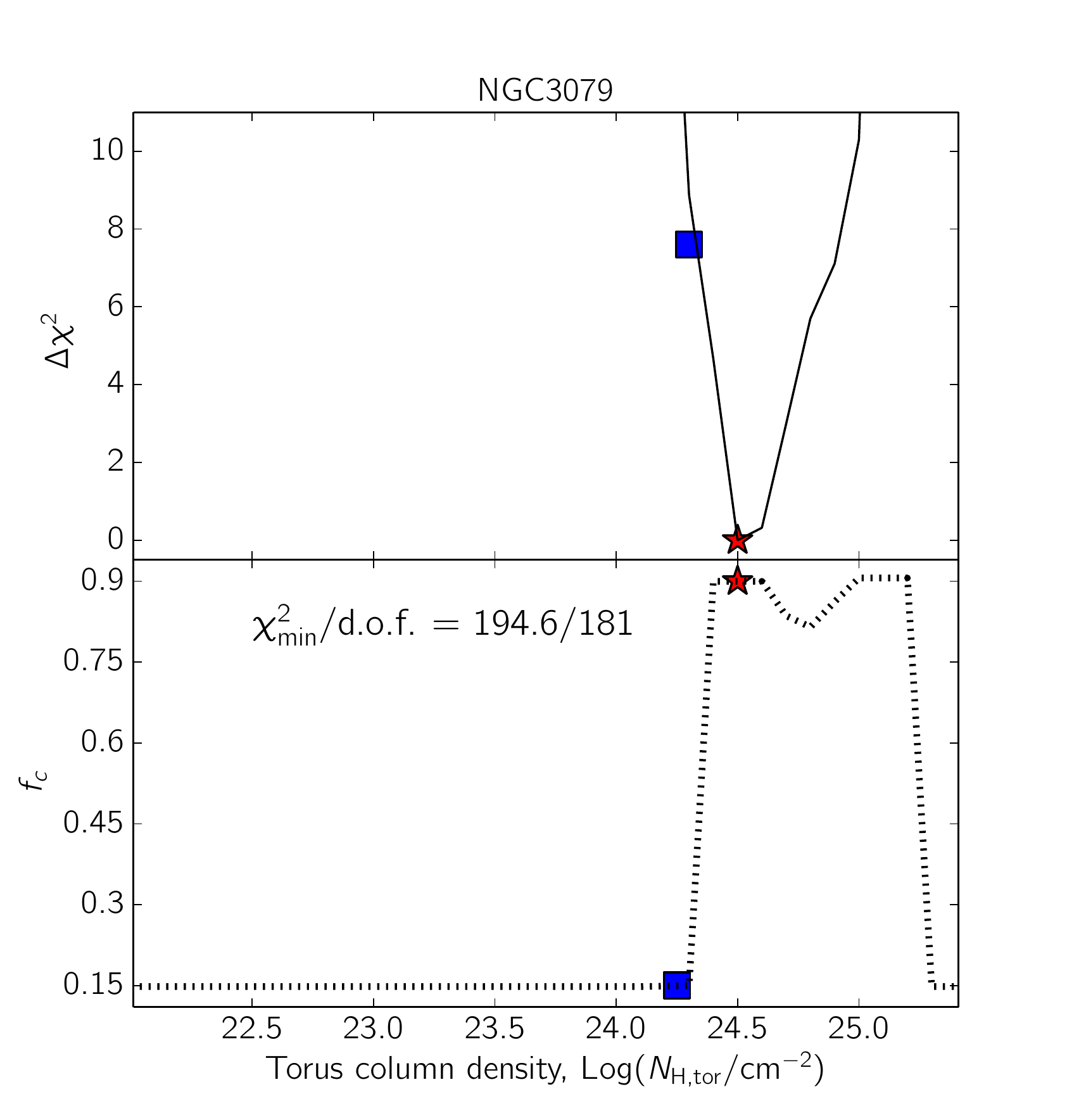}
  \end{minipage}
\caption{\normalsize Estimate of the torus covering factor in NGC 4945 (left) and NGC 3079 (right). Top panel: difference between the best-fit, minimum $\chi^2$ ($\chi^2_{\rm min}$) and the $\chi^2$ associated to $\log$($N_{\rm H, tor}$), as a function of the torus average column density. Bottom panel: torus covering factor as a function of the torus average column density. In both panels, we plot as a red star the combination of parameters associated to $\chi^2_{\rm min}$, and as a blue square the potential alternative $f_c$ solution we discuss in the text.}\label{fig:logNH_vs_chi_cf}
\end{figure*}

In our analysis with \borus\ we find that the best-fit covering factor is pegged at the lower limit of the model, $f_c$=0.10$_{-0.00}^{+0.12}$, in excellement agreement with the previous results reported in the literature. In Figure \ref{fig:logNH_vs_chi_cf}, left panel, we show how both $f_c$ and the best-fit $\chi^2$ vary as a function of the torus average covering factor: for $\log$($N_{\rm H, tor}$)$\leq$24.1 a low--$f_c$ solution is preferred, the model best-fit $\chi^2$ regularly decreasing since reaching a minimum at $\log$$N_{\rm H, tor}$=24.1, where the best-fit statistic is $\chi_\nu$=$\chi^2$/d.o.f.=1404.1/1399=1.00. At $\log$$N_{\rm H, tor}$$\geq$24.2 we instead enter in the high--$f_c$ solution regime, which has a significantly worse best-fit statistics: in this regime, we find a local $\chi^2$ minimum at $\log$$N_{\rm H, tor}$=24.5, where $f_c$=0.78$_{-0.09}^{+0.13}$. However, while this potential high-$f_c$ solution has a reasonable reduced $\chi^2$,  $\chi_\nu$=$\chi^2$/d.o.f.=1433.6/1399=1.03, the difference in $\chi^2$ between the two solutions, $\Delta$$\chi^2$=29.5, suggests that a low--$f_c$ solution is favored by our data.

Since NGC 4945 is known to be variable in intrinsic luminosity, and therefore the covering factor measurement is partially dependent on the observation used for the analysis, in our analysis we fitted the \nustar\ data alone, following the approach adopted by \citet{brightman15}. Nonetheless, even multi-epoch studies with \borus\ (Balokovi{\'c} et al. 2019 in prep.) find a low--$f_c$ solution ($f_c$=0.3$\pm$0.1). In conclusion, our analysis favors a low-$f_c$ solution for NGC 4945, in agreement with several other works. This result also confirms that the issue in \texttt{BNTorus} reported by \citet{liu15}, i.e., the tendency to overestimate the strength of the reprocessed component in an edge--on configuration, has been properly taken into account in \borus.  

Another source, NGC 3079, was also part of the sample studied by \citet{brightman15}. One of their \texttt{BNTorus} best-fit solutions implies a small covering factor, $f_c$=0.18$_{-0.03}^{+0.16}$, but  they also reported a potential second solution with $f_c$$>$0.90, and fitting the data with the \texttt{sphere} model, i.e., assuming $f_c$=1, also led to a slightly improved best-fit statistics, from $\chi^2_\nu$=1.33 to $\chi^2_\nu$=1.31. In our work, we find that the best-fit covering factor, corresponding to a torus average column density $\log$($N_{\rm H, tor}$)=24.5, is $f_c$=0.90$_{-0.21}^{+0.01u}$, in good agreement with the second of the solutions reported in \citet{brightman15}; the reduced $\chi^2$ of this solution is $\chi^2_\nu$=194.2/181=1.07. However, as we show in Figure \ref{fig:logNH_vs_chi_cf}, right panel, assuming a slightly lower average torus column density, $\log$($N_{\rm H, tor}$)=24.3, leads to a small covering factor solution, $f_c$=0.15$_{-0.02l}^{+0.05}$; the reduced $\chi^2$ of this second solution is $\chi^2_\nu$=201.8/181=1.11, with a $\chi^2$ difference $\Delta$$\chi^2$=7.6.
Consequently, we find that a high covering factor solution for NGC 3079 is favored by our data, but a small covering factor cannot be ruled out at a $>$3$\sigma$ level.

Finally, the covering factor of NGC 7582 was recently measured by \citet{balokovic18}, using \borus\ and fitting the \nustar\ data alone. As expected, since they used the same model we used in our analysis, their results are in excellent agreement with ours: they find a best-fit covering factor $f_c$=0.9 and a torus average column density $\log$($N_{\rm H, tor}$)=24.5, while our best-fit result, obtained combining \nustar\ and \xrt\ data, is $f_c$=0.91$_{-0.08}^{+0.00u}$ for $\log$($N_{\rm H, tor}$)=24.2. We point out that even assuming $\log$($N_{\rm H, tor}$)=24.5, as reported in \citet{balokovic18}, leads to a high covering factor, $f_c$=0.9, although with a significantly worse best-fit $\chi^2$.

Notably, the torus covering factor can be measured also using dusty torus models based on the mid-infrared (mIR; $\sim$7.5--13.5\,$\mu$m) spectral energy distribution (SED) fitting. Most of these models assume a clumpy distribution of the obscuring material \citep[see, e.g., ][]{rowan95,nenkova02,nenkova08,honig10,garcia17}, thus differing from both \myt\ and \borus, where the obscuring material is assumed to be uniformly distributed. 

\citet{alonso11} used the \texttt{CLUMPY} model \citet{nenkova08} to fit the infrared SED and measure the torus properties of a sample of 13 nearby Seyfert galaxies, among which there are also two objects we study in this work, NGC 1068 and NGC 7582. We find that for both sources there is a remarkable agreement between the $f_c$ value measured from the X-ray spectral fitting and the one inferred from the IR SED-fitting: more in detail, NGC 1068 has $f_{c,X}$=0.40$\pm$0.01 and $f_{c,IR}$=0.30$^{+0.11}_{-0.08}$, while NGC 7582 has $f_{c,X}$=0.91$_{-0.08}^{+0.00u}$ and $f_{c,IR}$=0.83$^{+0.06}_{-0.14}$.

\subsection{Covering factor trend with 2-10\,keV luminosity}
\begin{figure*}
\begin{minipage}[b]{.5\textwidth}
  \centering
  \includegraphics[width=1.05\textwidth]{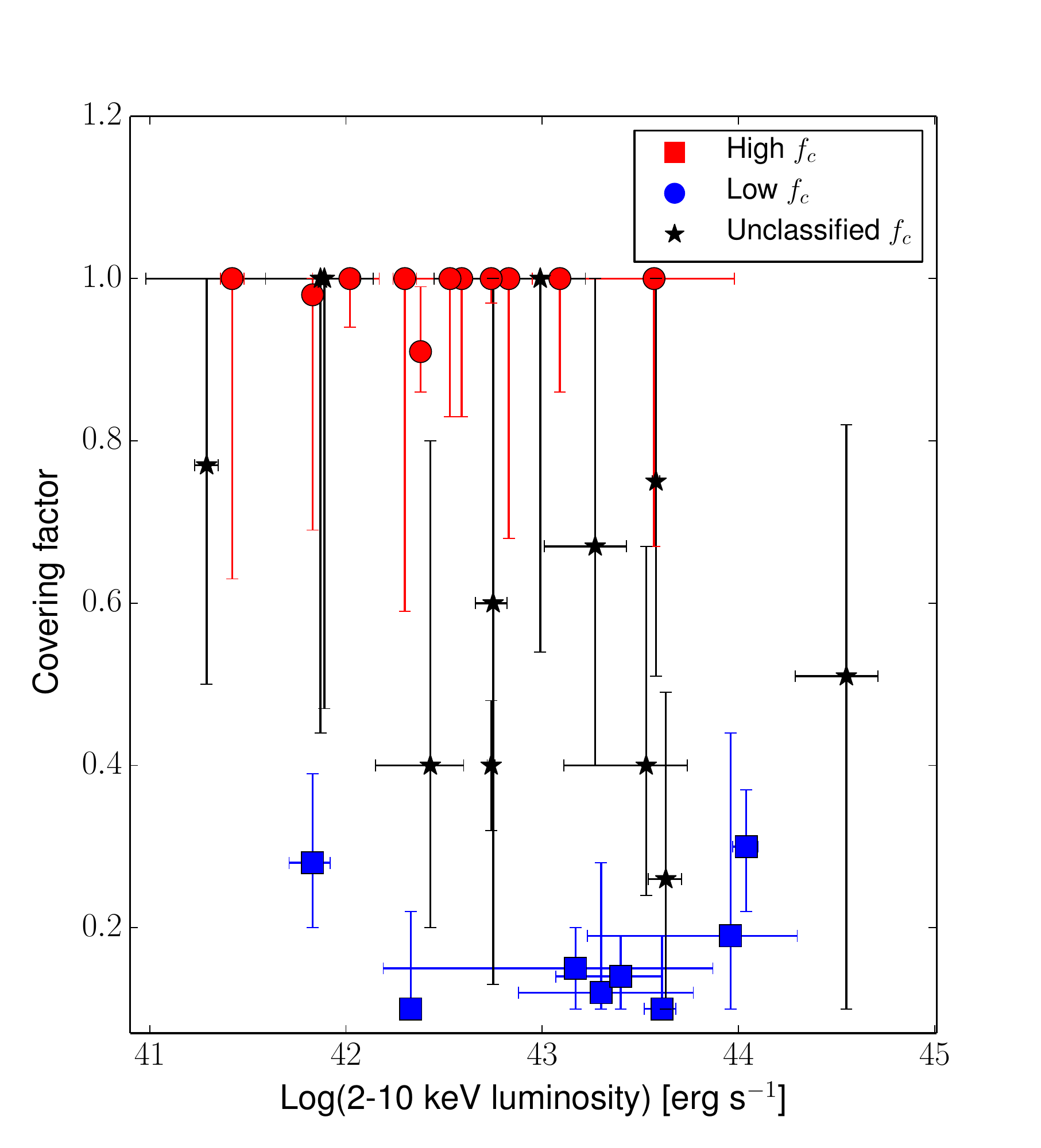}
  \end{minipage}
\begin{minipage}[b]{.5\textwidth}
  \centering
  \includegraphics[width=1.06\textwidth]{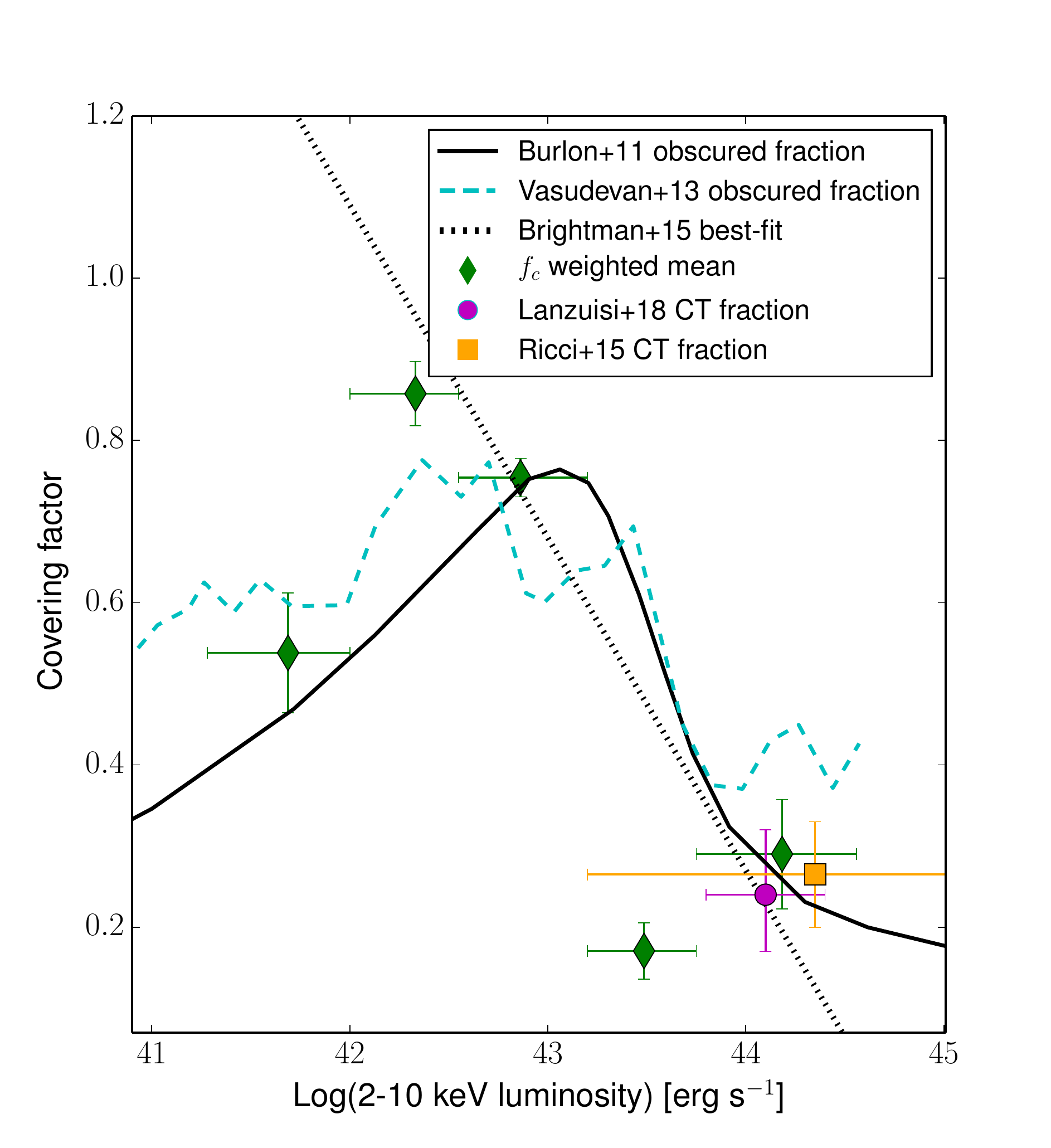}
  \end{minipage}
\caption{\normalsize \textit{Left}: torus covering factor, $f_c$, as a function of the intrinsic, absorption-corrected 2--10\,keV luminosity for the 31 sources in our sample having $\log$($N_{\rm H, z}$)$\geq$23. Objects with covering factor 90\% confidence lower boundary $>$0.55 are plotted as red circles, sources with with covering factor 90\% confidence upper boundary $<$0.45 are plotted as blue squares, and sources not belonging to either of the other two classes are plotted as black stars.
\textit{Right}: same as left, but this time plotting $f_c$ in five different luminosity bins (green diamonds). The best-fit to the same relation computed by \citet{brightman15} for a sample of 8 CT-AGNs is plotted as black dotted line. Measurements of the obscured fraction of AGNs with respect to the total population as computed by \citet[][black solid line]{burlon11} and \citet[][cyan dashed line]{vasudevan13} in samples of BAT-selected AGNs in the nearby Universe are also plotted for comparison, together with the CT fraction measurements made by \citet[][magenta circle]{lanzuisi18} and \citet[][orange square]{ricci15}
}\label{fig:lx_vs_cf}
\end{figure*}

In Figure \ref{fig:lx_vs_cf}, left panel, we plot the torus covering factor as a function of the de-absorbed 2--10\,keV luminosity for the 31 objects in our sample with $\log$($N_{\rm H, z}$)$\geq$23. As in Figure \ref{fig:nh_vs_cf}, the sources are divided in high--$f_c$ objects (red circles), low--$f_c$ objects (blue squares) and undefined--$f_c$ objects (black stars). Since no clear trend between the two quantities is immediately visible in our sample, we measure the difference in average L$_{\rm 2-10keV}$ between the high-- and the low--$f_c$ sample. The eleven high--$f_c$ sources with $\log$($N_{\rm H, z}$)$\geq$23 have average 2--10\,keV luminosity $\langle \log(L_{\rm 2-10keV,H-f_c})\rangle$=42.84, with standard deviation $\sigma_{\rm \log(L2-10),Hf_c}$=0.57. The eight low--$f_c$ sources with $\log$($N_{\rm H, z}$)$\geq$23, instead, have $\langle \log(L_{\rm 2-10keV,L-f_c})\rangle$=43.58, with $\sigma_{\rm \log(L2-10),Hf_c}$=0.72. When we perform a KS-test, we find that the hypothesis that the two $L_{\rm 2-10keV}$ samples are drawn from the same population can be rejected at the $\sim$2.5\,$\sigma$ level (p-value=0.018), a (marginal) evidence of the existence of different luminosity trends in high-- and low--$f_c$ sources.

While the overall distribution has large dispersion and only a marginal visible trend, when we compute the weighted average of $f_c$ in five different bins we find significant evidence of anti-correlation in our sample: as can be seen in Figure \ref{fig:lx_vs_cf}, right panel, at $\log$(L$_{\rm 2-10keV}$)$<$43 the average covering factor (green diamonds) value is $f_c$$\sim$0.6--0.8, while at $\log$(L$_{\rm 2-10keV}$)$>$43 the average covering factor value drops to $f_c$$\sim$0.2. At $\log$(L$_{\rm 2-10keV}$)$>$42.5, these results are in reasonable agreement with the trend reported in \citet{brightman15} using a sample of 8 CT-AGNs fitted with \texttt{BNTorus}, and plotted as a black dotted line. At lower luminosities, instead, we find an average covering factor $\langle f_c \rangle$$\sim$0.7, in significant disagreement with the expected $f_c$$\sim$1 reported by \citet{brightman15}.
We remind that our results are obtained with a sample 3.5 times larger than the \citet{brightman15} one, and that we made use of the \borus\ model, rather than the \texttt{BNTorus} one.

In Figure \ref{fig:lx_vs_cf}, right panel, we also show the fraction of obscured AGN, $f_{\rm obs}$=N$_{\rm obs}$/N$_{\rm tot}$, as a function of L$_{\rm 2-10keV}$, computed by \citet[][black solid line]{burlon11} and \citet[][magenta dashed line]{vasudevan13}. The \citet{burlon11} obscured fraction has been computed dividing the 15--55\,keV luminosity function (XLF) of the obscured AGN by the overall XLF, while the \citet{vasudevan13} one is derived directly counting the number of obscured AGN with respect to the whole population, in each bin of luminosity. As can be seen, the two curves show a fair agreement over the whole luminosity range, peaking at log(L$_{\rm 2-10keV}\sim$42.5--43 and declining significantly both at lower and at higher luminosities. A trend similar to the one observed in these two works has also been recently observed by \citet{ricci17b}, using a complete sample of  731 AGNs from the 70--month \swi\ catalog. The existence of a luminosity-dependent covering factor in CT-AGNs was also mentioned in \citet{boorman18}, as a potential cause for the observed anti-correlation between the Fe K$\alpha$ equivalent width and the AGN bolometric luminosity. 

Both curves have been computed from a sample of bright BAT-selected AGNs in the local Universe, and can therefore be compared with our results, since our sample has been selected in the same way. Furthermore, the $f_c$ of a source is also an indicator of the probability to observe that source as obscured (the higher the covering factor, the higher the probability), therefore $f_c$ and $f_{\rm obs}$ may in principle be directly compared. 
As for the anti-correlation between $f_c$ and $L_{\rm 2-10keV}$, we find some tentative agreement between our weighted average data and the measured fractions of obscured AGNs, particularly with the \citet{vasudevan13} one.

Finally, in Figure \ref{fig:lx_vs_cf}, right panel, we also plot the intrinsic fraction of CT-AGNs derived in two different works: the magenta circle is the measurement in the redshift range 0.04$<$$z$$<$1 made by \citet{lanzuisi18}, which computed the intrinsic CT fraction using the \cha\ \leg \citep{civano16,marchesi16a} AGN sample; the orange square is instead the one derived by \citet{ricci15} using the 70-month BAT catalog and assuming a torus opening angle $\theta_{\rm OA}$=60\degree. As can be seen, both measurements are in good agreement with both our result and the obscured AGN fractions measured by \citet{burlon11} and \citet{vasudevan13}.

\vspace{0.4cm}

\section{Conclusions}\label{sec:concl}
In this work, we analyzed the combined 2--100\,keV spectra of 35 AGNs selected in the 100-month BAT catalog. These objects have been selected among those candidate CT sources in the 100-month BAT catalog having an archival \nustar\ observation. In this work, we used only single-epoch \nustar\ observations. 30 out of 35 sources were already analyzed in \citet{marchesi18} using \myt. We report the main results of our analysis.

\begin{enumerate}
\item Among the five sources not studied in M18, three (ESO 116-G018, NGC 1358 and NGC 7479) are confirmed to be CT-AGNs at a $>$3\,$\sigma$ level. A fourth object, Mrk 3, is found to have best-fit line-of-sight column density $N_{\rm H, z}$=(7.8$\pm$0.1)$\times$10$^{23}$\,cm$^{-2}$, slightly below the CT threshold: this source is known to be highly variable, having line-of-sight column density varying in the range $N_{\rm H, z}$=[0.75--0.94]\,$\times$10$^{24}$\,cm$^{-2}$ in a timespan of seven months. Finally, we find that MCG-01-30-041, a candidate CT-AGN reported by \citet{vasudevan13}, is in fact an unobscured AGN ($N_{\rm H, z}$$<$10$^{22}$\,cm$^{-2}$). This discrepancy is likely caused by the low quality of the \xrt\ and \swi\ data fitted by \citet{vasudevan13}.
\item For all the 35 sources in our sample, we compared the best-fit line-of-sight column density obtained using \myt\ with those obtained using the new \borus\ with the same geometrical configuration of \myt, i.e., fixing the torus covering factor to $f_c$=0.5. We find that there is a general excellent agreement between the $N_{\rm H, z}$ value obtained using \borus\ and the one obtained using \myt\ (see Figure \ref{fig:trends}, left panel). While this trend is driven by the non-CT population, the weaker, albeit existent correlation observed in CT sources is not unexpected, given the increasing complexity in properly constrain the AGN spectral parameters in the CT-regime since above the CT threshold the $N_{\rm H, z}$. Nonetheless, 11 out of 16 sources confirmed CT-AGNs have $N_{\rm H, z}$ values in agreement at the 90\% confidence level.
\item We find an overall remarkable agreement between the photon indeces obtained using \myt\ and those measured with \borus\ in the same geometrical configuration of \myt\ (see Figure \ref{fig:trends}, right panel).
29 out of 35 sources have $\Gamma_{\rm Borus}$ consistent with $\Gamma_{\rm MyT}$ within the 90\% confidence uncertainty.
\item After validating \borus\ showing its excellent agreement with \myt, we used it to measure the torus covering factor and average column density for the 35 objects in our sample. We find 12 high--$f_c$ sources, i.e.,  objects having 90\% confidence lower boundary $>$0.55; 8 low--$f_c$ sources, i.e., objects having 90\% confidence lower boundary $<$0.45; and 15 undefined--$f_c$ sources, i.e., objects do not belonging to either of the two groups.
\item We find a tentative evidence of different trends between $f_c$ and the difference between the average torus column density and the line-of-sight column density:  the offset is larger in low--$f_c$ objects
, where the average column density is always smaller than the line-of-sight column density, than in high--$f_c$ objects.
These results are consistent with a scenario where low--$f_c$ AGNs  are more likely to have a patchy torus, while high--$f_c$ AGNs are more likely to be obscured by a more uniform distribution of gas.
\item In 6 out of 35 sources, leaving $f_c$ to vary leads to a significant variation in the line-of-sight column density measurement, and a corresponding significant improvement in the best-fit $\chi^2$ value. Interestingly, three of these objects (MCG+08-03-018, ESO 201-IG004 and CGCG 420--15), which were found to have $\log$($N_{\rm H, z, l.o.s.}$)$<$24 using \myt, are now re-classified as CT-AGN on the basis of the \borus\ modelling. Overall, 19 out of 35 candidate CT-AGNs (54\% of the sources in our sample) are confirmed CT-AGNs.
\item We find that our data favors a low--$f_c$ solution for NGC 4945  ($f_c$$<$0.22), in agreement with previous results based on the strength of the reprocessed component and the observed strong variability at energies $>$10\,keV. 
\item We find potential evidence of an inverse trend between the torus covering factor and the AGN 2--10\,keV luminosity (Figure \ref{fig:lx_vs_cf}, right panel), i.e., sources with higher $f_c$ values have on average lower luminosities, although the $f_c$ dispersion in sources in the same range of luminosity is large. Our results partially disagree, in the low-luminosity regime,  with the findings of \citet{brightman15}, since we observe a flattening in the anti-correlation at Log(L$_{\rm 2-10keV}$)$<$42.5, while they observed a more regular trend, implying that the vast majority of CT-AGN should have $f_c$$\sim$1 at Log(L$_{\rm 2-10keV}$)$<$42.
However, their analysis was performed using a sample of only 8 sources, while we have 31 objects with $\log$($N_{\rm H, z}$)$\geq$23. The trend we observe needs however to be validated using a larger sample of sources, all observed with \nustar\ to properly constrain both $N_{\rm H, z}$ and L$_{\rm 2-10keV}$.
\end{enumerate}

\subsection*{Acknowledgements}
We thank an anonymous referee for the useful comments, which helped in improving the paper.

S.M., M.A., and X.Z. acknowledge funding under NASA contract 80NSSC17K0635. Mi.Ba. acknowledges support from the Black Hole Initiative at Harvard University, which is funded by a grant from the John Templeton Foundation. This work made use of data supplied by the UK Swift Science Data Centre at the University of Leicester, as well as of the TOPCAT software \citep{taylor05} for the analysis of data tables.

\appendix

\section{A. \borus\ best-fit spectra}\label{app:borus_spectra}
\begin{figure*}
\begin{minipage}[b]{.5\textwidth}
  \centering
  \includegraphics[width=0.78\textwidth,angle=-90]{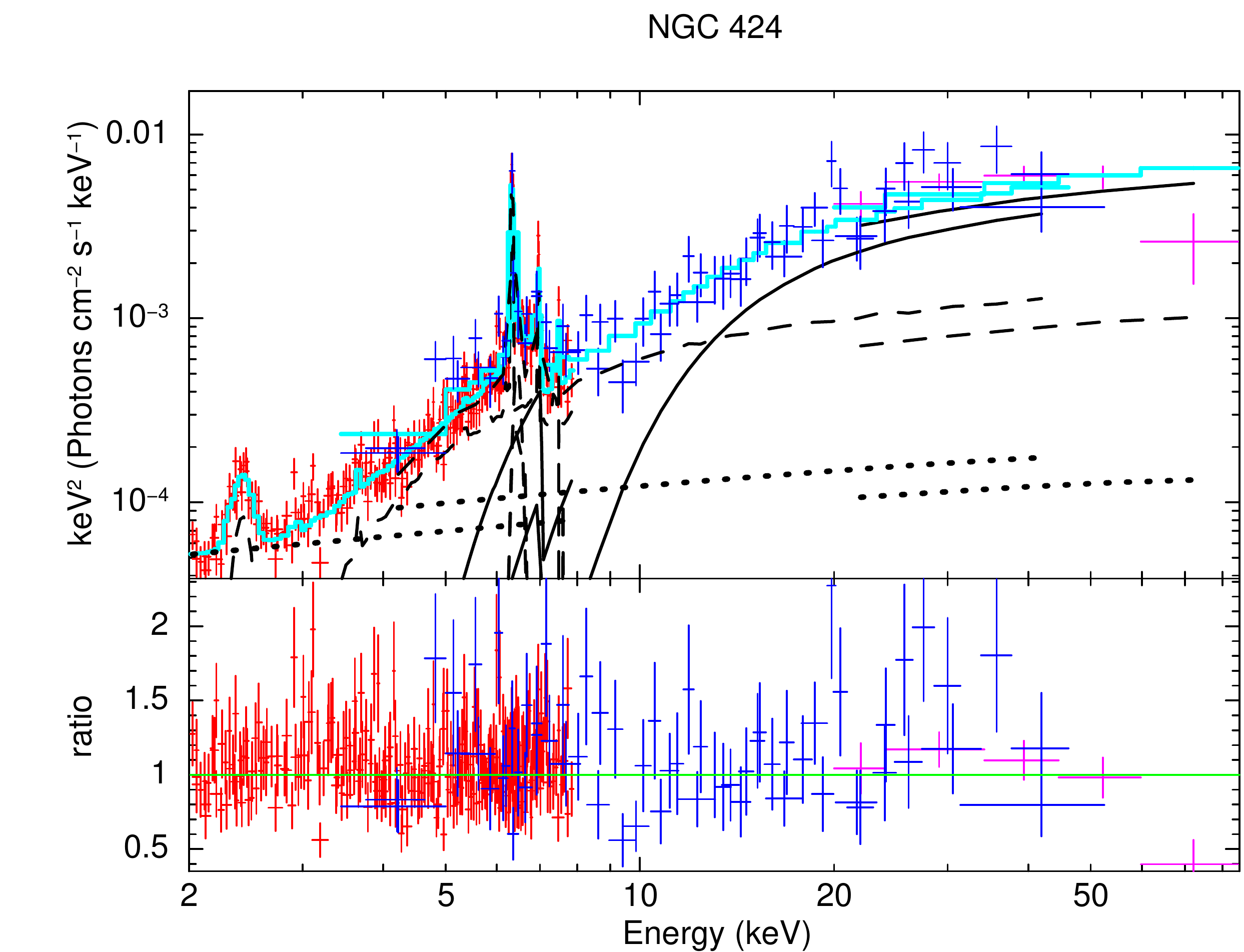}
  \end{minipage}
\begin{minipage}[b]{.5\textwidth}
  \centering
  \includegraphics[width=0.78\textwidth,angle=-90]{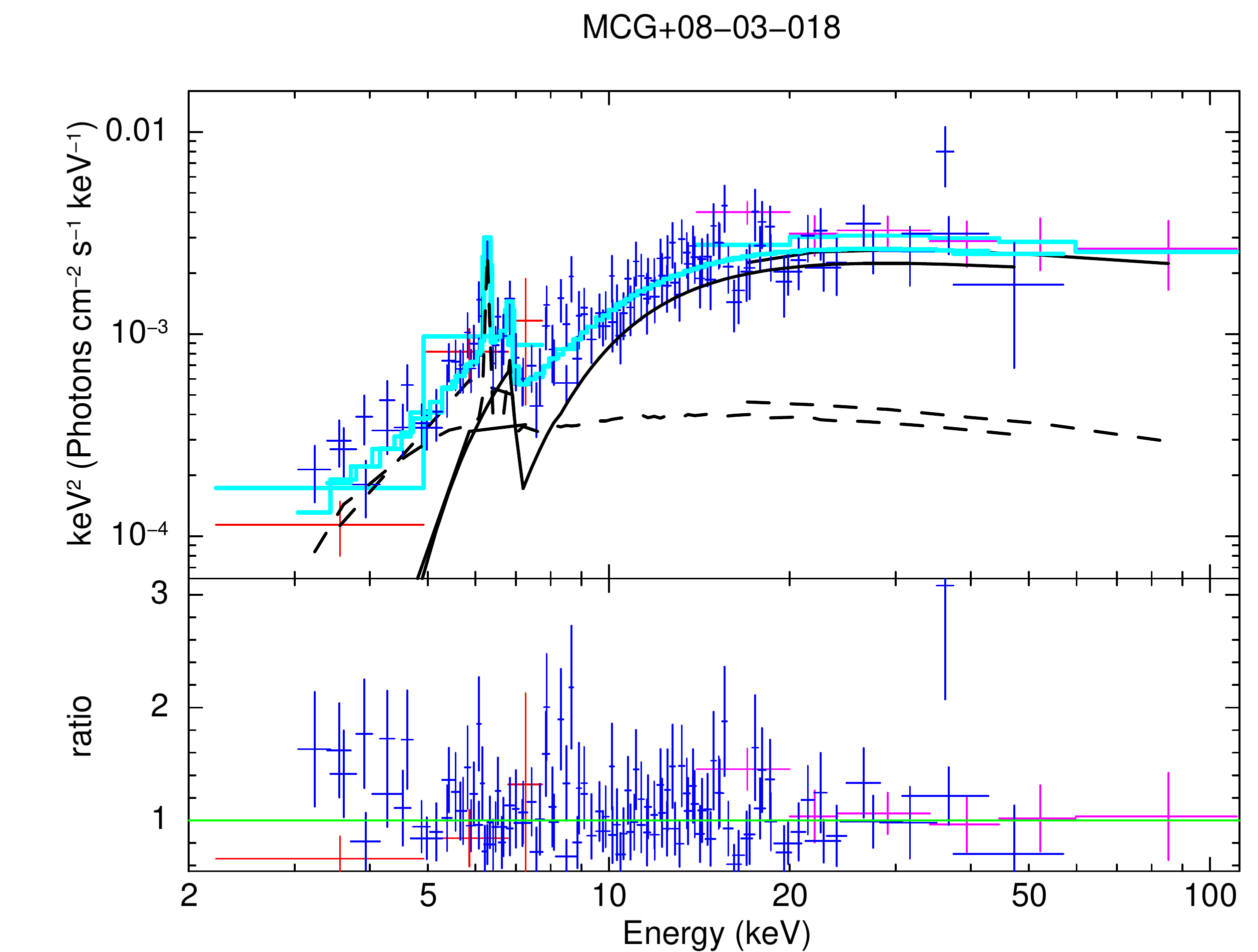}
  \end{minipage}
\begin{minipage}[b]{.5\textwidth}
  \centering
  \includegraphics[width=0.78\textwidth,angle=-90]{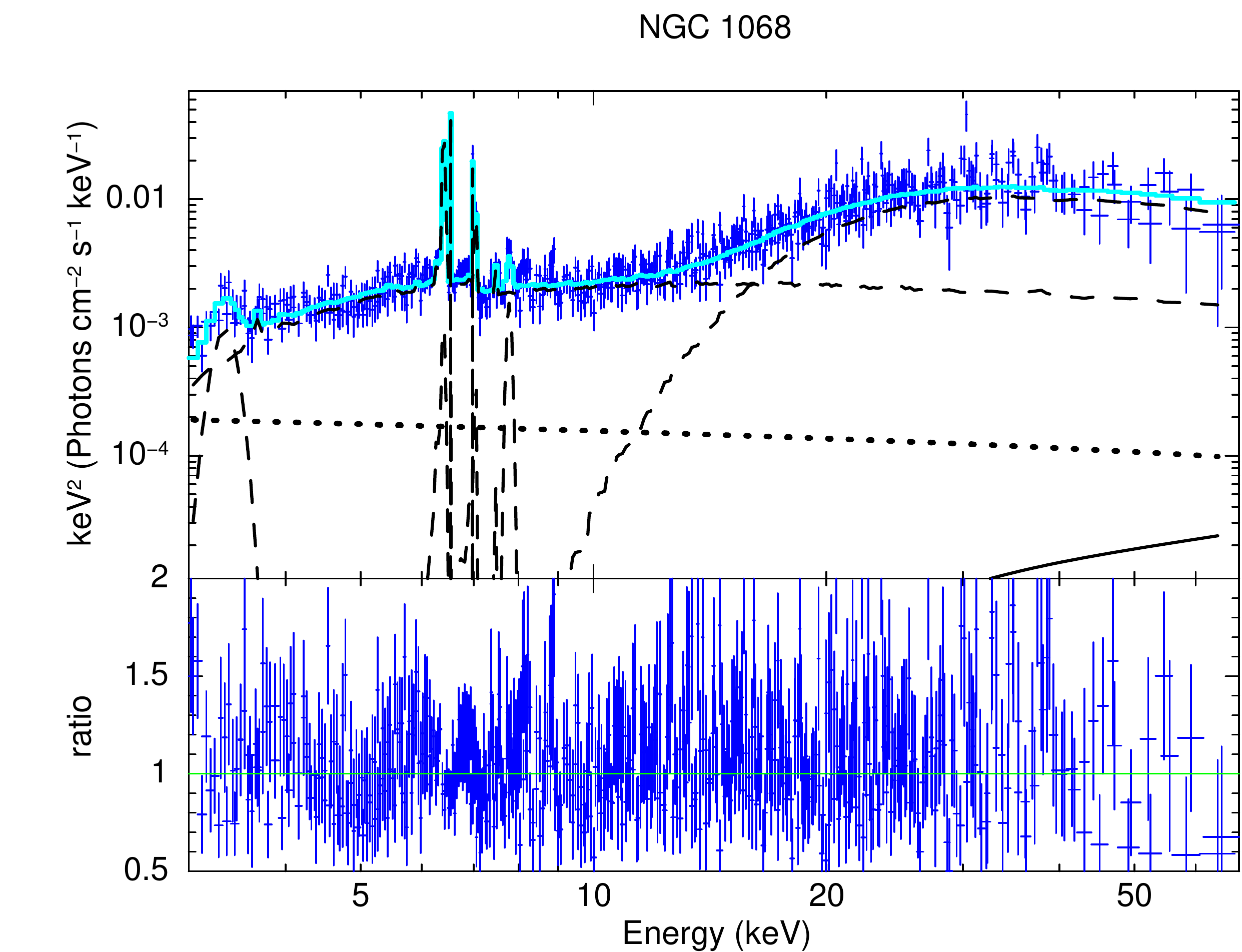}
  \end{minipage}
\begin{minipage}[b]{.5\textwidth}
  \centering
  \includegraphics[width=0.78\textwidth,angle=-90]{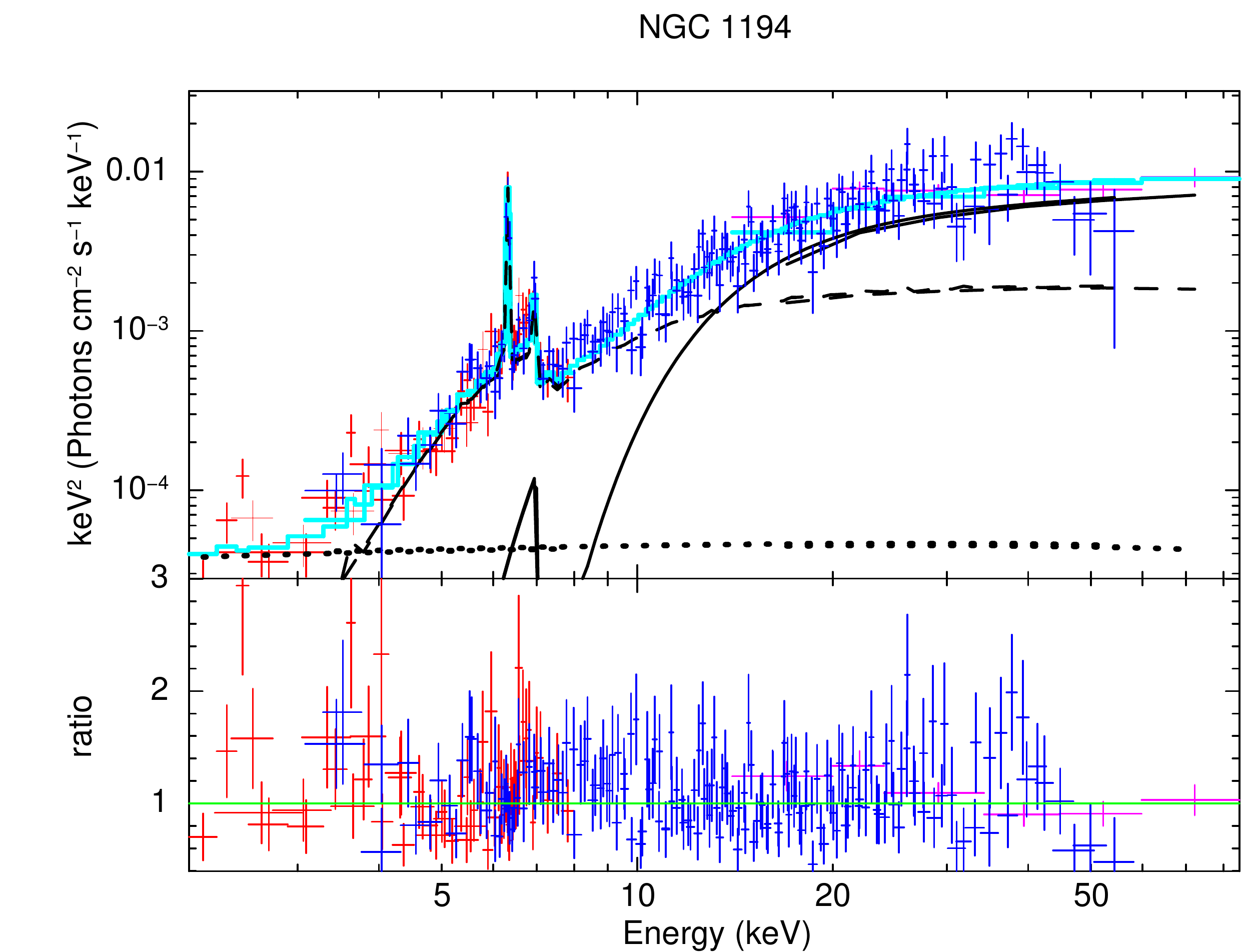}
  \end{minipage}
\begin{minipage}[b]{.5\textwidth}
  \centering
  \includegraphics[width=0.78\textwidth,angle=-90]{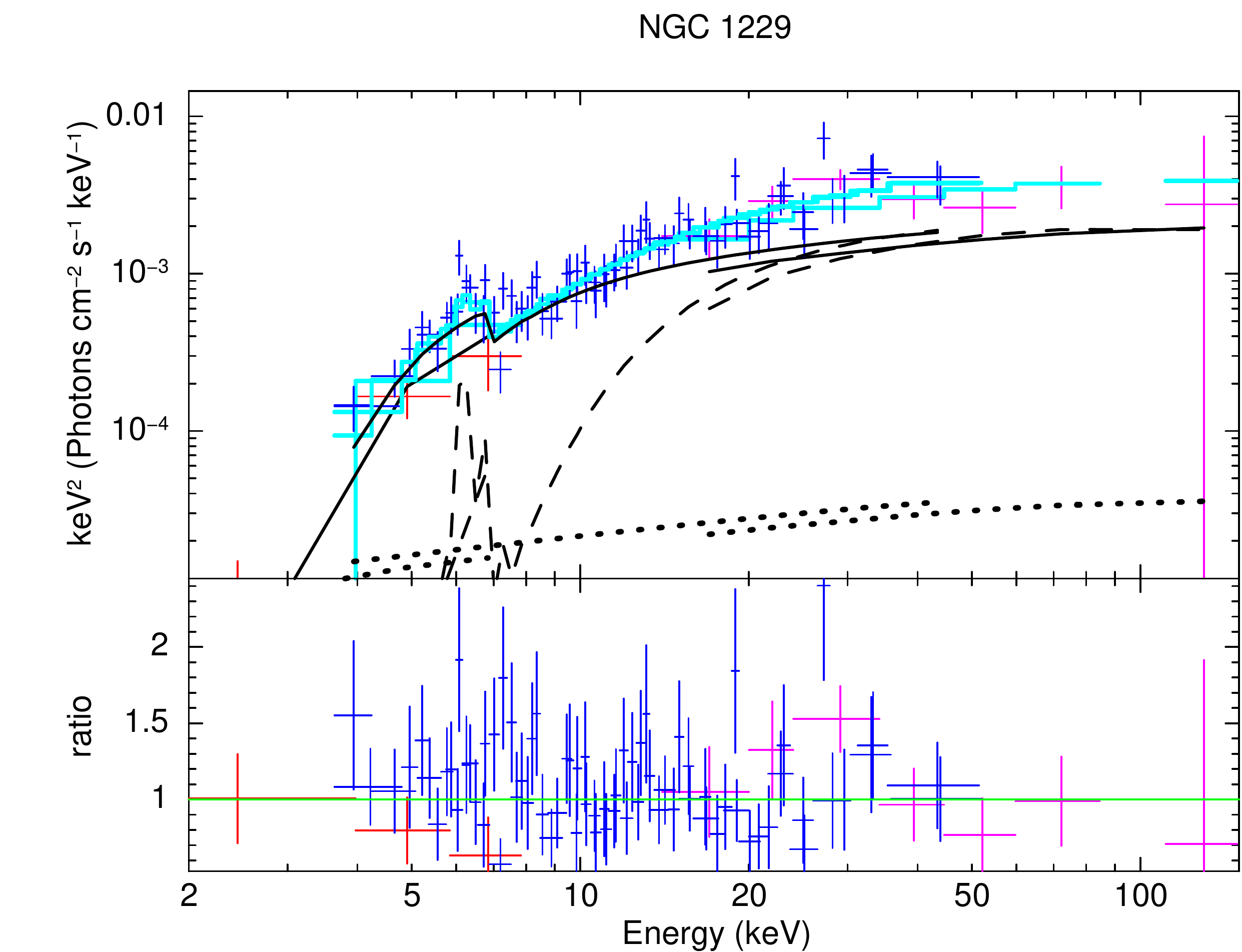}
  \end{minipage}
  \begin{minipage}[b]{.5\textwidth}
  \centering
  \includegraphics[width=0.78\textwidth,angle=-90]{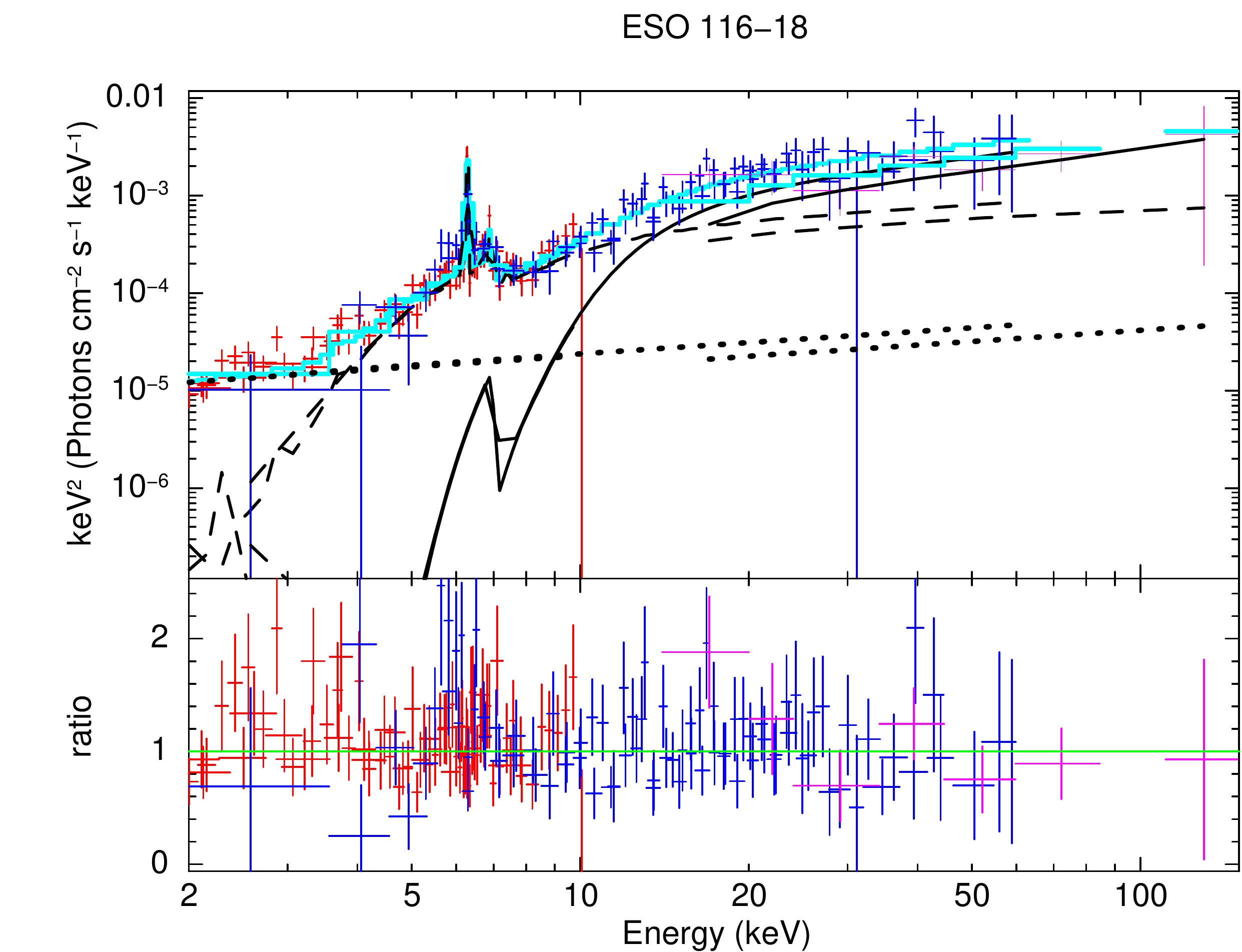}
  \end{minipage}
\caption{\normalsize \normalsize Background-subtracted spectra (top panel) and data-to-model ratio (bottom) of the CT-AGNs analyzed in this work using \borus\ with the covering factor as a free parameter. 2--10\,keV data are plotted in red, \nustar\ data in blue and \swi\ data in magenta. The best-fitting model is plotted as a cyan solid line, the AGN main continuum is plotted as a black solid line, while the reprocessed component modelled by \borus\ and other additional emission lines are plotted as a black dashed line. Finally, the main power law component scattered, rather than absorbed, by the torus is plotted as a black dotted line.}\label{fig:spectra_borus}
\end{figure*}

\begin{figure*}
\begin{minipage}[b]{.5\textwidth}
  \centering
  \includegraphics[width=0.78\textwidth,angle=-90]{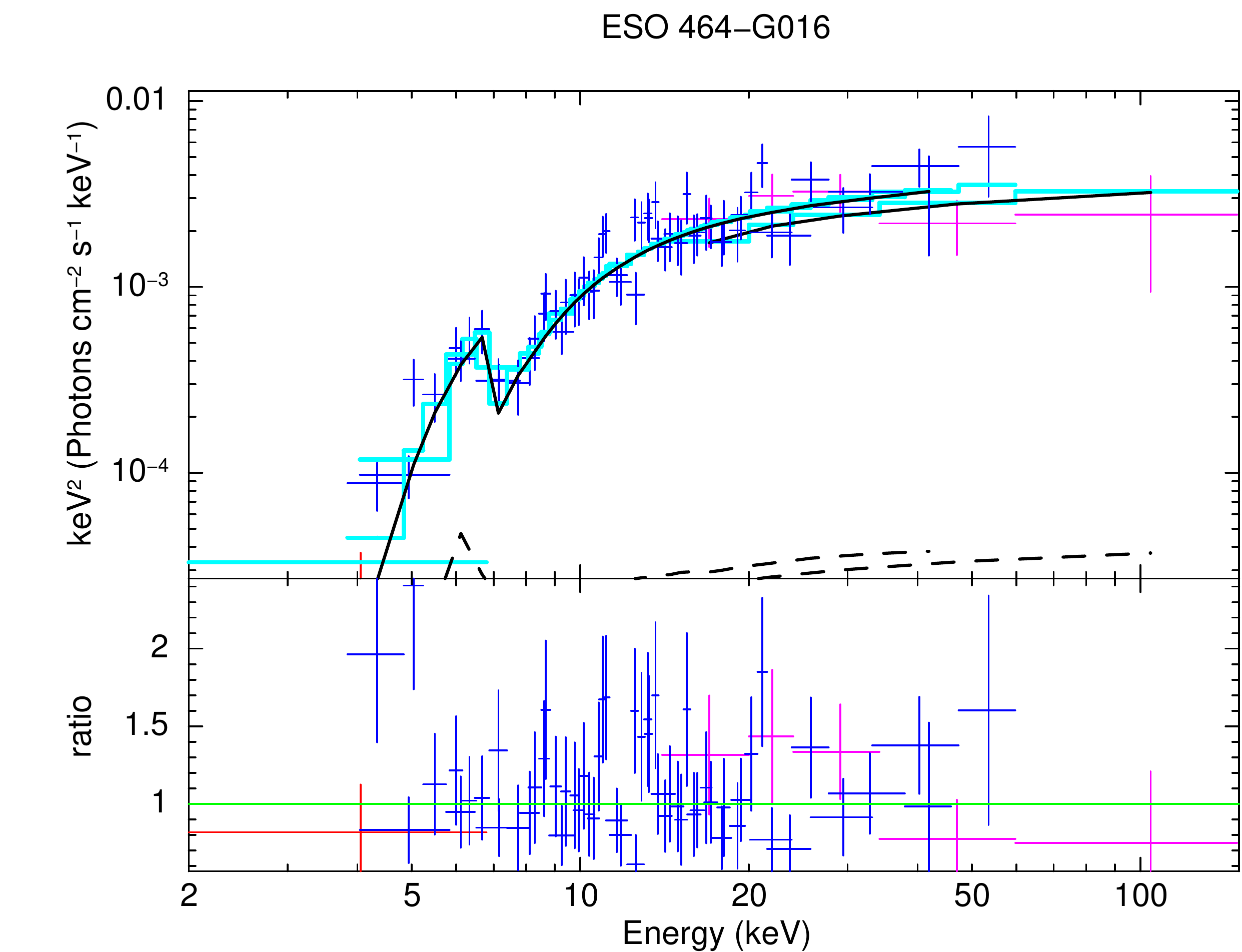}
  \end{minipage}
\begin{minipage}[b]{.5\textwidth}
  \centering
  \includegraphics[width=0.78\textwidth,angle=-90]{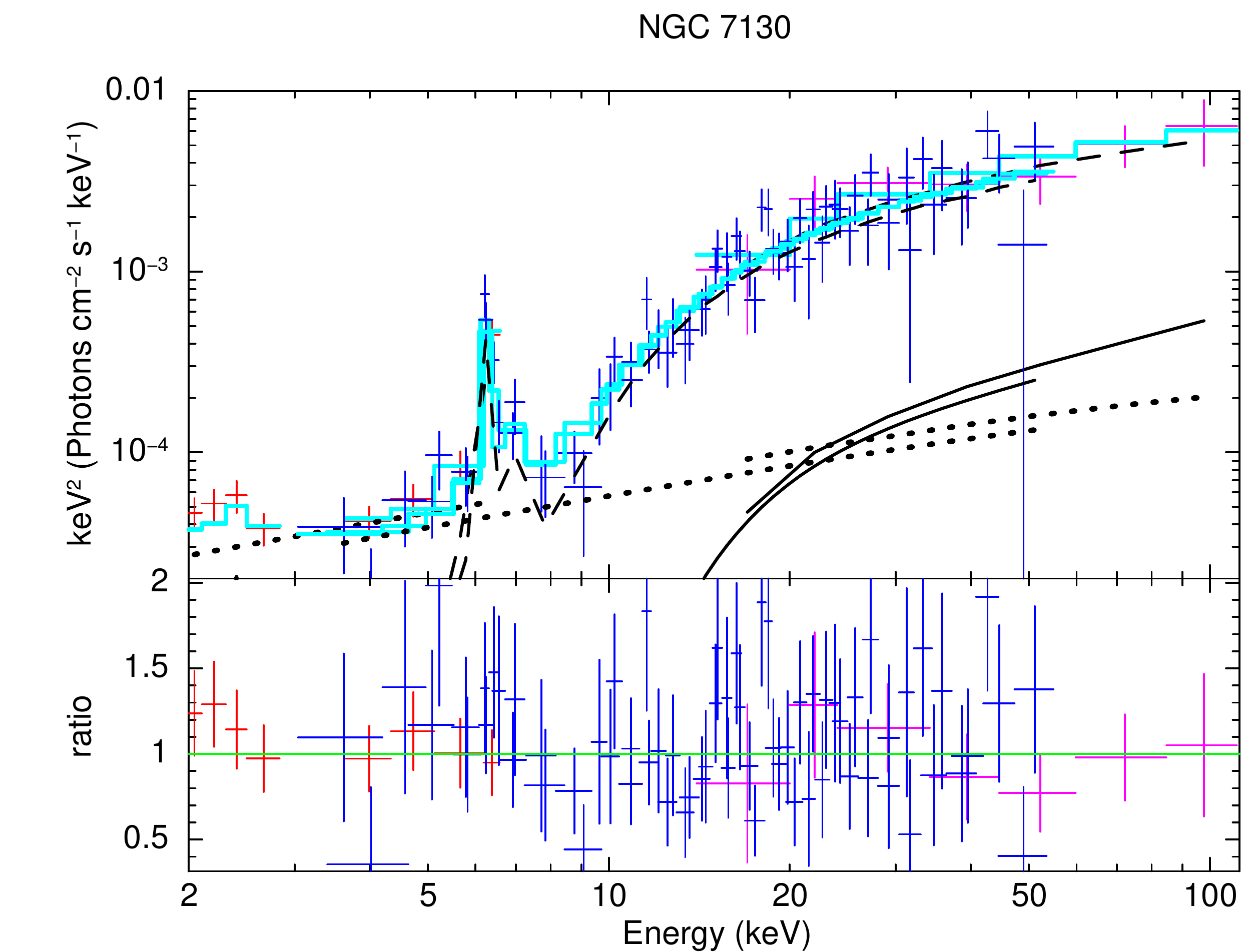}
  \end{minipage}
\begin{minipage}[b]{.5\textwidth}
  \centering
  \includegraphics[width=0.78\textwidth,angle=-90]{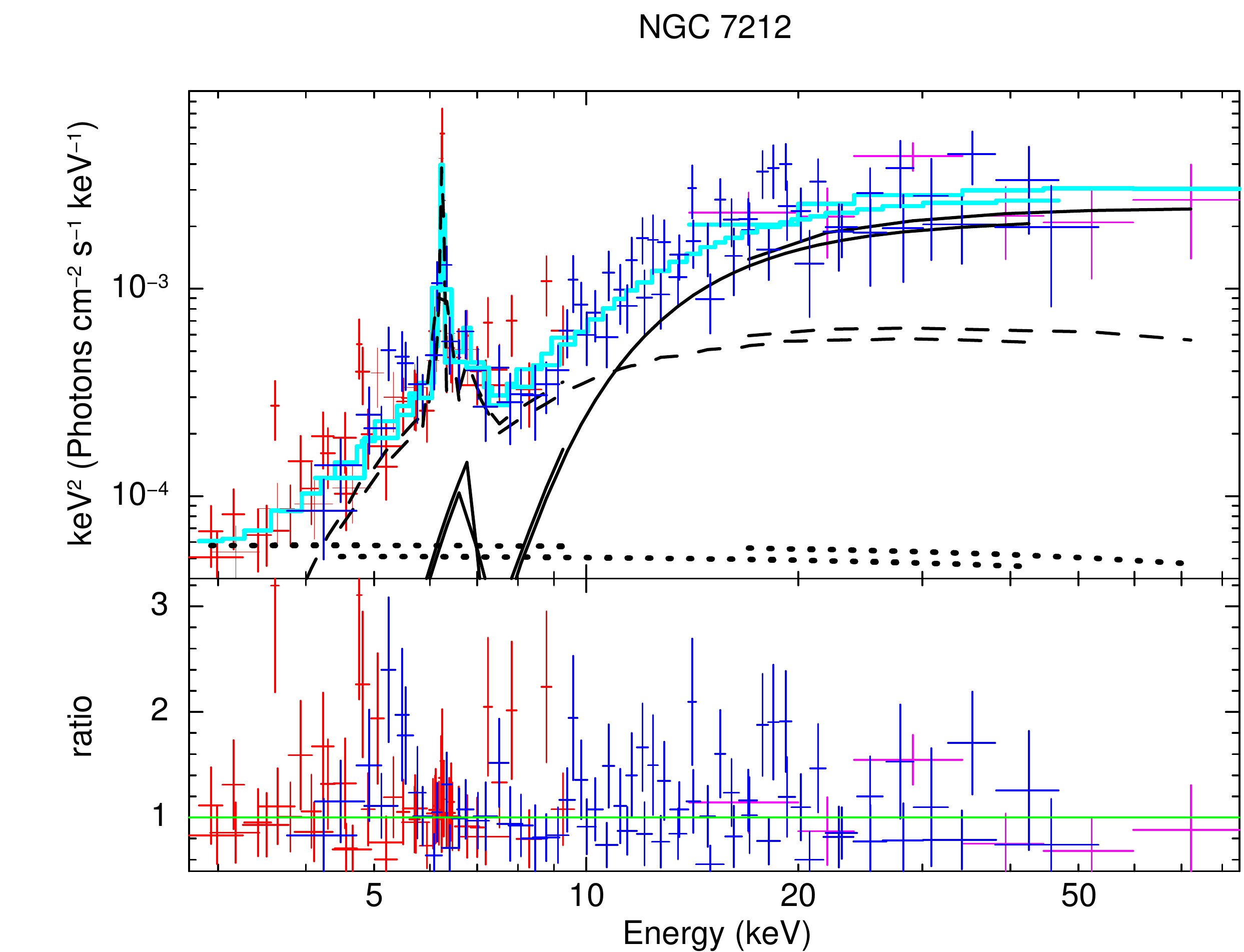}
  \end{minipage}
\begin{minipage}[b]{.5\textwidth}
  \centering
  \includegraphics[width=0.78\textwidth,angle=-90]{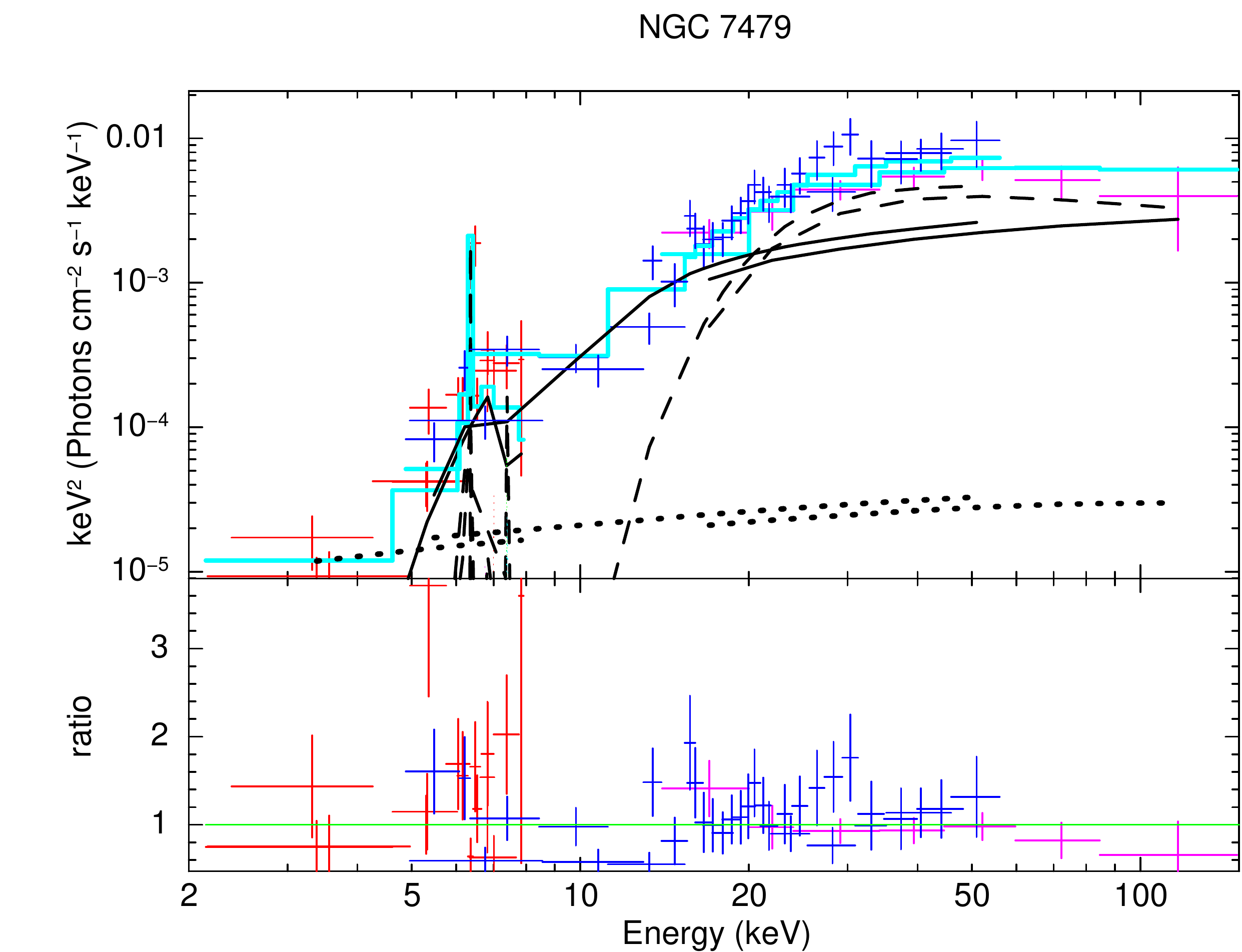}
  \end{minipage}
\begin{minipage}[b]{.5\textwidth}
  \centering
  \includegraphics[width=0.78\textwidth,angle=-90]{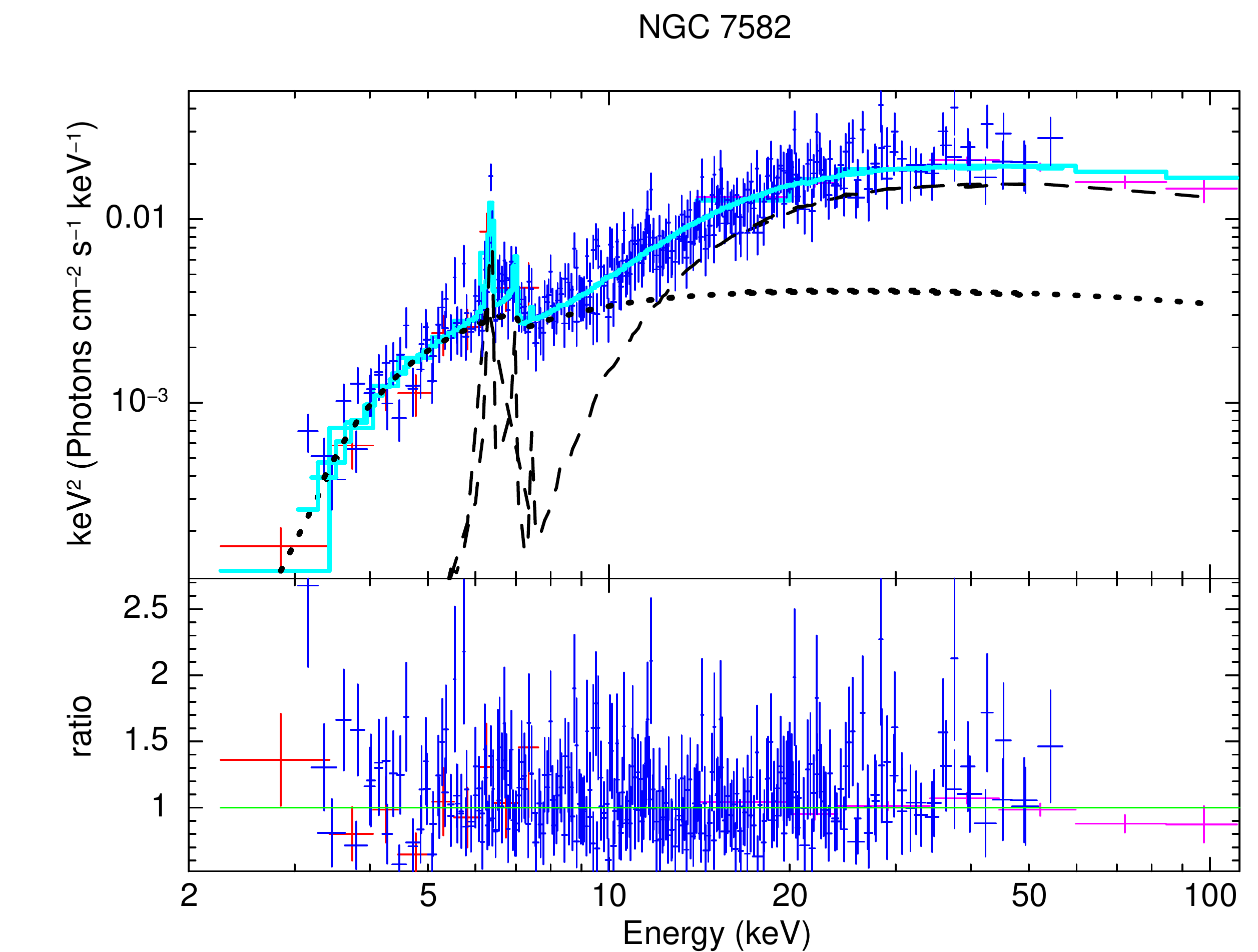}
  \end{minipage}
\caption{\normalsize \normalsize Background-subtracted spectra (top panel) and data-to-model ratio (bottom) of the CT-AGNs analyzed in this work using \borus\ with the covering factor as a free parameter. 2--10\,keV data are plotted in red, \nustar\ data in blue and \swi\ data in magenta. The best-fitting model is plotted as a cyan solid line, the AGN main continuum is plotted as a black solid line, while the reprocessed component modelled by \borus\ and other additional emission lines are plotted as a black dashed line. Finally, the main power law component scattered, rather than absorbed, by the torus is plotted as a black dotted line.}\label{fig:spectra_borus_last}
\end{figure*}

We report in Figures \ref{fig:spectra_borus}--\ref{fig:spectra_borus_last} the unfolded spectra and data-to-model ratios of all the 35 sources in our sample. The best-fit models are those obtained using \borus\ with the covering factor left free to vary: the best-fit parameters of these spectra are reported in Table \ref{tab:cf}.

\begin{figure*}
\begin{minipage}[b]{.5\textwidth}
  \centering
  \includegraphics[width=0.78\textwidth,angle=-90]{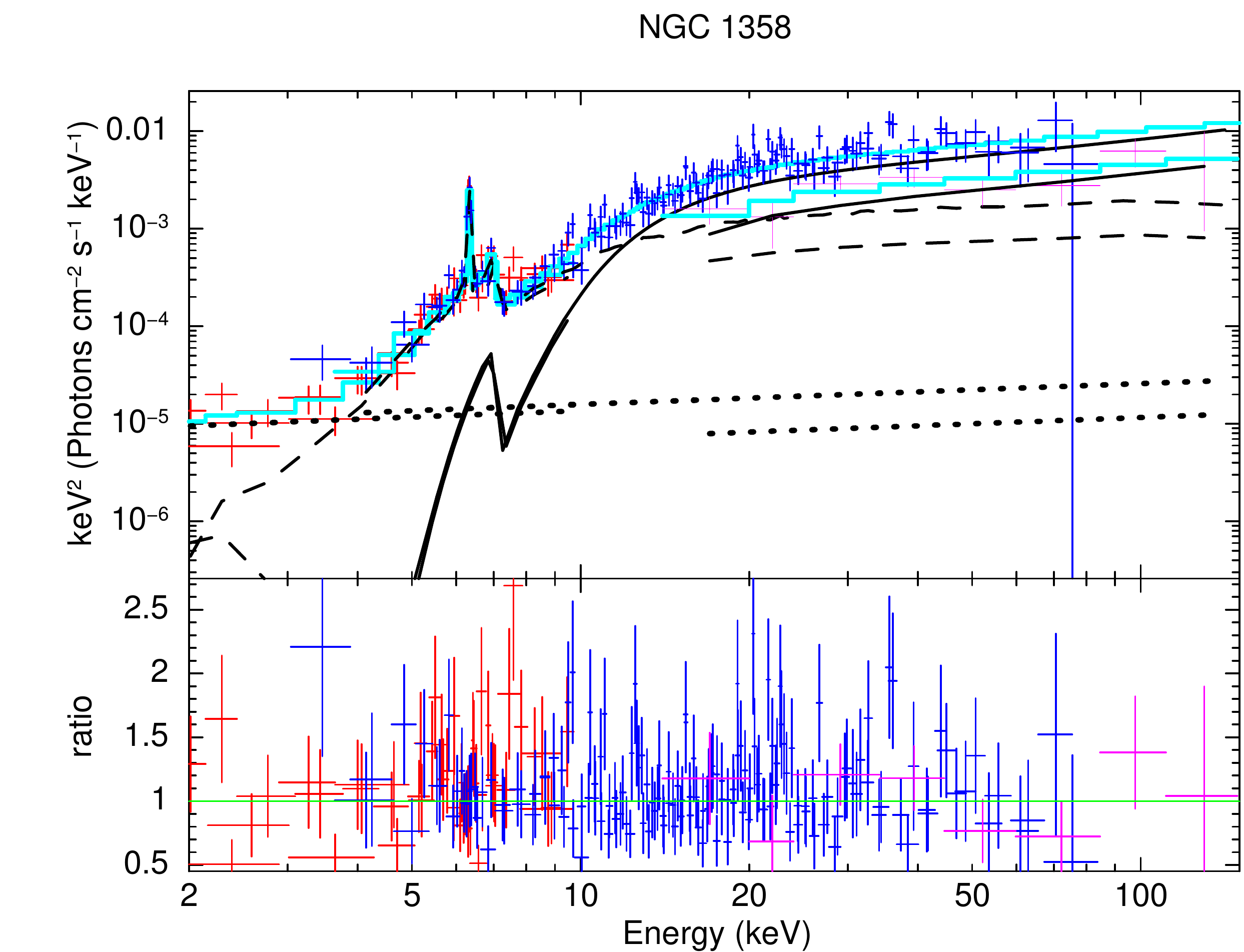}
  \end{minipage}
\begin{minipage}[b]{.5\textwidth}
  \centering
  \includegraphics[width=0.78\textwidth,angle=-90]{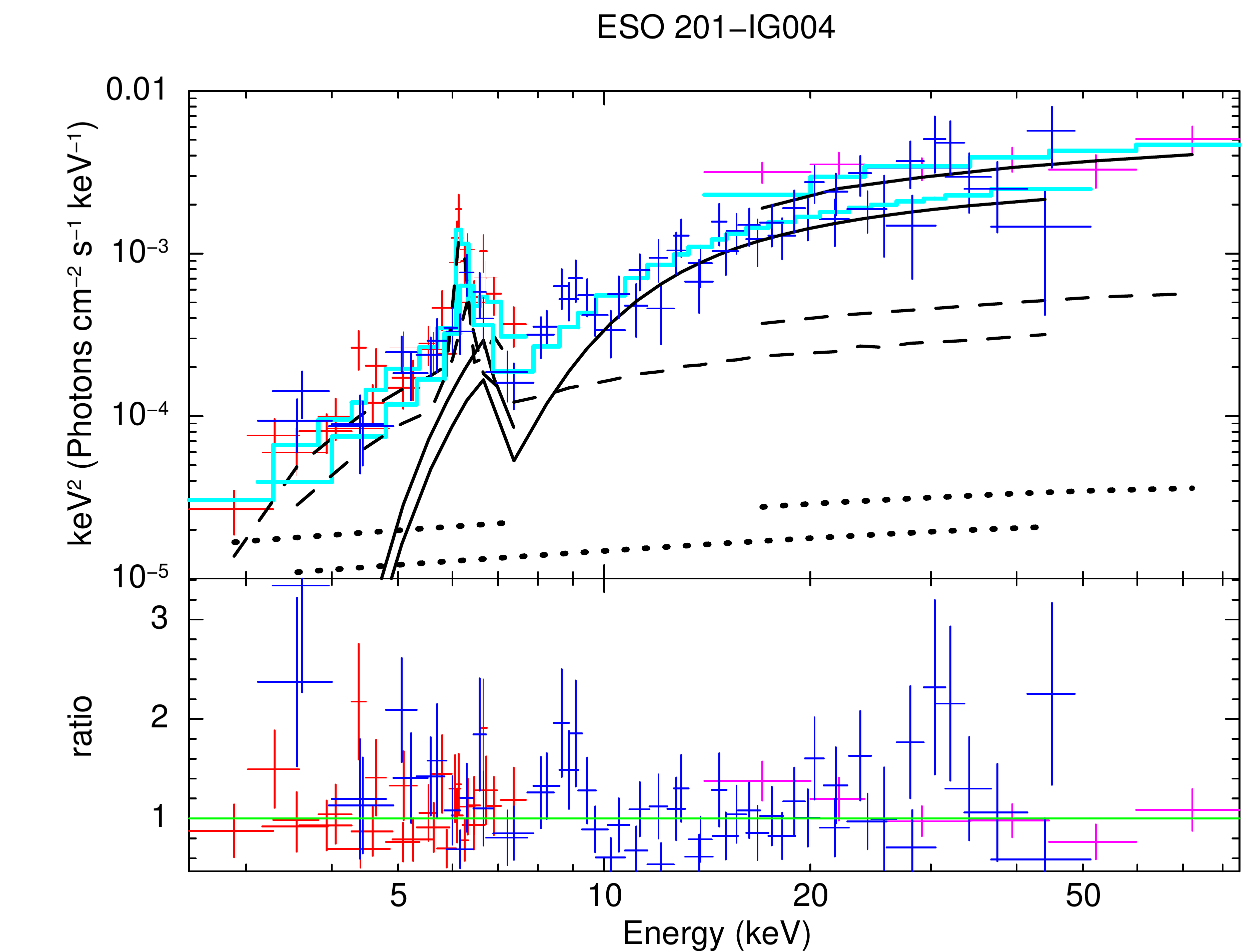}
  \end{minipage}
\begin{minipage}[b]{.5\textwidth}
  \centering
  \includegraphics[width=0.78\textwidth,angle=-90]{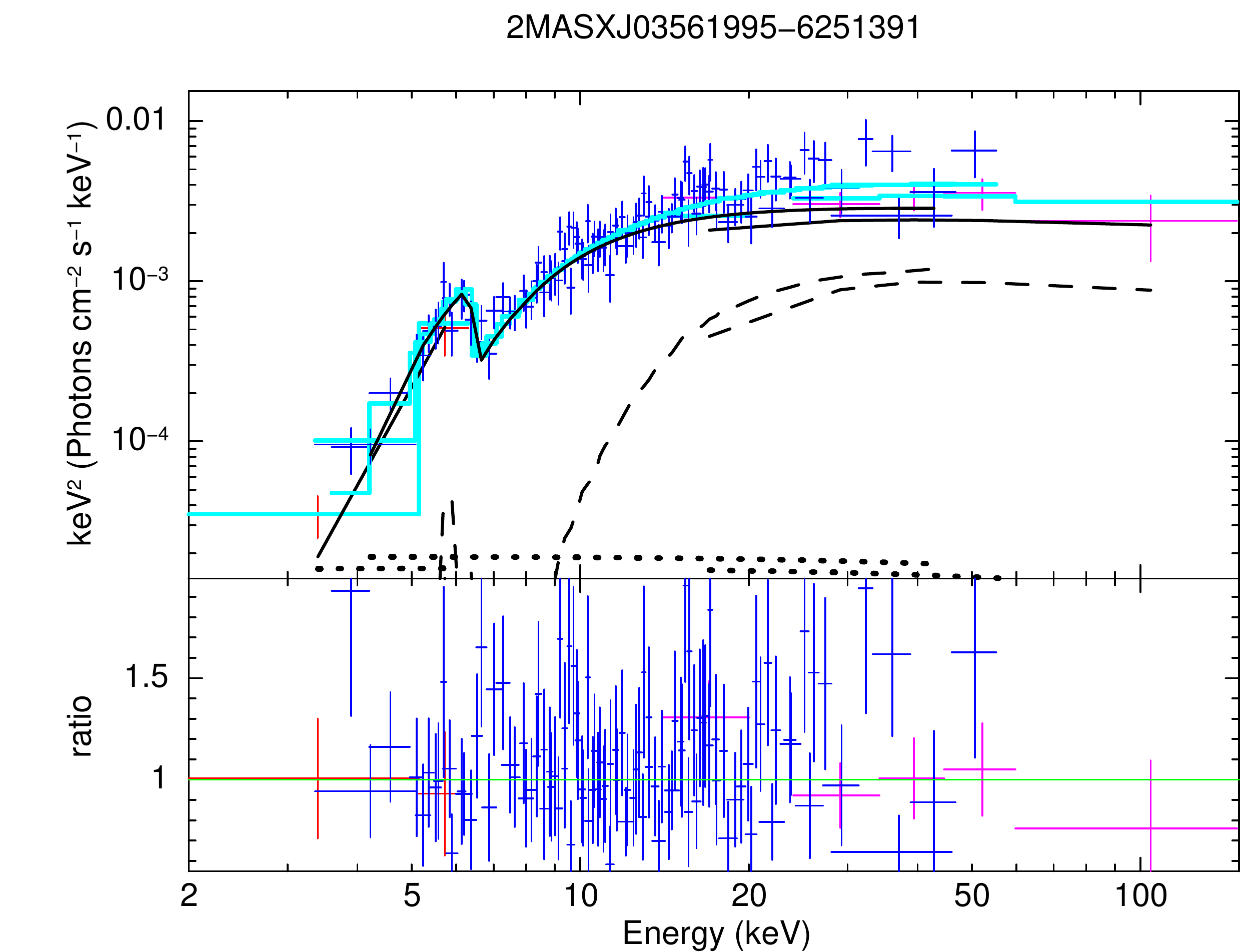}
  \end{minipage}
\begin{minipage}[b]{.5\textwidth}
  \centering
  \includegraphics[width=0.78\textwidth,angle=-90]{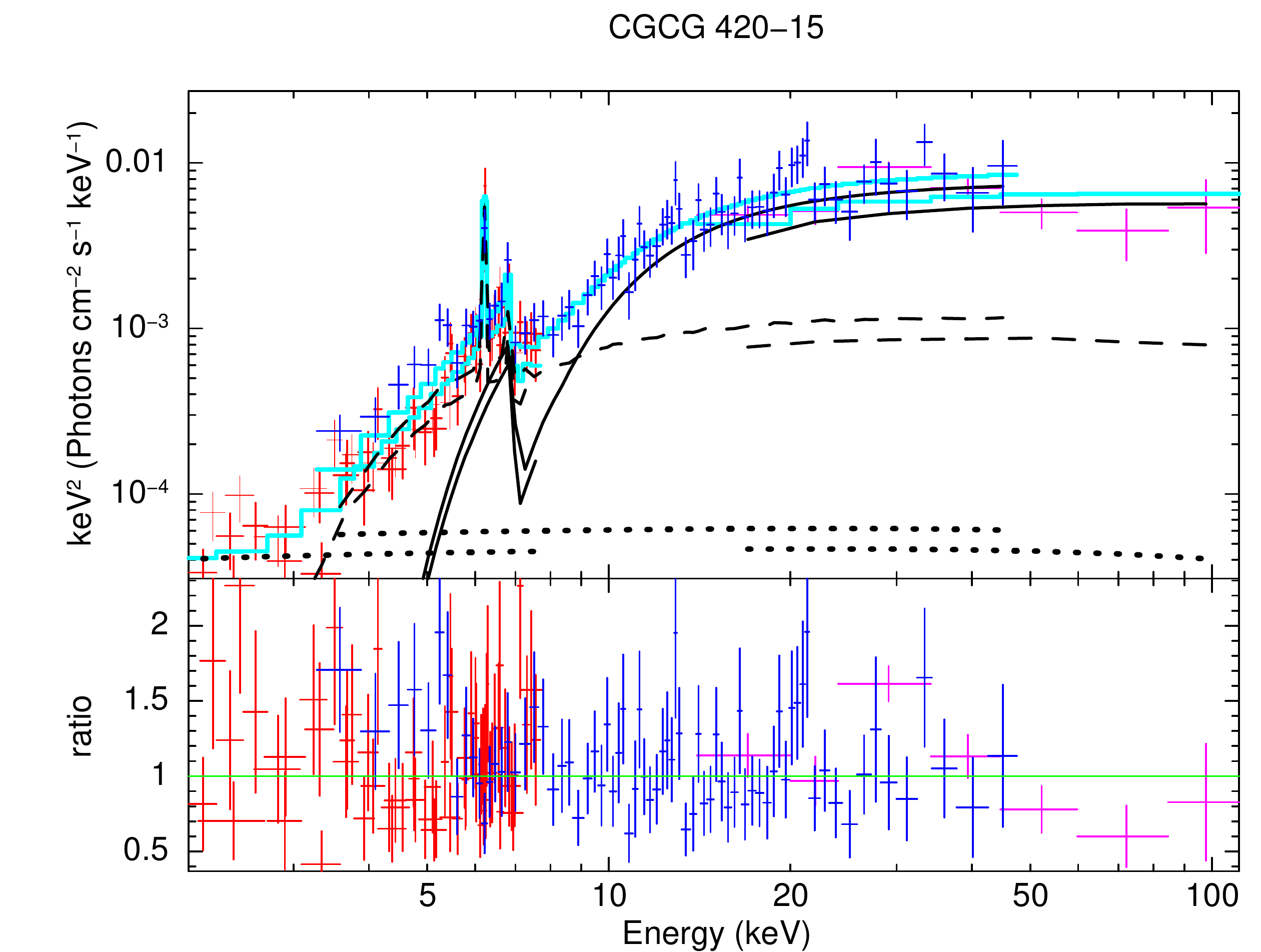}
  \end{minipage}
\begin{minipage}[b]{.5\textwidth}
  \centering
  \includegraphics[width=0.78\textwidth,angle=-90]{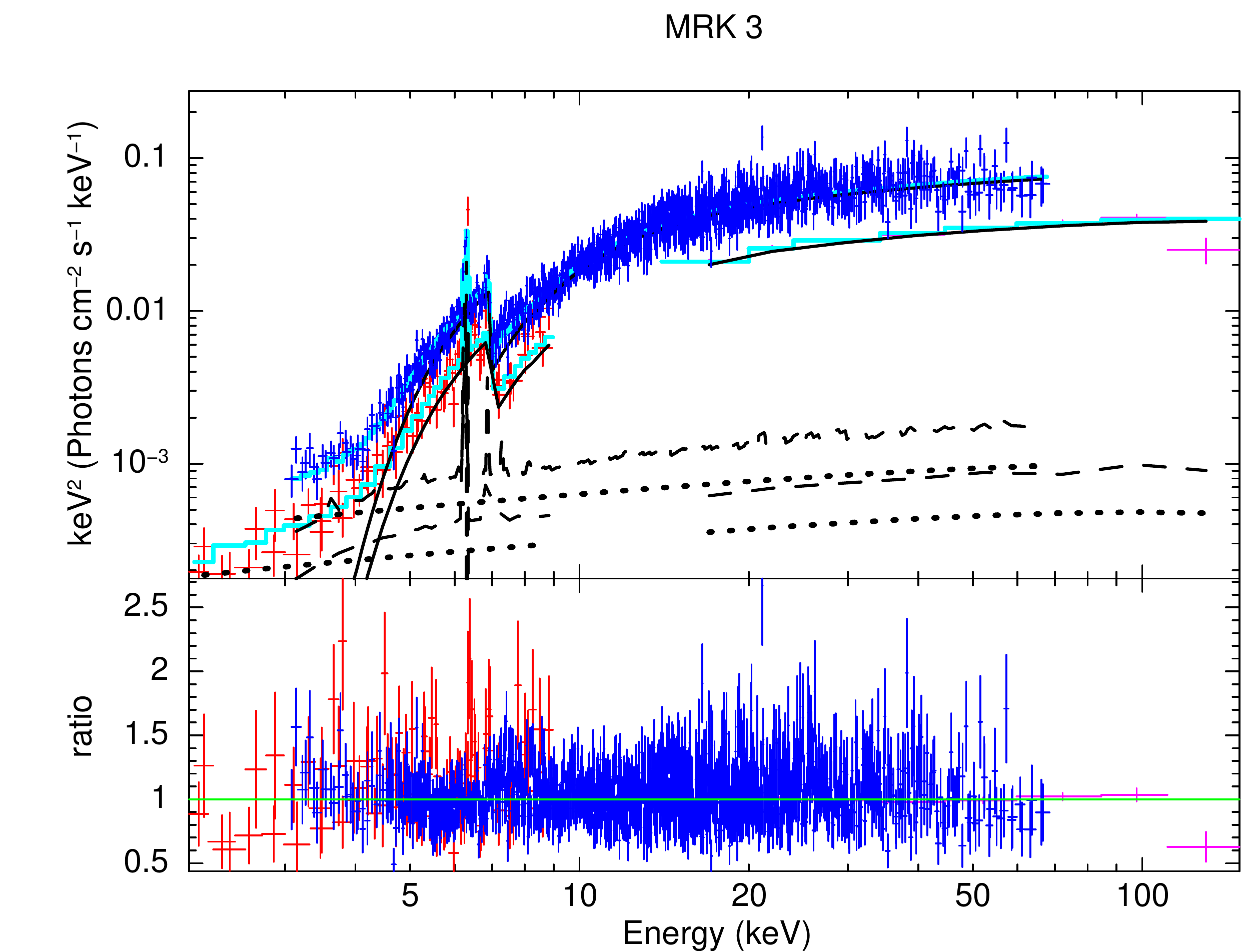}
  \end{minipage}
  \begin{minipage}[b]{.5\textwidth}
  \centering
  \includegraphics[width=0.78\textwidth,angle=-90]{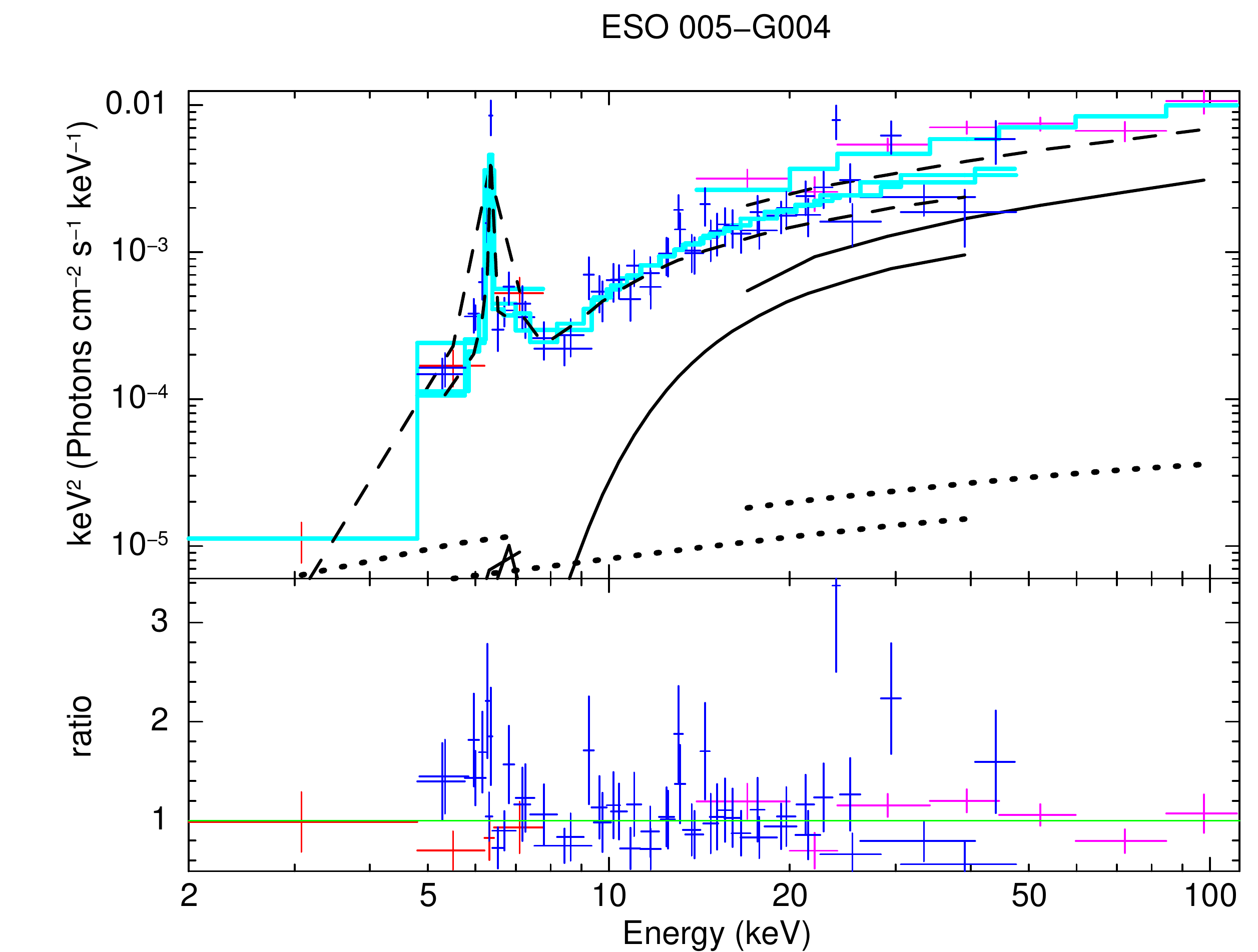}
  \end{minipage}
\caption{\normalsize \normalsize Background-subtracted spectra (top panel) and data-to-model ratio (bottom) of the CT-AGNs analyzed in this work using \borus\ with the covering factor as a free parameter. 2--10\,keV data are plotted in red, \nustar\ data in blue and \swi\ data in magenta. The best-fitting model is plotted as a cyan solid line, the AGN main continuum is plotted as a black solid line, while the reprocessed component modelled by \borus\ and other additional emission lines are plotted as a black dashed line. Finally, the main power law component scattered, rather than absorbed, by the torus is plotted as a black dotted line.}
\end{figure*}

\begin{figure*}
\begin{minipage}[b]{.5\textwidth}
  \centering
  \includegraphics[width=0.78\textwidth,angle=-90]{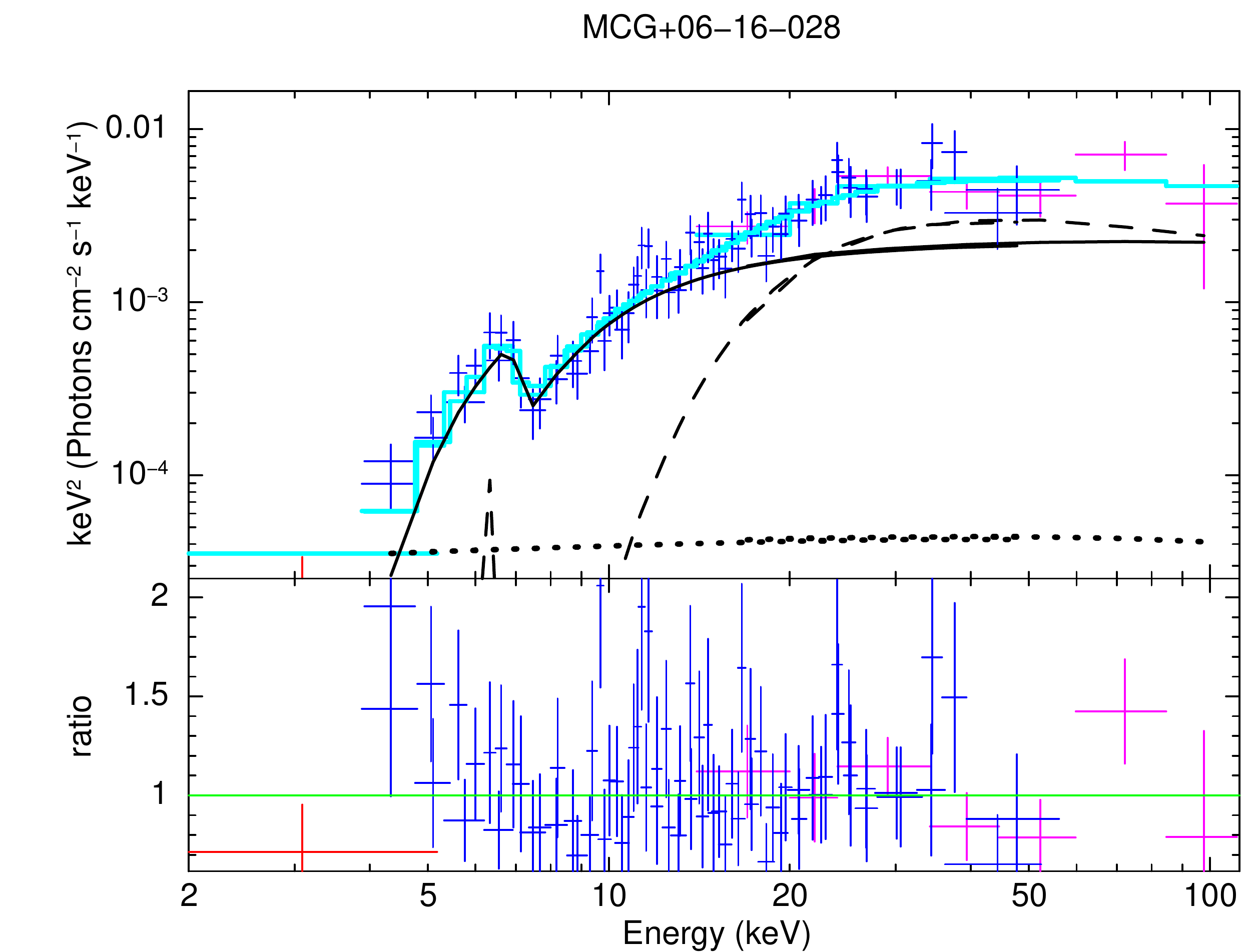}
  \end{minipage}
\begin{minipage}[b]{.5\textwidth}
  \centering
  \includegraphics[width=0.78\textwidth,angle=-90]{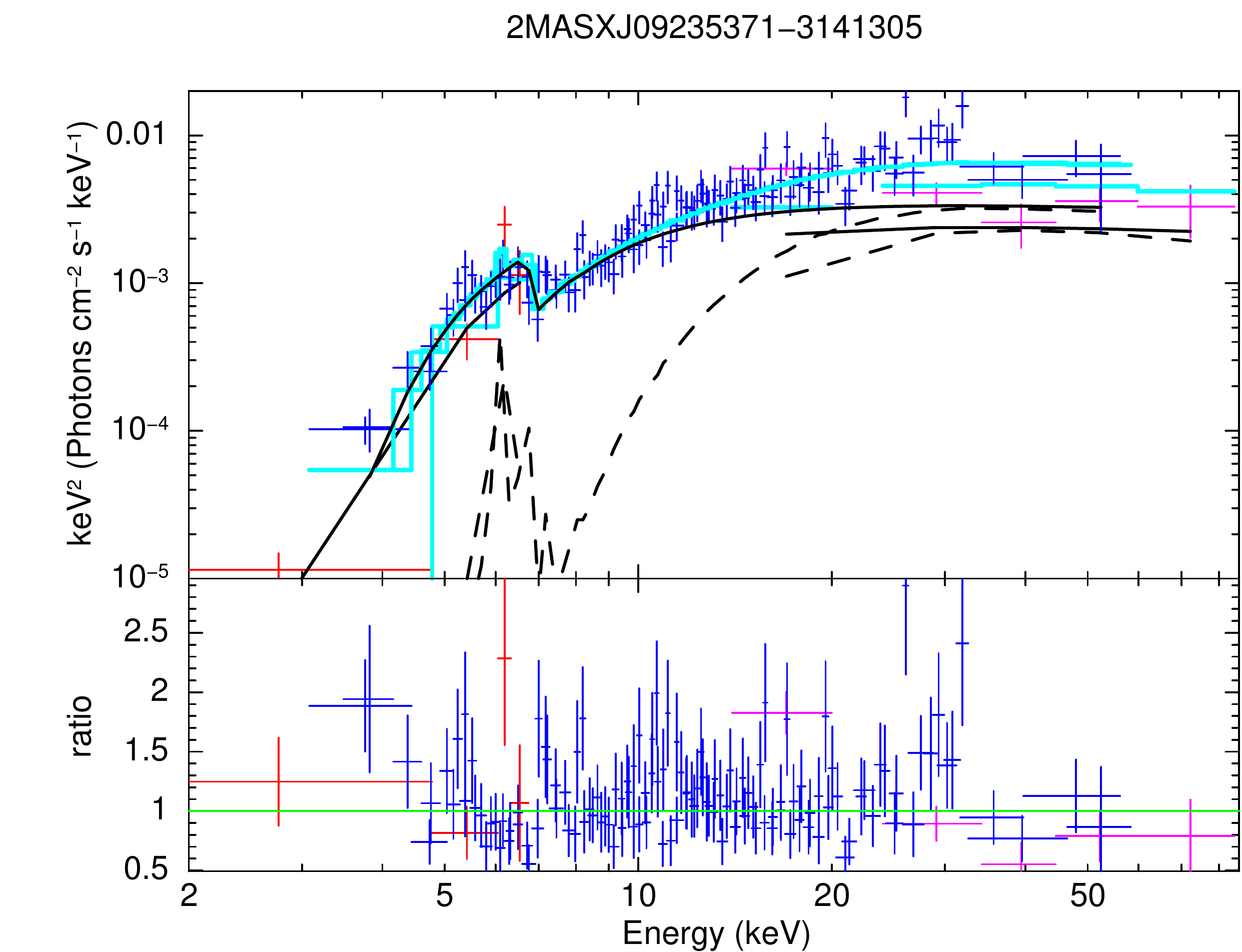}
  \end{minipage}
\begin{minipage}[b]{.5\textwidth}
  \centering
  \includegraphics[width=0.78\textwidth,angle=-90]{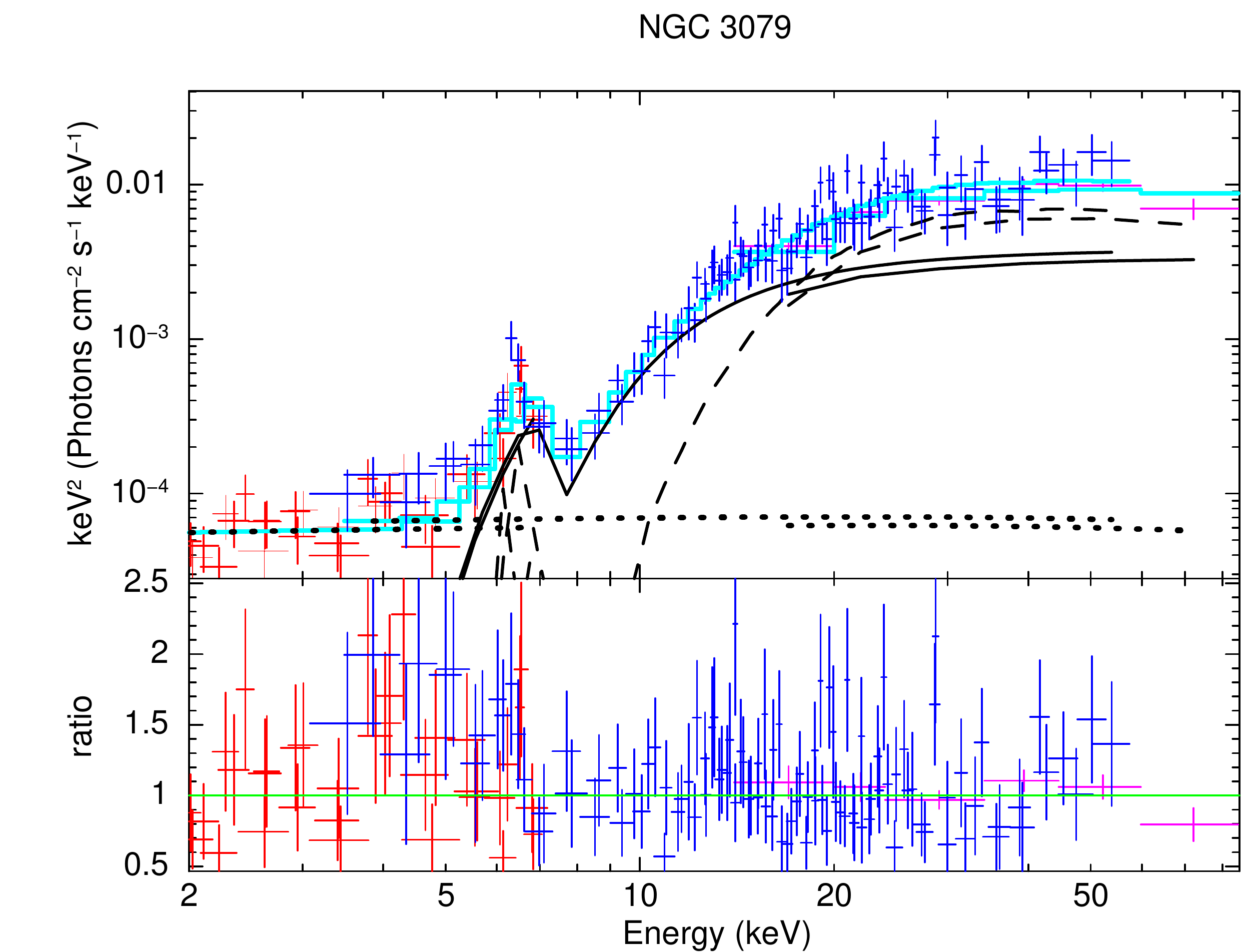}
  \end{minipage}
\begin{minipage}[b]{.5\textwidth}
  \centering
  \includegraphics[width=0.78\textwidth,angle=-90]{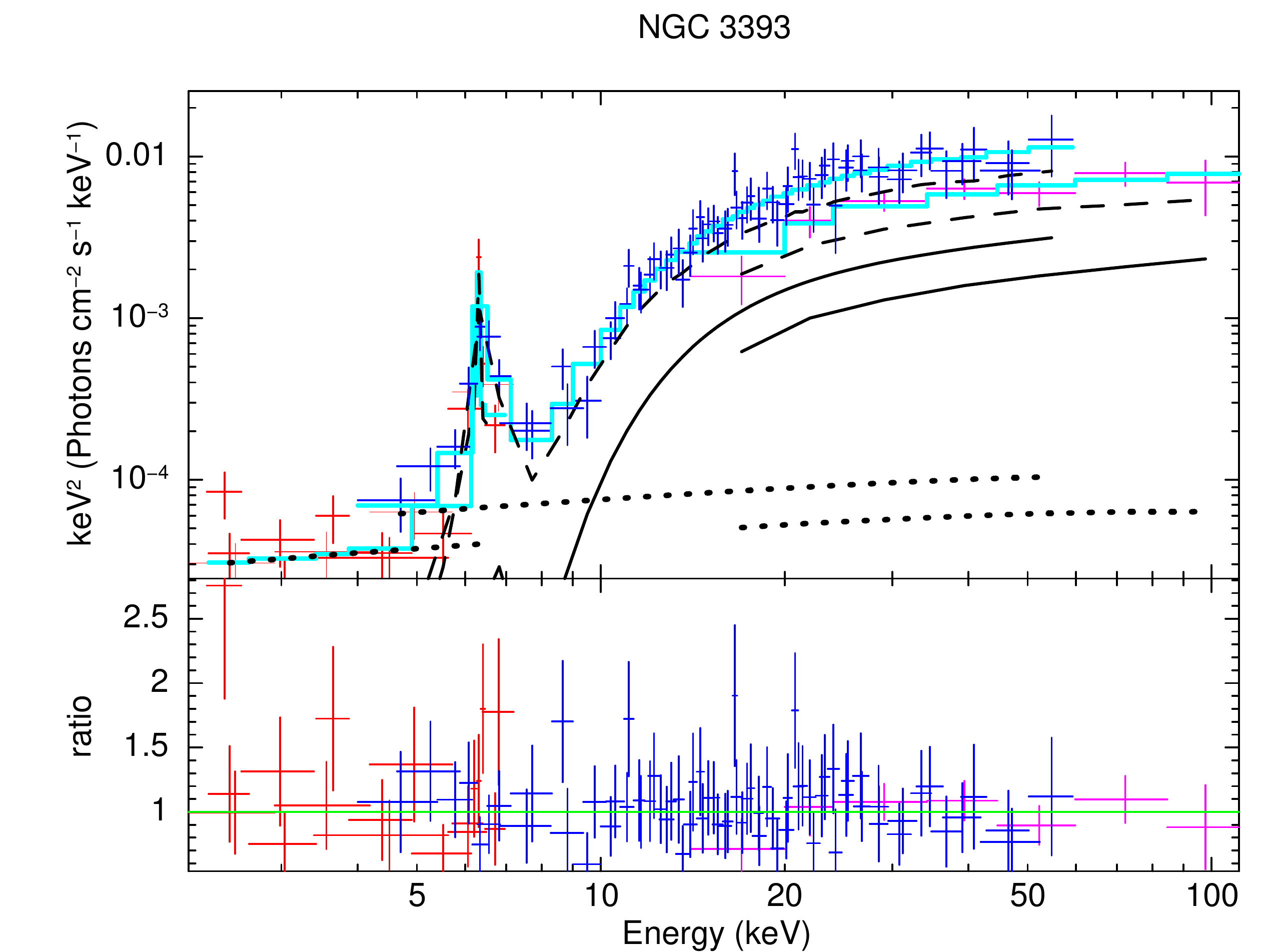}
  \end{minipage}
\begin{minipage}[b]{.5\textwidth}
  \centering
  \includegraphics[width=0.78\textwidth,angle=-90]{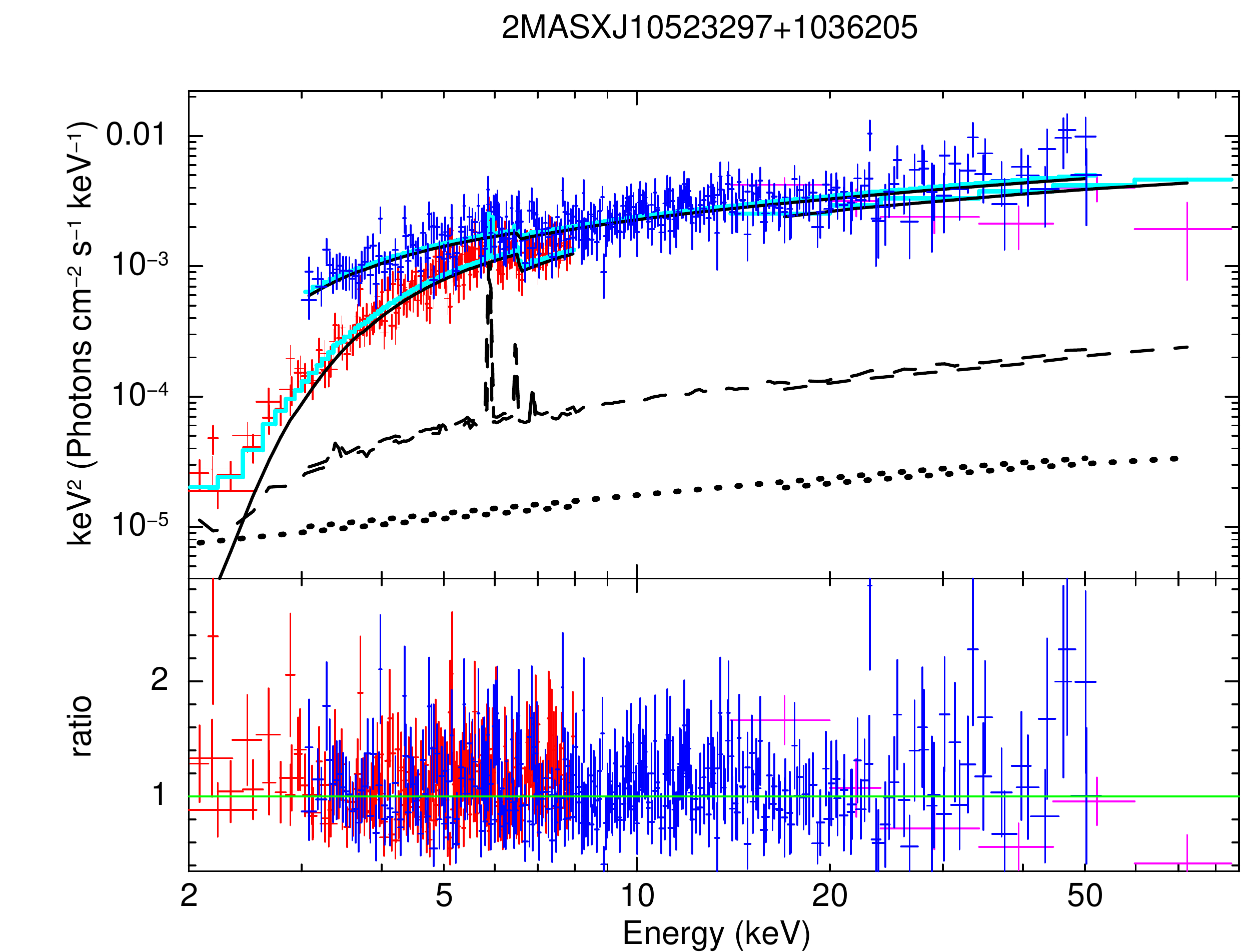}
  \end{minipage}
  \begin{minipage}[b]{.5\textwidth}
  \centering
  \includegraphics[width=0.78\textwidth,angle=-90]{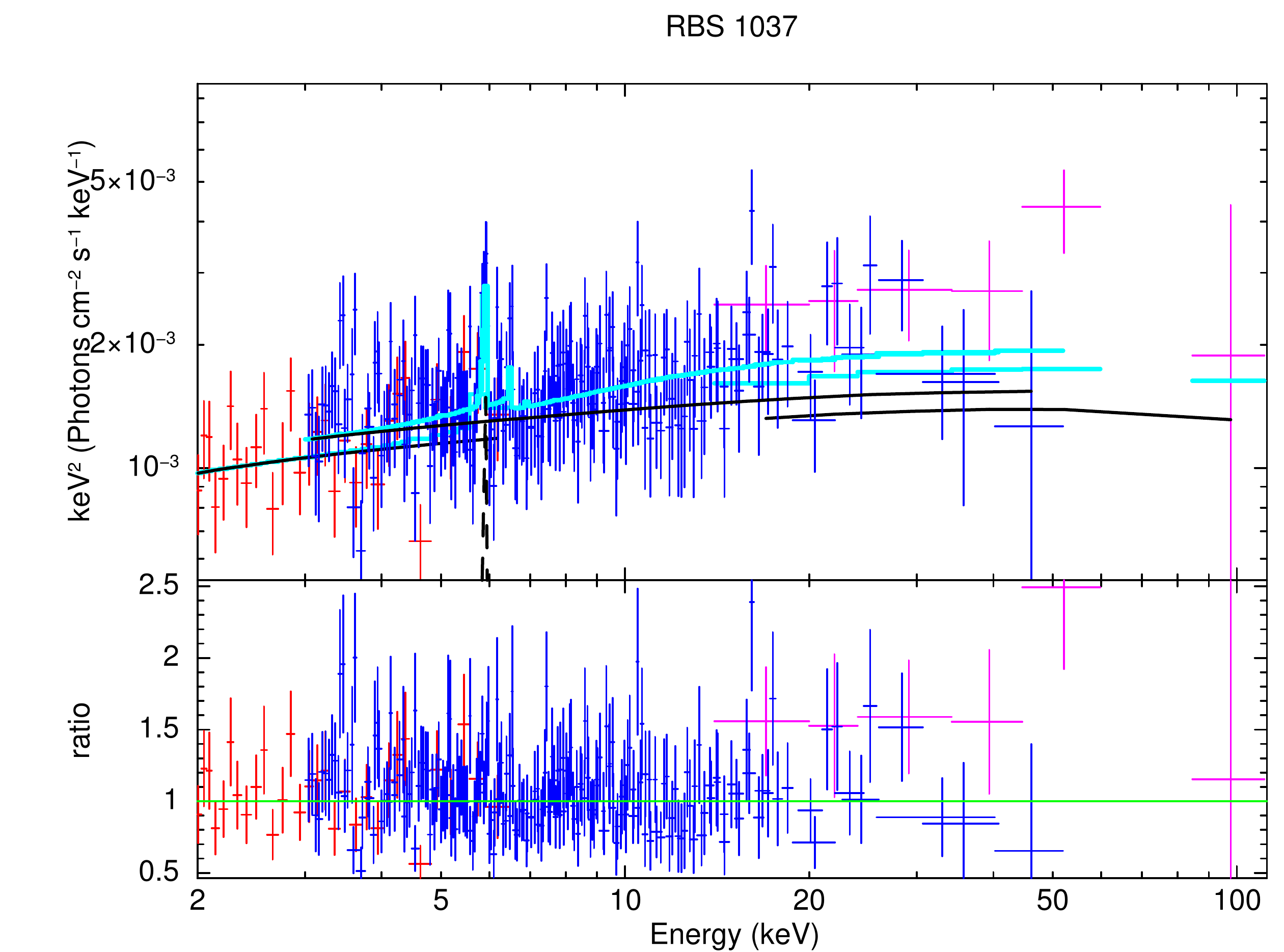}
  \end{minipage}
\caption{\normalsize \normalsize Background-subtracted spectra (top panel) and data-to-model ratio (bottom) of the CT-AGNs analyzed in this work using \borus\ with the covering factor as a free parameter. 2--10\,keV data are plotted in red, \nustar\ data in blue and \swi\ data in magenta. The best-fitting model is plotted as a cyan solid line, the AGN main continuum is plotted as a black solid line, while the reprocessed component modelled by \borus\ and other additional emission lines are plotted as a black dashed line. Finally, the main power law component scattered, rather than absorbed, by the torus is plotted as a black dotted line.}
\end{figure*}

\begin{figure*}
\begin{minipage}[b]{.5\textwidth}
  \centering
  \includegraphics[width=0.78\textwidth,angle=-90]{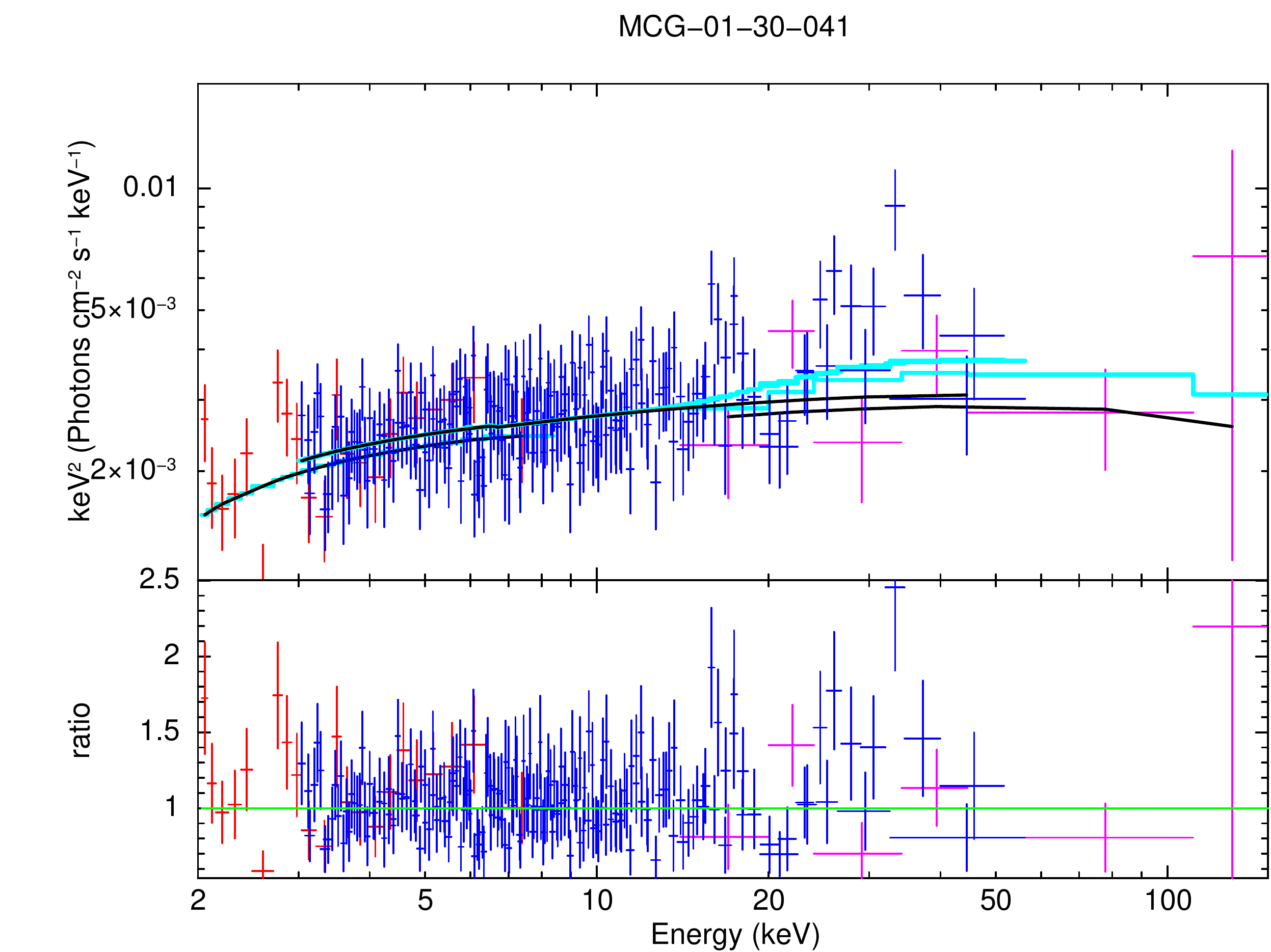}
  \end{minipage}
\begin{minipage}[b]{.5\textwidth}
  \centering
  \includegraphics[width=0.78\textwidth,angle=-90]{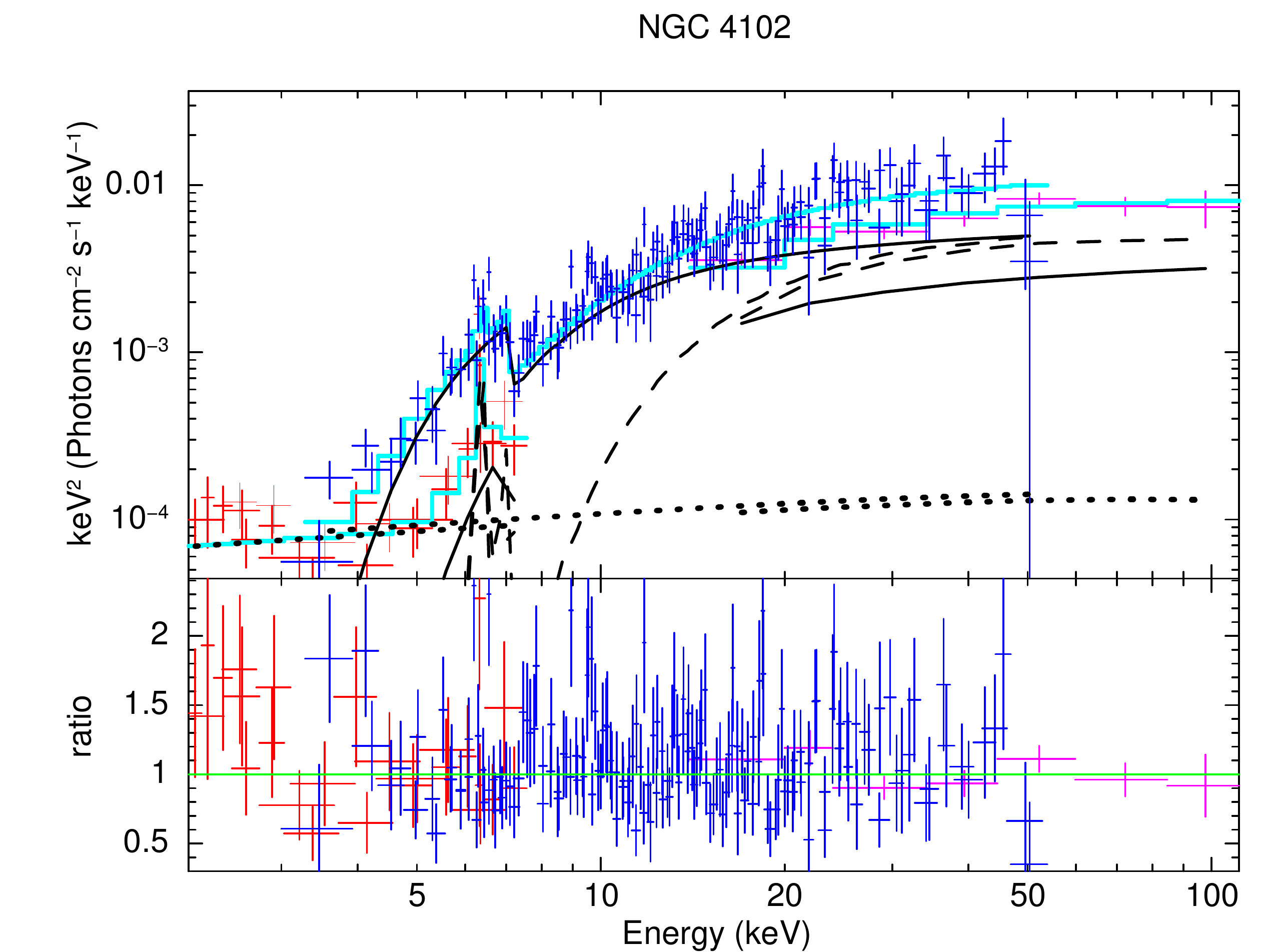}
  \end{minipage}
\begin{minipage}[b]{.5\textwidth}
  \centering
  \includegraphics[width=0.78\textwidth,angle=-90]{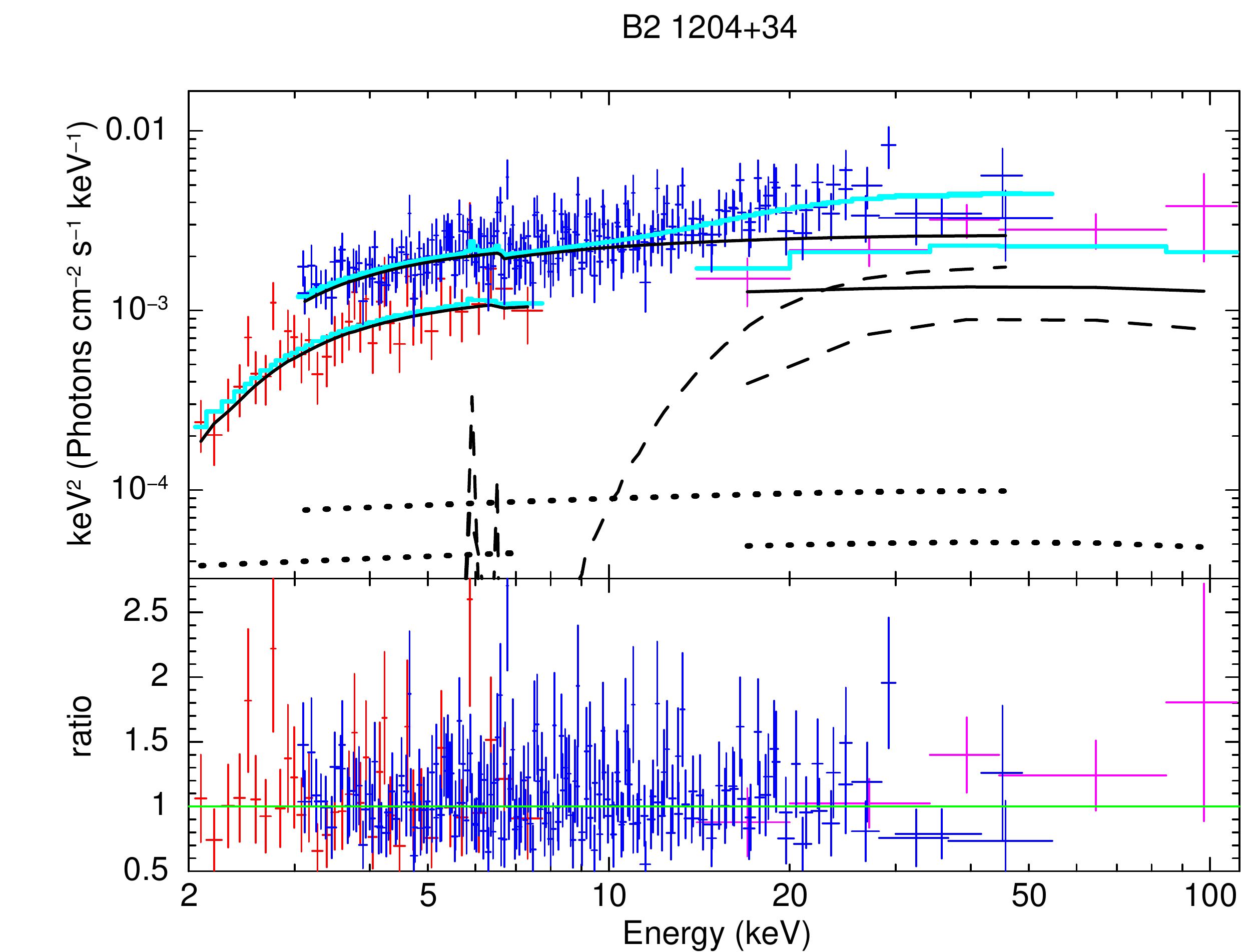}
  \end{minipage}
\begin{minipage}[b]{.5\textwidth}
  \centering
  \includegraphics[width=0.78\textwidth,angle=-90]{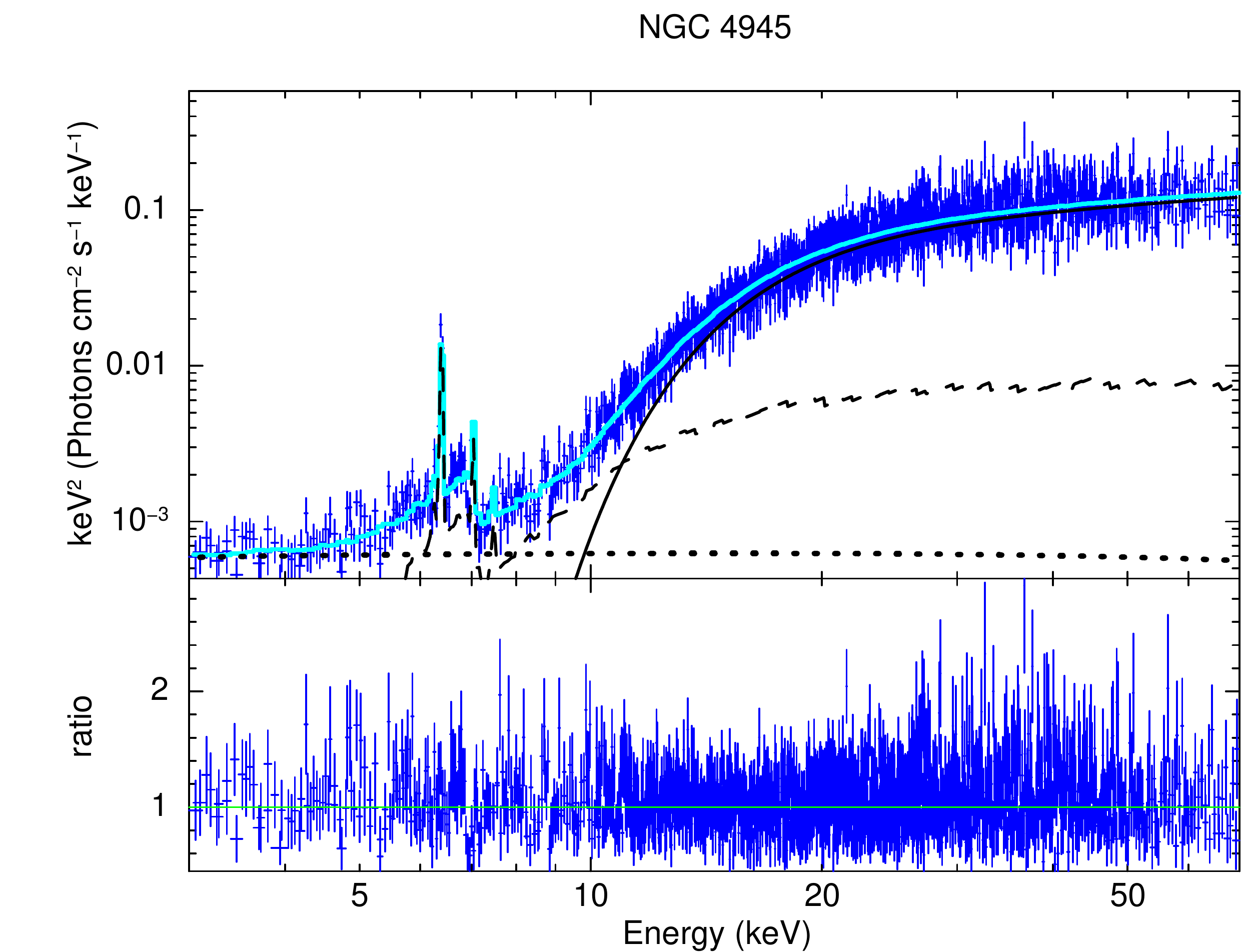}
  \end{minipage}
\begin{minipage}[b]{.5\textwidth}
  \centering
  \includegraphics[width=0.78\textwidth,angle=-90]{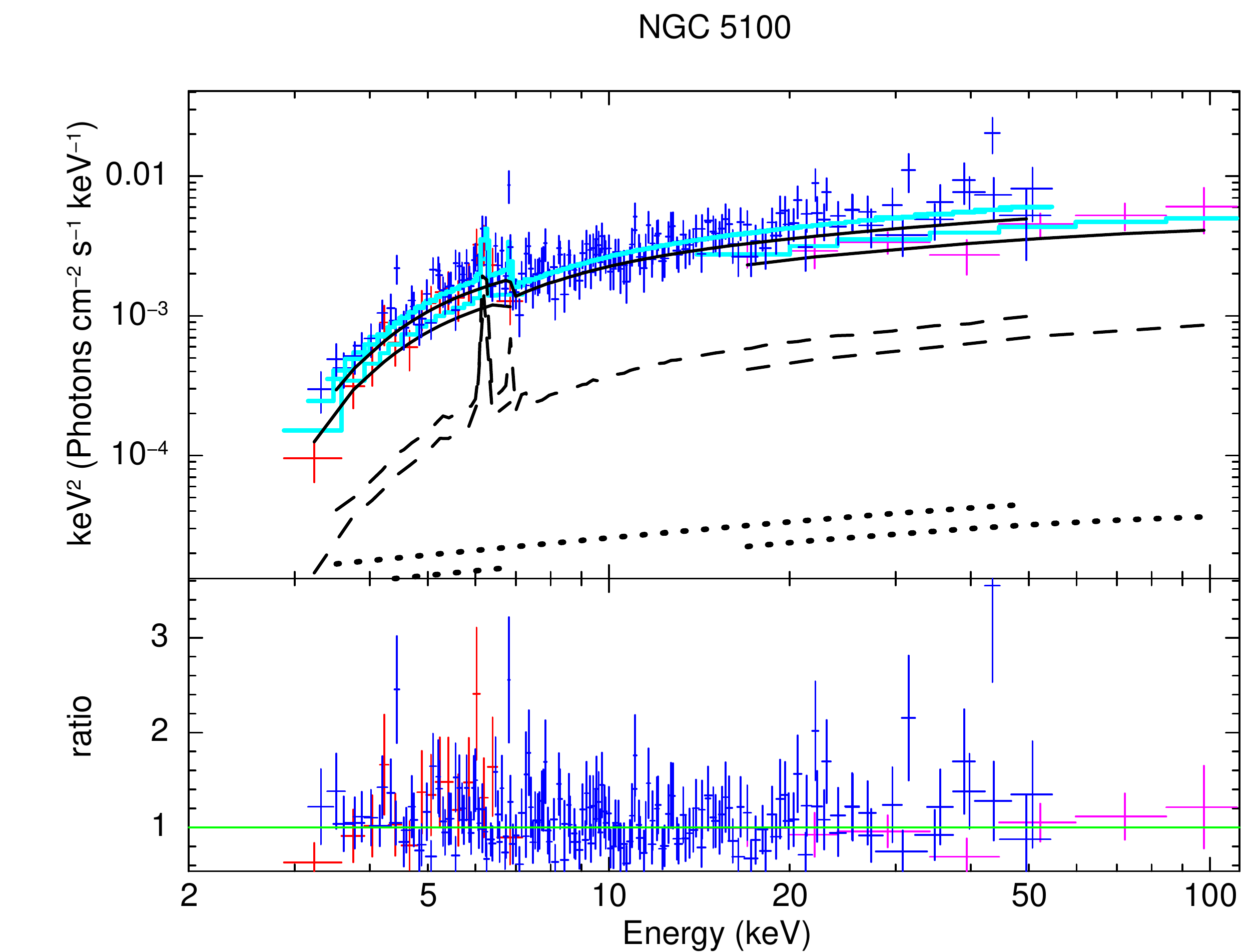}
  \end{minipage}
  \begin{minipage}[b]{.5\textwidth}
  \centering
  \includegraphics[width=0.78\textwidth,angle=-90]{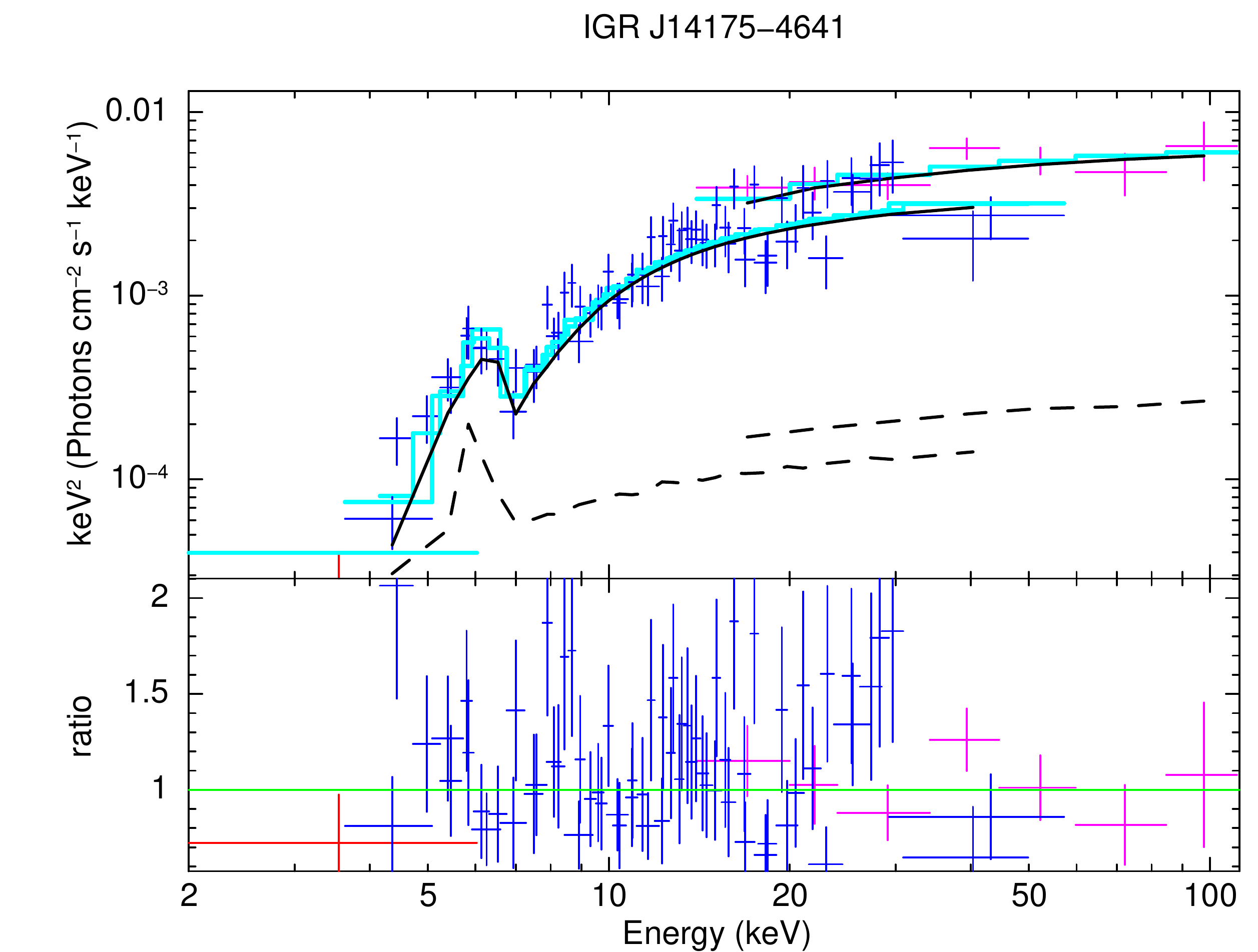}
  \end{minipage}
\caption{\normalsize \normalsize Background-subtracted spectra (top panel) and data-to-model ratio (bottom) of the CT-AGNs analyzed in this work using \borus\ with the covering factor as a free parameter. 2--10\,keV data are plotted in red, \nustar\ data in blue and \swi\ data in magenta. The best-fitting model is plotted as a cyan solid line, the AGN main continuum is plotted as a black solid line, while the reprocessed component modelled by \borus\ and other additional emission lines are plotted as a black dashed line. Finally, the main power law component scattered, rather than absorbed, by the torus is plotted as a black dotted line.}
\end{figure*}

\begin{figure*}
\begin{minipage}[b]{.5\textwidth}
  \centering
  \includegraphics[width=0.78\textwidth,angle=-90]{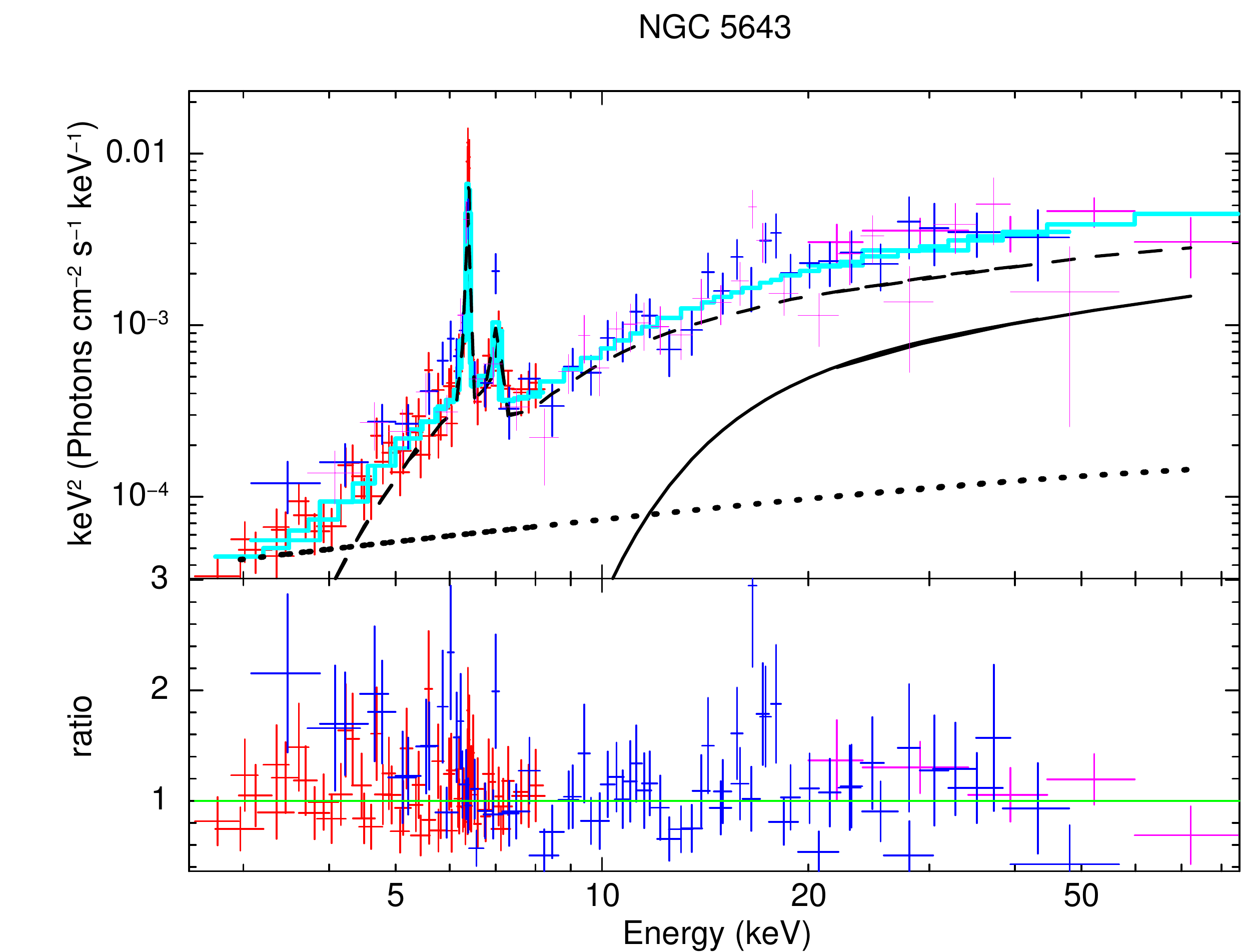}
  \end{minipage}
\begin{minipage}[b]{.5\textwidth}
  \centering
  \includegraphics[width=0.78\textwidth,angle=-90]{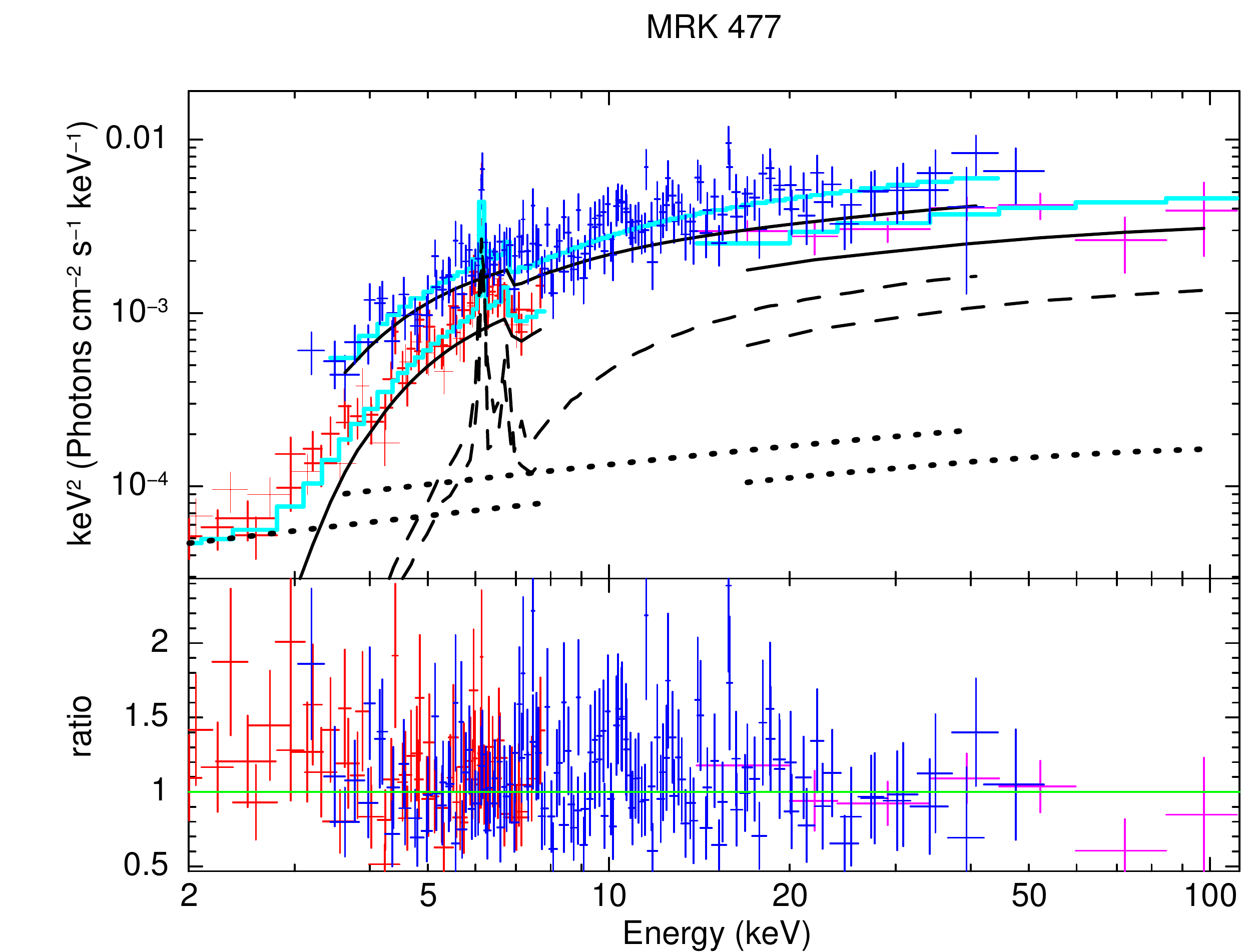}
  \end{minipage}
\begin{minipage}[b]{.5\textwidth}
  \centering
  \includegraphics[width=0.78\textwidth,angle=-90]{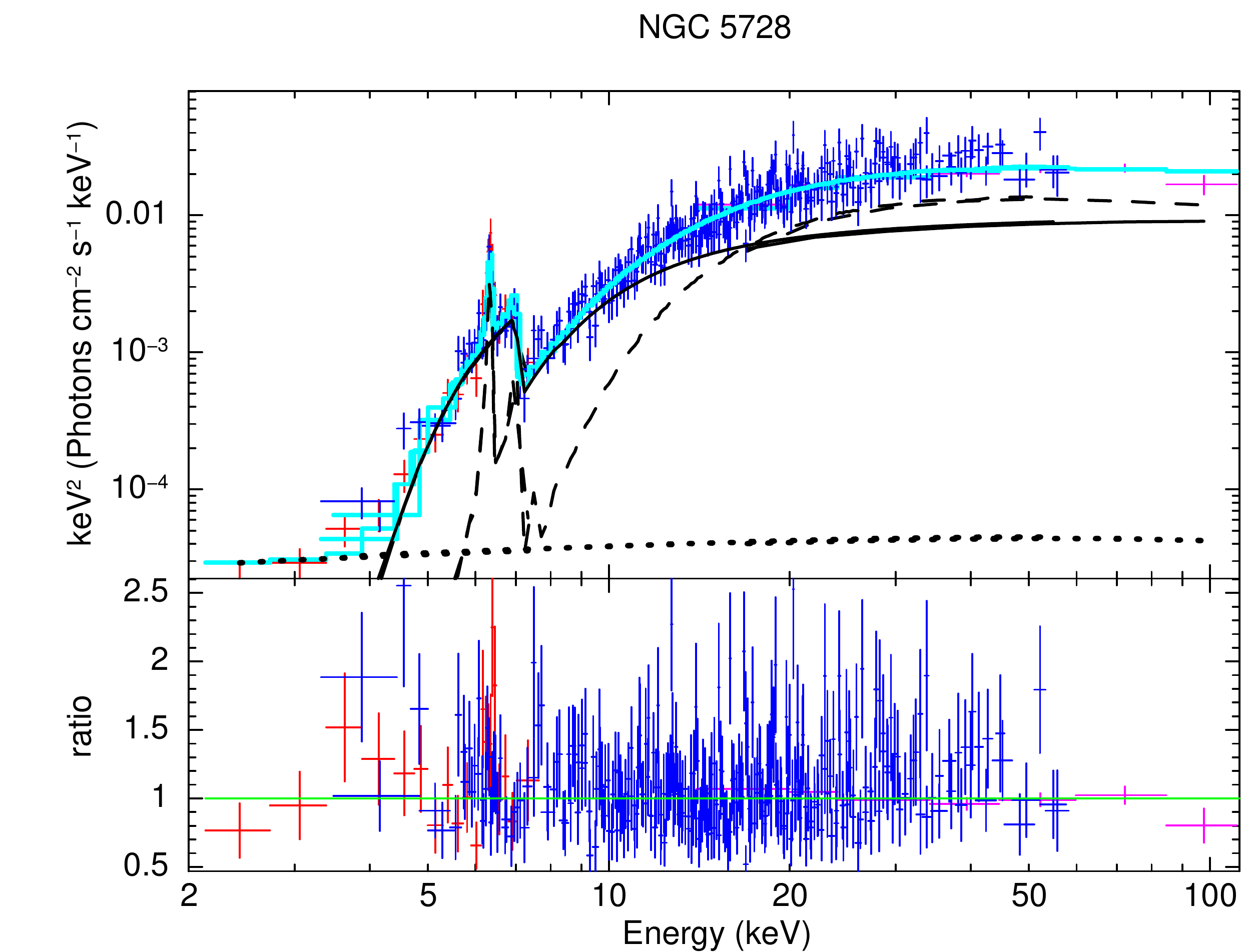}
  \end{minipage}
\begin{minipage}[b]{.5\textwidth}
  \centering
  \includegraphics[width=0.78\textwidth,angle=-90]{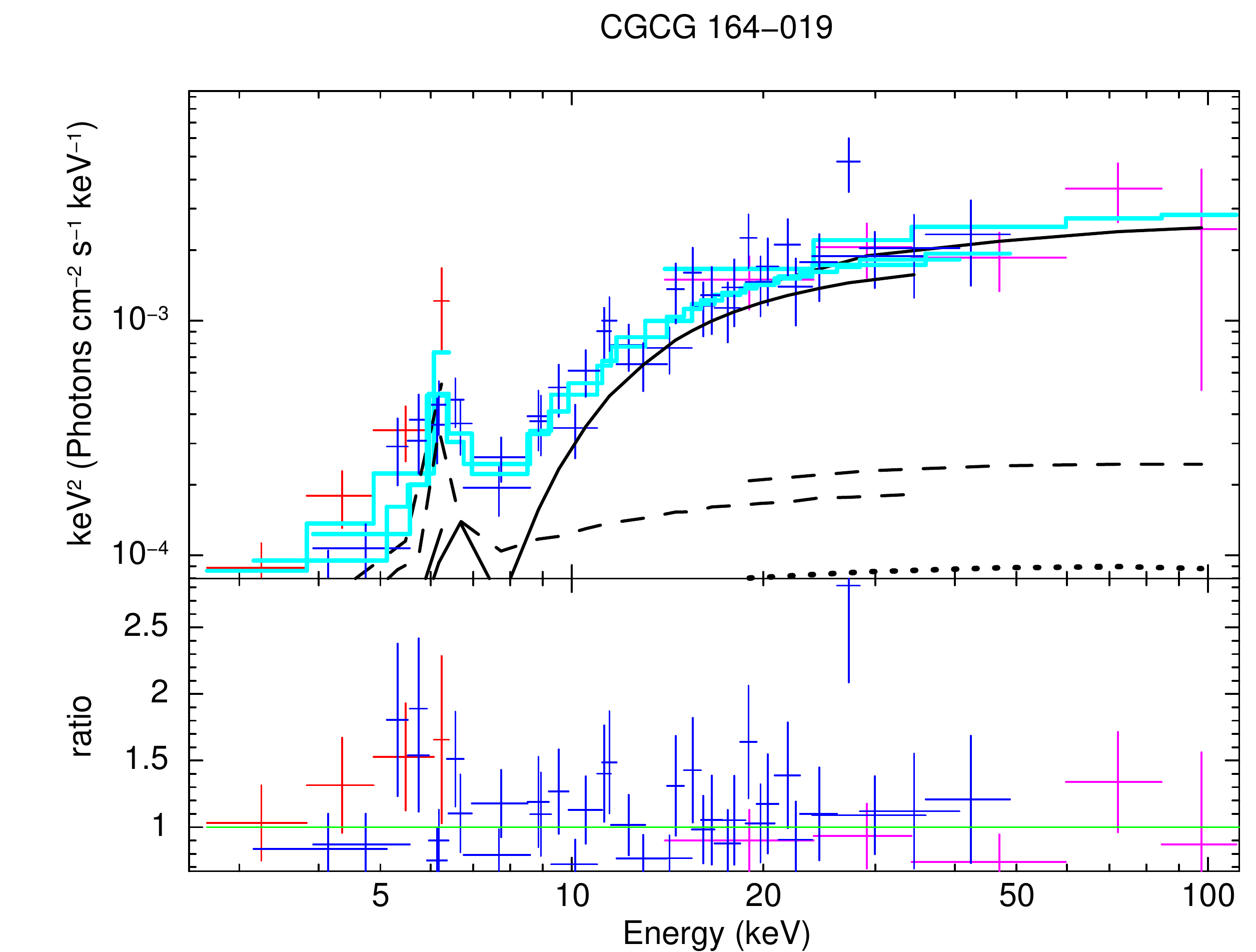}
  \end{minipage}
\begin{minipage}[b]{.5\textwidth}
  \centering
  \includegraphics[width=0.78\textwidth,angle=-90]{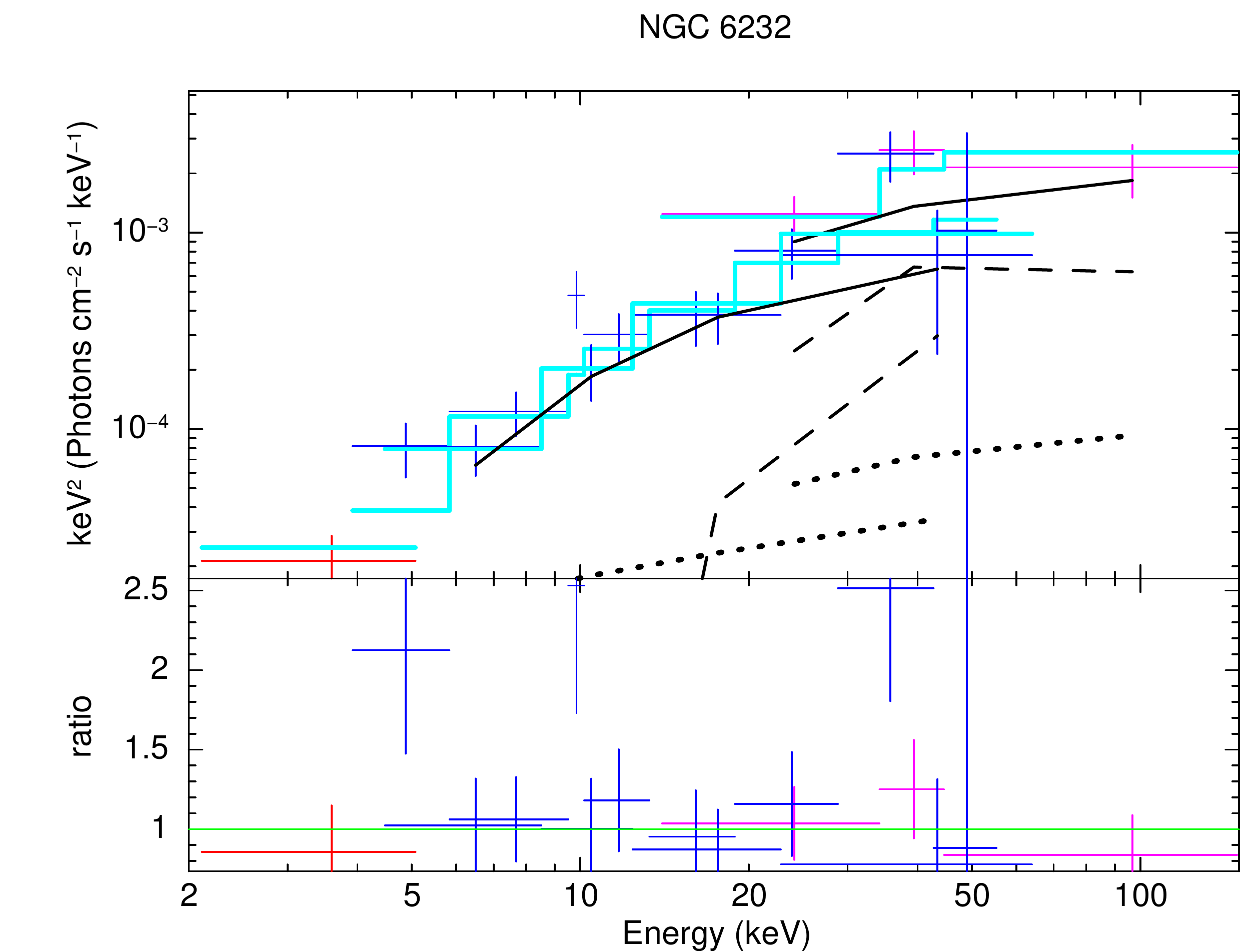}
  \end{minipage}
  \begin{minipage}[b]{.5\textwidth}
  \centering
  \includegraphics[width=0.78\textwidth,angle=-90]{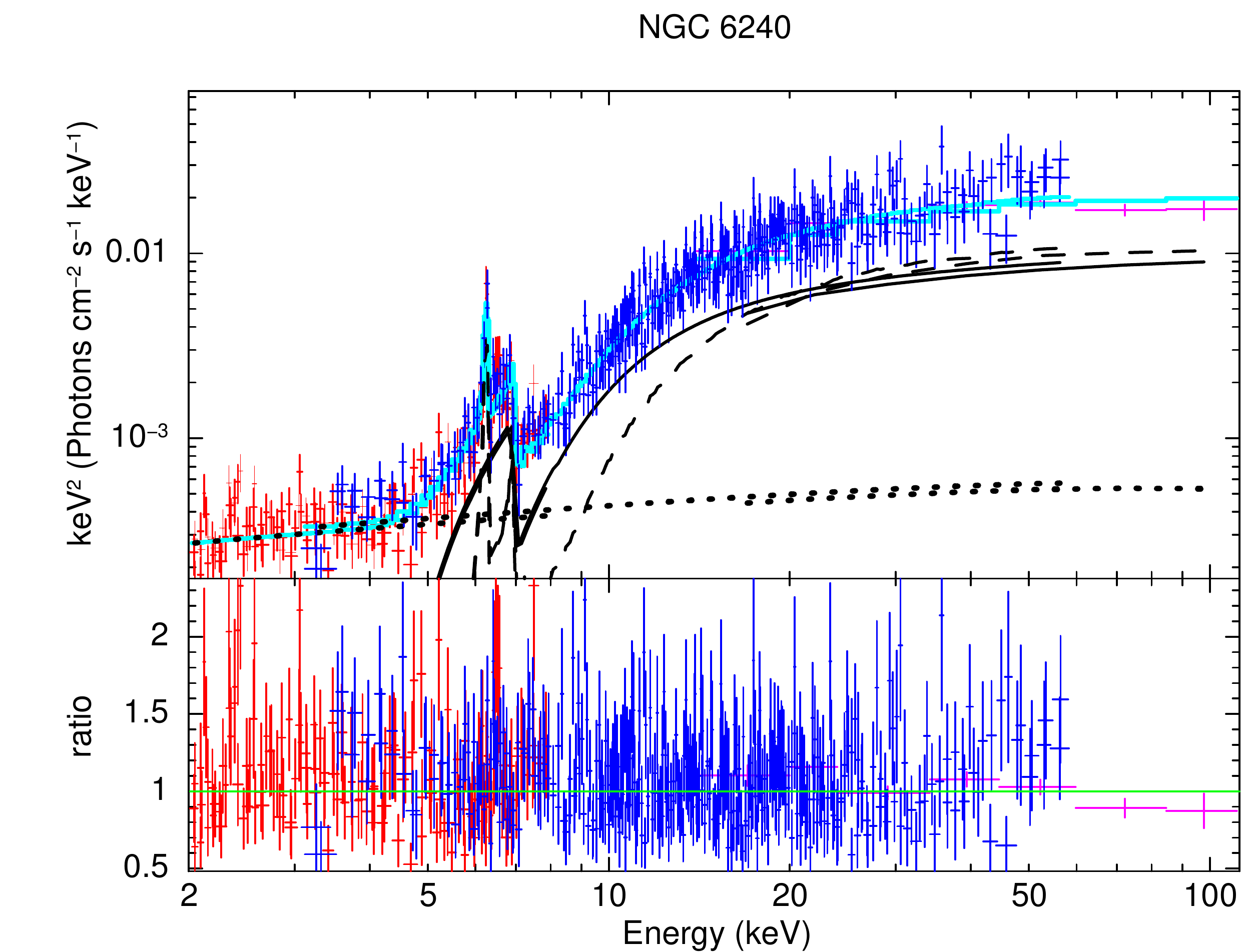}
  \end{minipage}
\caption{\normalsize \normalsize Background-subtracted spectra (top panel) and data-to-model ratio (bottom) of the CT-AGNs analyzed in this work using \borus\ with the covering factor as a free parameter. 2--10\,keV data are plotted in red, \nustar\ data in blue and \swi\ data in magenta. The best-fitting model is plotted as a cyan solid line, the AGN main continuum is plotted as a black solid line, while the reprocessed component modelled by \borus\ and other additional emission lines are plotted as a black dashed line. Finally, the main power law component scattered, rather than absorbed, by the torus is plotted as a black dotted line.}
\end{figure*}

\section{B. Confidence contours of the torus covering factor  versus the line-of-sight column density}\label{app:nhz_vs_fc}
\begin{figure*}
\begin{minipage}[b]{.5\textwidth}
  \centering
  \includegraphics[width=0.78\textwidth,angle=-90]{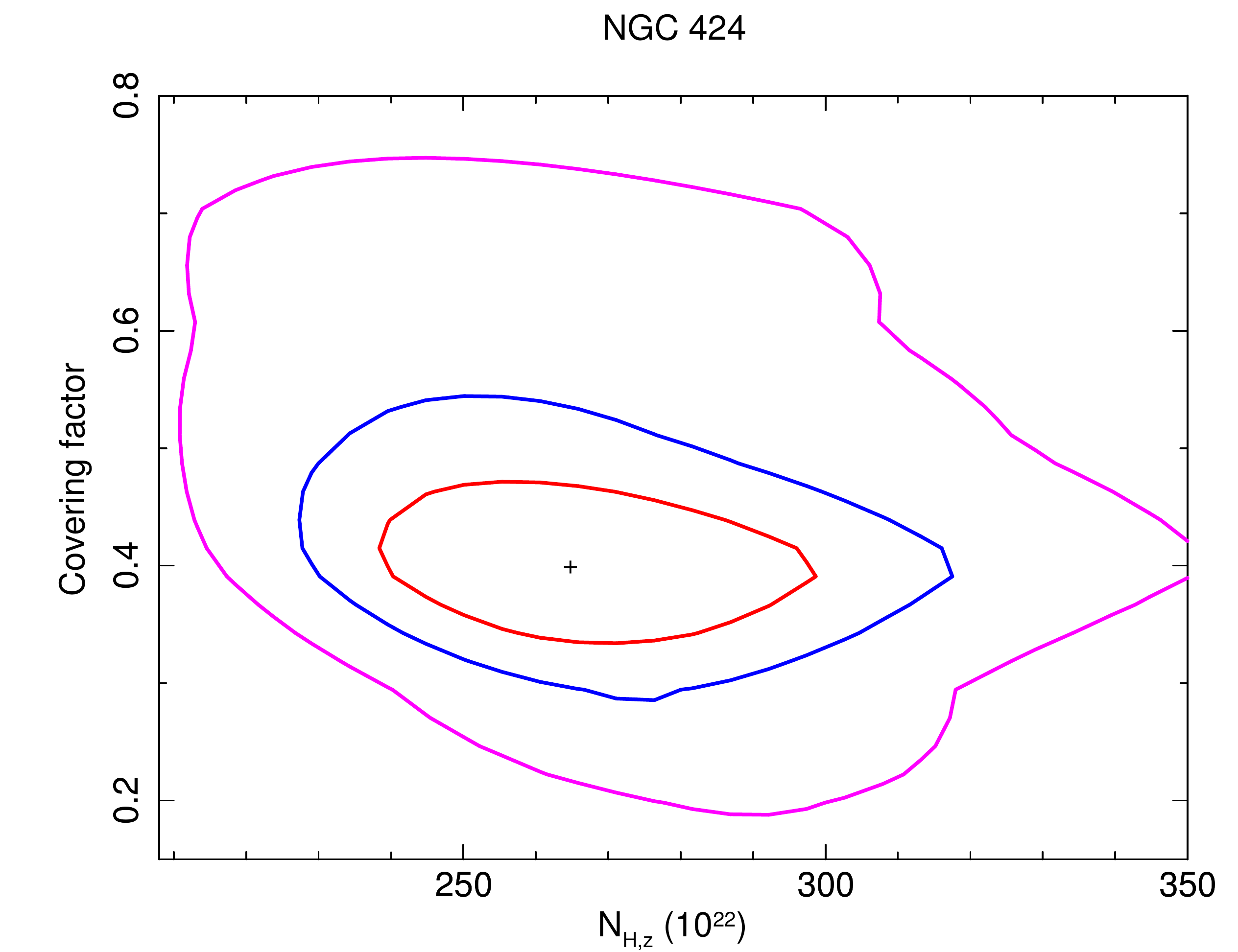}
  \end{minipage}
\begin{minipage}[b]{.5\textwidth}
  \centering
  \includegraphics[width=0.78\textwidth,angle=-90]{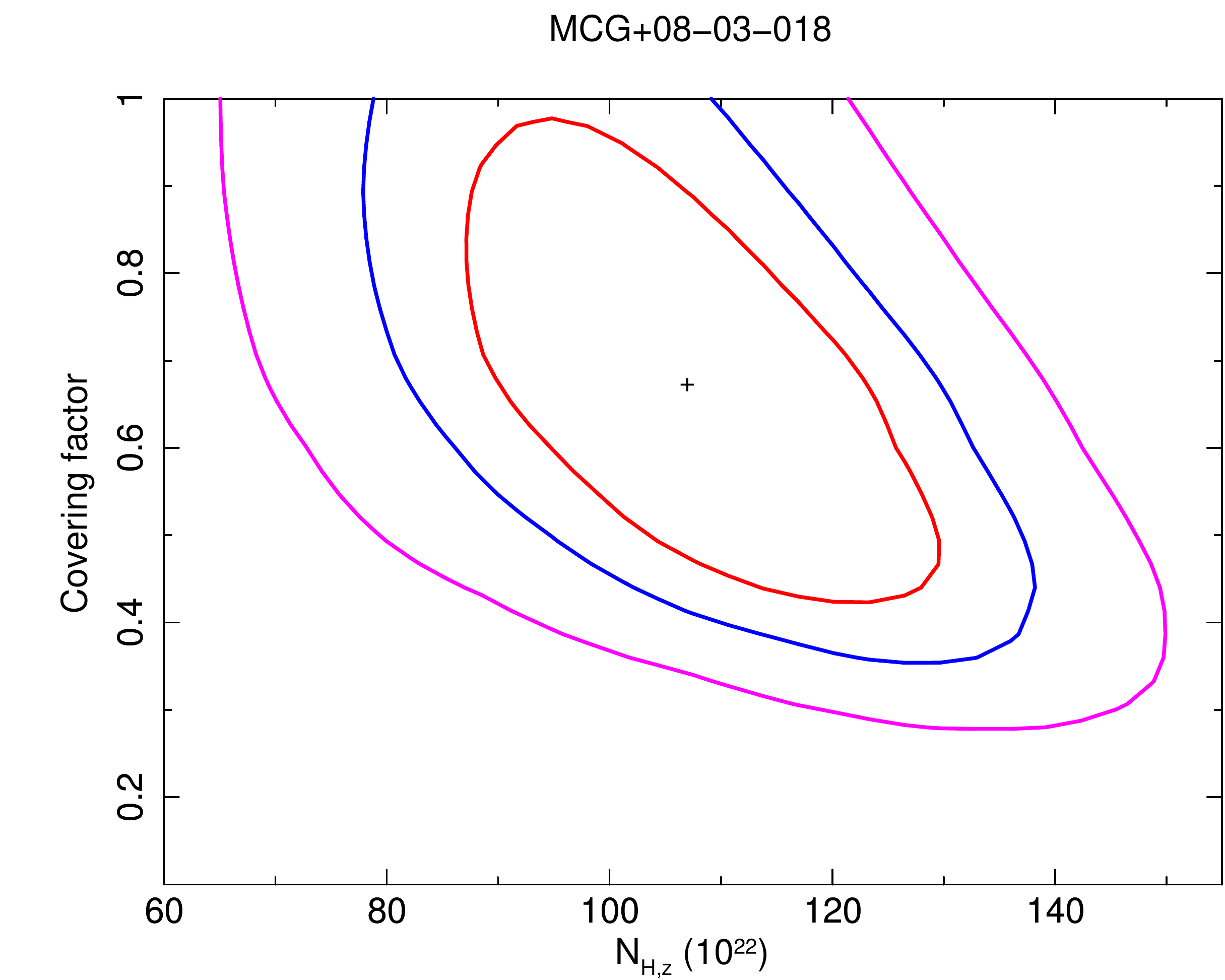}
  \end{minipage}
\begin{minipage}[b]{.5\textwidth}
  \centering
  \includegraphics[width=0.78\textwidth,angle=-90]{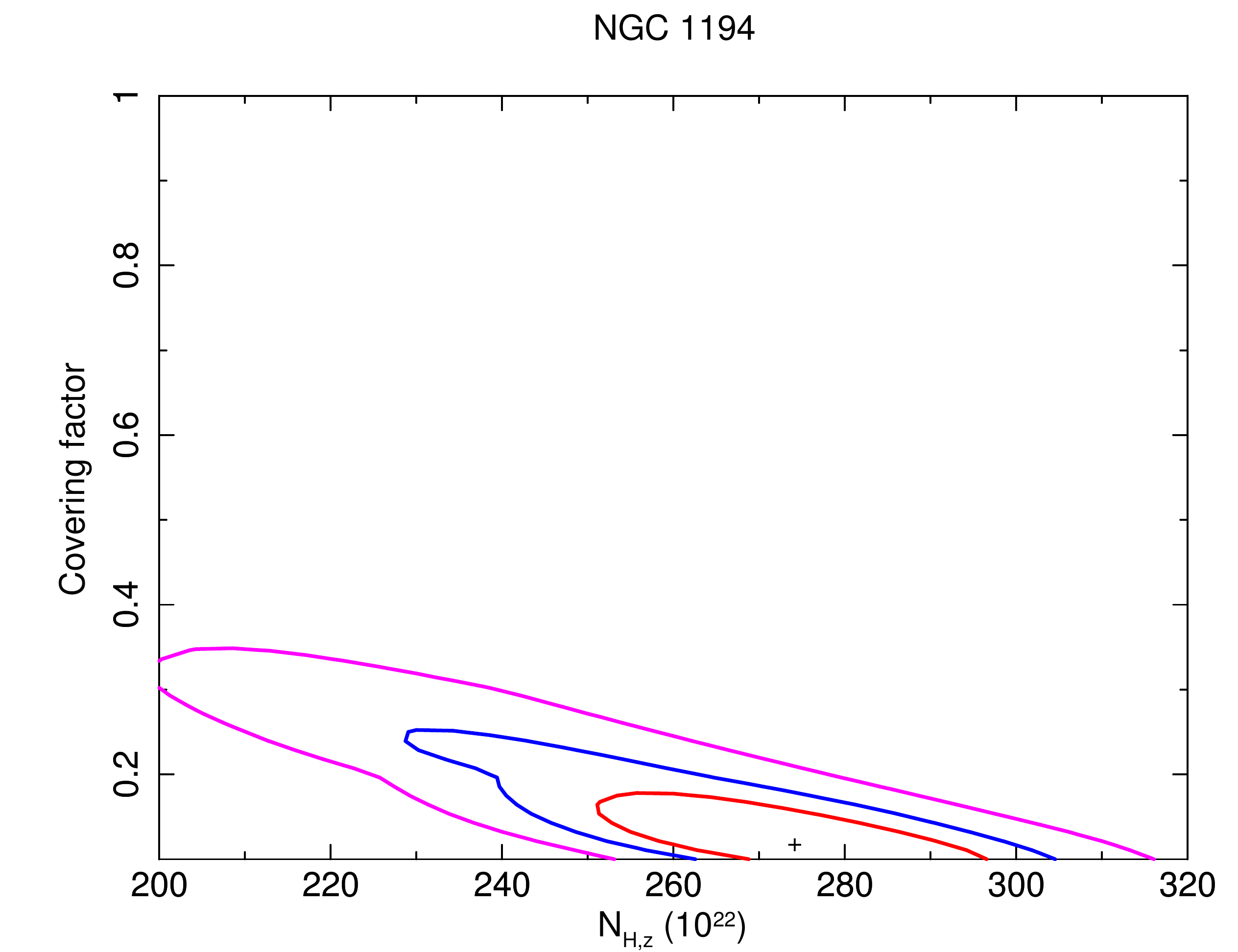}
  \end{minipage}
\begin{minipage}[b]{.5\textwidth}
  \centering
  \includegraphics[width=0.78\textwidth,angle=-90]{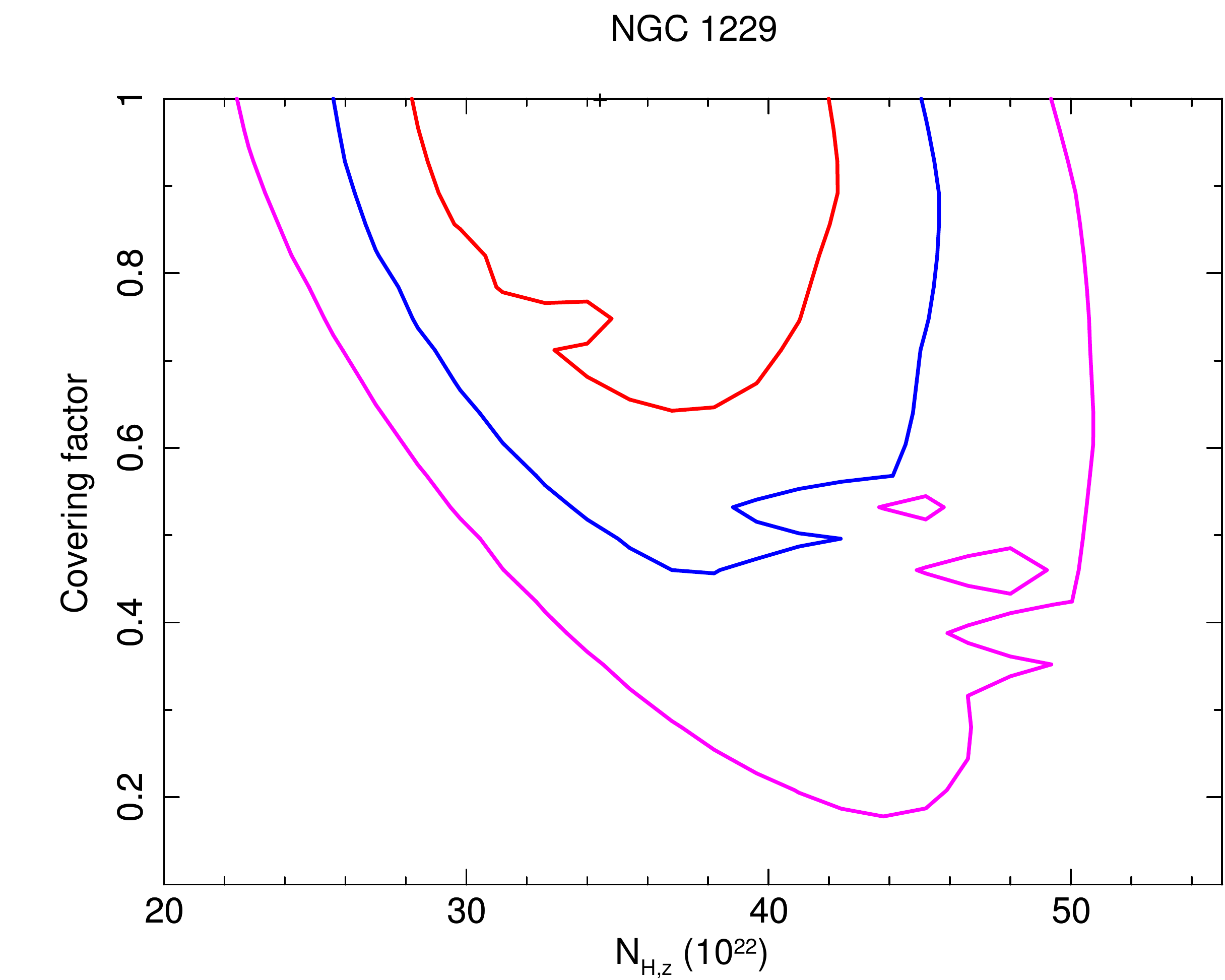}
  \end{minipage}
  \begin{minipage}[b]{.5\textwidth}
  \centering
  \includegraphics[width=0.78\textwidth,angle=-90]{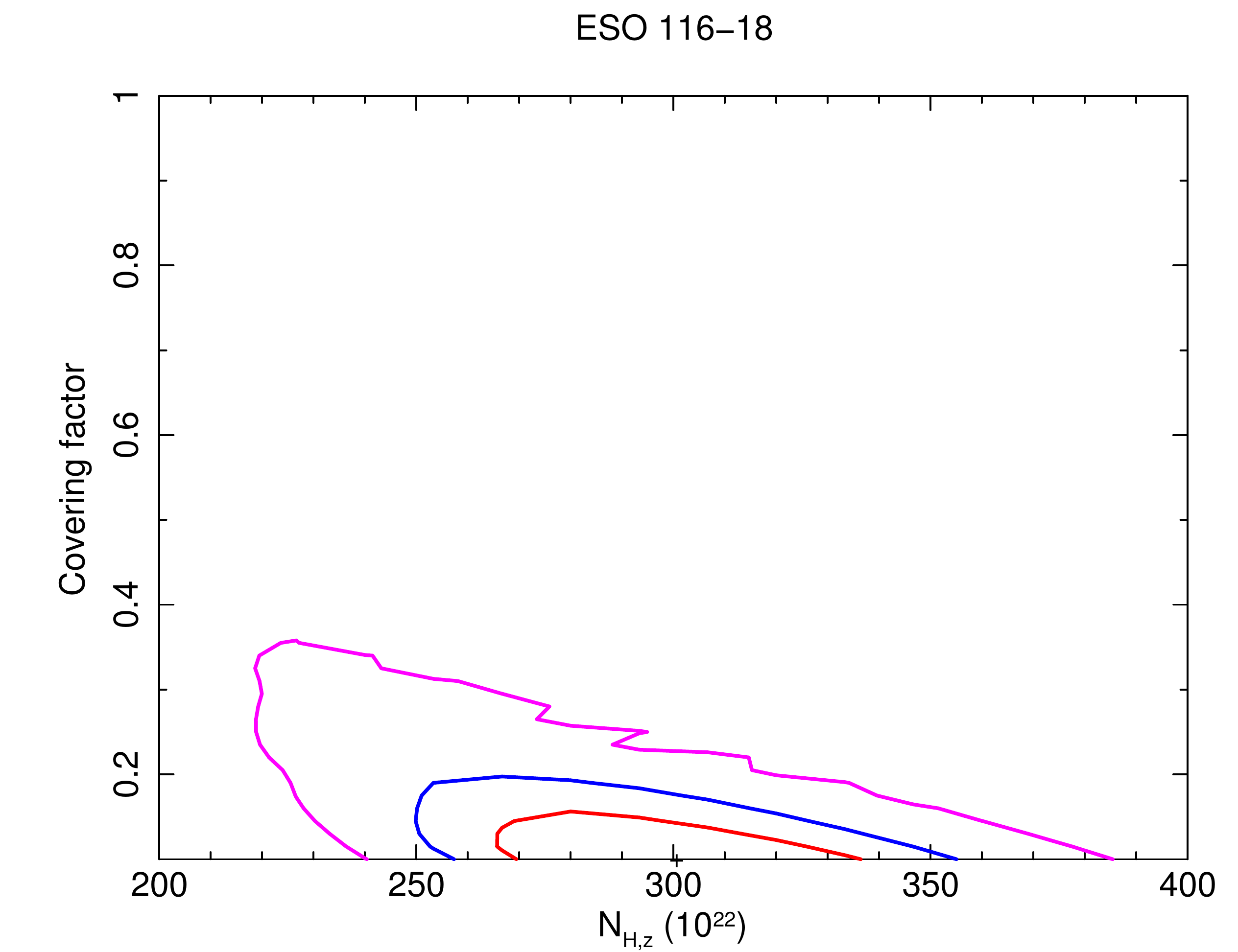}
  \end{minipage}
  \begin{minipage}[b]{.5\textwidth}
  \centering
  \includegraphics[width=0.78\textwidth,angle=-90]{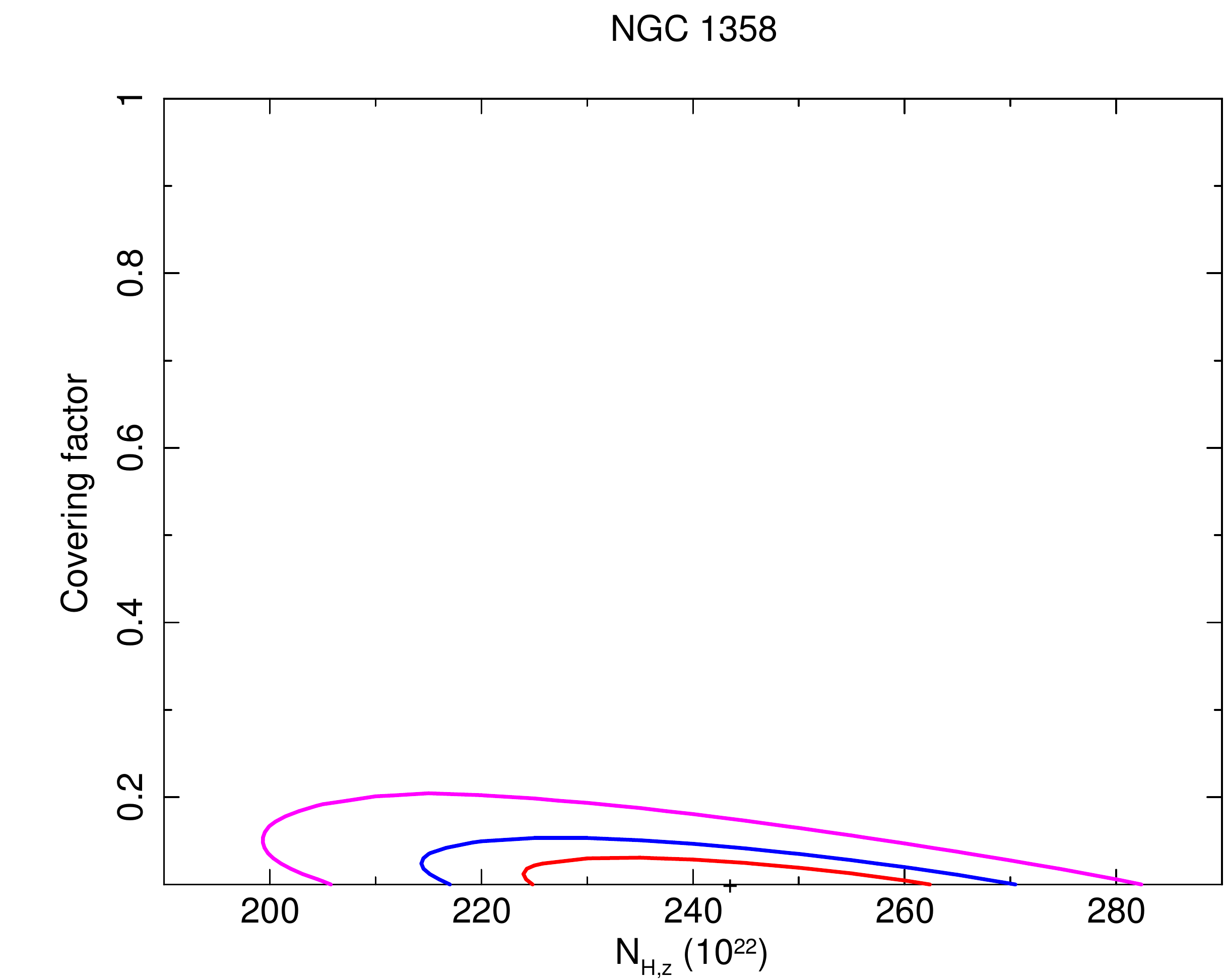}
  \end{minipage}
\caption{\normalsize \normalsize Confidence contours at 68, 90 and 99\% confidence level for the line-of-sight column density, $N_{\rm H, z}$, and the torus covering factor, $f_c$, for six of the 35 sources analyzed in this work.}\label{fig:NHz_vs_fc}
\end{figure*}

\begin{figure*}
\begin{minipage}[b]{.5\textwidth}
  \centering
  \includegraphics[width=0.78\textwidth,angle=-90]{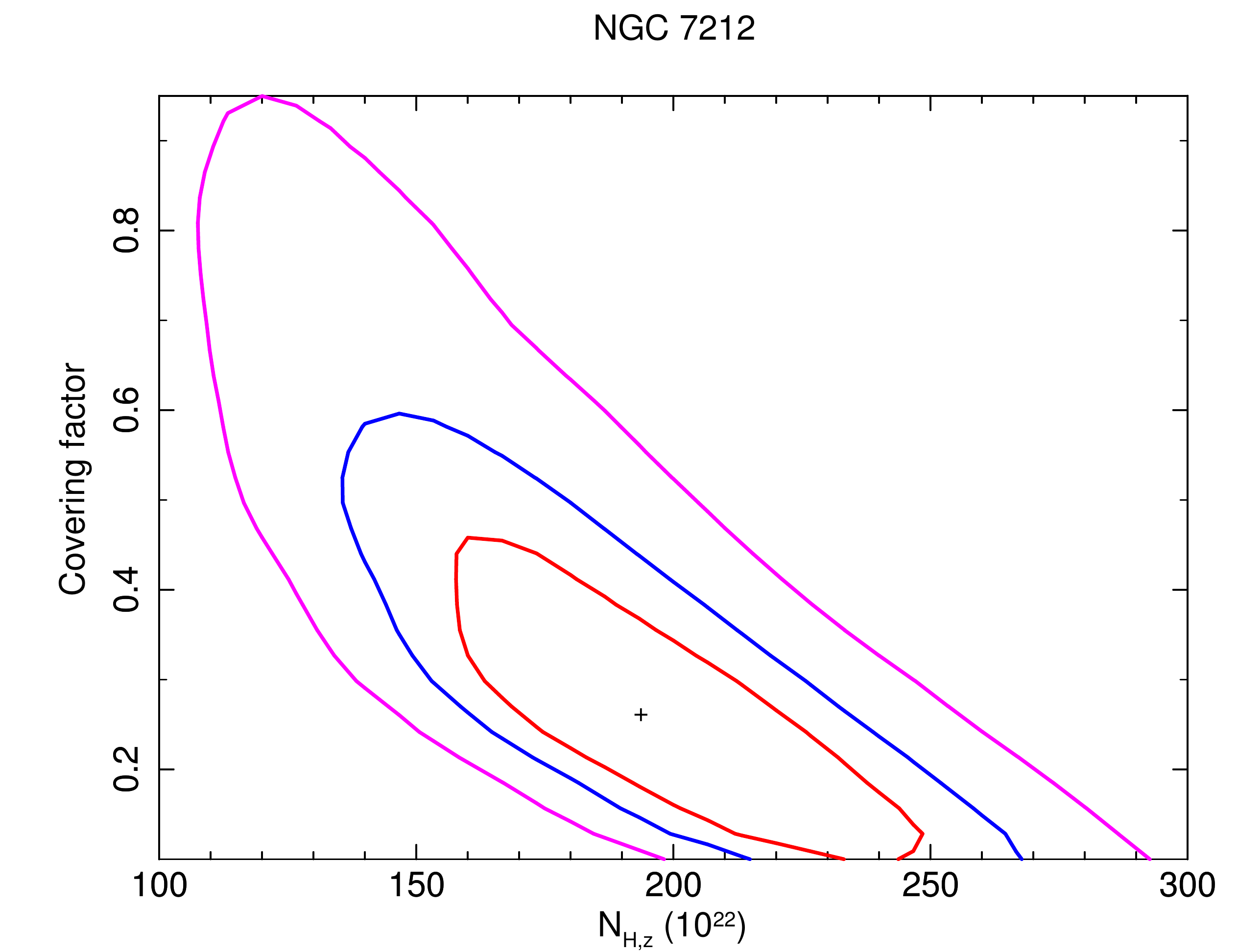}
  \end{minipage}
\begin{minipage}[b]{.5\textwidth}
  \centering
  \includegraphics[width=0.78\textwidth,angle=-90]{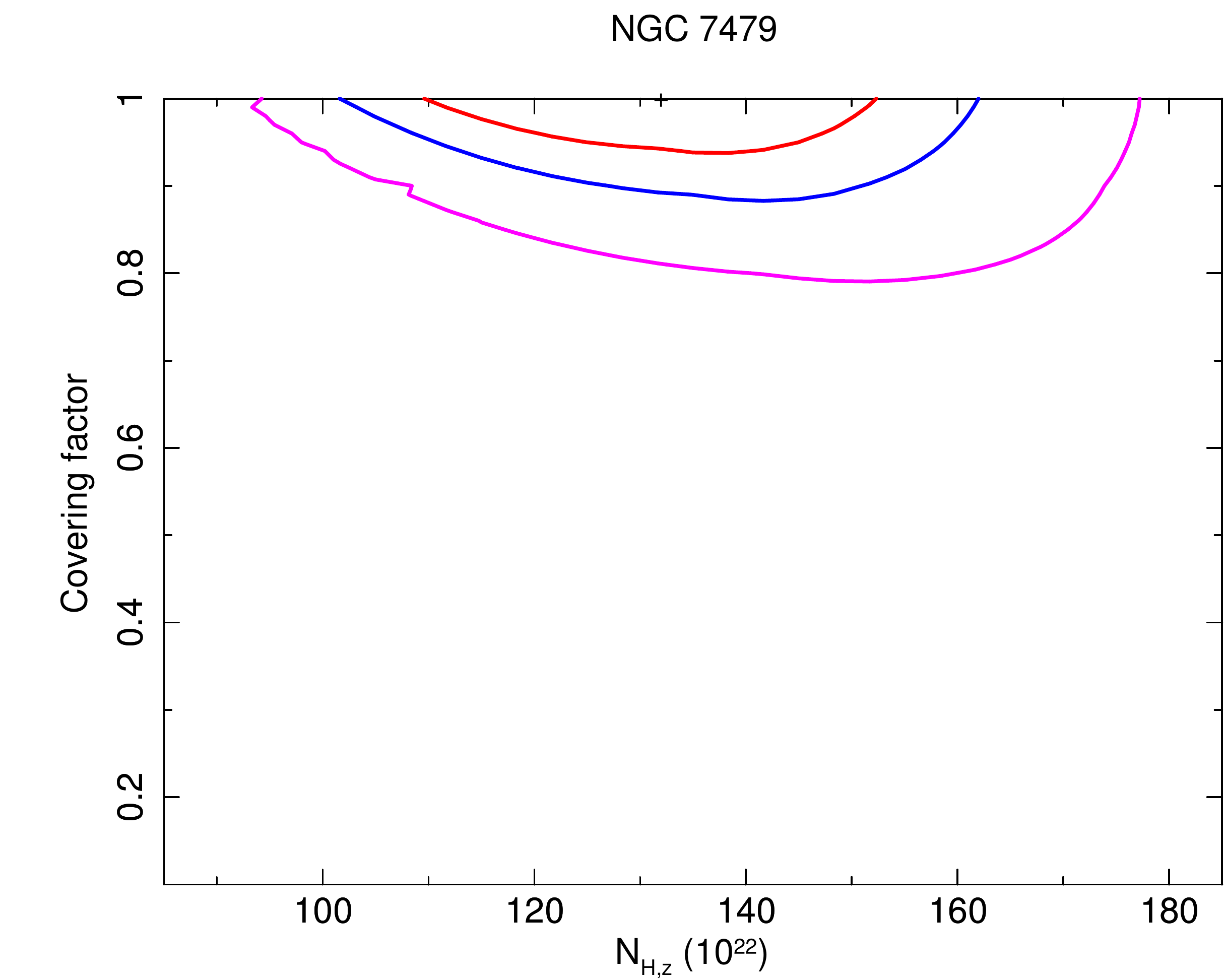}
  \end{minipage}
\caption{\normalsize \normalsize Confidence contours at 68, 90 and 99\% confidence level for the line-of-sight column density, $N_{\rm H, z}$, and the torus covering factor, $f_c$, for two of the 35 sources analyzed in this work.}\label{fig:NHz_vs_fc_last}
\end{figure*}

We report in Figures \ref{fig:NHz_vs_fc}--\ref{fig:NHz_vs_fc_last} the confidence contours of the covering factor, $f_c$, versus the line-of-sight column density, $N_{\rm H, z}$, for 32 sources out of the 35 in our sample. We do not report the contours of NGC 1068 and NGC 7582, where we fix the line-of-sight column density to $N_{\rm H, z}$=10$^{25}$ cm$^{-2}$, and of RBS 1037, which is an unobscured AGN where $f_c$ is unconstrained.

\begin{figure*}
\begin{minipage}[b]{.5\textwidth}
  \centering
  \includegraphics[width=0.78\textwidth,angle=-90]{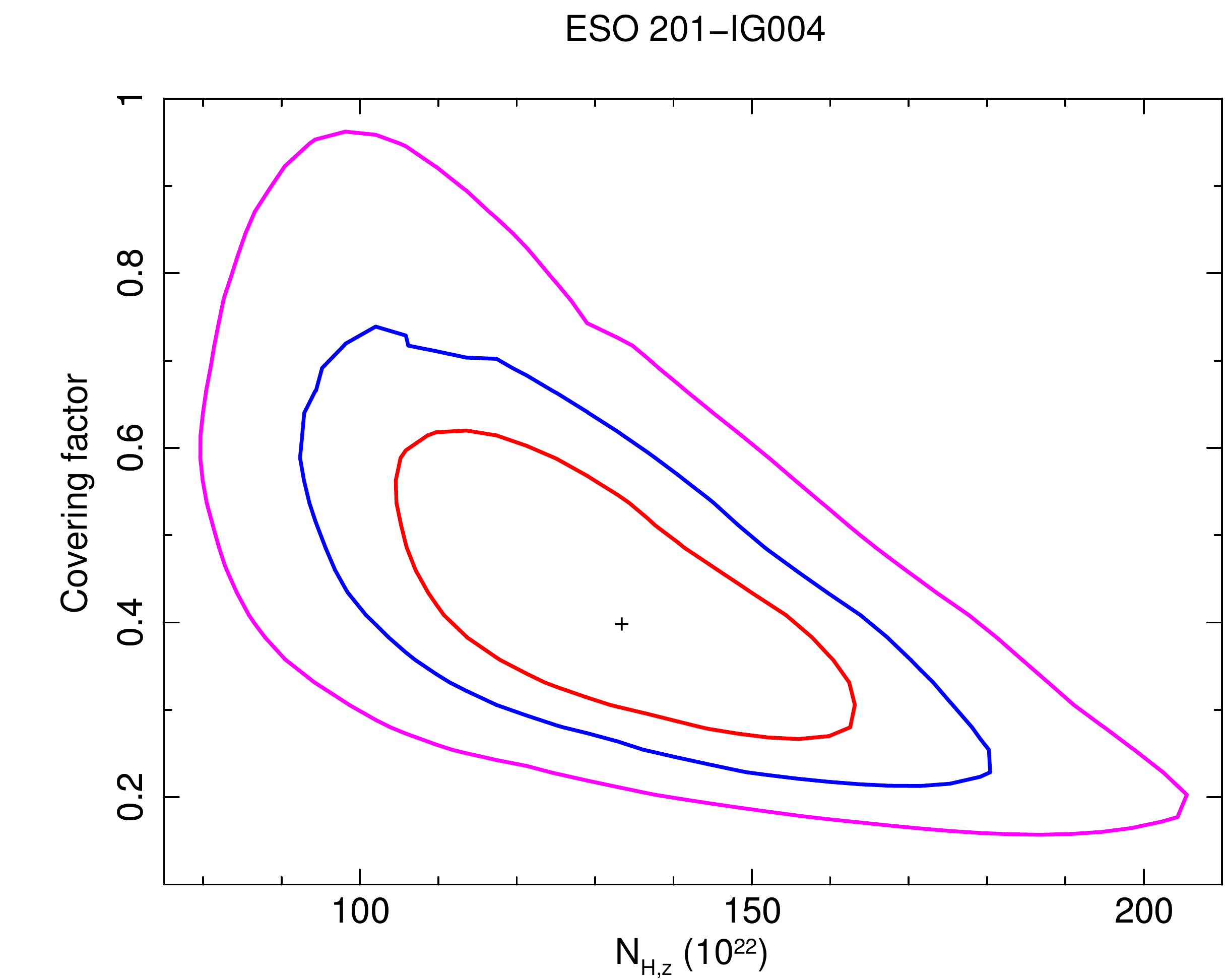}
  \end{minipage}
\begin{minipage}[b]{.5\textwidth}
  \centering
  \includegraphics[width=0.78\textwidth,angle=-90]{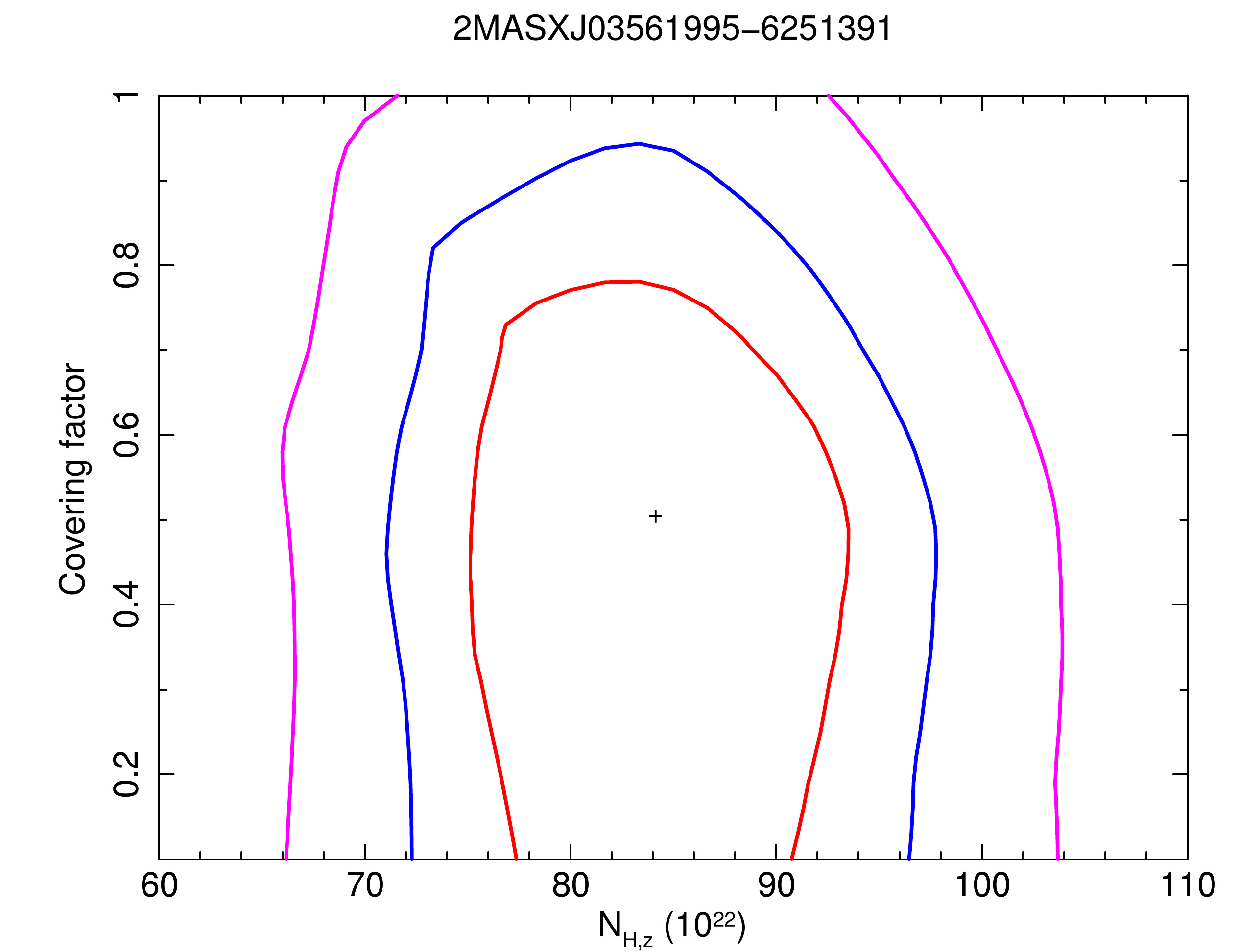}
  \end{minipage}
\begin{minipage}[b]{.5\textwidth}
  \centering
  \includegraphics[width=0.78\textwidth,angle=-90]{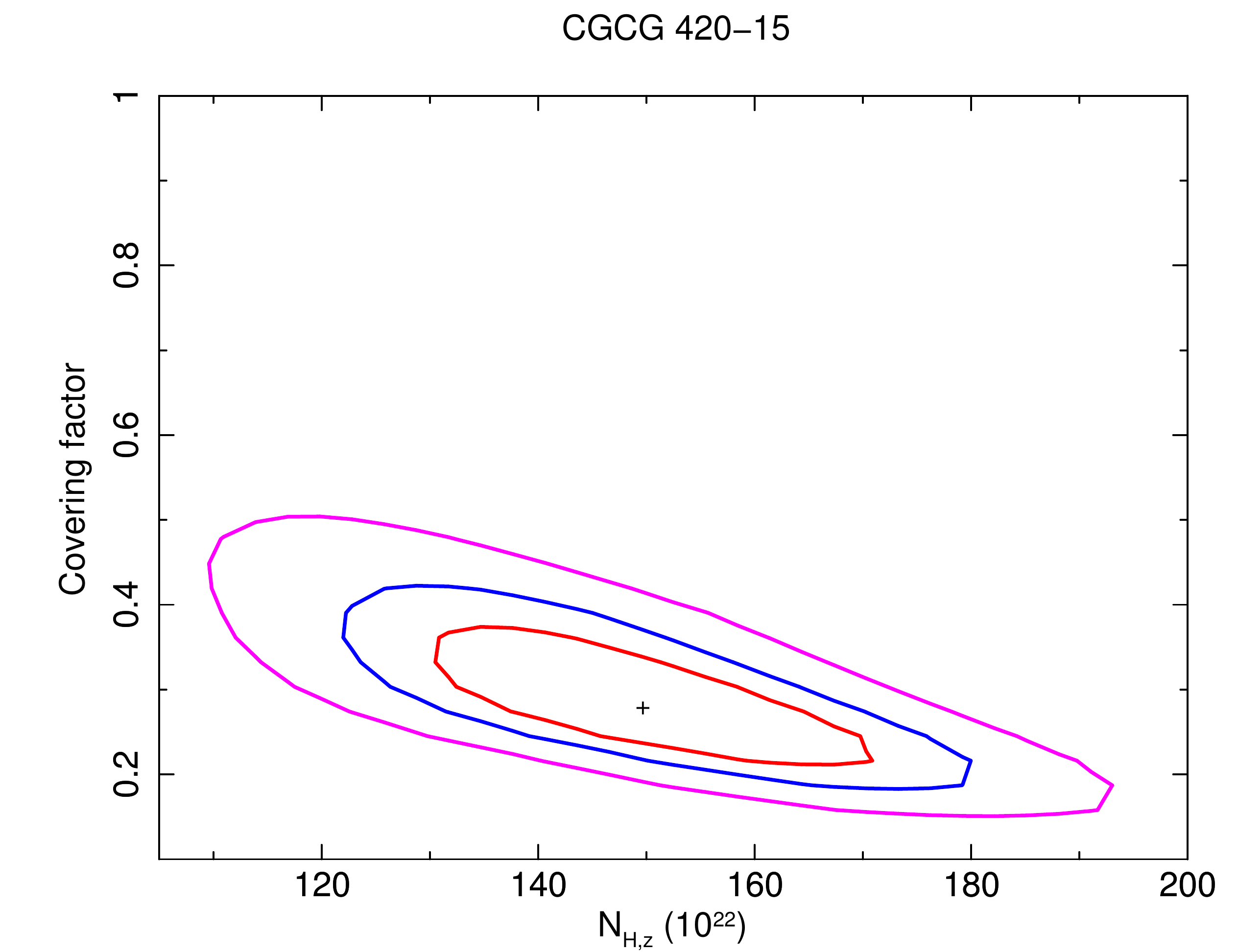}
  \end{minipage}
\begin{minipage}[b]{.5\textwidth}
  \centering
  \includegraphics[width=0.78\textwidth,angle=-90]{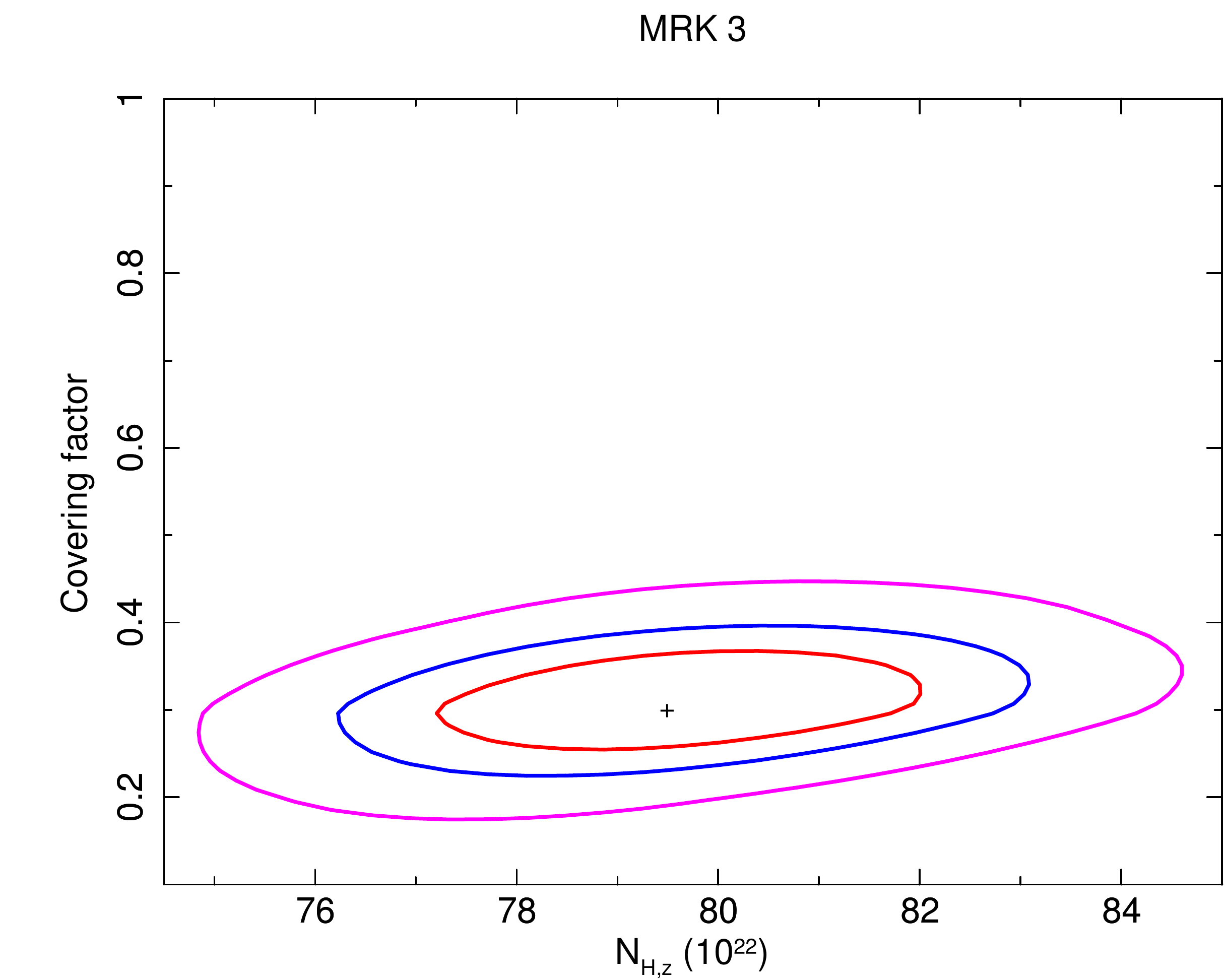}
  \end{minipage}
  \begin{minipage}[b]{.5\textwidth}
  \centering
  \includegraphics[width=0.78\textwidth,angle=-90]{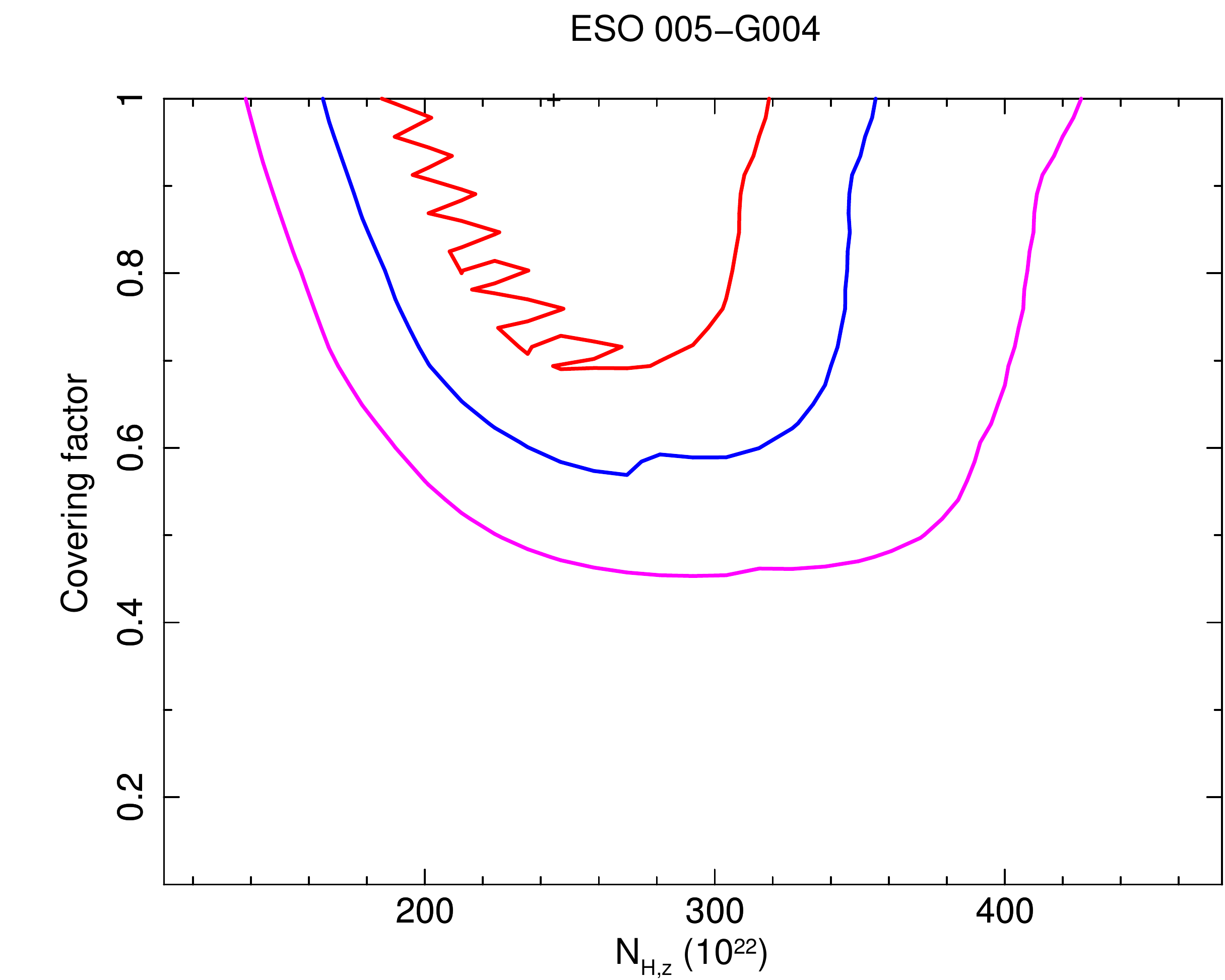}
  \end{minipage}
  \begin{minipage}[b]{.5\textwidth}
  \centering
  \includegraphics[width=0.78\textwidth,angle=-90]{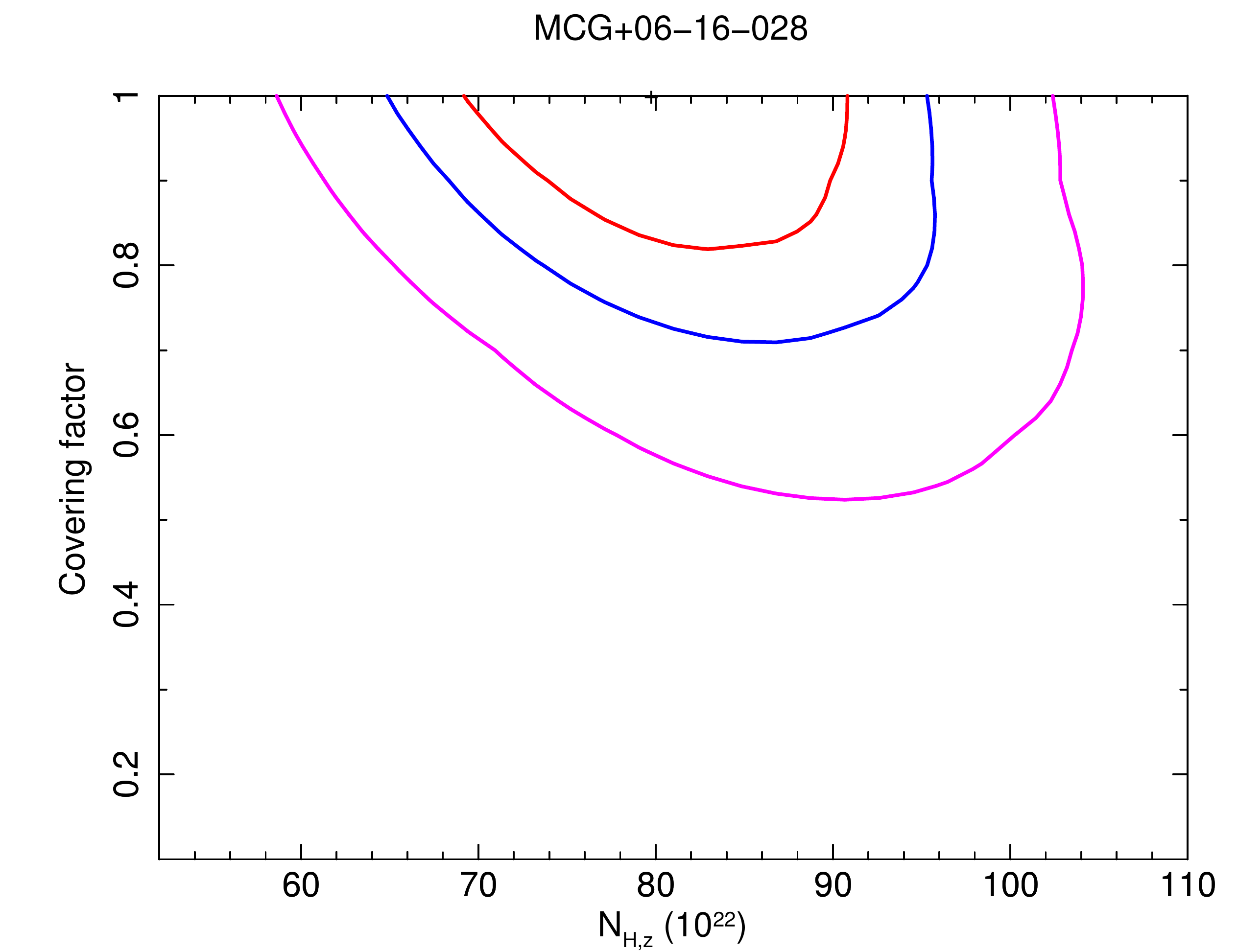}
  \end{minipage}
\caption{\normalsize \normalsize Confidence contours at 68, 90 and 99\% confidence level for the line-of-sight column density, $N_{\rm H, z}$, and the torus covering factor, $f_c$, for six of the 35 sources analyzed in this work.}
\end{figure*}

\begin{figure*}
\begin{minipage}[b]{.5\textwidth}
  \centering
  \includegraphics[width=0.78\textwidth,angle=-90]{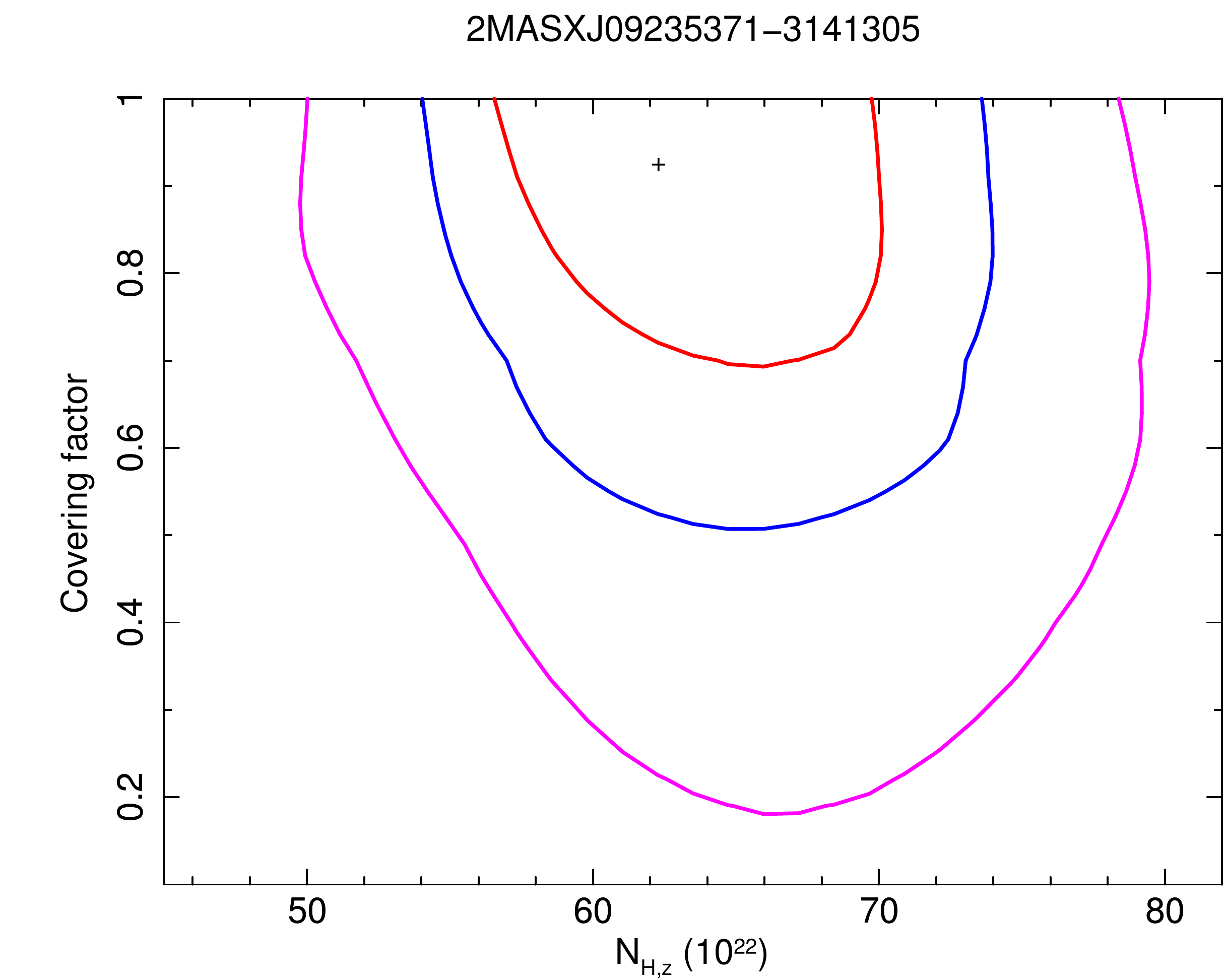}
  \end{minipage}
\begin{minipage}[b]{.5\textwidth}
  \centering
  \includegraphics[width=0.78\textwidth,angle=-90]{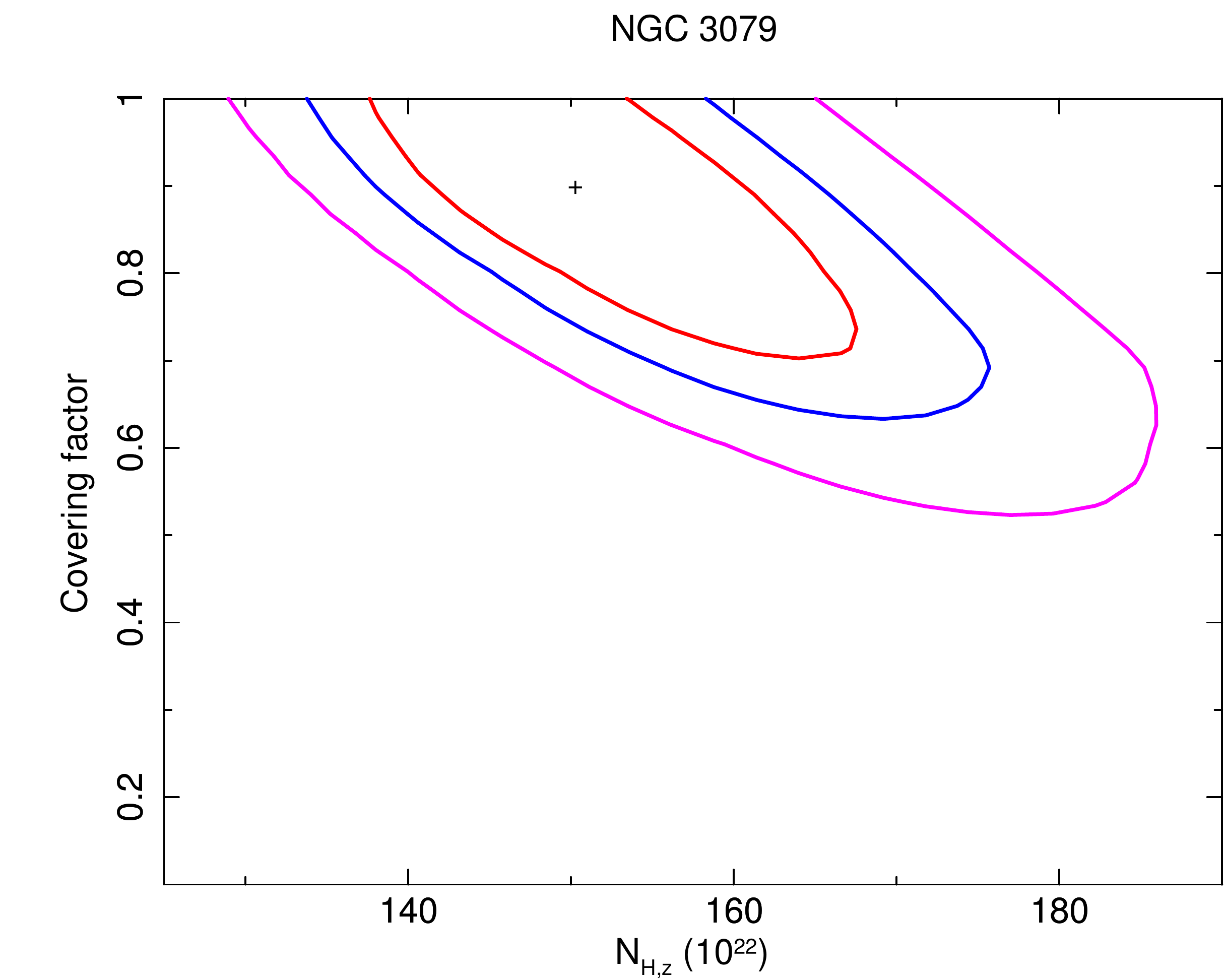}
  \end{minipage}
\begin{minipage}[b]{.5\textwidth}
  \centering
  \includegraphics[width=0.78\textwidth,angle=-90]{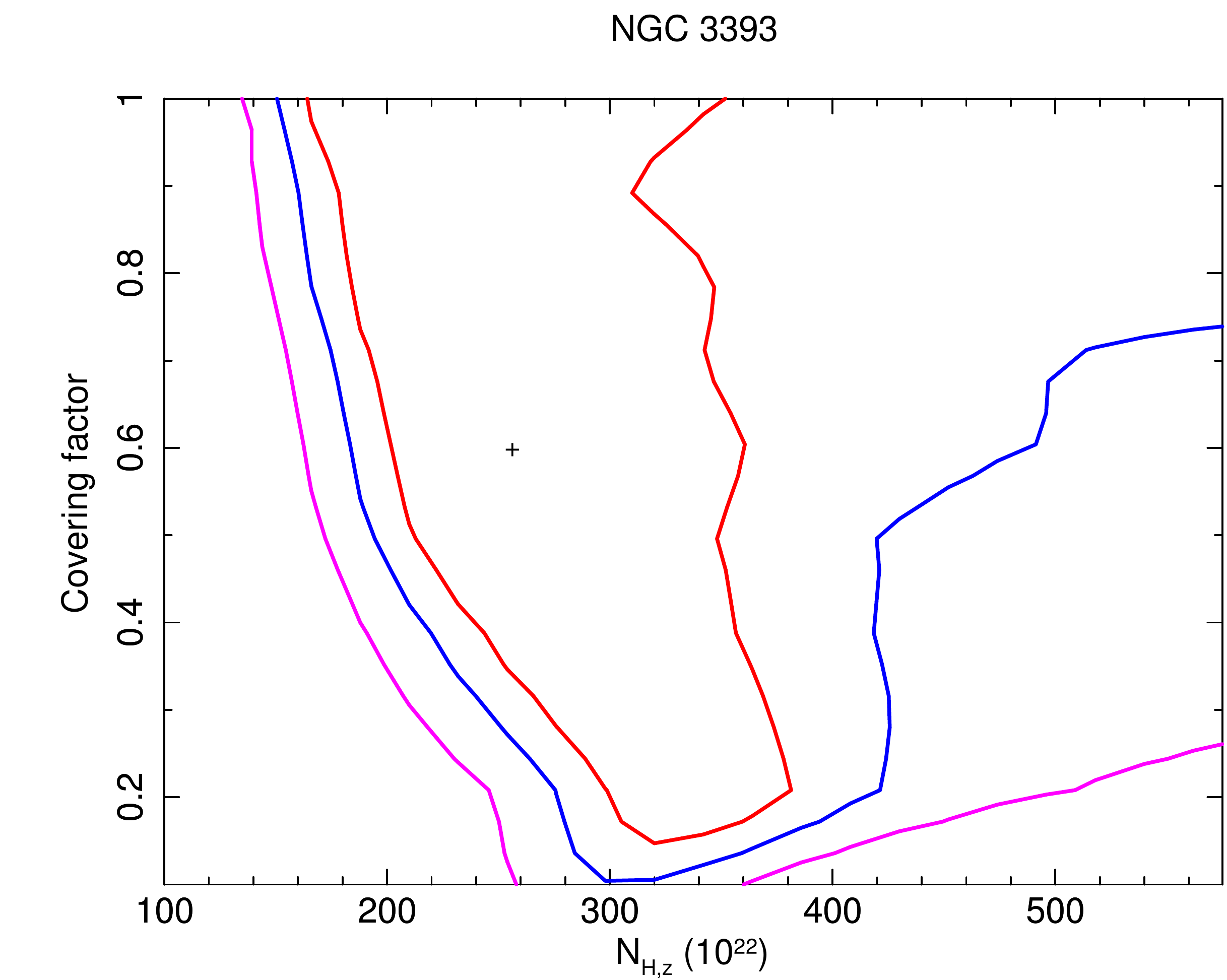}
  \end{minipage}
\begin{minipage}[b]{.5\textwidth}
  \centering
  \includegraphics[width=0.78\textwidth,angle=-90]{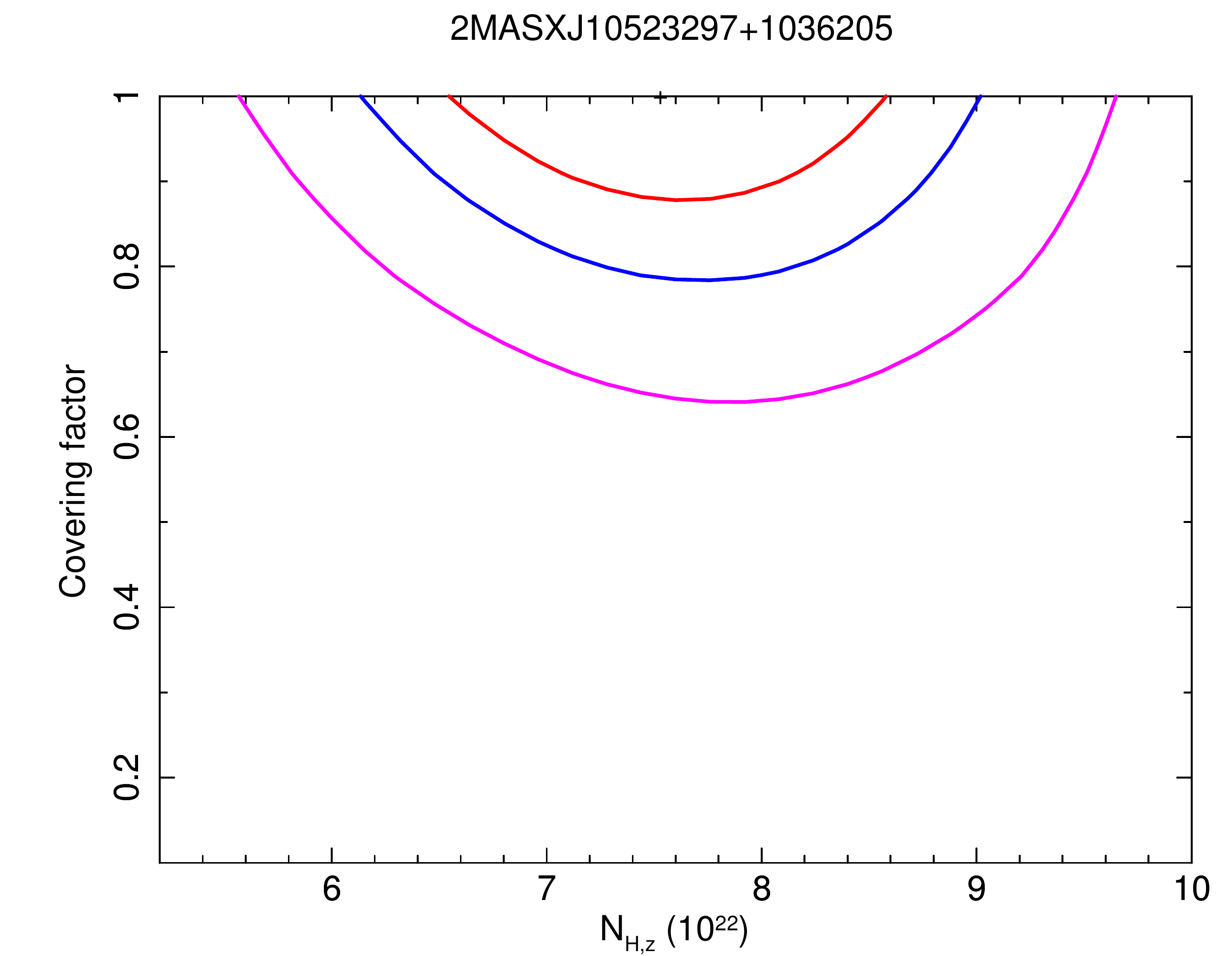}
  \end{minipage}
  \begin{minipage}[b]{.5\textwidth}
  \centering
  \includegraphics[width=0.78\textwidth,angle=-90]{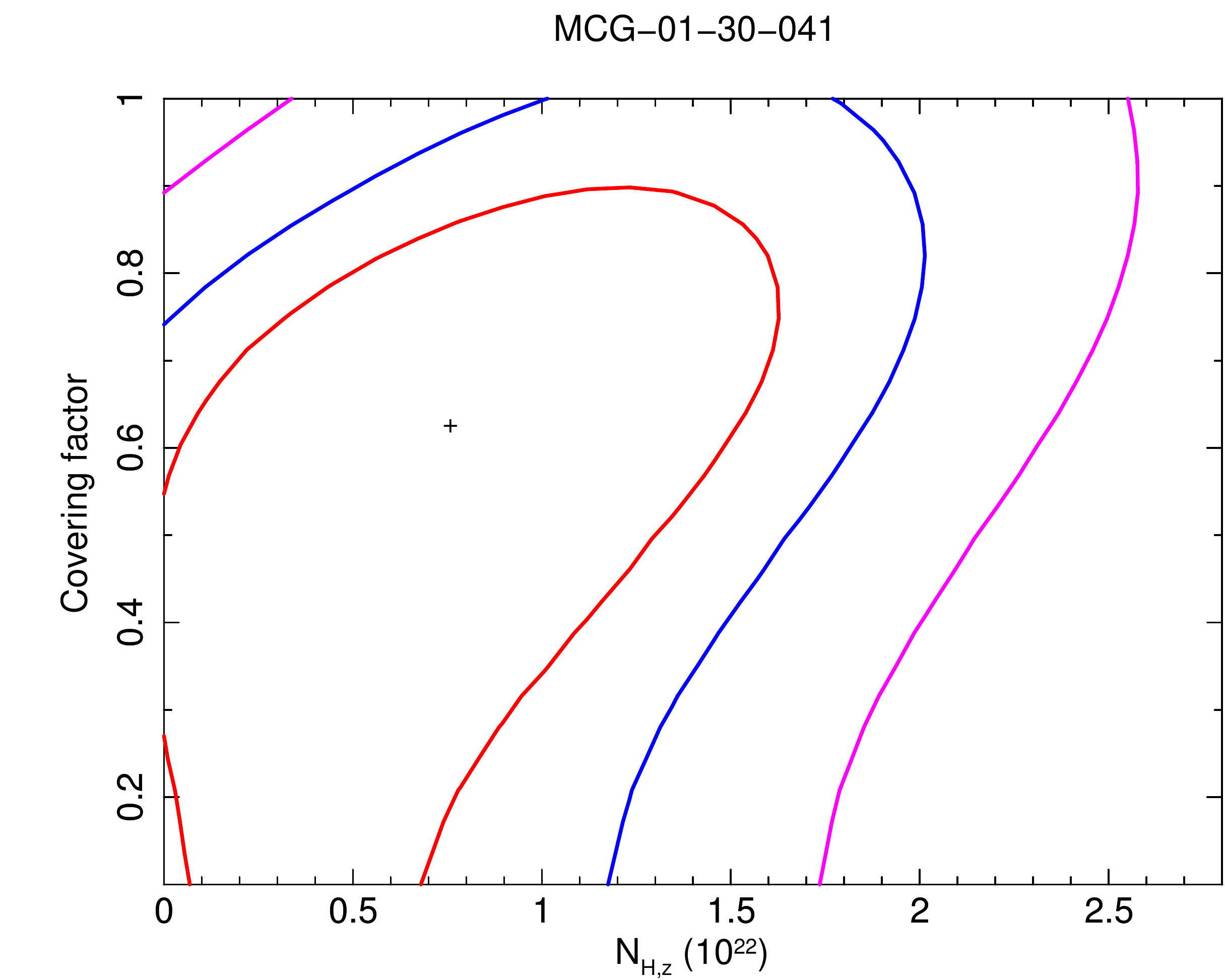}
  \end{minipage}
\begin{minipage}[b]{.5\textwidth}
  \centering
  \includegraphics[width=0.78\textwidth,angle=-90]{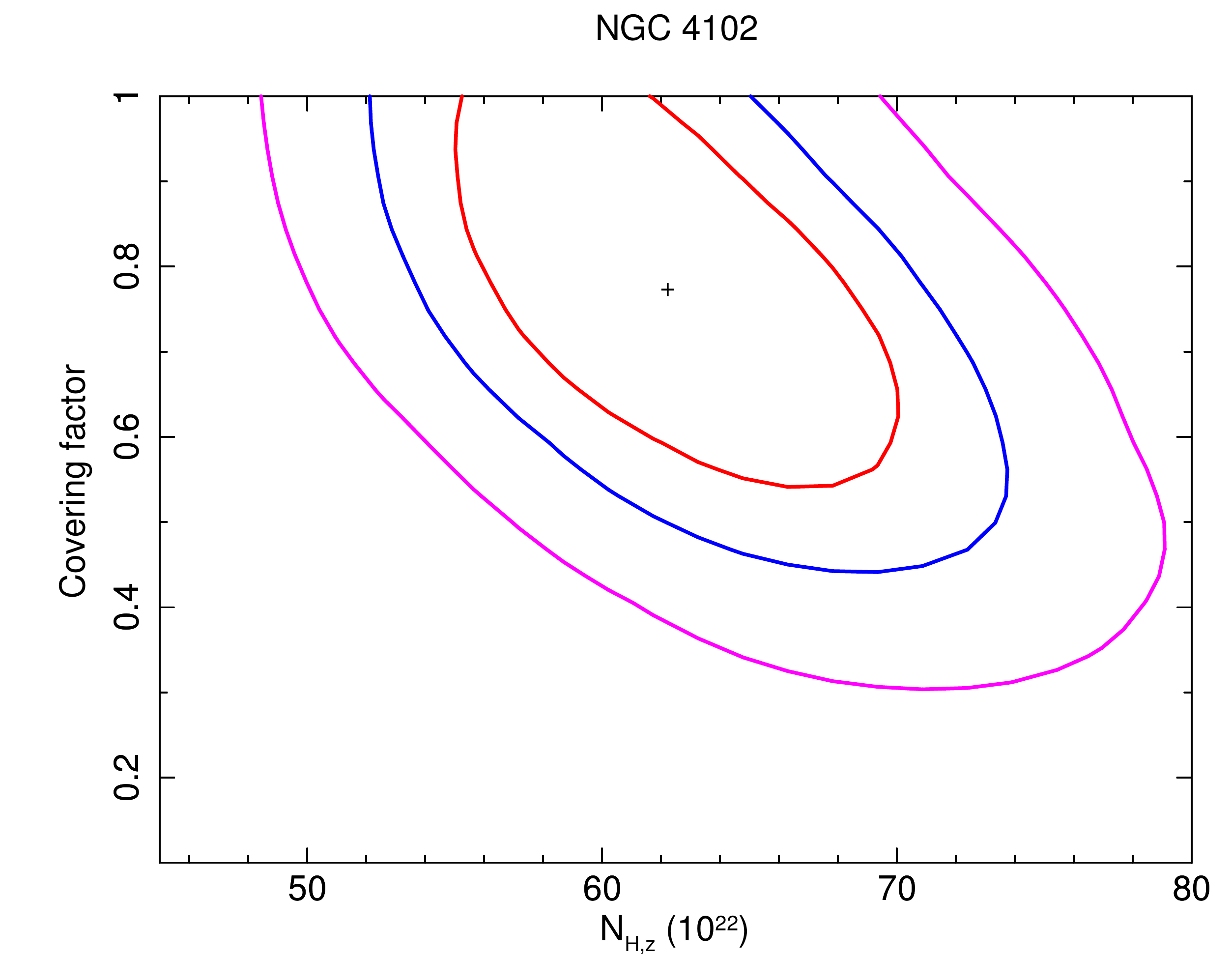}
  \end{minipage}
\caption{\normalsize \normalsize Confidence contours at 68, 90 and 99\% confidence level for the line-of-sight column density, $N_{\rm H, z}$, and the torus covering factor, $f_c$, for six of the 35 sources analyzed in this work.}
\end{figure*}

\begin{figure*}
\begin{minipage}[b]{.5\textwidth}
  \centering
  \includegraphics[width=0.78\textwidth,angle=-90]{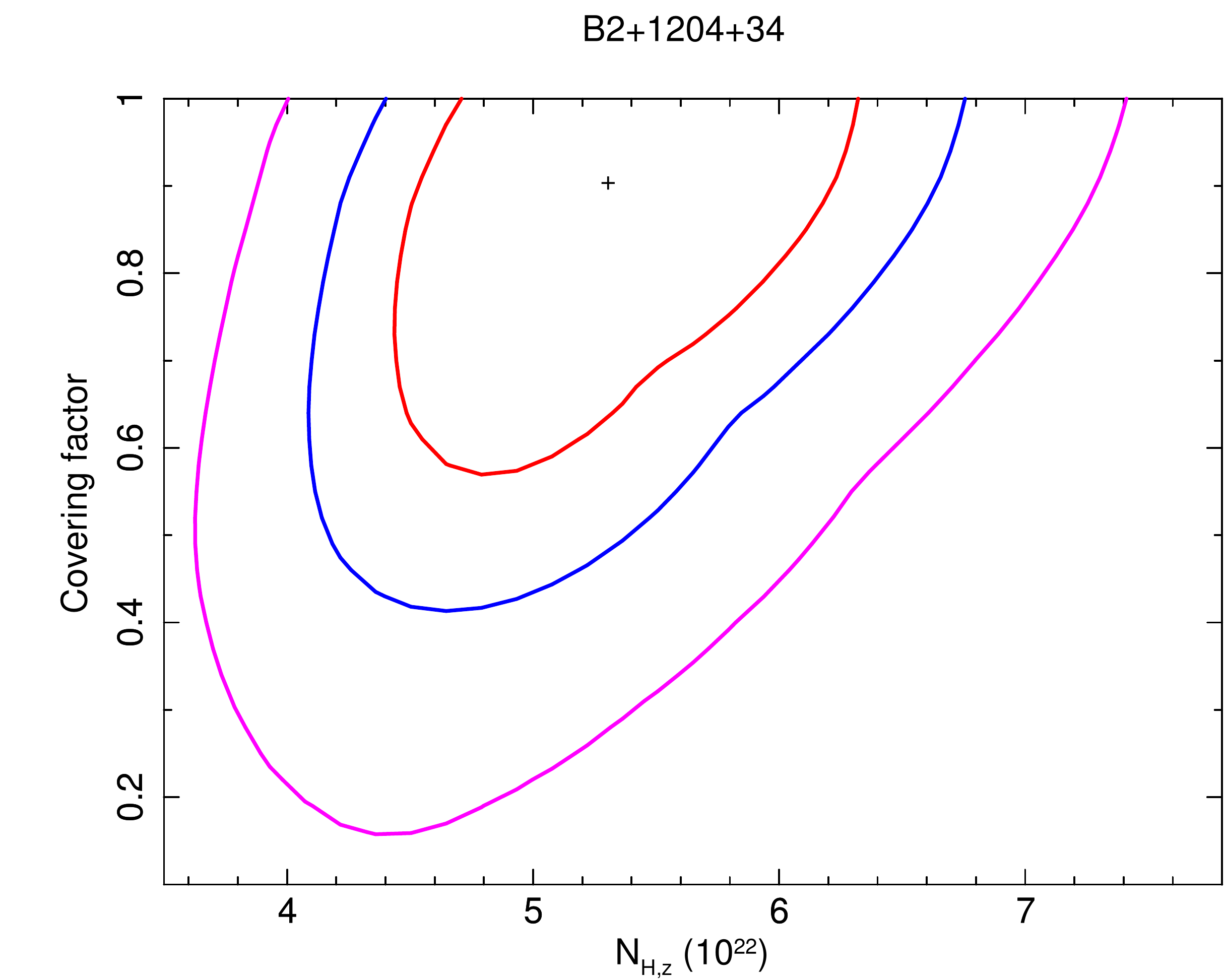}
  \end{minipage}
\begin{minipage}[b]{.5\textwidth}
  \centering
  \includegraphics[width=0.78\textwidth,angle=-90]{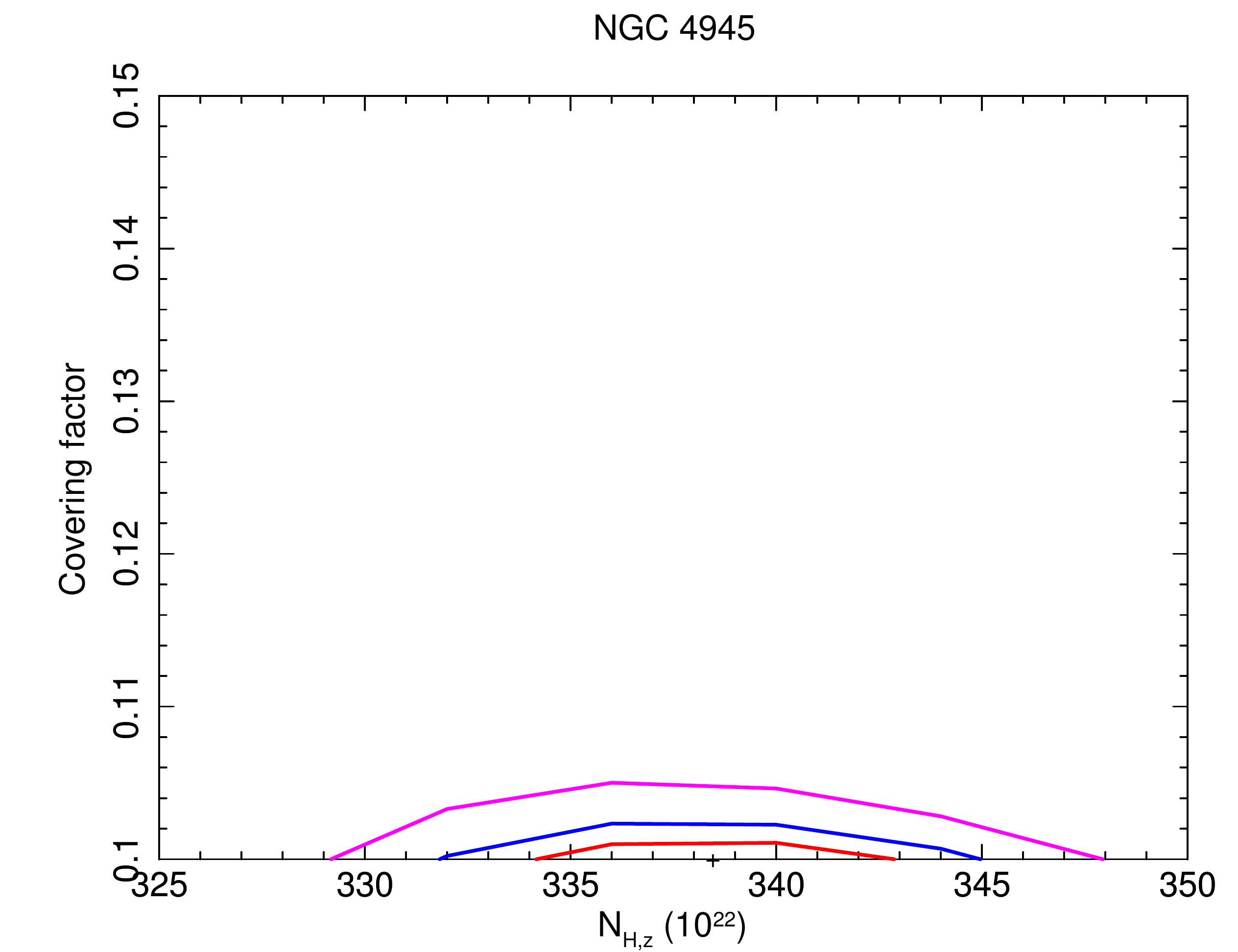}
  \end{minipage}
\begin{minipage}[b]{.5\textwidth}
  \centering
  \includegraphics[width=0.78\textwidth,angle=-90]{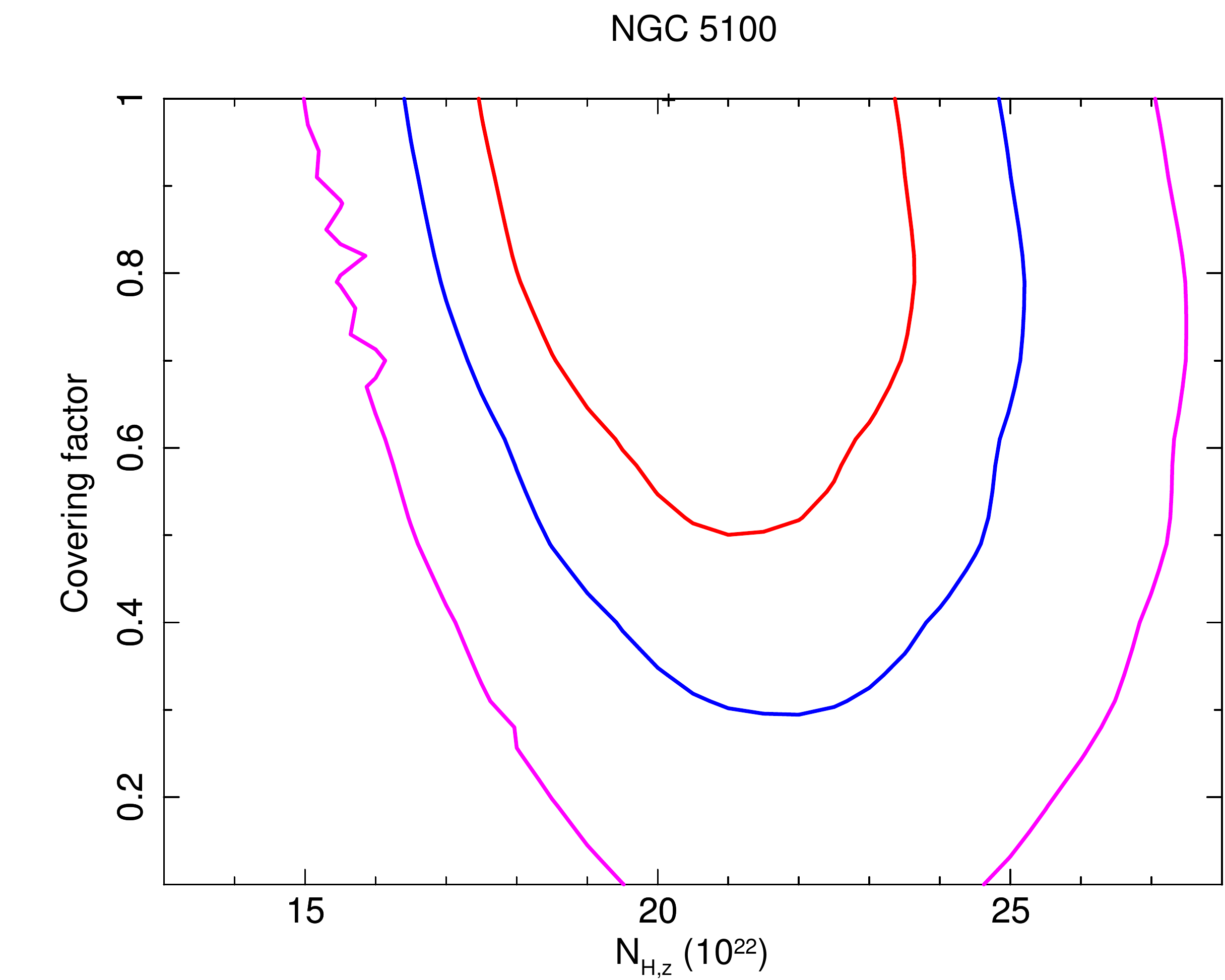}
  \end{minipage}
  \begin{minipage}[b]{.5\textwidth}
  \centering
  \includegraphics[width=0.78\textwidth,angle=-90]{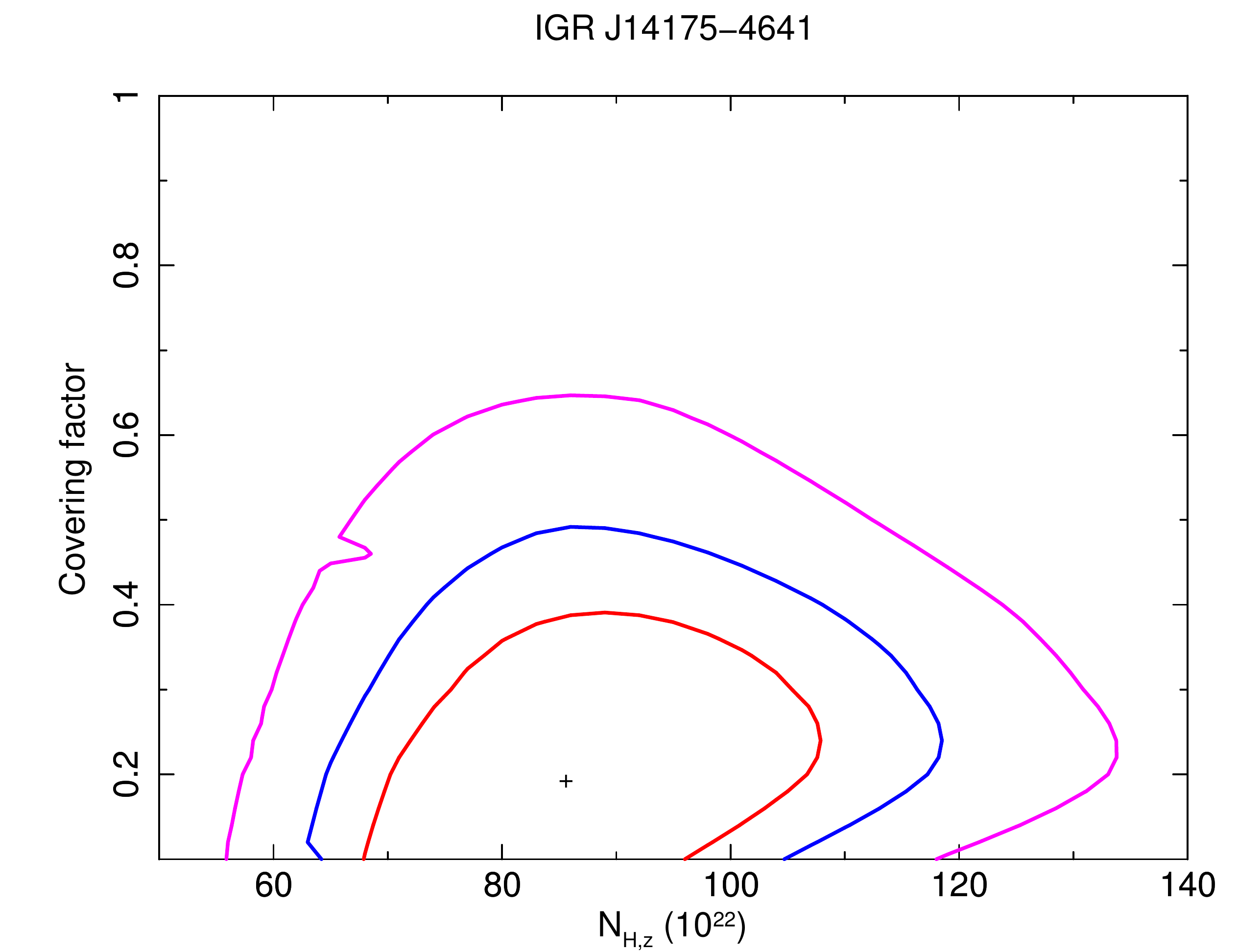}
  \end{minipage}
  \begin{minipage}[b]{.5\textwidth}
  \centering
  \includegraphics[width=0.78\textwidth,angle=-90]{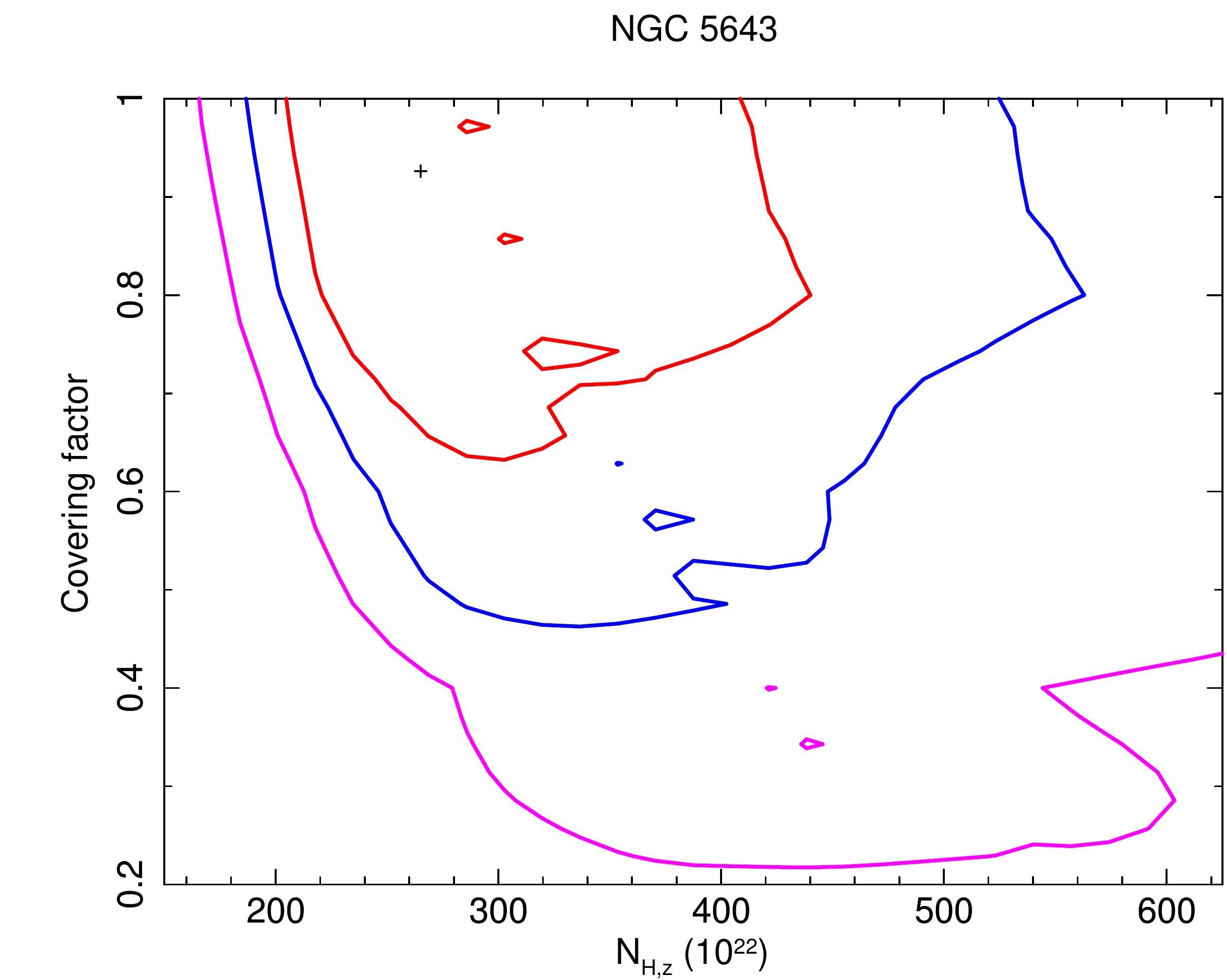}
  \end{minipage}
\begin{minipage}[b]{.5\textwidth}
  \centering
  \includegraphics[width=0.78\textwidth,angle=-90]{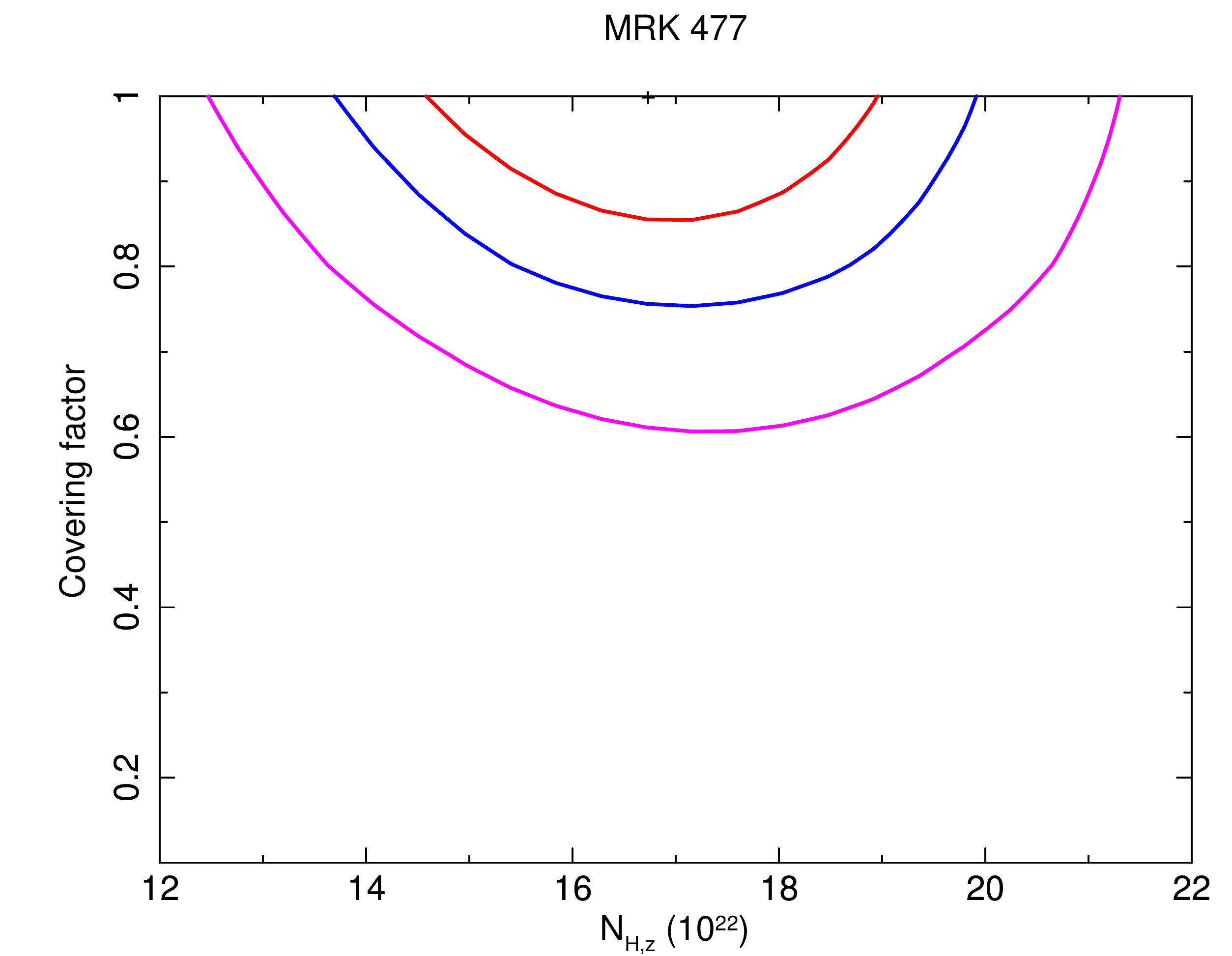}
  \end{minipage}
\caption{\normalsize \normalsize Confidence contours at 68, 90 and 99\% confidence level for the line-of-sight column density, $N_{\rm H, z}$, and the torus covering factor, $f_c$, for six of the 35 sources analyzed in this work.}
\end{figure*}

\begin{figure*}
\begin{minipage}[b]{.5\textwidth}
  \centering
  \includegraphics[width=0.78\textwidth,angle=-90]{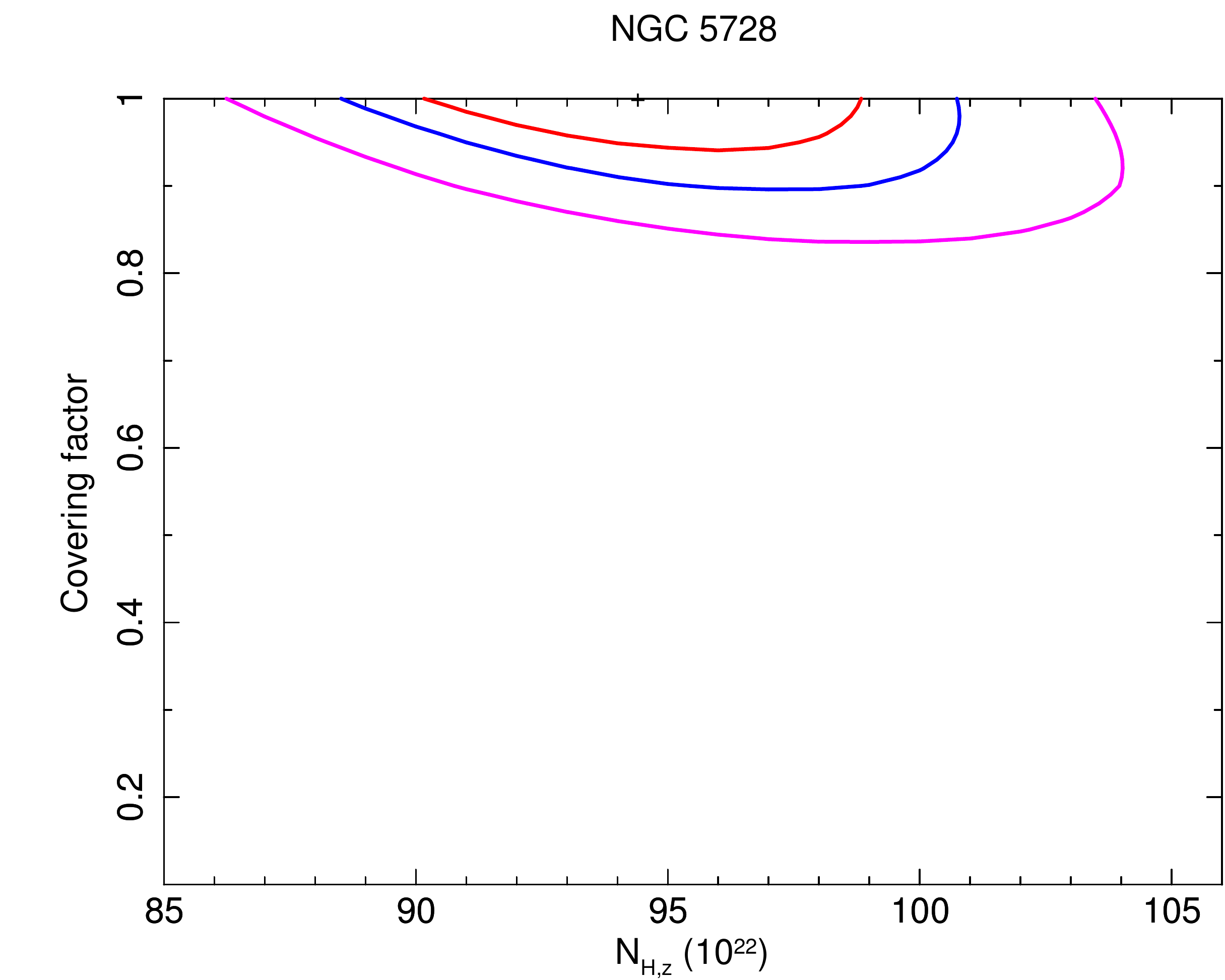}
  \end{minipage}
\begin{minipage}[b]{.5\textwidth}
  \centering
  \includegraphics[width=0.78\textwidth,angle=-90]{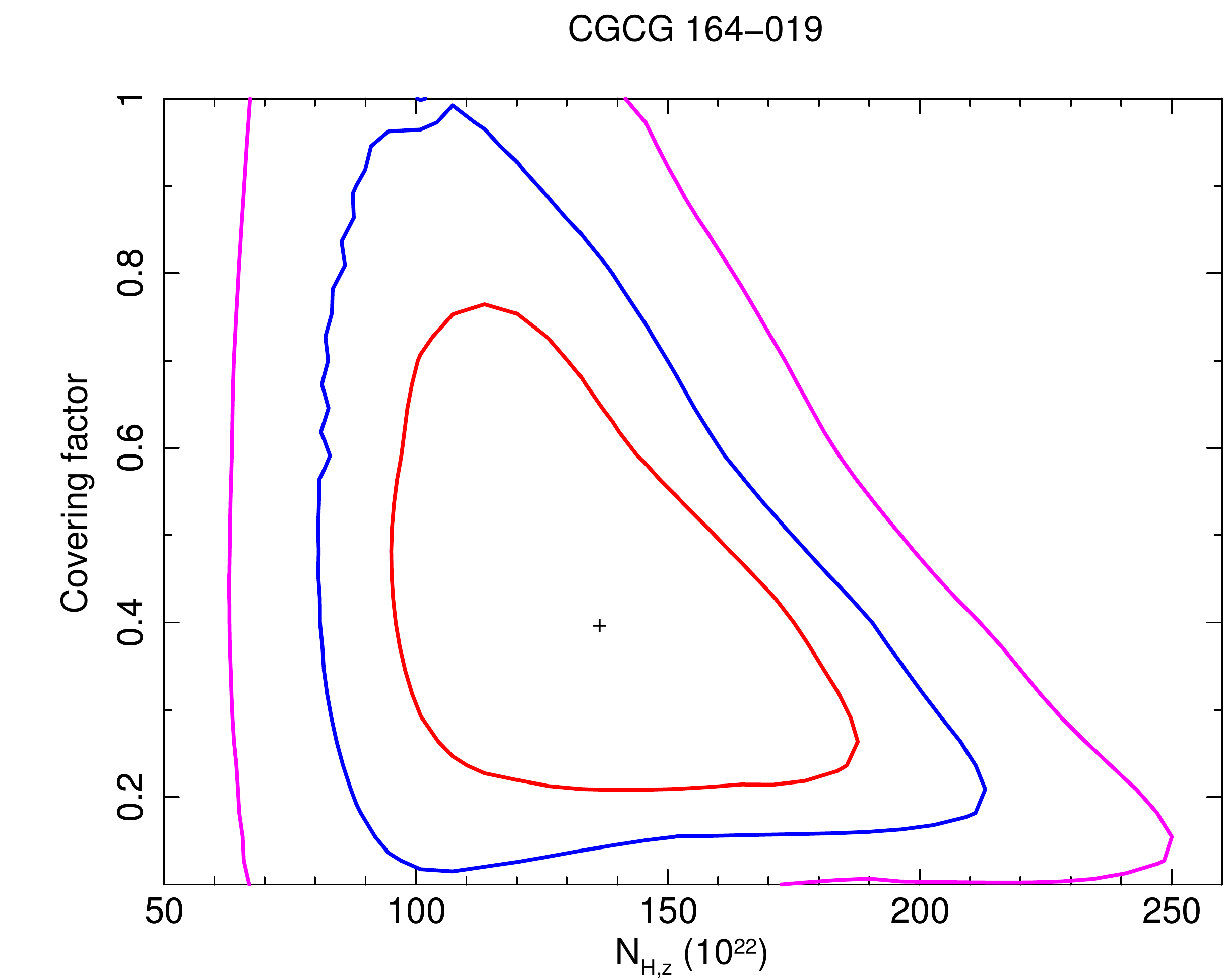}
  \end{minipage}
\begin{minipage}[b]{.5\textwidth}
  \centering
  \includegraphics[width=0.78\textwidth,angle=-90]{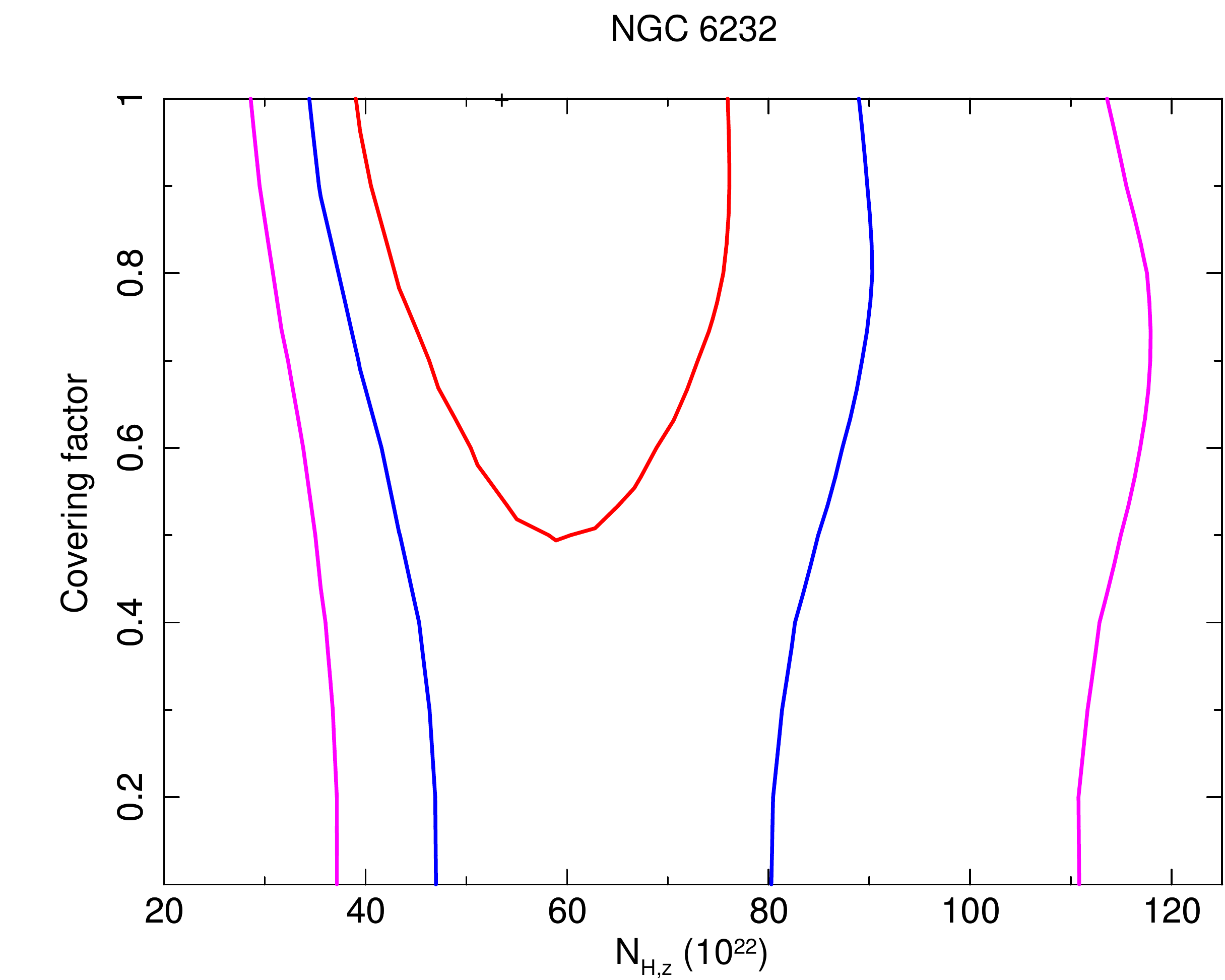}
  \end{minipage}
  \begin{minipage}[b]{.5\textwidth}
  \centering
  \includegraphics[width=0.78\textwidth,angle=-90]{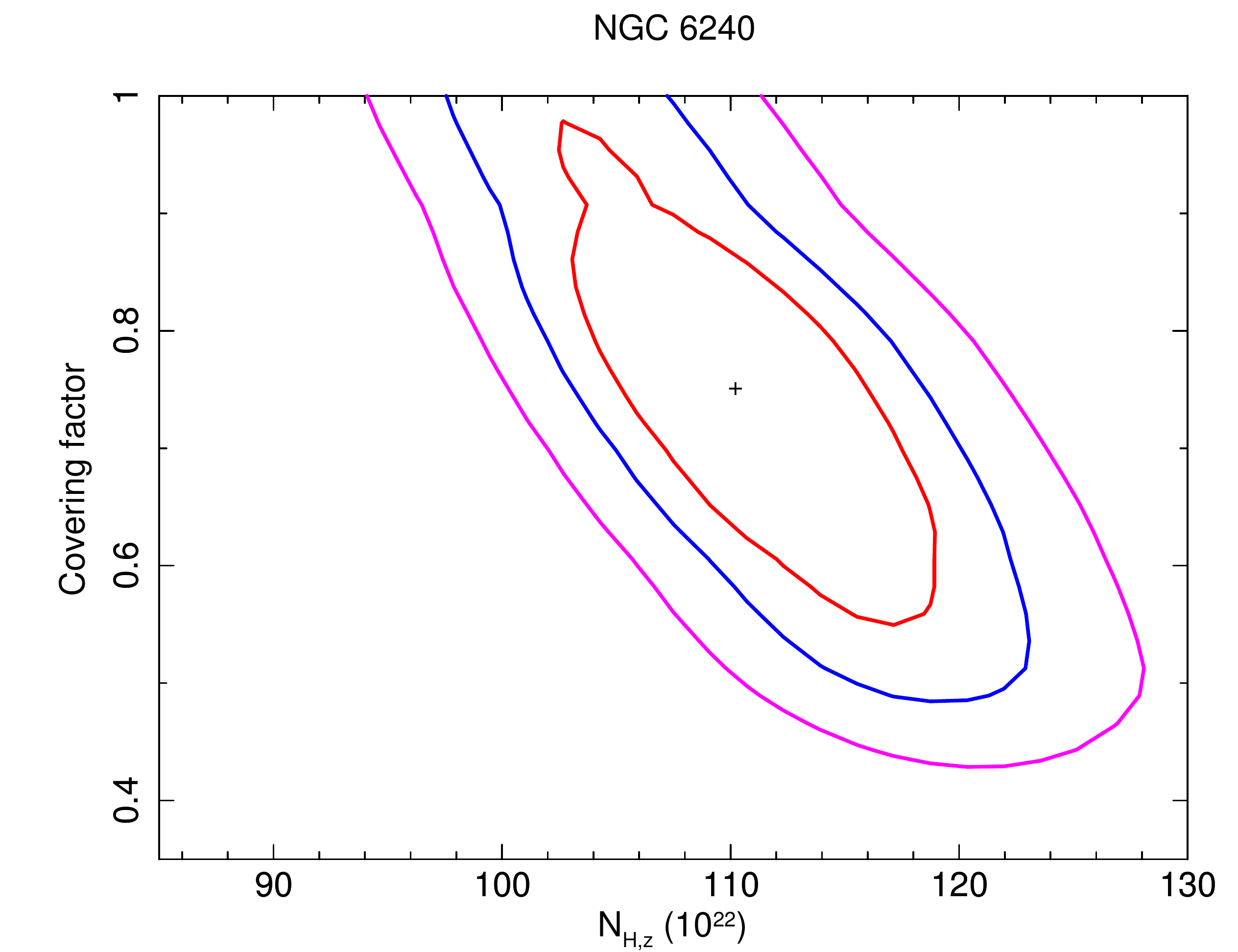}
  \end{minipage}
  \begin{minipage}[b]{.5\textwidth}
  \centering
  \includegraphics[width=0.78\textwidth,angle=-90]{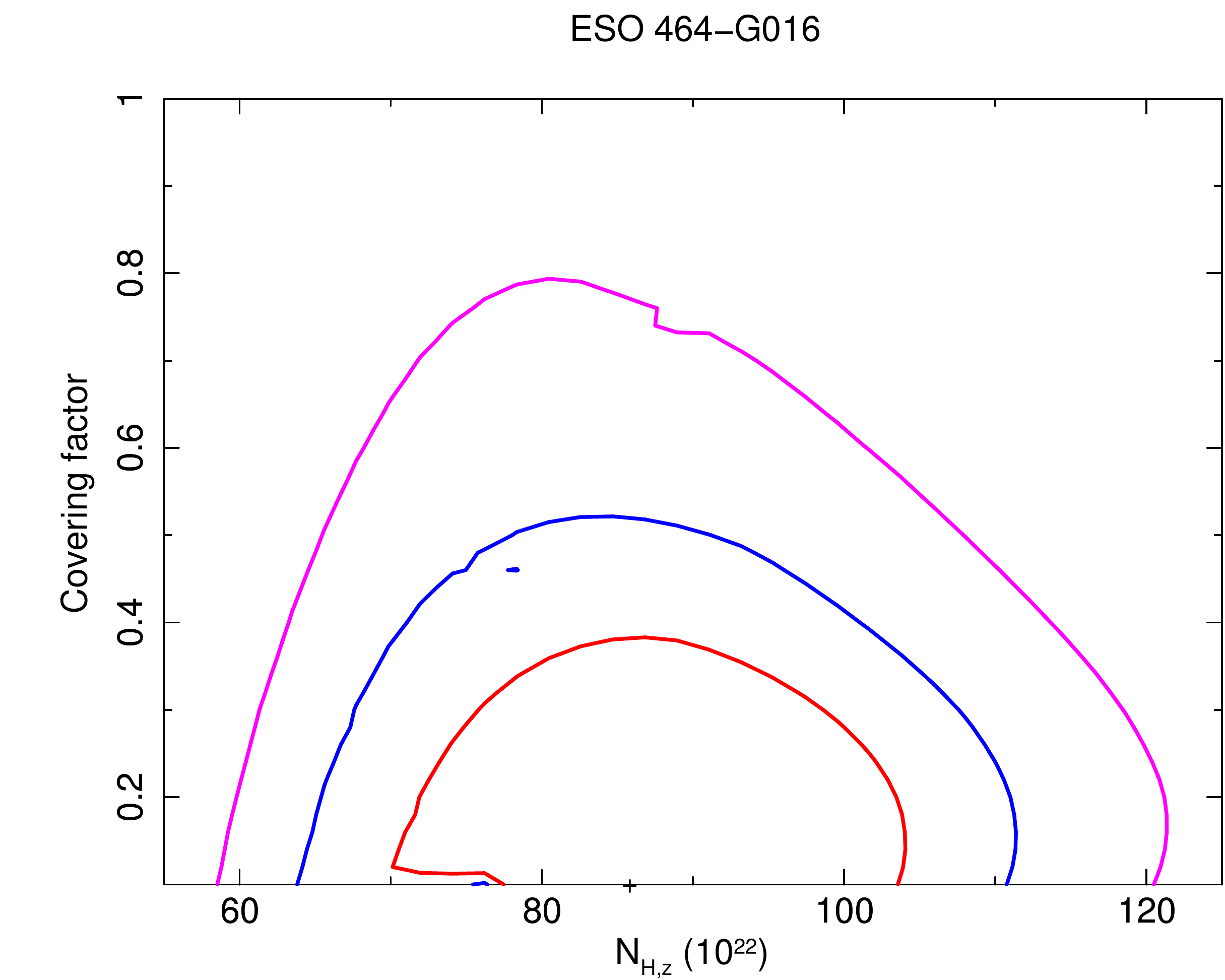}
  \end{minipage}
\begin{minipage}[b]{.5\textwidth}
  \centering
  \includegraphics[width=0.78\textwidth,angle=-90]{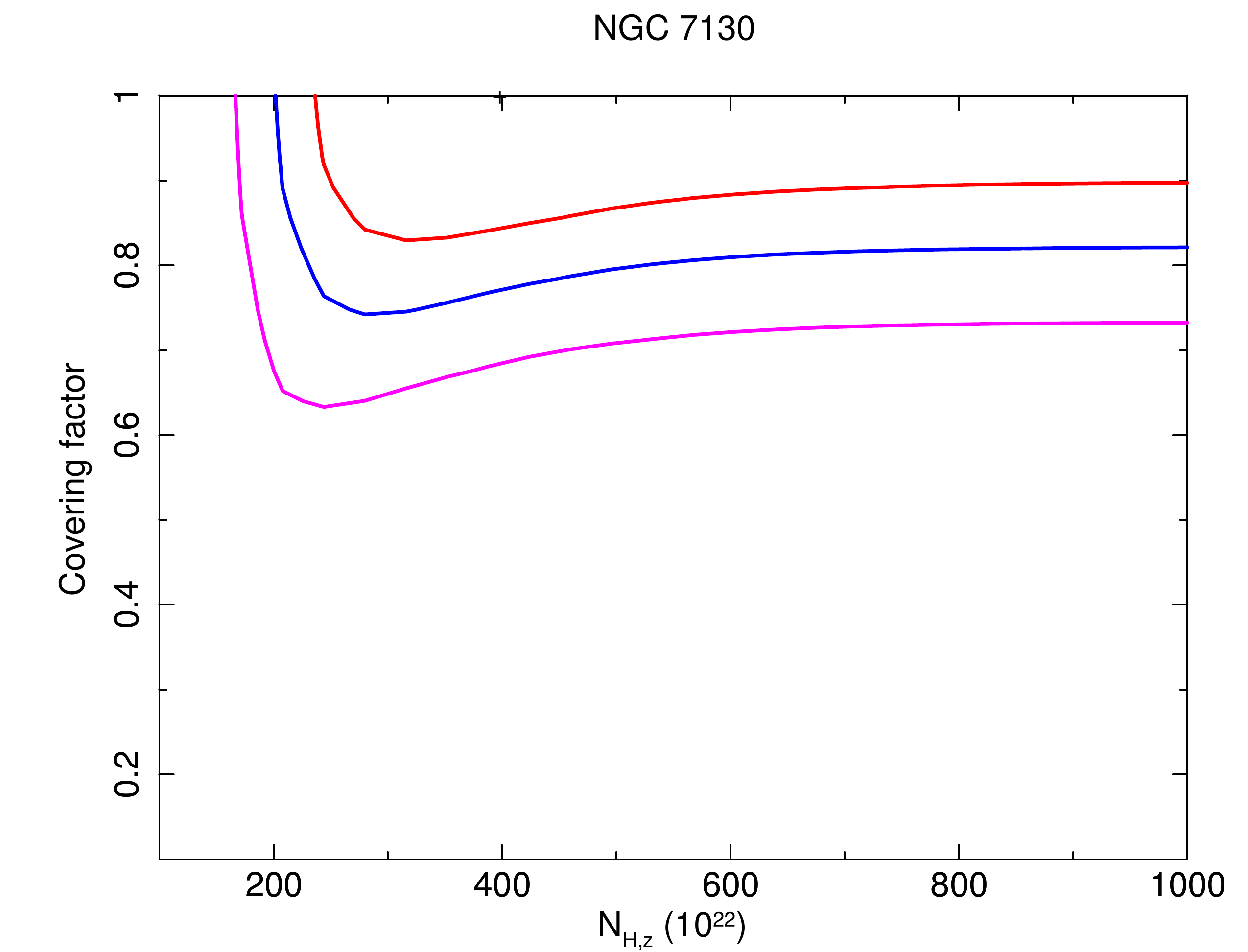}
  \end{minipage}
\caption{\normalsize \normalsize Confidence contours at 68, 90 and 99\% confidence level for the line-of-sight column density, $N_{\rm H, z}$, and the torus covering factor, $f_c$, for six of the 35 sources analyzed in this work.}
\end{figure*}

\section{C. Confidence contours of the torus covering factor versus the torus average column density}\label{app:nhtor_vs_fc}
\begin{figure*}
\begin{minipage}[b]{.5\textwidth}
  \centering
  \includegraphics[width=0.78\textwidth,angle=-90]{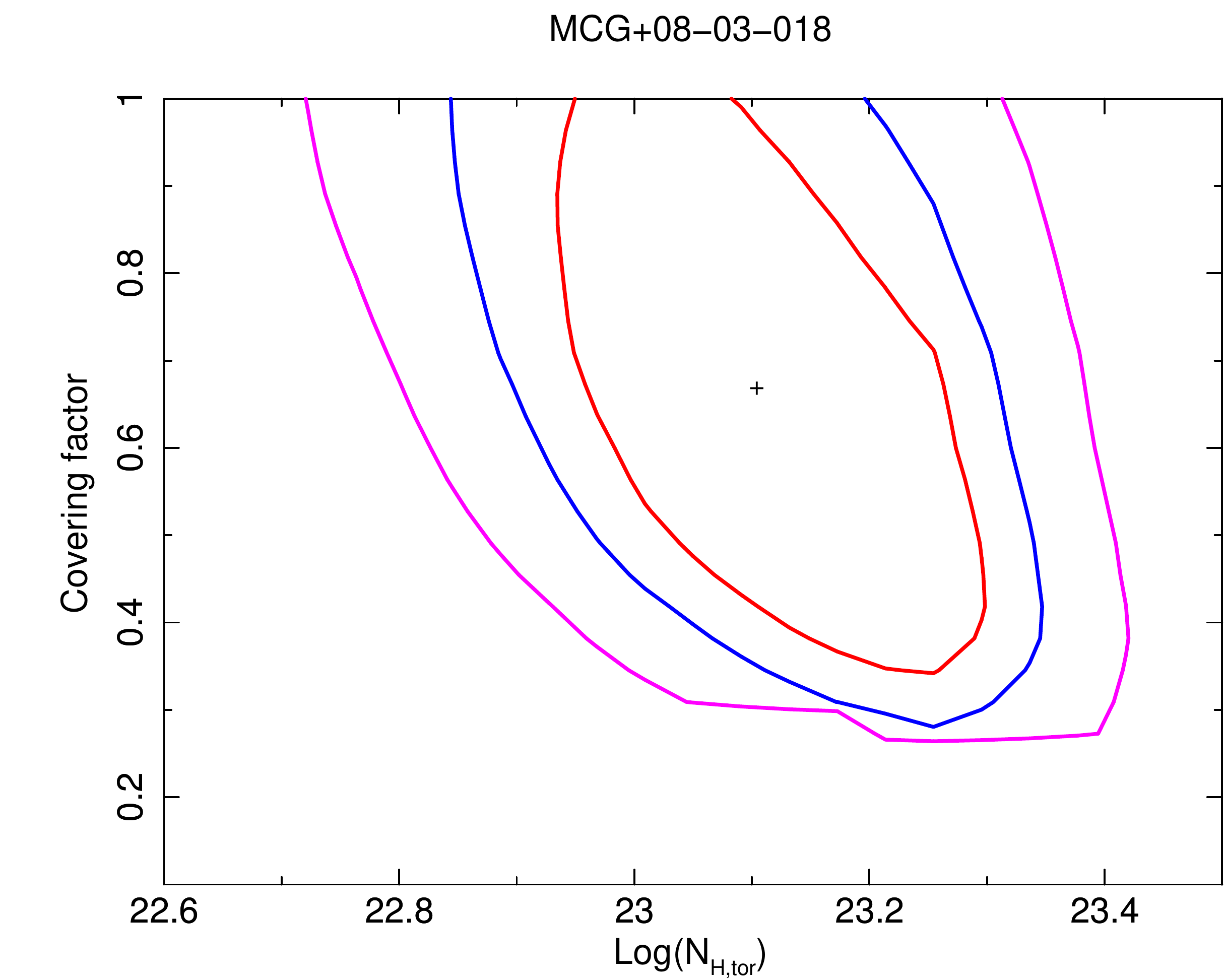}
  \end{minipage}
\begin{minipage}[b]{.5\textwidth}
  \centering
  \includegraphics[width=0.78\textwidth,angle=-90]{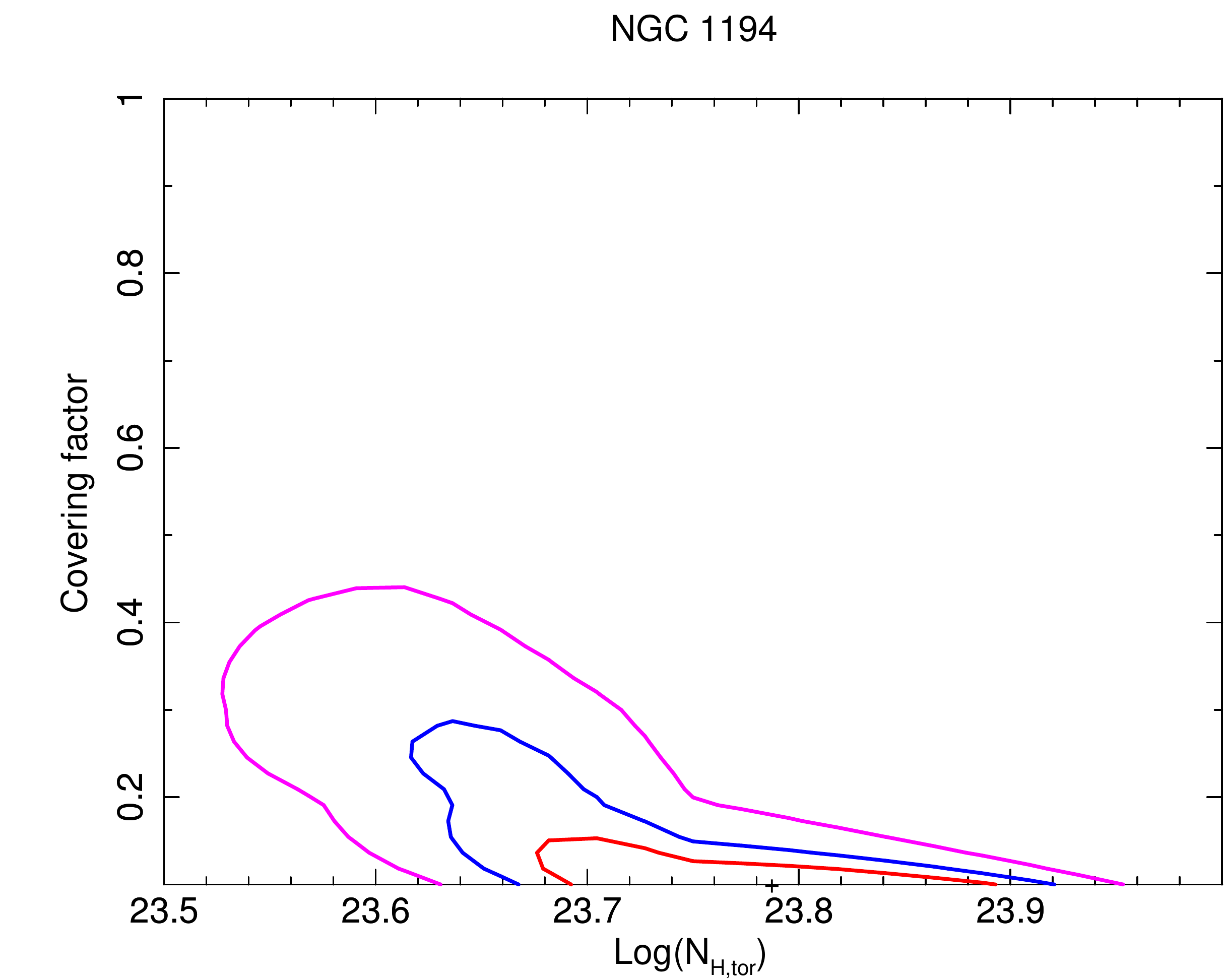}
  \end{minipage}
\begin{minipage}[b]{.5\textwidth}
  \centering
  \includegraphics[width=0.78\textwidth,angle=-90]{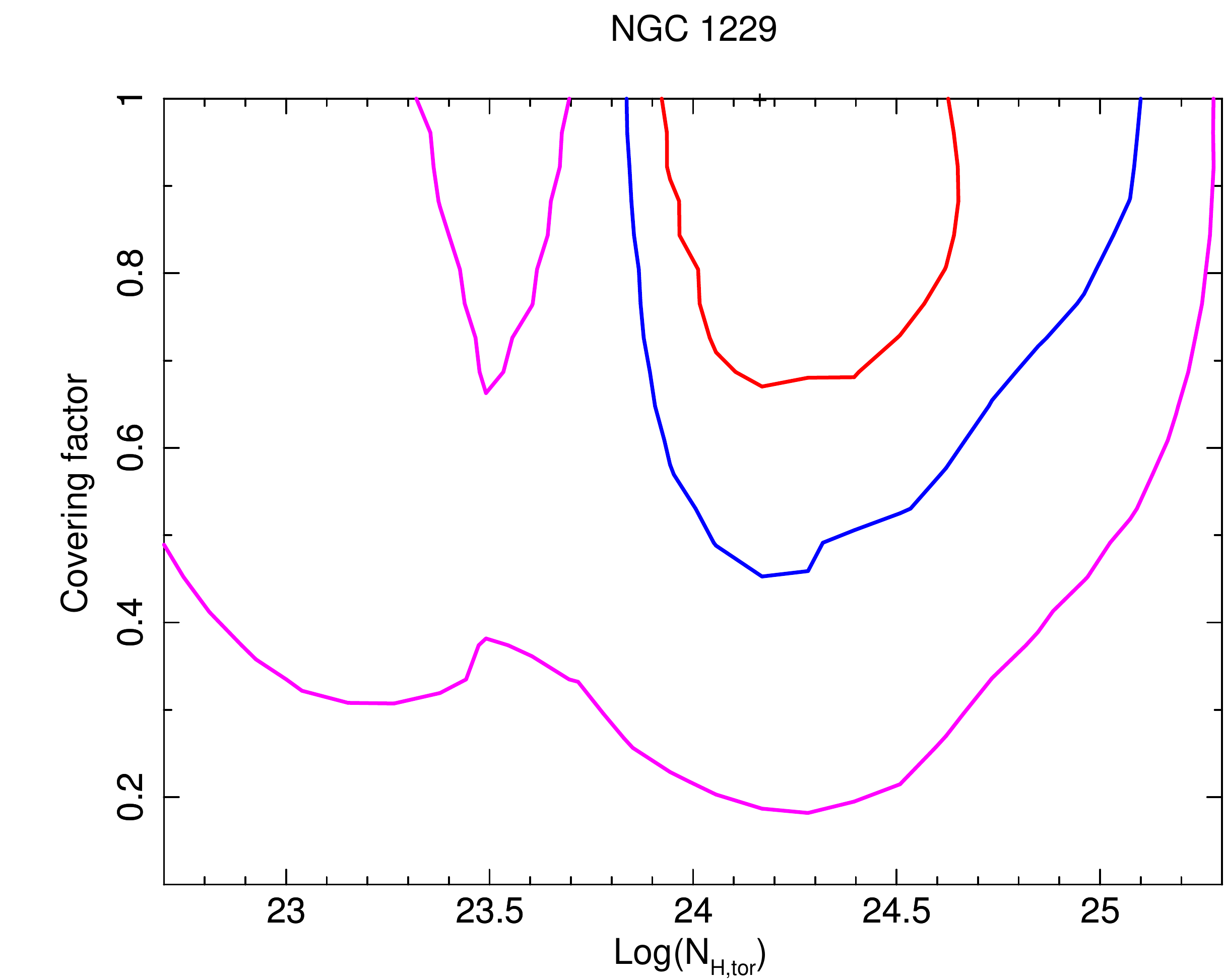}
  \end{minipage}
  \begin{minipage}[b]{.5\textwidth}
  \centering
  \includegraphics[width=0.78\textwidth,angle=-90]{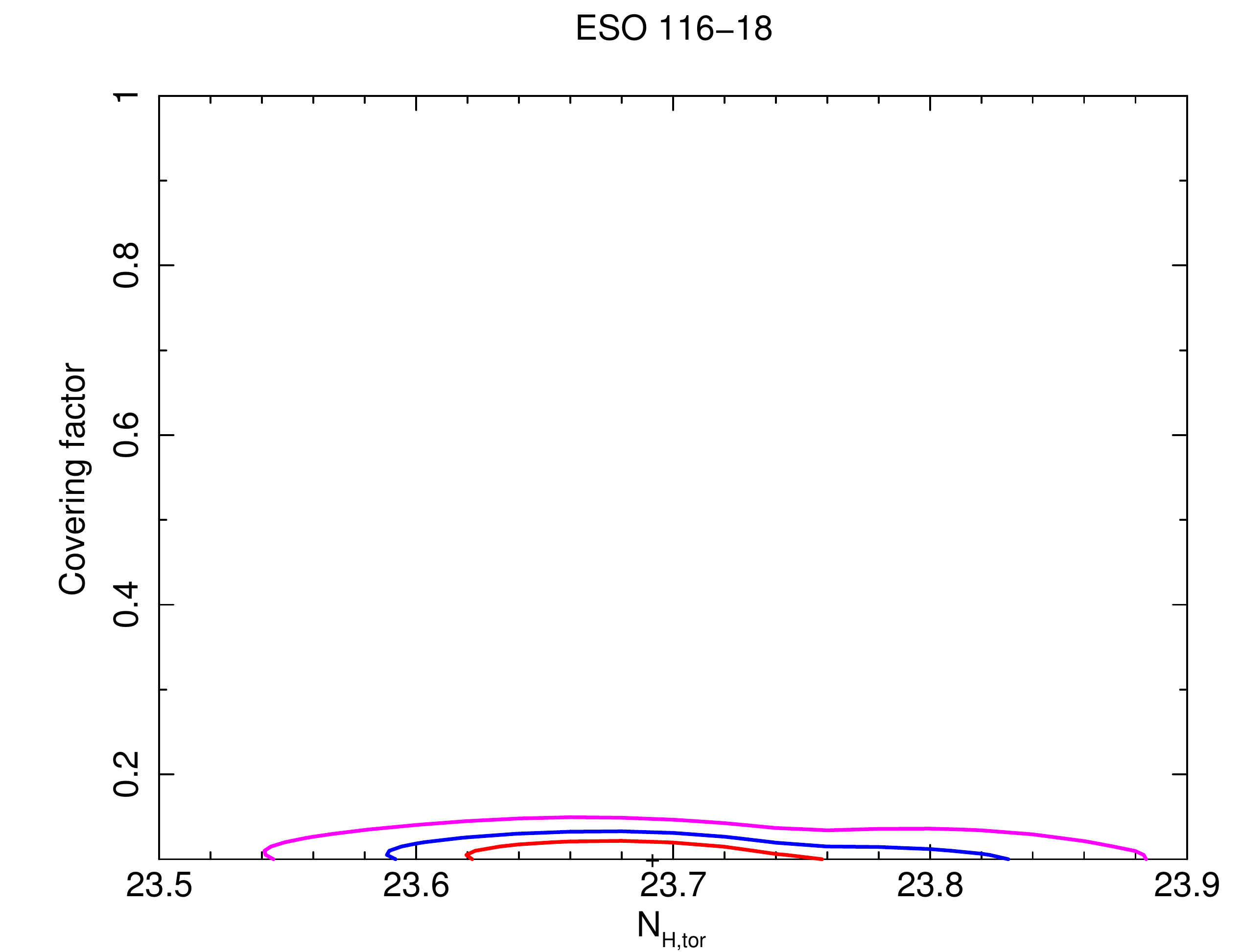}
  \end{minipage}
  \begin{minipage}[b]{.5\textwidth}
  \centering
  \includegraphics[width=0.78\textwidth,angle=-90]{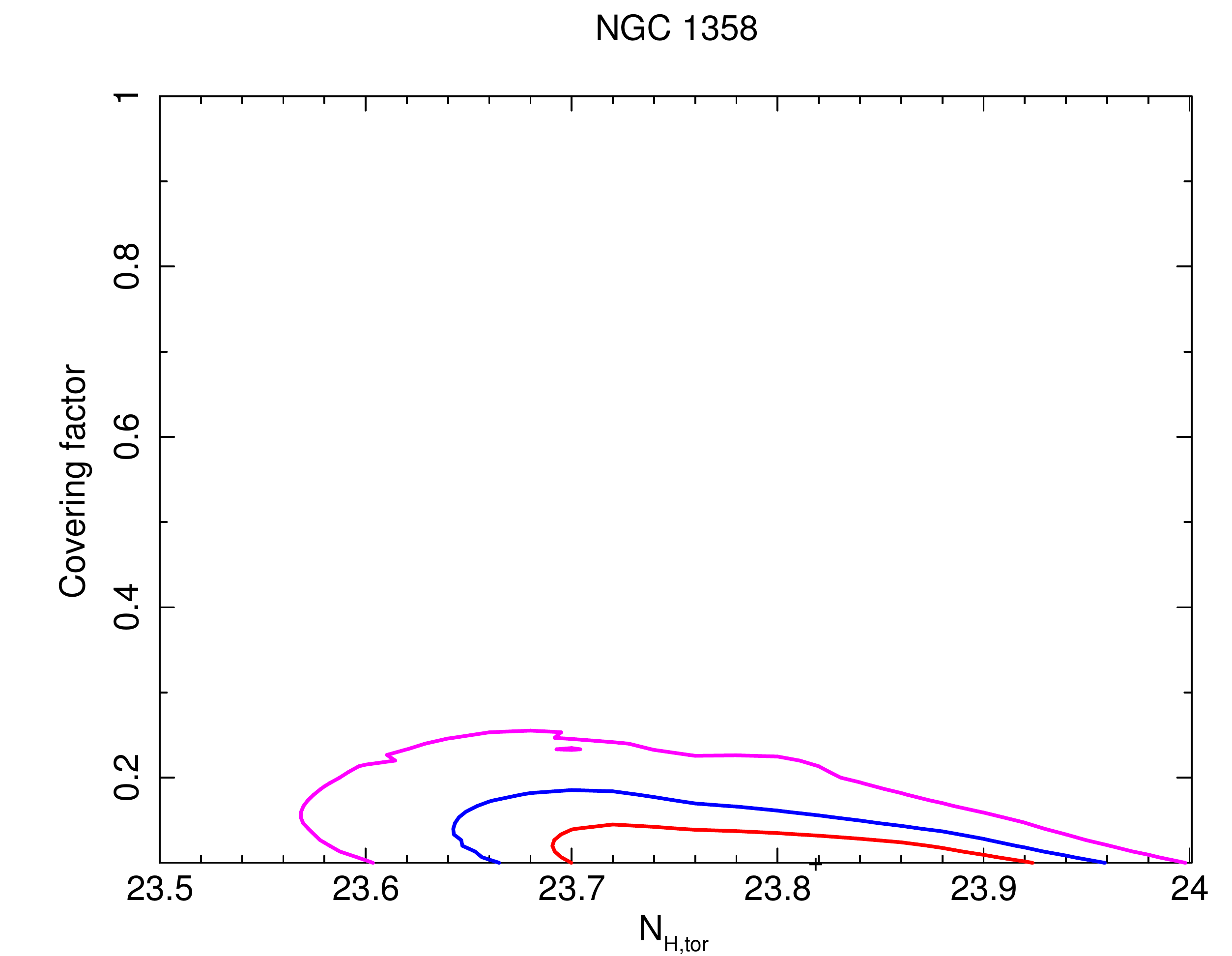}
  \end{minipage}
  \begin{minipage}[b]{.5\textwidth}
  \centering
  \includegraphics[width=0.78\textwidth,angle=-90]{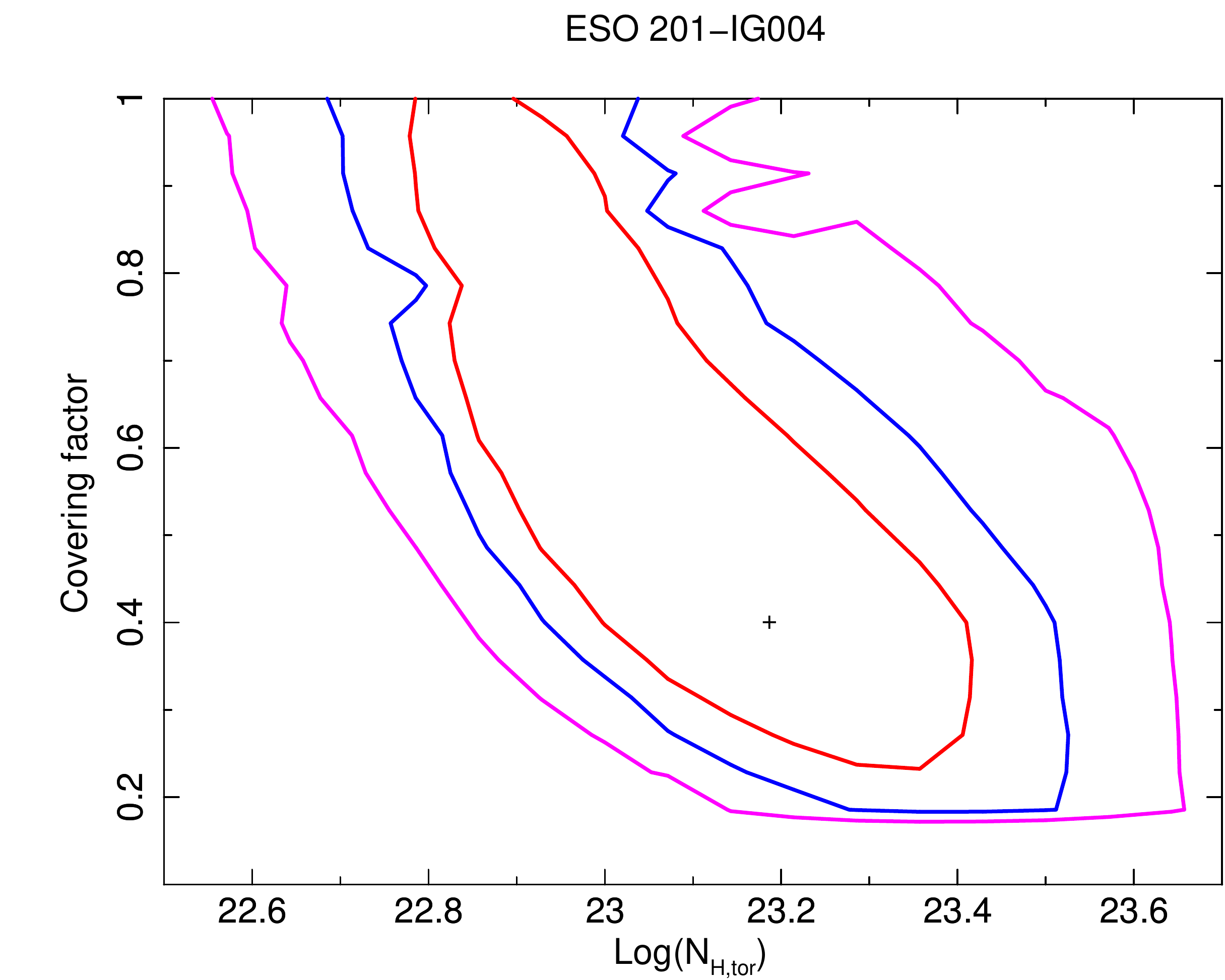}
  \end{minipage}
\caption{\normalsize \normalsize Confidence contours at 68, 90 and 99\% confidence level for the torus average column density, $N_{\rm H, tor}$, and the torus covering factor, $f_c$, for six of the 35 sources analyzed in this work.}\label{fig:NHtor_vs_fc}
\end{figure*}

\begin{figure*}
\begin{minipage}[b]{.5\textwidth}
  \centering
  \includegraphics[width=0.78\textwidth,angle=-90]{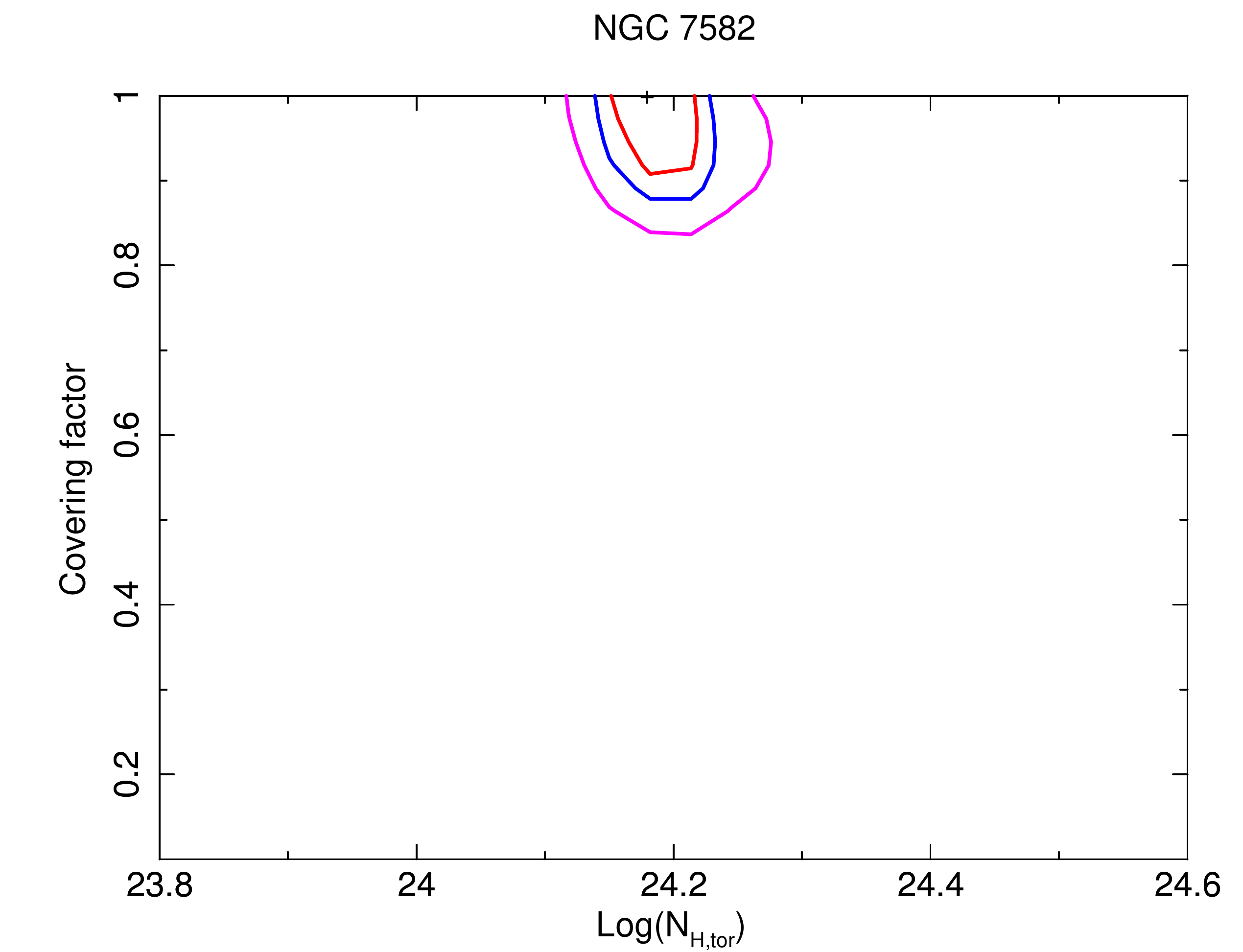}
  \end{minipage}
\caption{\normalsize \normalsize Confidence contours at 68, 90 and 99\% confidence level for the line-of-sight column density, $N_{\rm H, z}$, and the torus covering factor, $f_c$, for NGC 7582, one of the 35 sources analyzed in this work.}\label{fig:NHtor_vs_fc_last}
\end{figure*}

We report in Figure \ref{fig:NHtor_vs_fc}--\ref{fig:NHtor_vs_fc_last} the confidence contours of the covering factor, $f_c$, versus the logarithm of the torus average density, Log($N_{\rm H, tor}$), for 31 sources out of the 35 in our sample. We do not report the contours of NGC 424 and NGC 1068, which are best-fitted by a multi-reprocessed component and therefore have more than one best-fit  Log($N_{\rm H, tor}$) (see Table \ref{tab:cf}), and those of MCG-01-30-041 and RBS 1037, the two unobscured AGNs in our sample, because $f_c$ and/or  Log($N_{\rm H, tor}$) are unconstrained.

\begin{figure*}
\begin{minipage}[b]{.5\textwidth}
  \centering
  \includegraphics[width=0.78\textwidth,angle=-90]{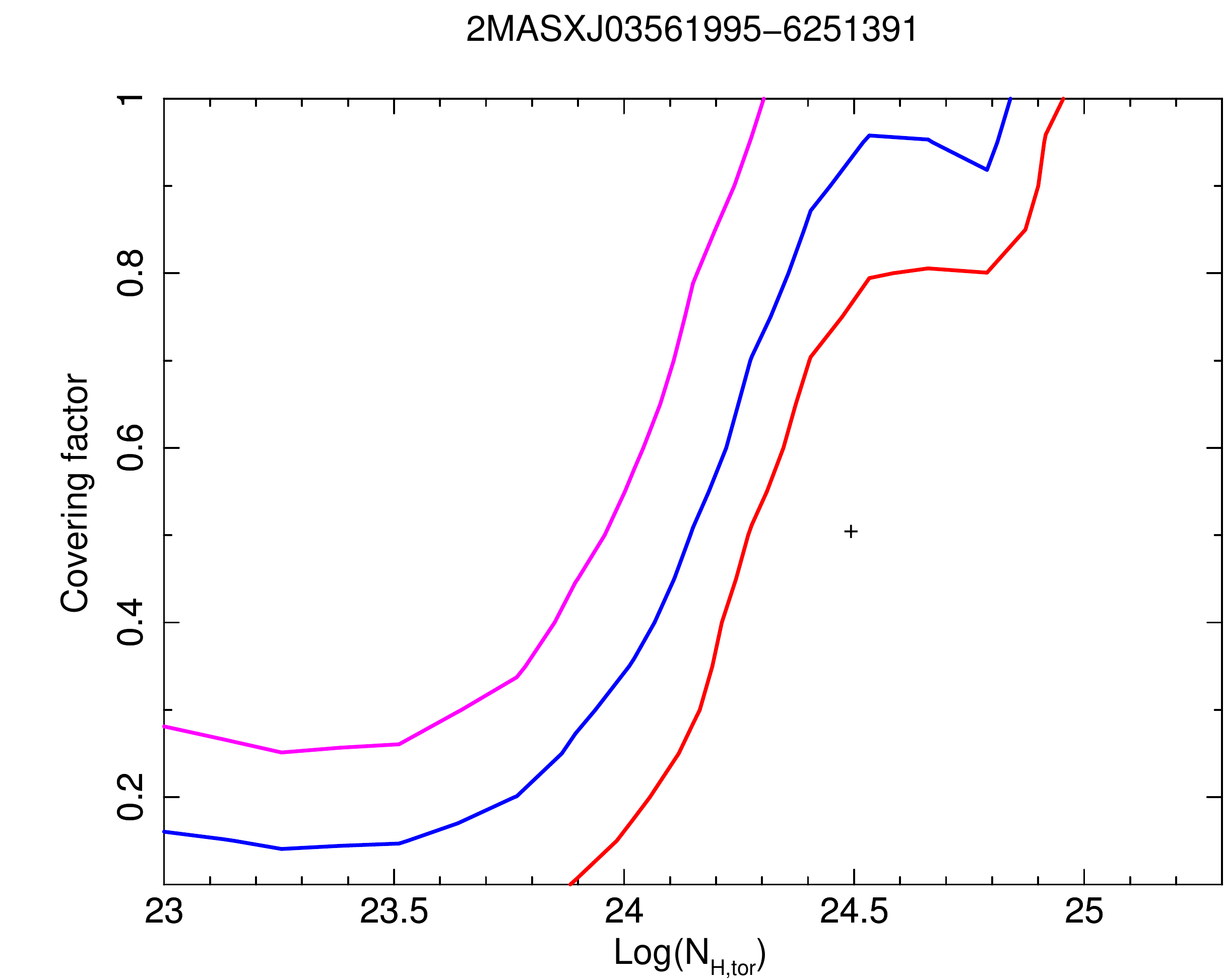}
  \end{minipage}
\begin{minipage}[b]{.5\textwidth}
  \centering
  \includegraphics[width=0.78\textwidth,angle=-90]{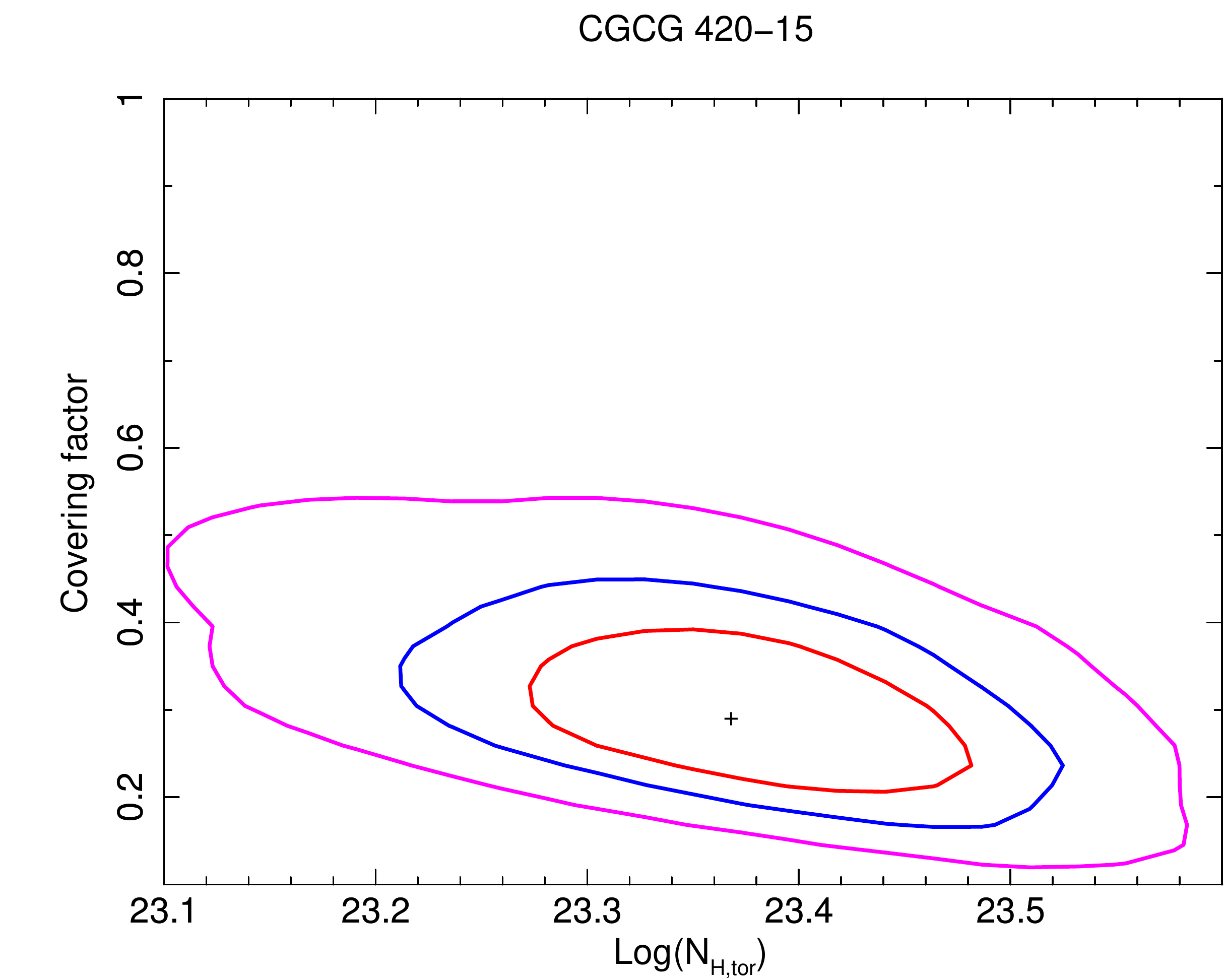}
  \end{minipage}
\begin{minipage}[b]{.5\textwidth}
  \centering
  \includegraphics[width=0.78\textwidth,angle=-90]{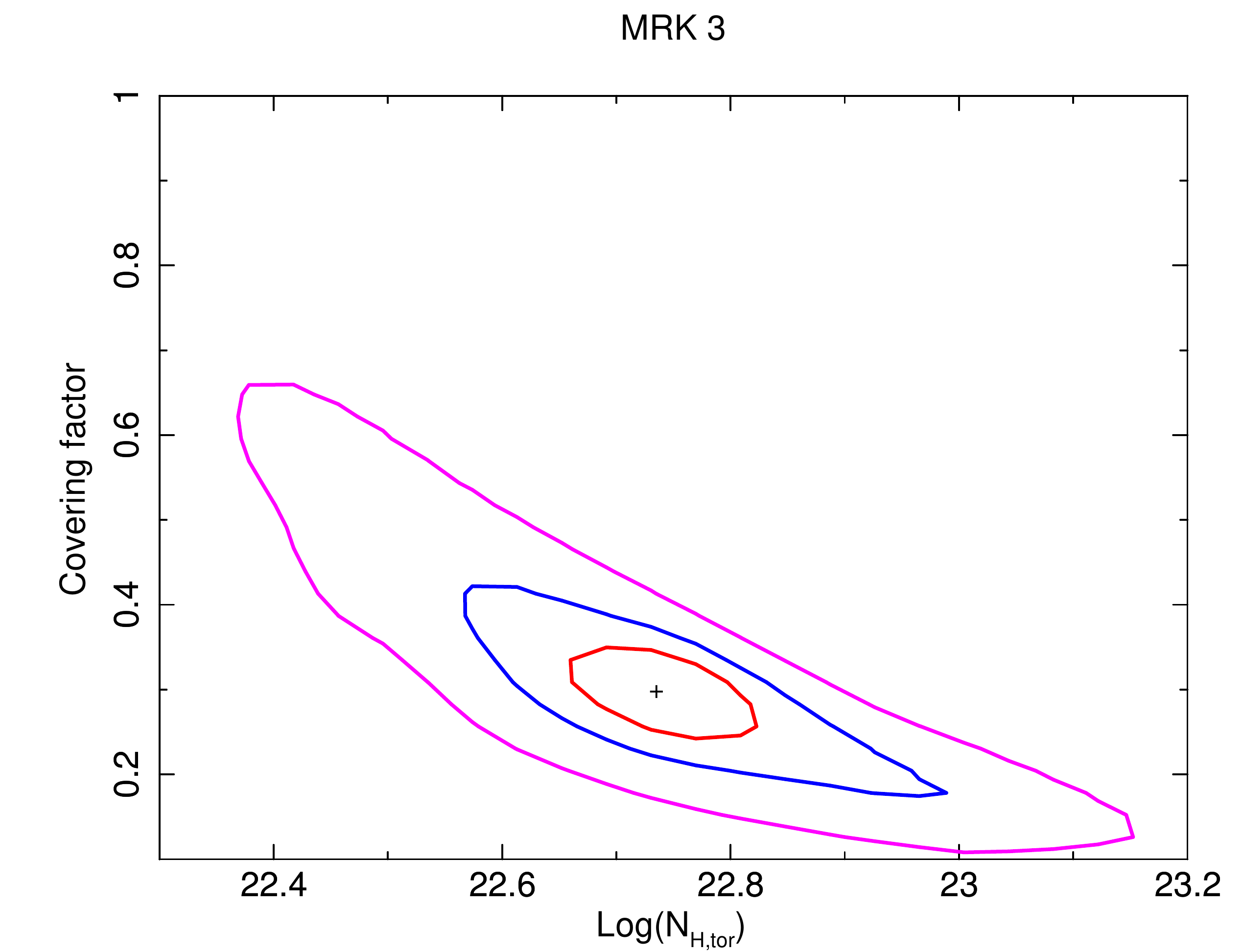}
  \end{minipage}
  \begin{minipage}[b]{.5\textwidth}
  \centering
  \includegraphics[width=0.78\textwidth,angle=-90]{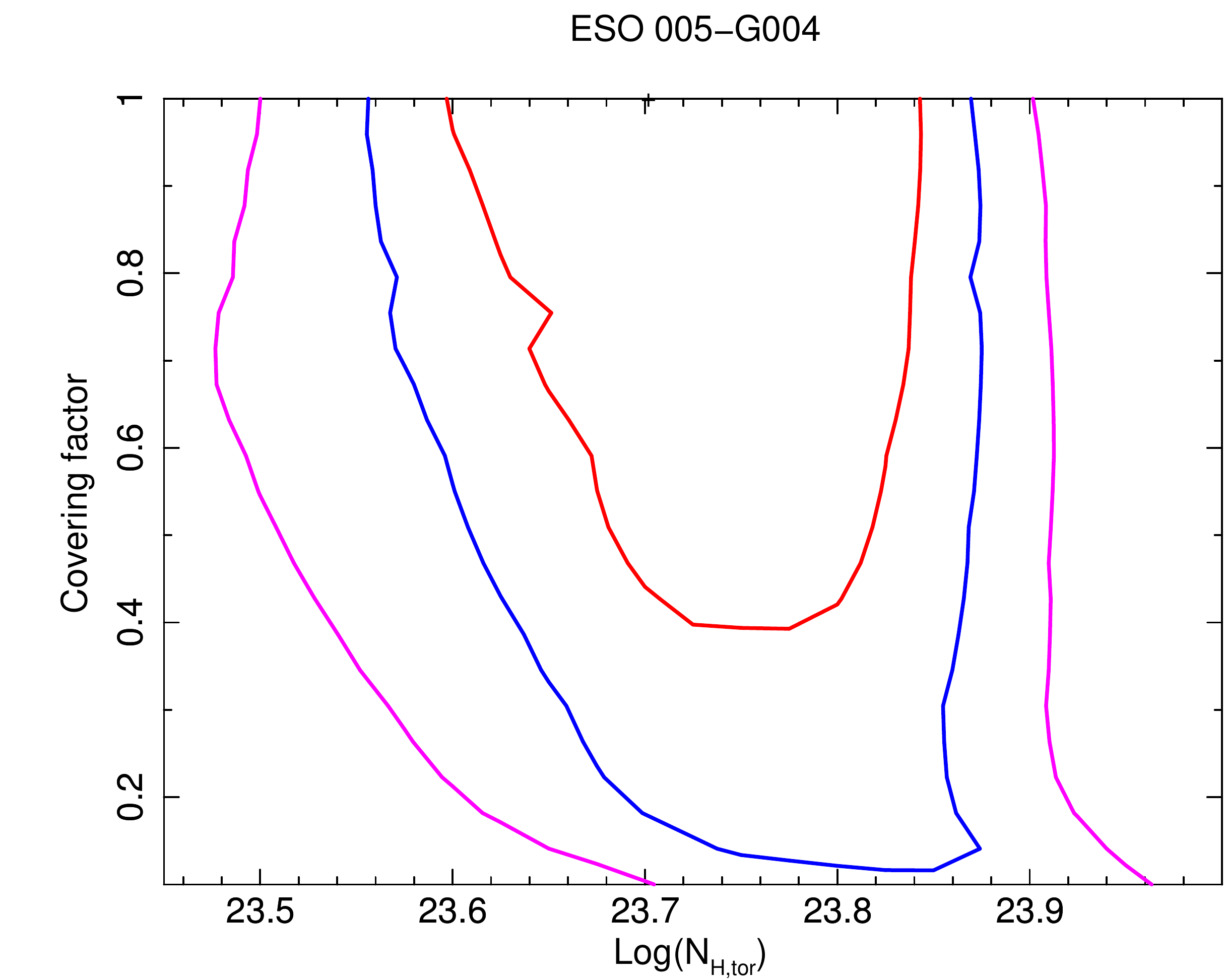}
  \end{minipage}
  \begin{minipage}[b]{.5\textwidth}
  \centering
  \includegraphics[width=0.78\textwidth,angle=-90]{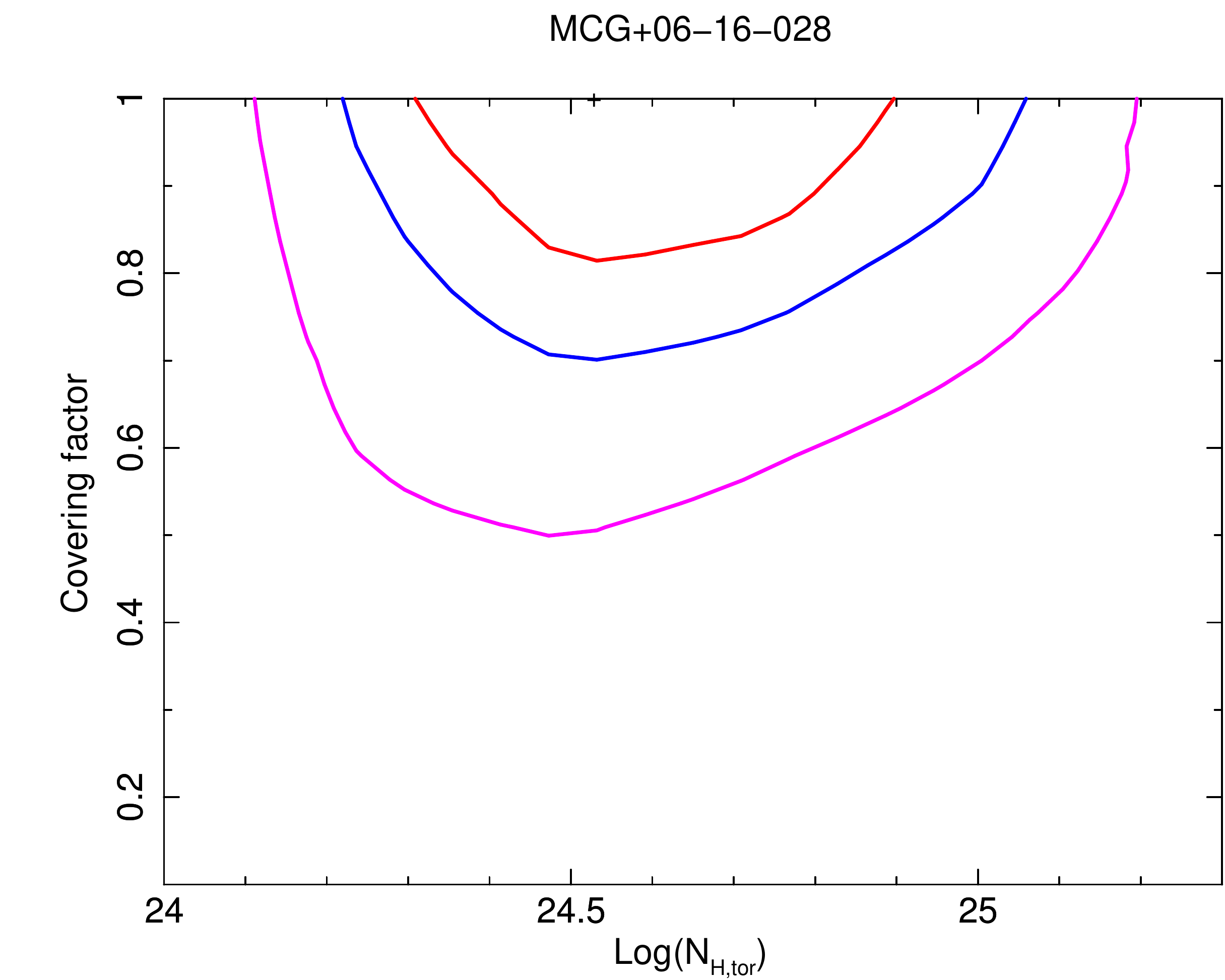}
  \end{minipage}
  \begin{minipage}[b]{.5\textwidth}
  \centering
  \includegraphics[width=0.78\textwidth,angle=-90]{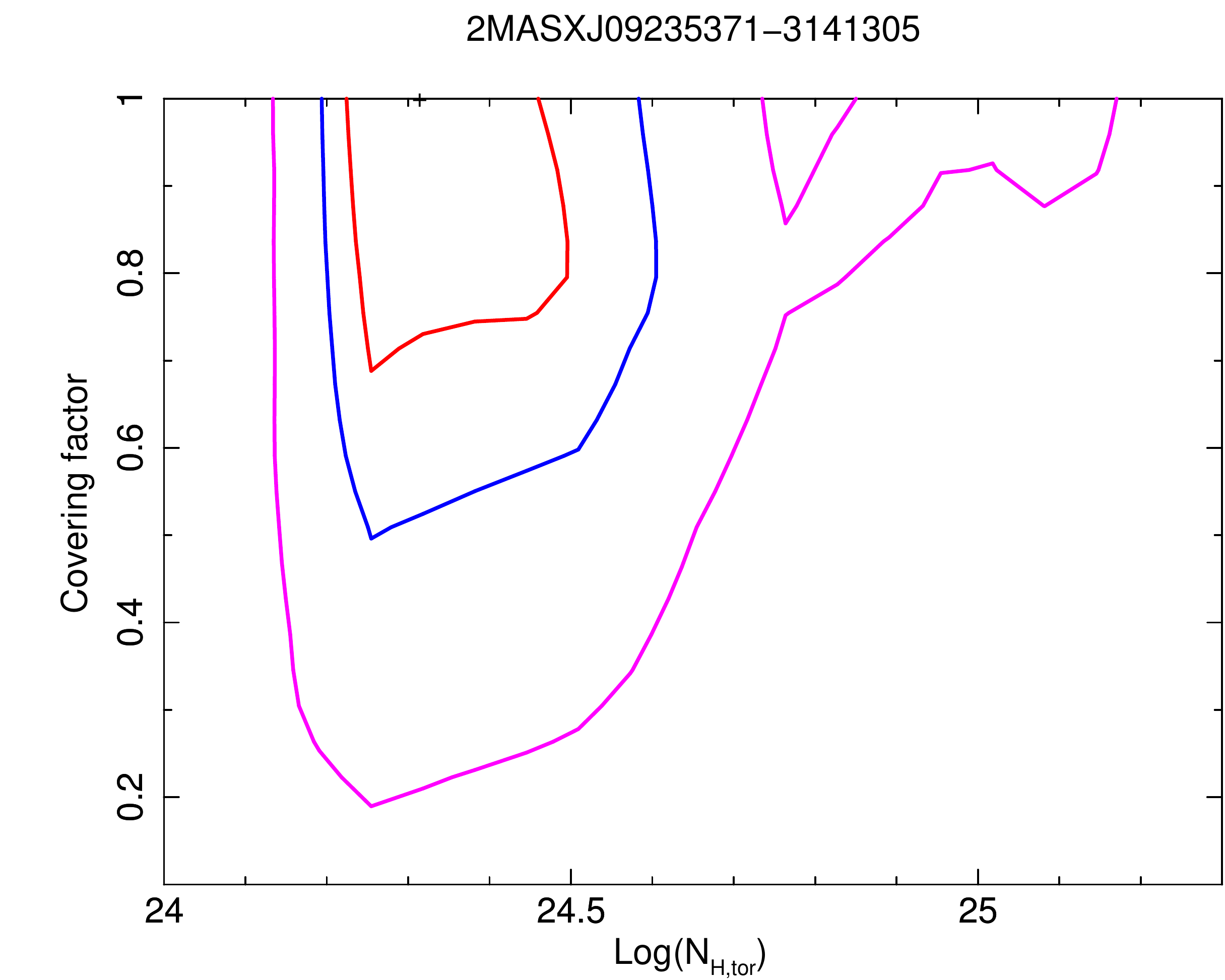}
  \end{minipage}
\caption{\normalsize \normalsize Confidence contours at 68, 90 and 99\% confidence level for the line-of-sight column density, $N_{\rm H, z}$, and the torus covering factor, $f_c$, for six of the 35 sources analyzed in this work.}
\end{figure*}

\begin{figure*}
\begin{minipage}[b]{.5\textwidth}
  \centering
  \includegraphics[width=0.78\textwidth,angle=-90]{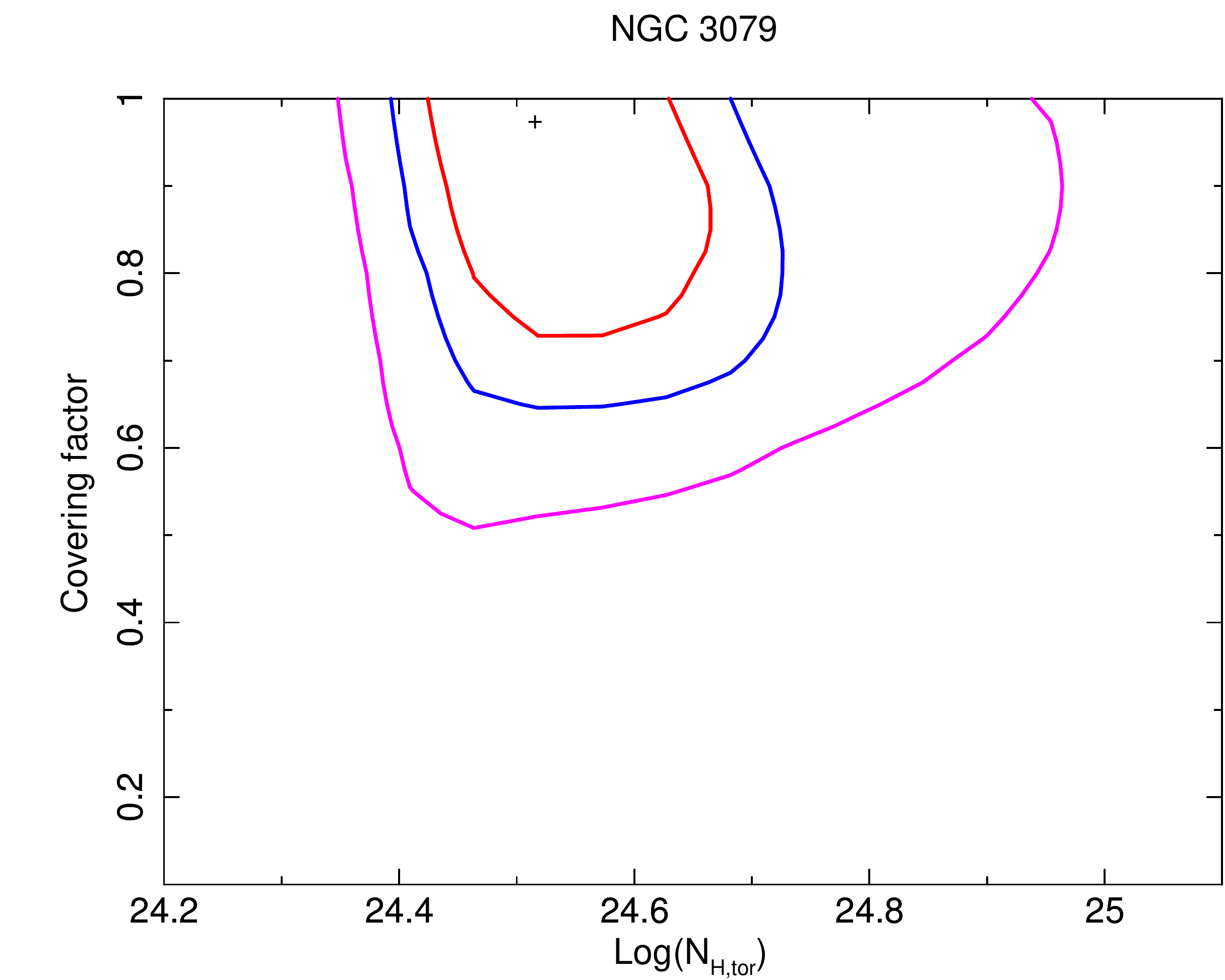}
  \end{minipage}
\begin{minipage}[b]{.5\textwidth}
  \centering
  \includegraphics[width=0.78\textwidth,angle=-90]{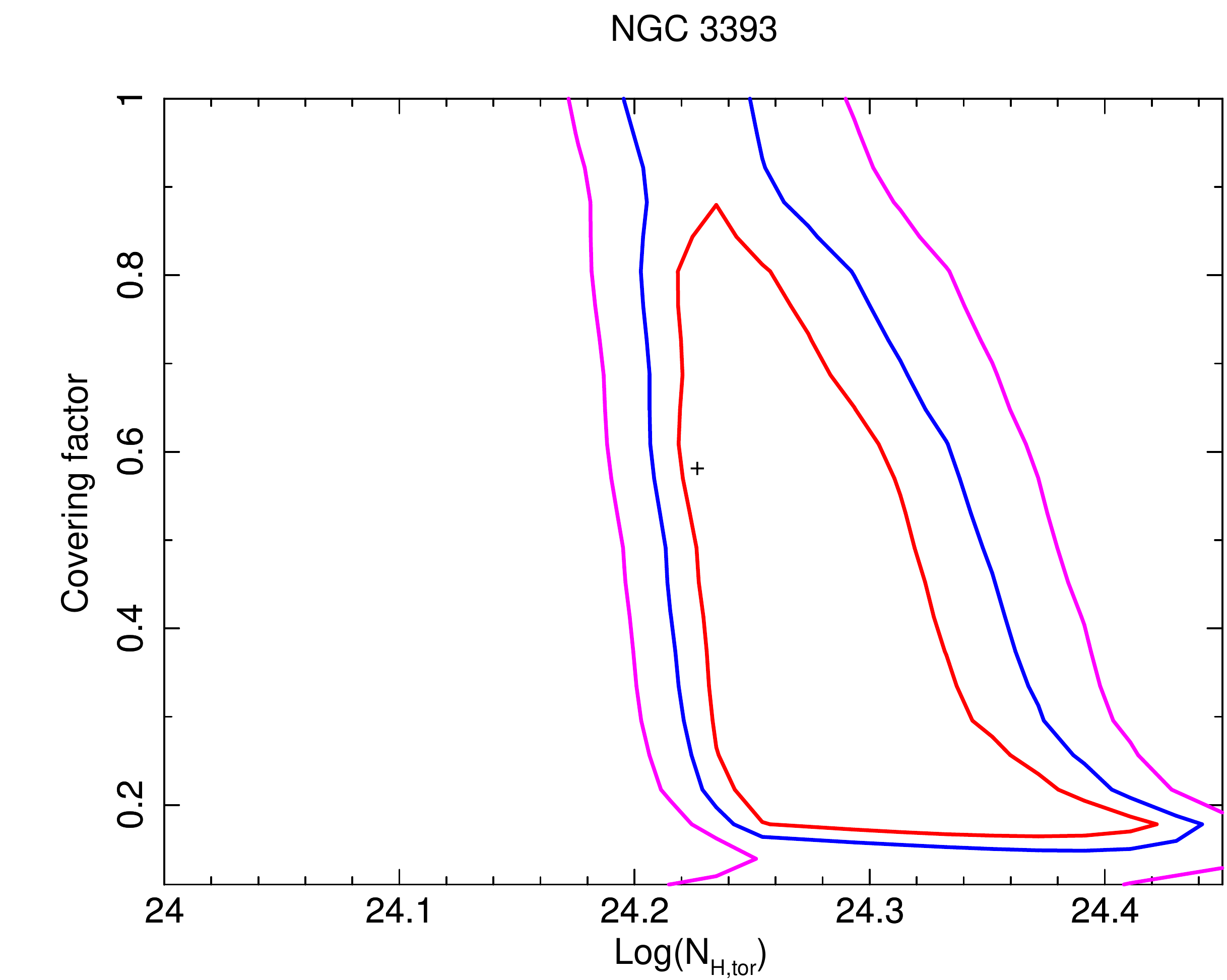}
  \end{minipage}
\begin{minipage}[b]{.5\textwidth}
  \centering
  \includegraphics[width=0.78\textwidth,angle=-90]{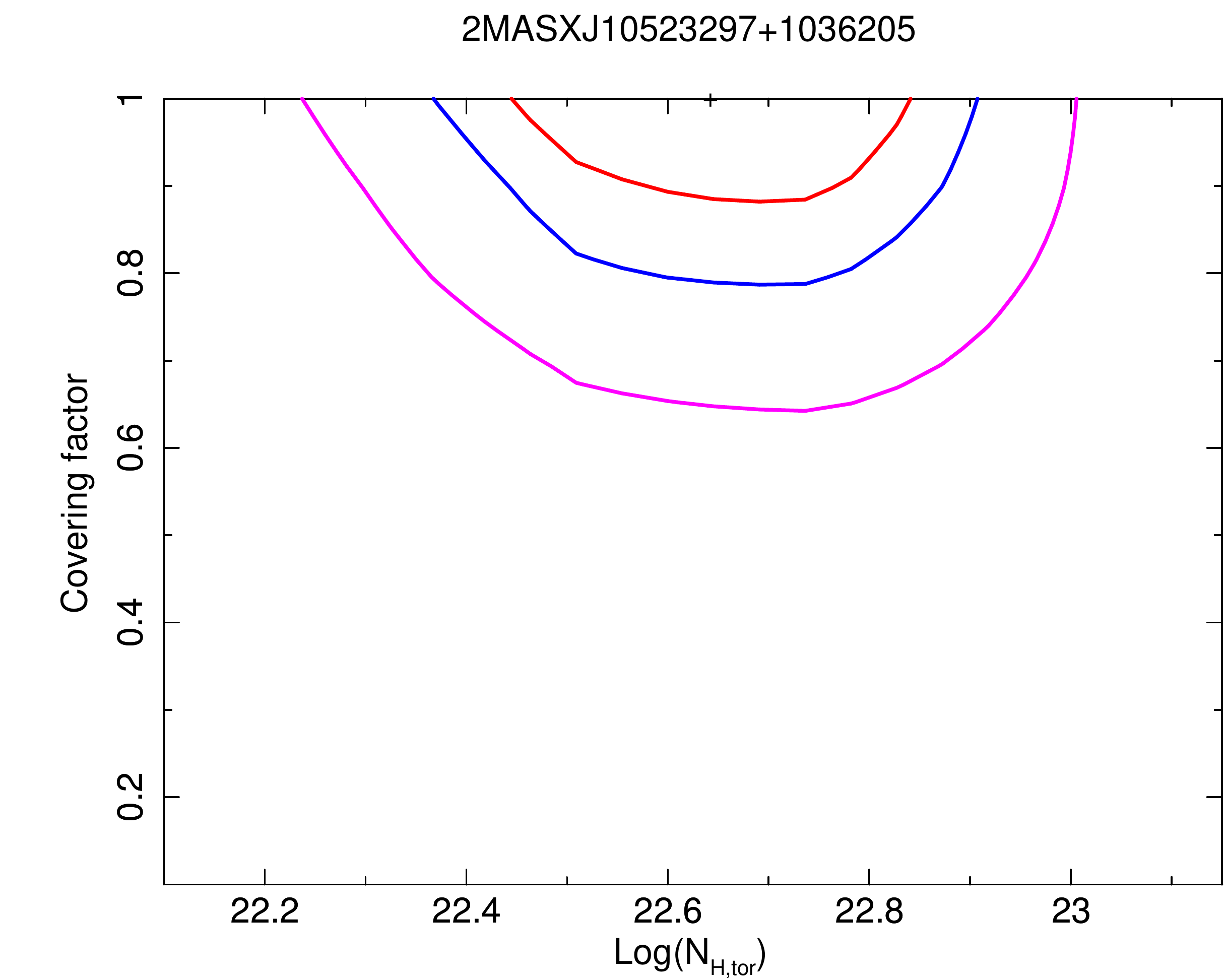}
  \end{minipage}
\begin{minipage}[b]{.5\textwidth}
  \centering
  \includegraphics[width=0.78\textwidth,angle=-90]{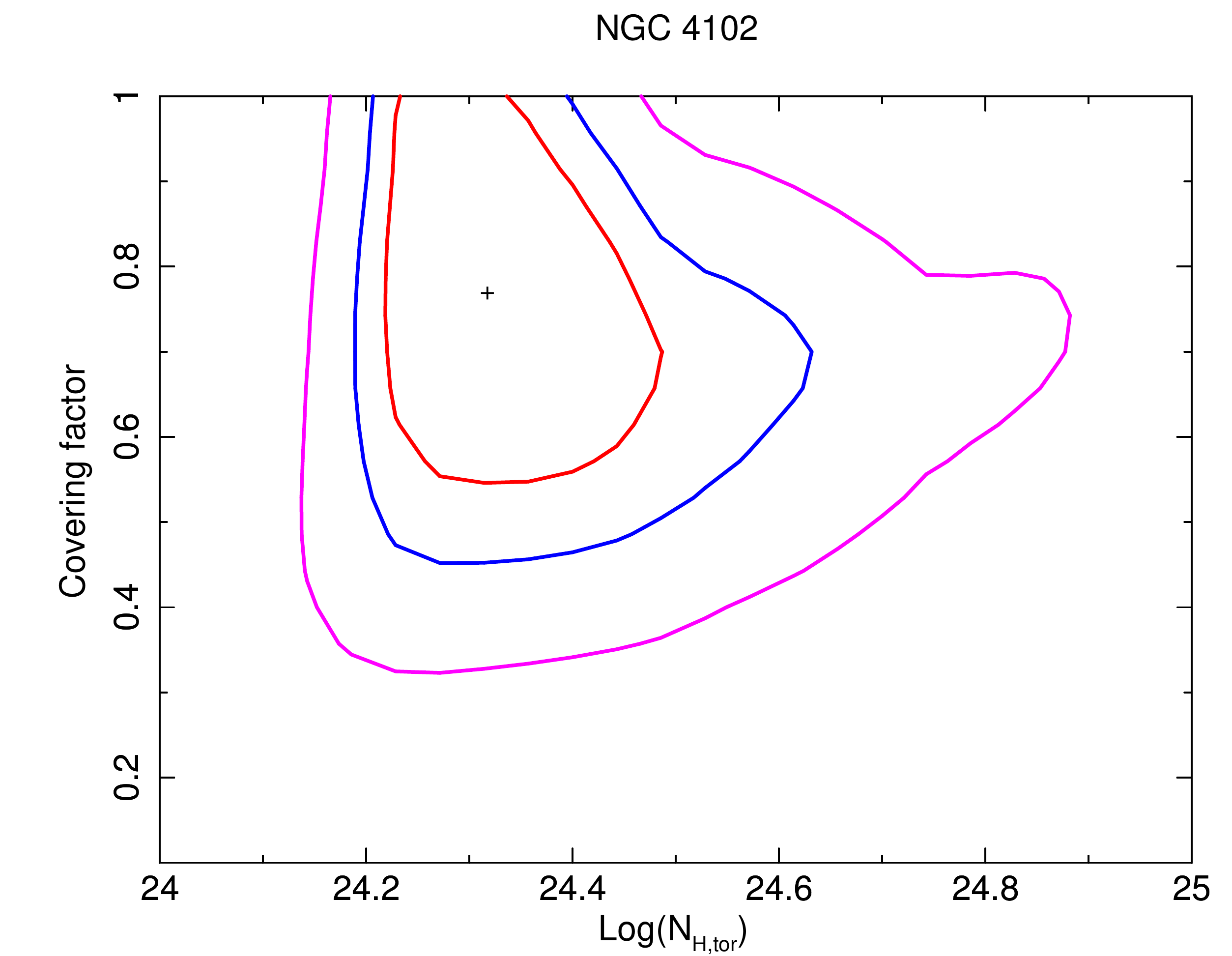}
  \end{minipage}
  \begin{minipage}[b]{.5\textwidth}
  \centering
  \includegraphics[width=0.78\textwidth,angle=-90]{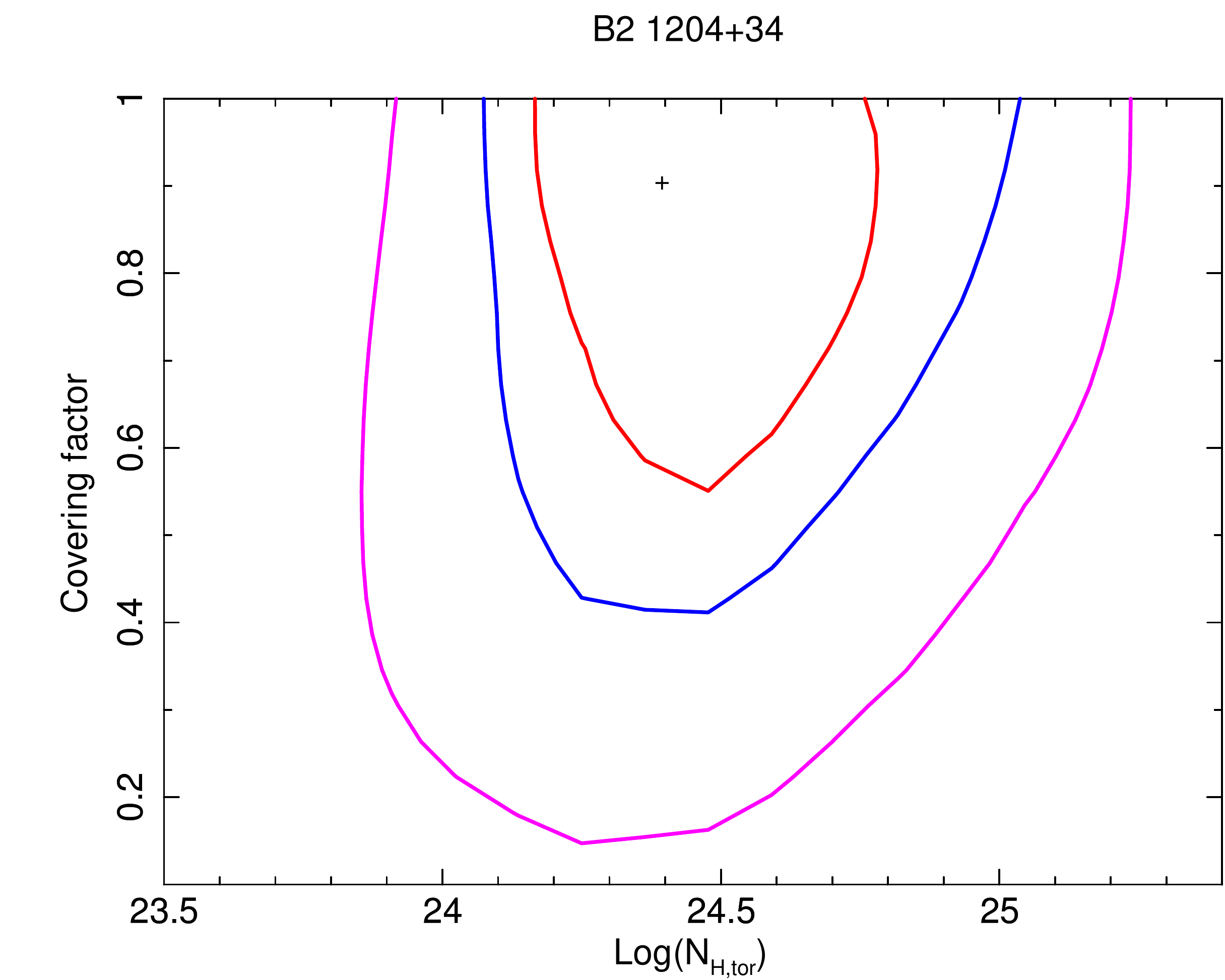}
  \end{minipage}
\begin{minipage}[b]{.5\textwidth}
  \centering
  \includegraphics[width=0.78\textwidth,angle=-90]{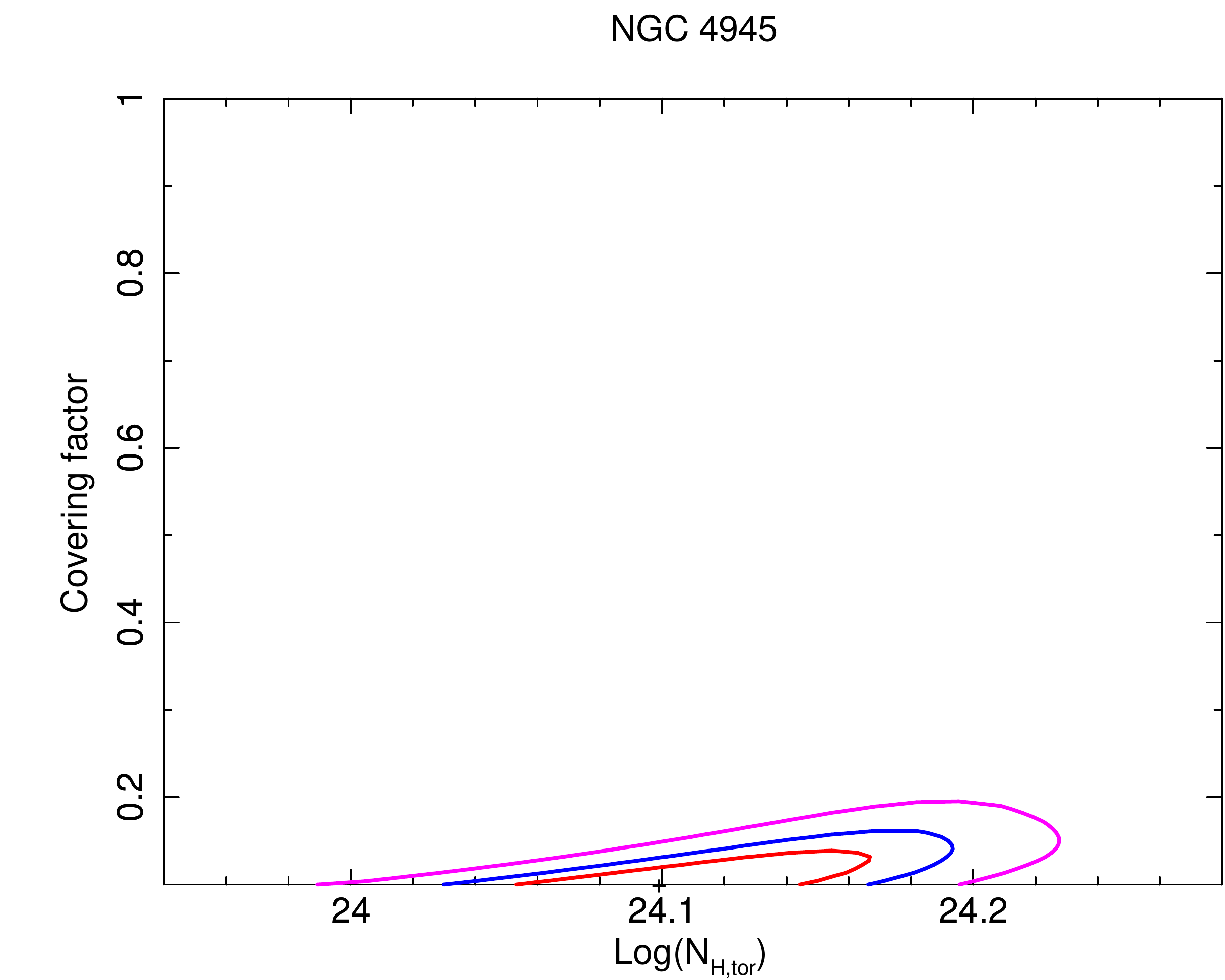}
  \end{minipage}
\caption{\normalsize \normalsize Confidence contours at 68, 90 and 99\% confidence level for the line-of-sight column density, $N_{\rm H, z}$, and the torus covering factor, $f_c$, for six of the 35 sources analyzed in this work.}
\end{figure*}

\begin{figure*}
\begin{minipage}[b]{.5\textwidth}
  \centering
  \includegraphics[width=0.78\textwidth,angle=-90]{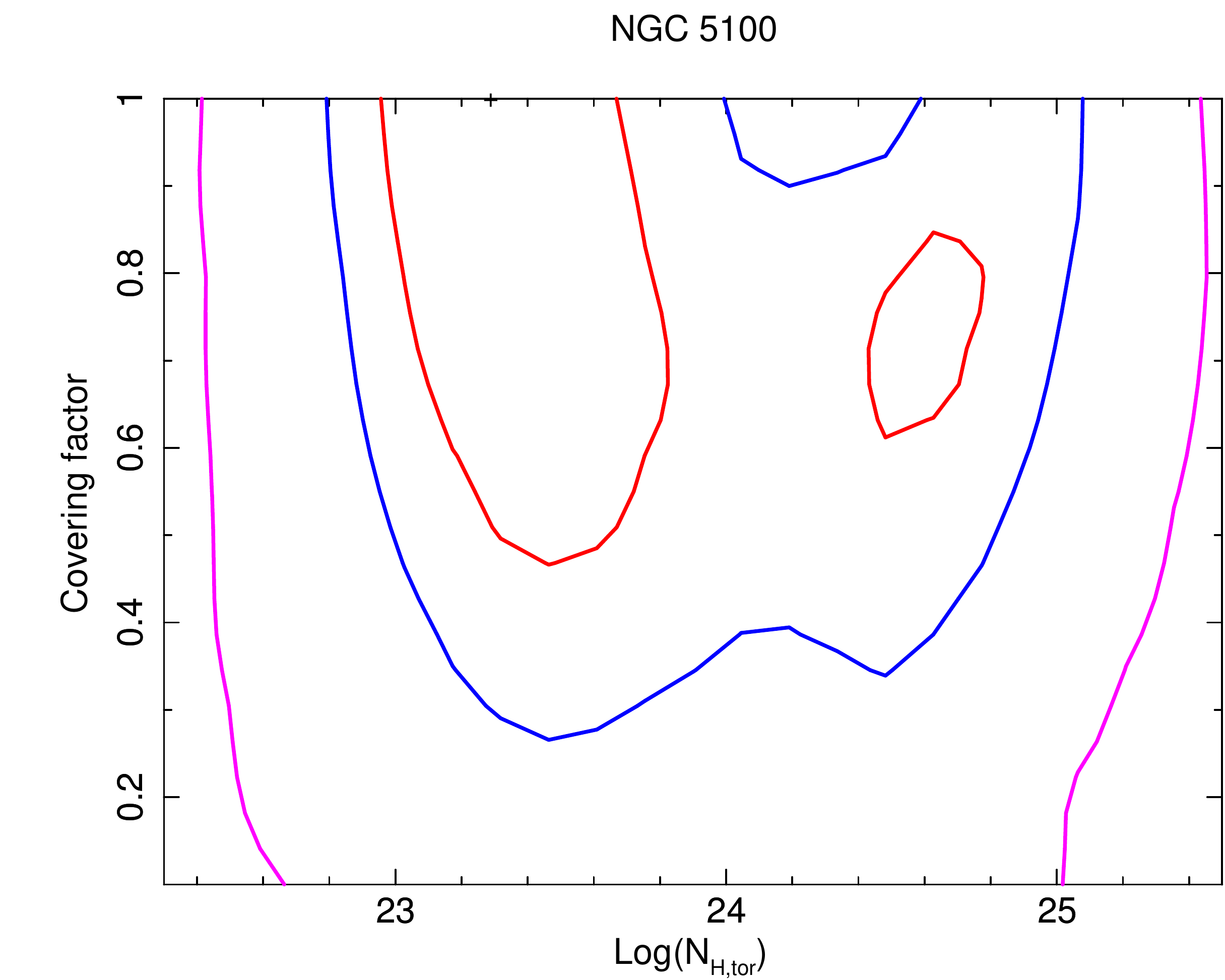}
  \end{minipage}
  \begin{minipage}[b]{.5\textwidth}
  \centering
  \includegraphics[width=0.78\textwidth,angle=-90]{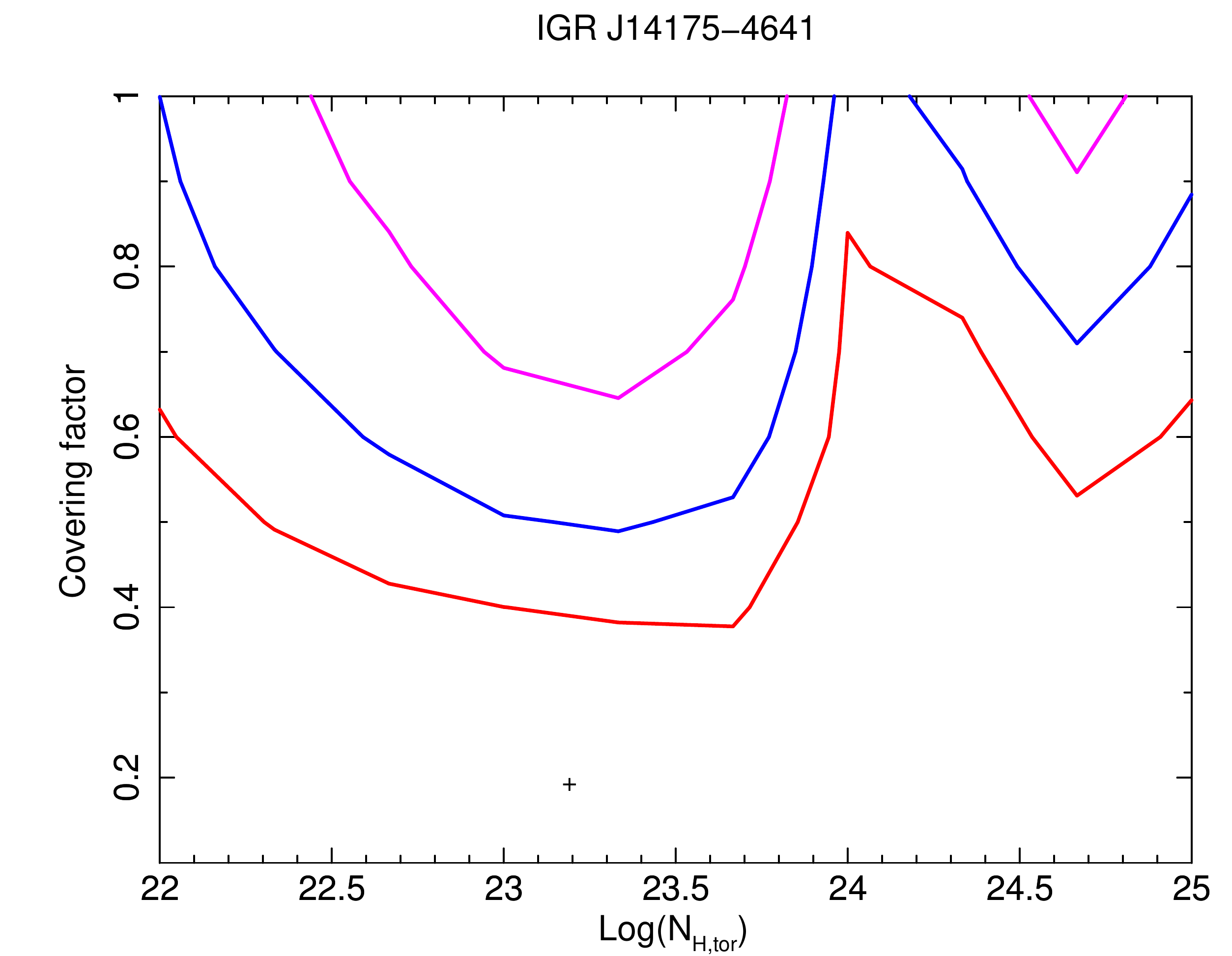}
  \end{minipage}
  \begin{minipage}[b]{.5\textwidth}
  \centering
  \includegraphics[width=0.78\textwidth,angle=-90]{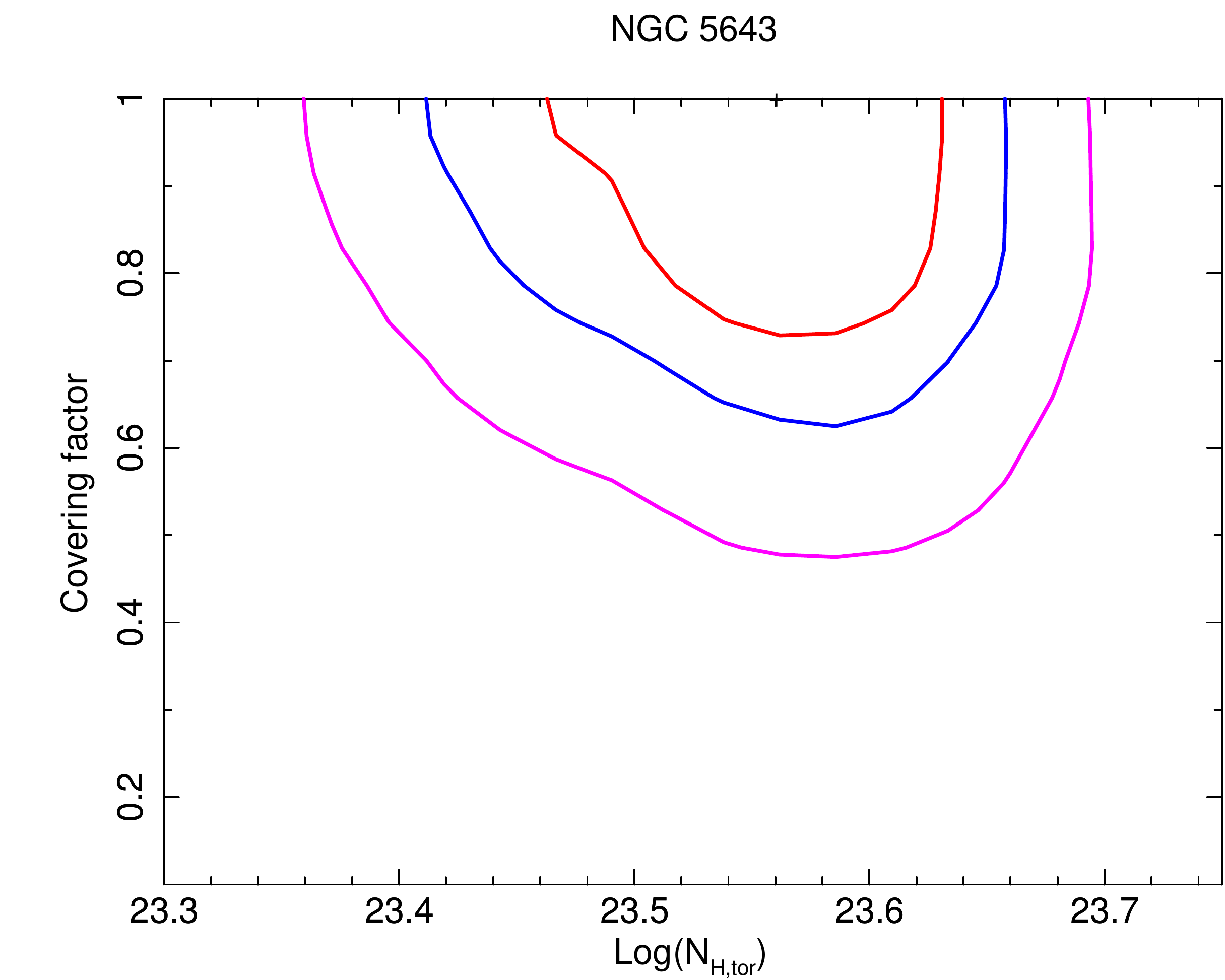}
  \end{minipage}
\begin{minipage}[b]{.5\textwidth}
  \centering
  \includegraphics[width=0.78\textwidth,angle=-90]{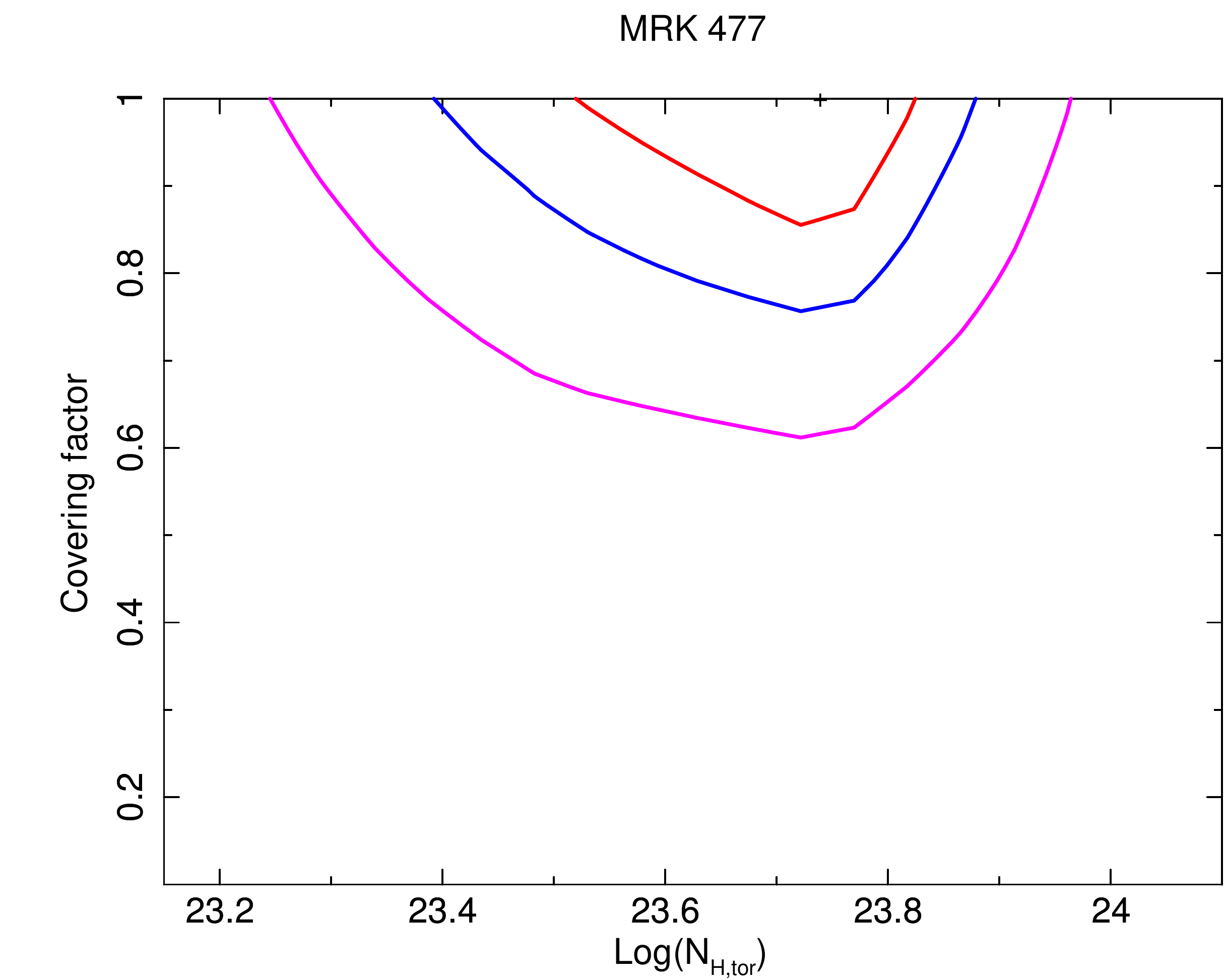}
  \end{minipage}
  \begin{minipage}[b]{.5\textwidth}
  \centering
  \includegraphics[width=0.78\textwidth,angle=-90]{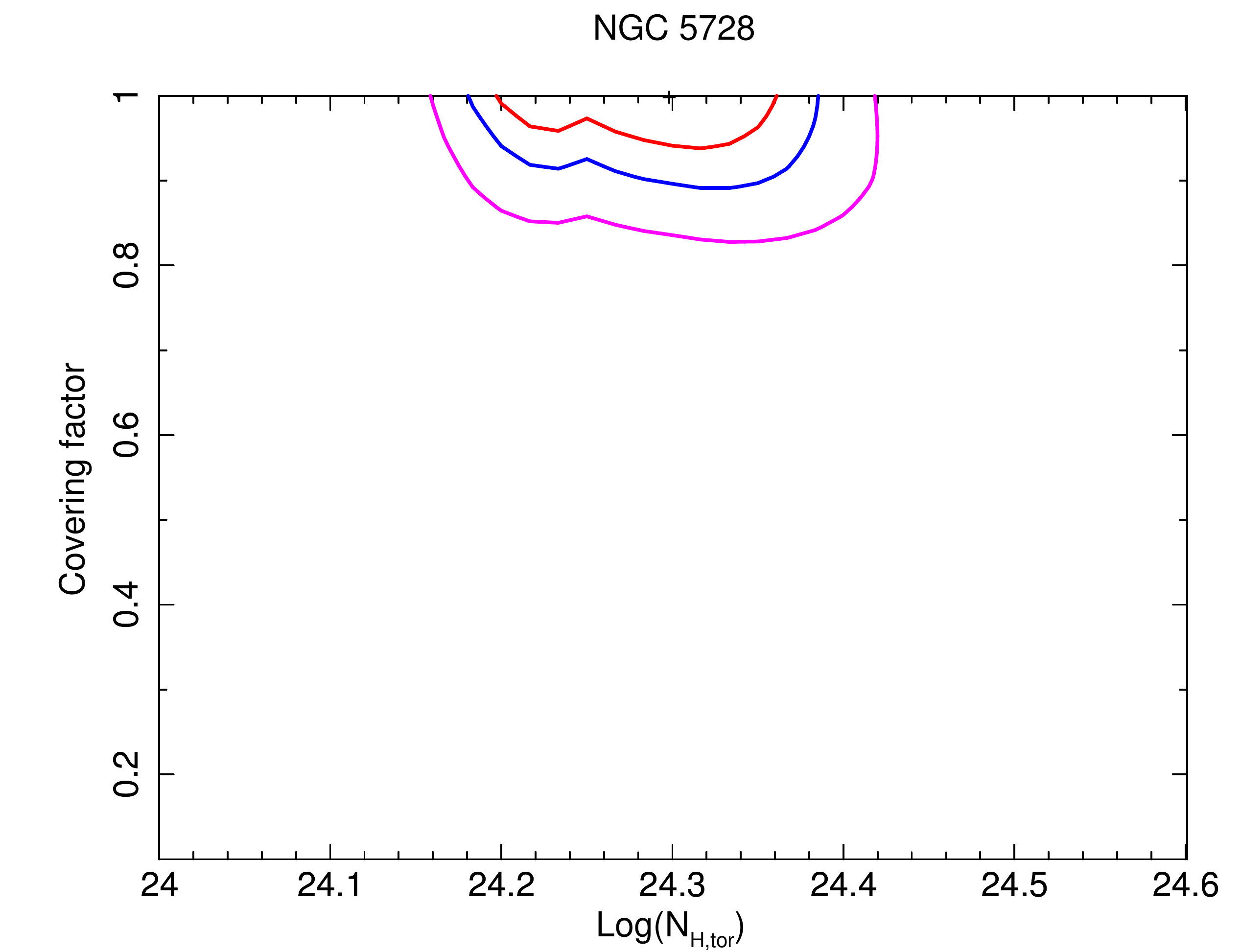}
  \end{minipage}
\begin{minipage}[b]{.5\textwidth}
  \centering
  \includegraphics[width=0.78\textwidth,angle=-90]{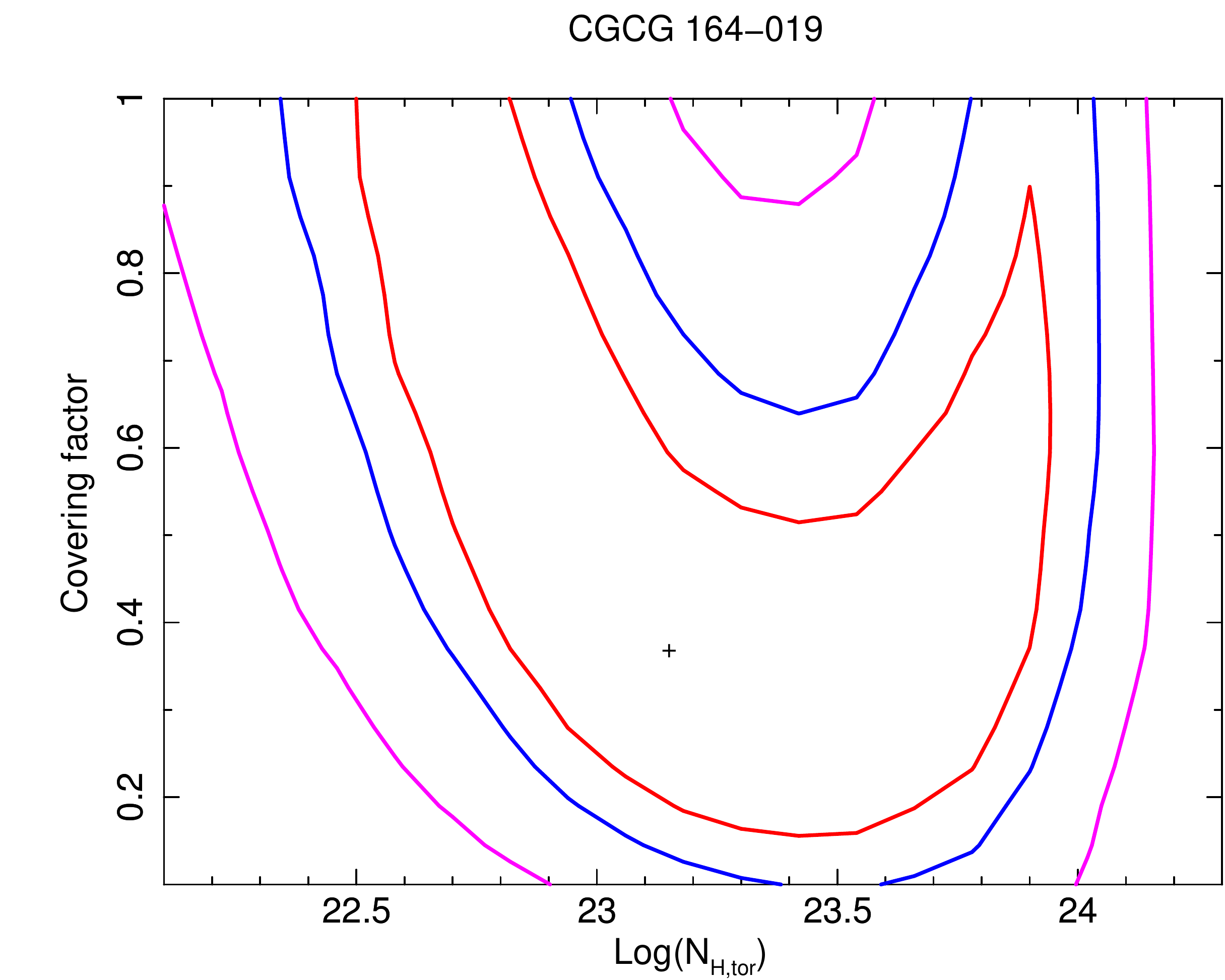}
  \end{minipage}
\caption{\normalsize \normalsize Confidence contours at 68, 90 and 99\% confidence level for the line-of-sight column density, $N_{\rm H, z}$, and the torus covering factor, $f_c$, for six of the 35 sources analyzed in this work.}
\end{figure*}

\begin{figure*}
\begin{minipage}[b]{.5\textwidth}
  \centering
  \includegraphics[width=0.78\textwidth,angle=-90]{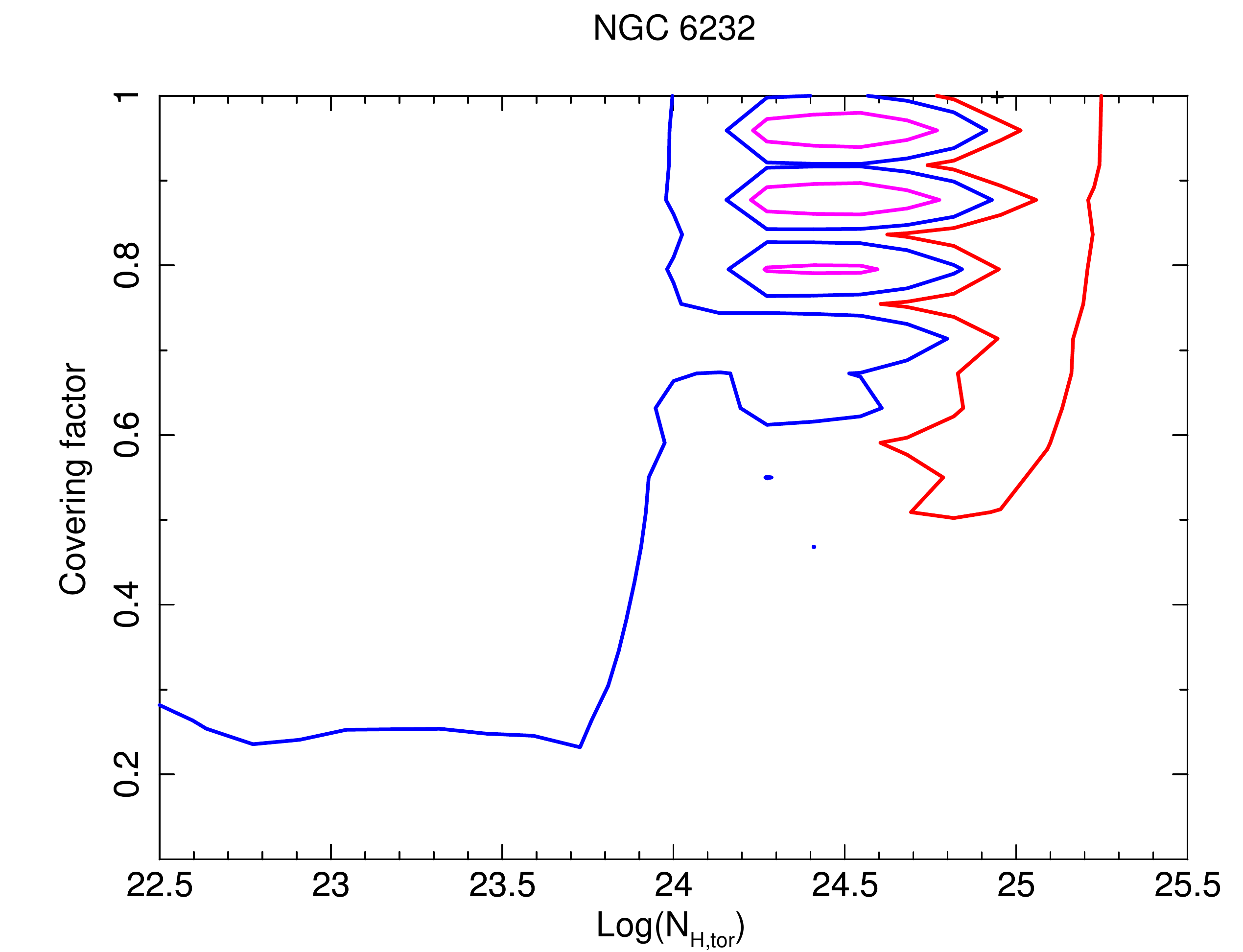}
  \end{minipage}
  \begin{minipage}[b]{.5\textwidth}
  \centering
  \includegraphics[width=0.78\textwidth,angle=-90]{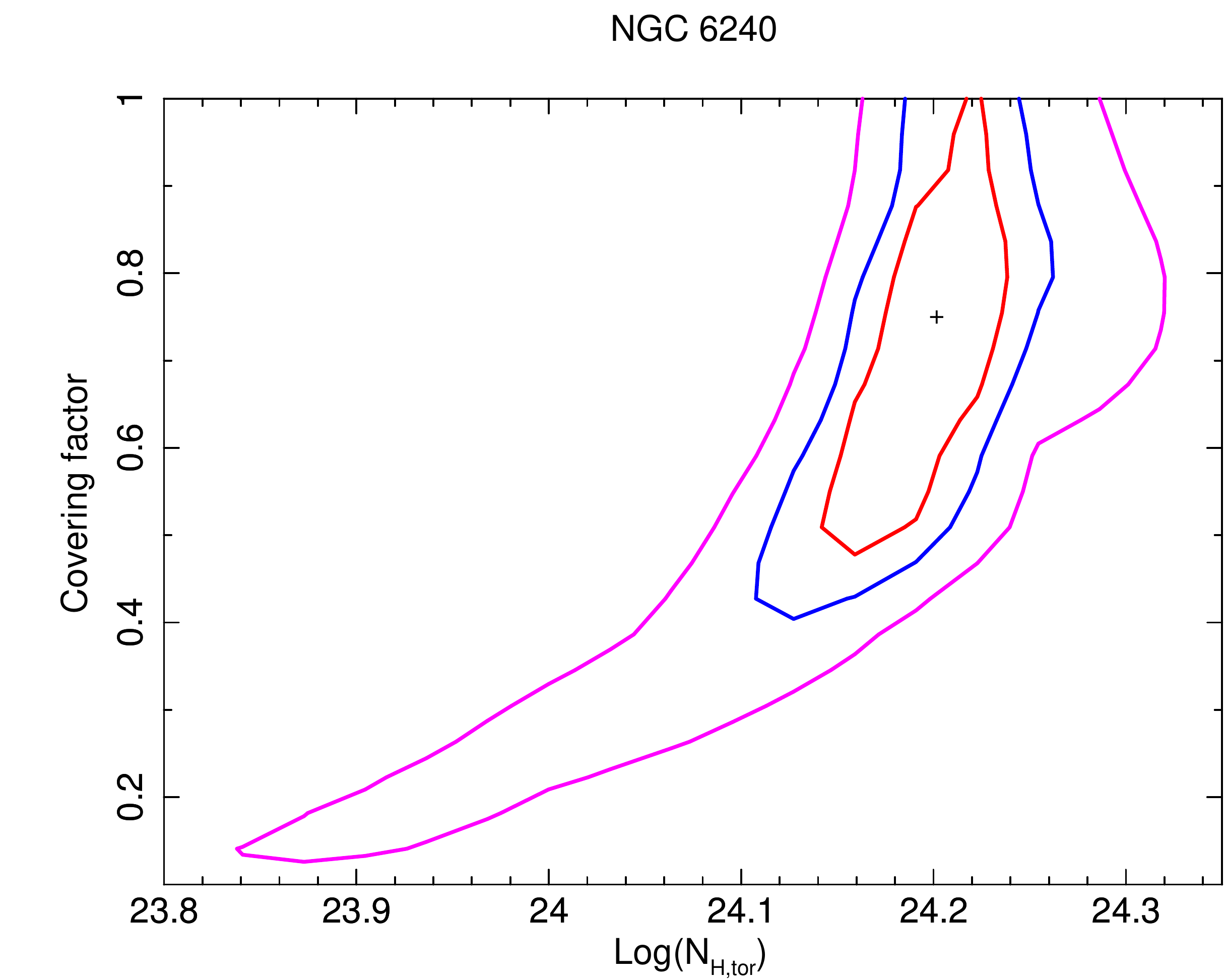}
  \end{minipage}
  \begin{minipage}[b]{.5\textwidth}
  \centering
  \includegraphics[width=0.78\textwidth,angle=-90]{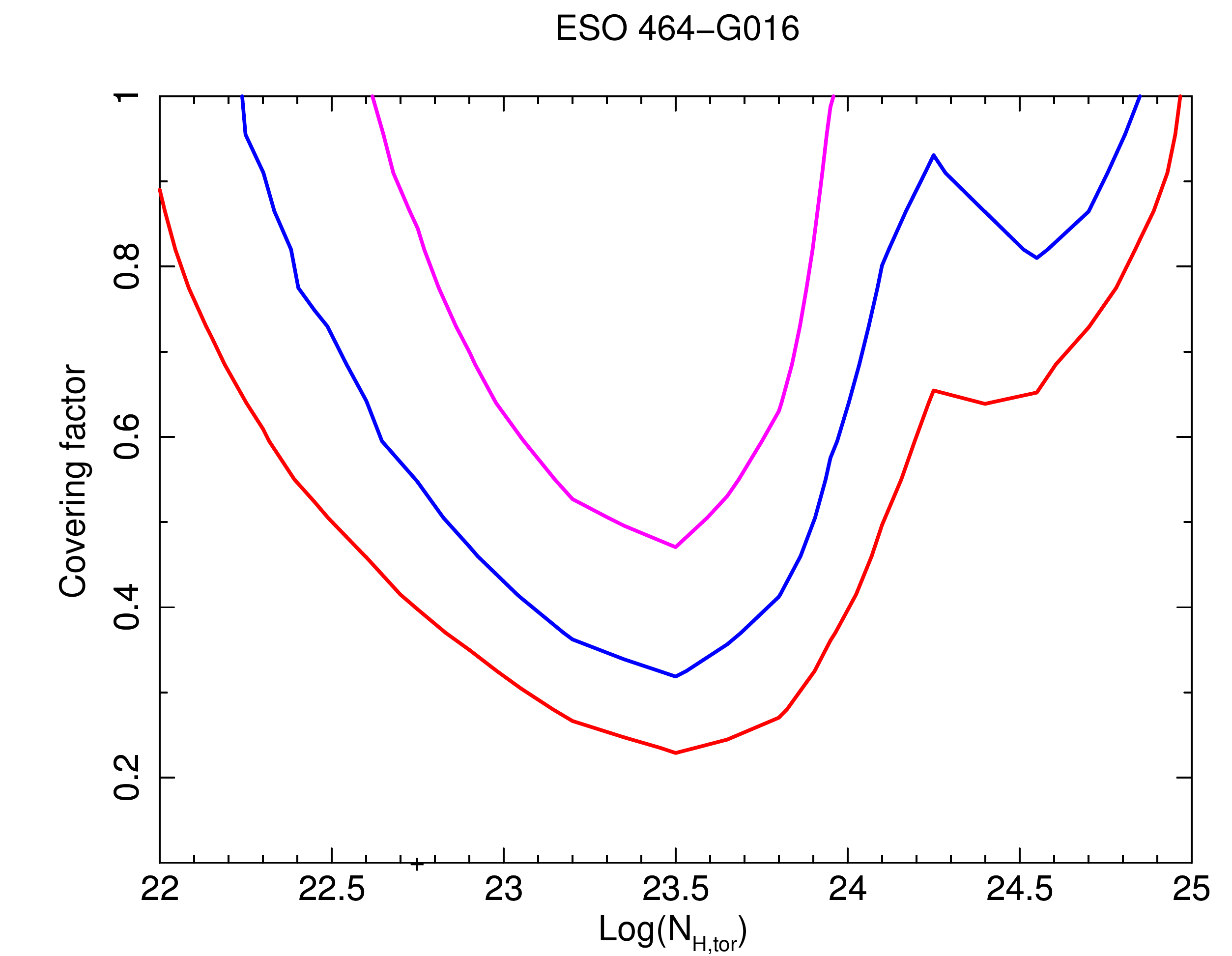}
  \end{minipage}
\begin{minipage}[b]{.5\textwidth}
  \centering
  \includegraphics[width=0.78\textwidth,angle=-90]{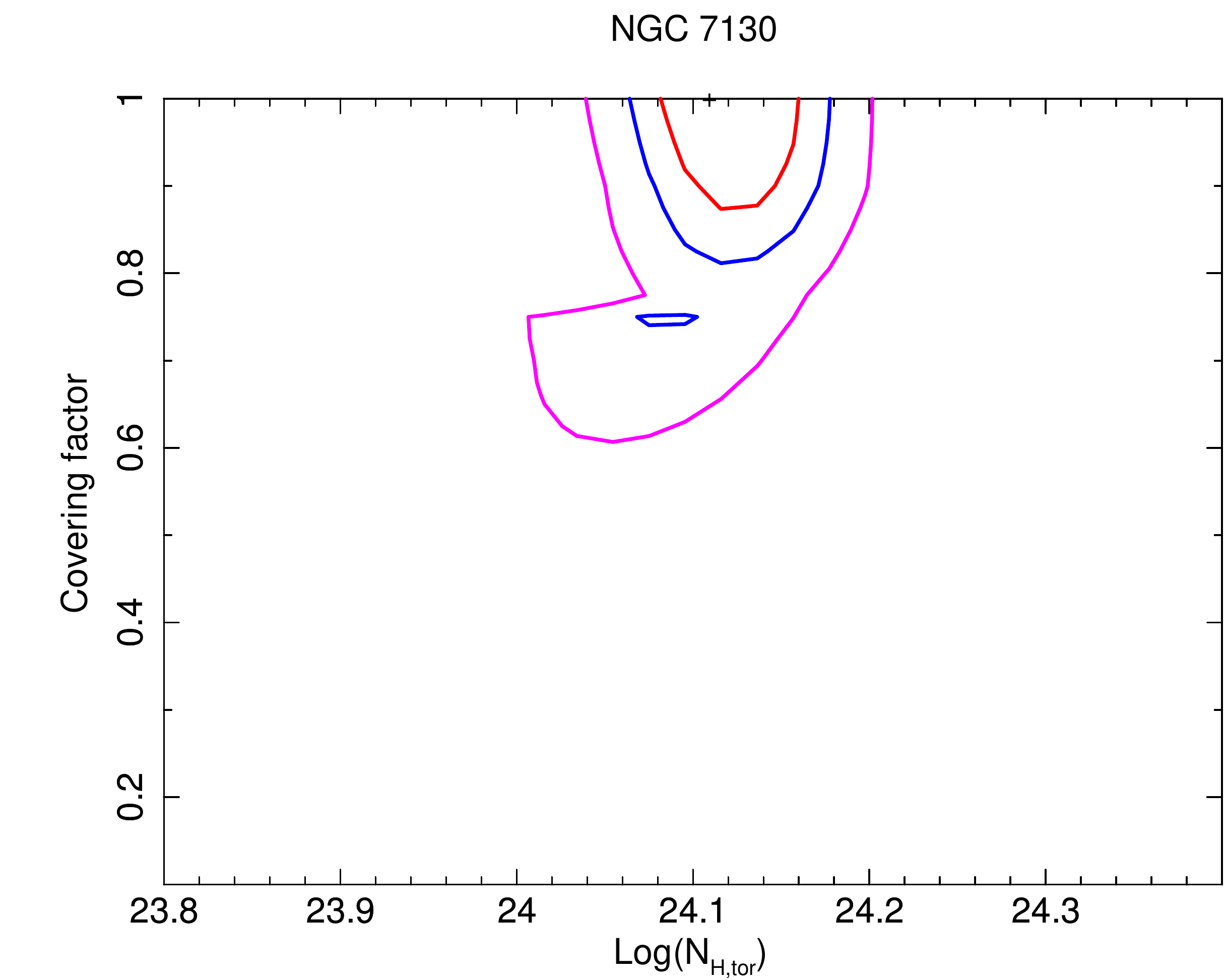}
  \end{minipage}
  \begin{minipage}[b]{.5\textwidth}
  \centering
  \includegraphics[width=0.78\textwidth,angle=-90]{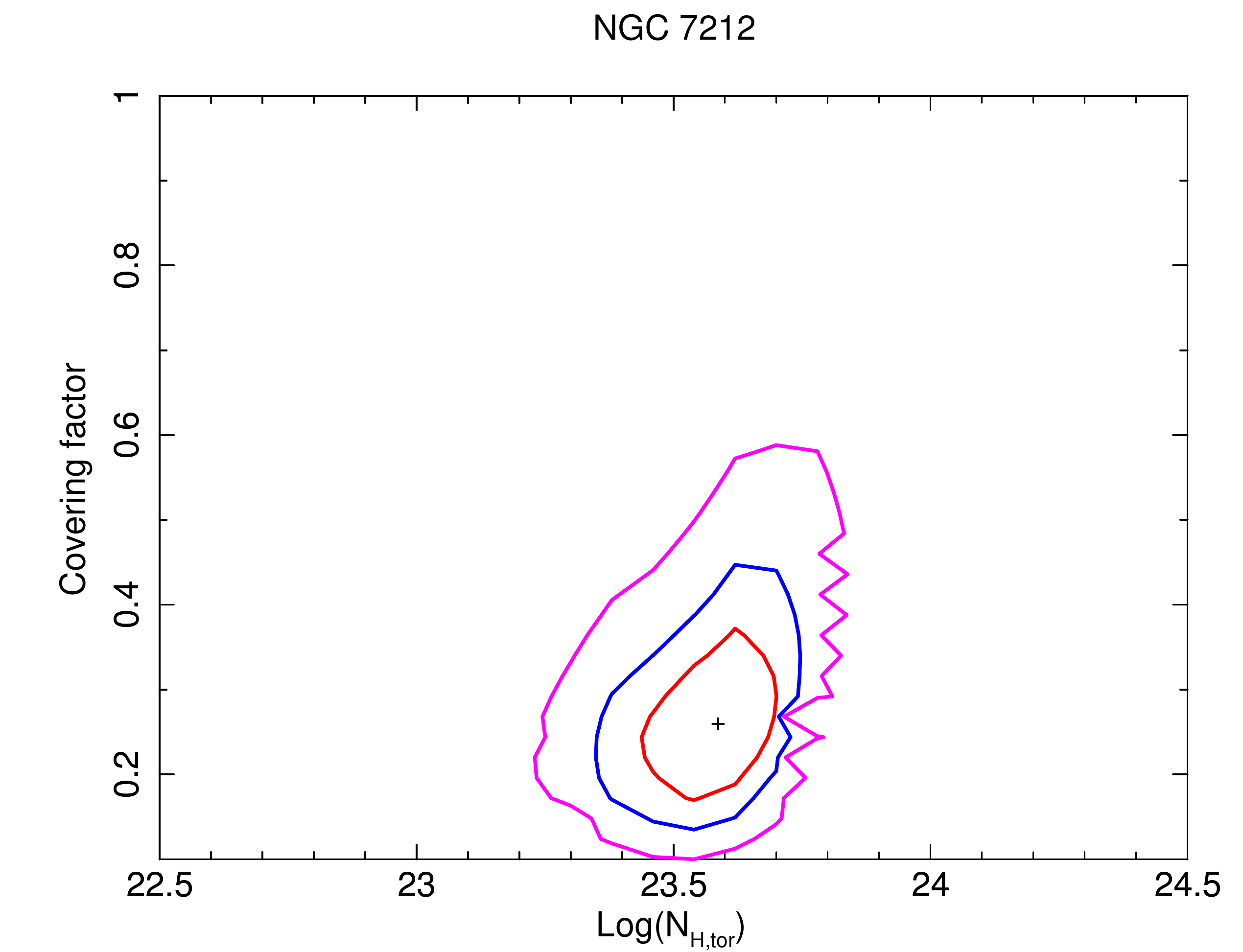}
  \end{minipage}
    \begin{minipage}[b]{.5\textwidth}
  \centering
  \includegraphics[width=0.78\textwidth,angle=-90]{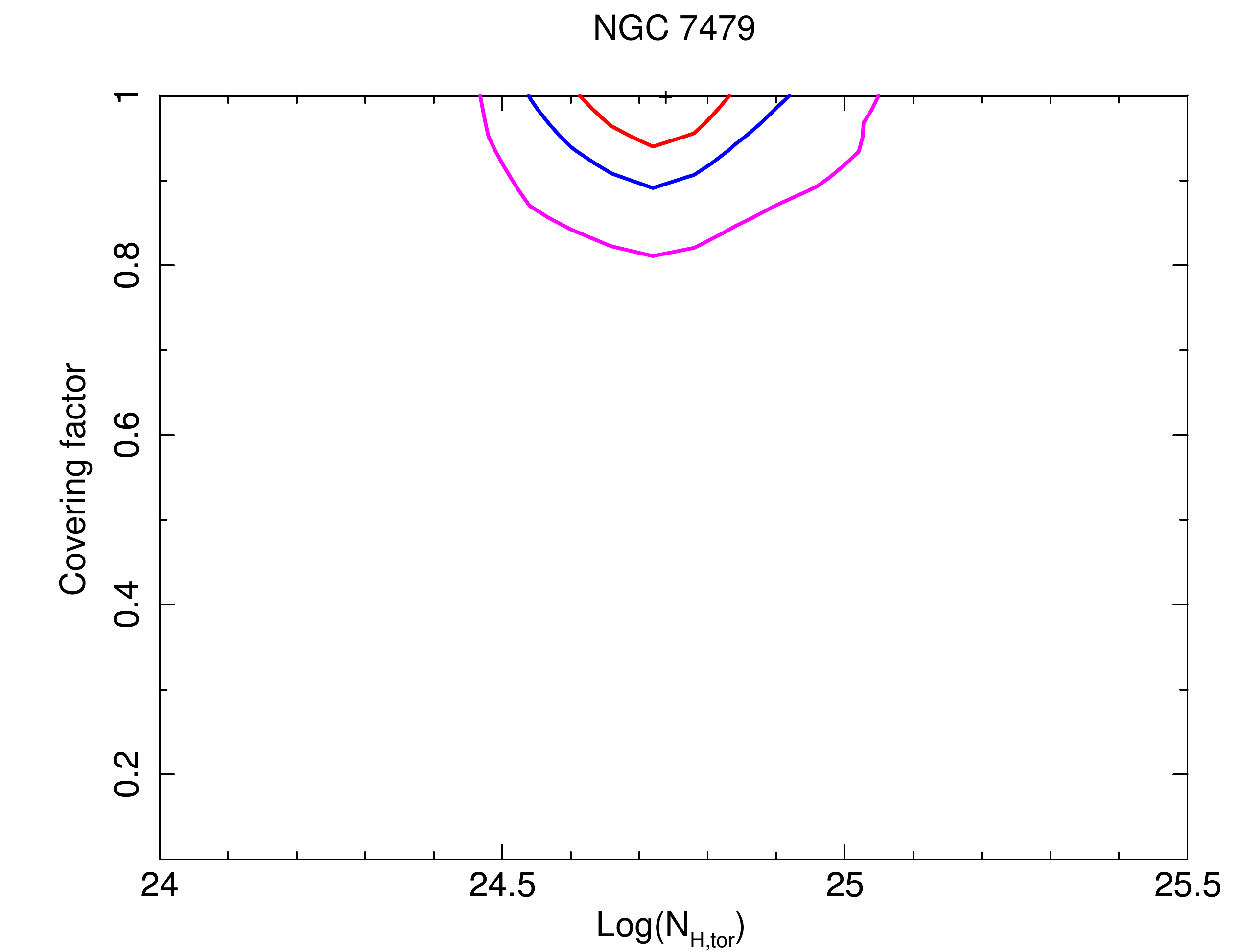}
  \end{minipage}
\caption{\normalsize \normalsize Confidence contours at 68, 90 and 99\% confidence level for the line-of-sight column density, $N_{\rm H, z}$, and the torus covering factor, $f_c$, for six of the 35 sources analyzed in this work.}
\end{figure*}

\bibliographystyle{aa}
\bibliography{covering_factor_nustar_arxiv}

\end{document}